\pdfoutput=1

\documentclass[11pt,twoside,a4paper,cmspaper,final,collab]{cms-tdr}
\def\svnVersion{459133P}\def\cmsCernNoTag{CERN-EP-2017-326}\def\cmsCernDate{\today}\def\cmsMessage{Published in the Journal of Instrumentation as \href{http://dx.doi.org/10.1088/1748-0221/13/05/P05011}{\doi{10.1088/1748-0221/13/05/P05011.}}}

\begin{document}\cmsNoteHeader{BTV-16-002}

\hyphenation{had-ron-i-za-tion}
\hyphenation{cal-or-i-me-ter}
\hyphenation{de-vices}
\RCS$HeadURL: svn+ssh://svn.cern.ch/reps/tdr2/papers/BTV-16-002/trunk/BTV-16-002.tex $
\RCS$Id: BTV-16-002.tex 459126 2018-05-08 18:23:52Z pvmulder $

\newcommand{\lm}{\ensuremath{\lambda_{M}}\xspace}
\newcommand{\ENN}[1]{N_{\mathrm{#1}}}
\newcommand{\EFF}[2]{\varepsilon_{\mathrm{#1}}^{\mathrm{#2}}}
\newcommand{\Wc}{\ensuremath{\PW + \cPqc}}
\newcommand{\PWmc}{\ensuremath{\PWm + \cPqc}}
\newcommand{\PWmac}{\ensuremath{\PWp + \cPaqc}}
\providecommand{\MT}{\ensuremath{M_\cmsSymbolFace{T}}\xspace}
\newcommand{\Zj}{\ensuremath{\cPZ + \text{jets}}}
\newcommand{\Wj}{\ensuremath{\PW + \text{jets}}}
\newcommand{\ZZ}{\ensuremath{\cPZ\cPZ}}
\providecommand{\NA}{\ensuremath{\text{---}}}
\hyphenation{brems-strah-lung}

\cmsNoteHeader{BTV-16-002}
\title{Identification of heavy-flavour jets with the CMS detector in pp collisions at 13\TeV}

\date{\today}

\abstract{
Many measurements and searches for physics beyond the standard model at the LHC rely on the efficient identification of heavy-flavour jets, \ie jets originating from bottom or charm quarks.
In this paper, the discriminating variables and the algorithms used for heavy-flavour jet identification during the first years of operation of the CMS experiment in proton-proton collisions at a centre-of-mass energy of 13\TeV, are presented. Heavy-flavour jet identification algorithms have been improved compared to those used previously at centre-of-mass energies of 7 and 8\TeV. For jets with transverse momenta in the range expected in simulated \ttbar events, these new developments result in an efficiency of 68\% for the correct identification of a {\cPqb} jet for a probability of 1\% of misidentifying a light-flavour jet. The improvement in relative efficiency at this misidentification probability is about 15\%, compared to previous CMS algorithms. In addition, for the first time algorithms have been developed to identify jets containing two {\cPqb} hadrons in Lorentz-boosted event topologies, as well as to tag {\cPqc} jets. The large data sample recorded in 2016 at a centre-of-mass energy of 13\TeV has also allowed the development of new methods to measure the efficiency and misidentification probability of heavy-flavour jet identification algorithms. The {\cPqb} jet identification efficiency is measured with a precision of a few per cent at moderate jet transverse momenta (between 30 and 300\GeV) and about 5\% at the highest jet transverse momenta (between 500 and 1000\GeV).
}

\hypersetup{%
pdfauthor={CMS Collaboration},%
pdftitle={Identification of heavy-flavour jets with the CMS detector at 13 TeV},%
pdfsubject={CMS},%
pdfkeywords={CMS, physics, heavy-flavour jet identification, b tagging, c tagging, boosted objects}}

\maketitle

\tableofcontents

\section{Introduction}
The success of the physics programme of the CMS experiment at the CERN LHC requires the particles created in the LHC collisions to be reconstructed and identified as accurately as possible. With the exception of the top quark, quarks and gluons produced in {\Pp\Pp} collisions develop a parton shower and eventually hadronize giving rise to jets of collimated particles observed in the CMS detector. Heavy-flavour jet identification techniques exploit the properties of the hadrons in the jet to discriminate between jets originating from {\cPqb} or {\cPqc} quarks (heavy-flavour jets) and those originating from light-flavour quarks or gluons (light-flavour jets). The CMS Collaboration presented in Ref.~\cite{BTV12001} a set of {\cPqb} jet identification techniques used in physics analyses performed on LHC Run 1 {\Pp\Pp} collision data, collected in 2011 and 2012 at centre-of-mass energies of 7 and 8\TeV. This paper presents a comprehensive summary of the newly developed and optimized techniques compared to our previous results. In particular, the larger recorded data set of {\Pp\Pp} collisions at a centre-of-mass energy of 13 TeV during Run 2 of the LHC in 2016, allows the study of rarer high-momentum topologies in which daughter jets from a Lorentz-boosted parent particle merge into a single jet. Examples of such topologies include the identification of boosted Higgs bosons decaying to two {\cPqb} quarks, and of {\cPqb} jets from boosted top quarks. The identification of {\cPqc} jets is also of significant interest, \eg for the study of Higgs boson decays to a pair of {\cPqc} quarks, and for top squark searches in the {\cPqc} quark plus neutralino final-state topology.

The paper is organized as follows. A brief summary of particle and jet reconstruction in the CMS detector is given in Section~\ref{sec:cms}. Details about the simulated proton-proton collision samples and the data-taking conditions are given in Section~\ref{sec:simulation}. The properties of heavy-flavour jets and the variables used to discriminate between these and other jets are discussed in Section~\ref{sec:vars}, while the algorithms are presented in Sections~\ref{sec:ak4algos} and~\ref{sec:boostedalgos}. For some physics processes, it is important to identify {\cPqb} jets at the trigger level. This topic is discussed in Section~\ref{sec:trigger}. The large recorded number of proton-proton ({\Pp\Pp}) collisions permits the exploration of new methods to measure the efficiency of the heavy-flavour jet identification algorithms using data. These new methods, as well as the techniques used during the Run 1, are summarized in Sections~\ref{sec:ak4eff} and~\ref{sec:boostedeff} for efficiency measurements in nonboosted and boosted event topologies, respectively.

\section{The CMS detector}
\label{sec:cms}
The central feature of the CMS apparatus is a superconducting solenoid of 6\unit{m} internal diameter and a magnetic field of 3.8\unit{T}. Within the solenoid volume are a silicon pixel and strip tracker, a lead tungstate crystal electromagnetic calorimeter (ECAL), and a brass and scintillator hadron calorimeter (HCAL), each composed of a barrel and two endcap sections, together providing coverage in pseudorapidity ($\eta$) up to $\abs{\eta}=3.0$. Forward calorimeters extend the coverage to $\abs{\eta} = 5.2$. Muons are detected in the pseudorapidity range $\abs{\eta} < 2.4$ using gas-ionization chambers embedded in the steel flux-return yoke outside the solenoid.

The silicon tracker measures charged particles within the range $\abs{\eta} < 2.5$. During the first two years of Run 2 operation at a centre-of-mass energy of 13 TeV, the silicon tracker setup did not change compared to the Run 1 of the LHC. The trajectories of charged particles are reconstructed from the hits in the silicon tracking system using an iterative procedure with a Kalman filter. The tracking efficiency is typically over 98\% for tracks with a transverse momentum (\pt) above 1\GeV. For nonisolated particles with $1 < \pt < 10\GeV$ and $\abs{\eta} < 1.4$, the track resolutions are typically 1.5\% in \pt and 25--90 (45--150)\mum in the transverse (longitudinal) impact parameter (IP)~\cite{TRK-11-001}. The {\Pp\Pp} interaction vertices are reconstructed by clustering tracks on the basis of their $z$ coordinates at their points of closest approach to the centre of the beam spot using a deterministic annealing algorithm~\cite{Rose98deterministicannealing}. The position of each vertex is estimated with an adaptive vertex fit~\cite{Fruhwirth:2007hz}. The resolution on the position is around 20\mum in the transverse plane and around 30\mum along the beam axis for primary vertices reconstructed using at least 50 tracks~\cite{TRK-11-001}.

The global event reconstruction, also called particle-flow (PF) event reconstruction~\cite{Sirunyan:2017ulk}, consists of reconstructing and identifying each individual particle with an optimized combination of all subdetector information. In this process, the identification of the particle type (photon, electron, muon, charged hadron, neutral hadron) plays an important role in the determination of the particle direction and energy. Photons, \eg coming from neutral pion decays or from electron bremsstrahlung, are identified as ECAL energy clusters not linked to the extrapolation of any charged-particle trajectory to the ECAL. Electrons, \eg coming from photon conversions in the tracker material or from heavy-flavour hadron semileptonic decays, are identified as combinations of charged-particle tracks reconstructed in the tracker and multiple ECAL energy clusters corresponding to both the passage of the electron through the ECAL plus any associated bremsstrahlung photons. Muons, \eg from the semileptonic decay of heavy-flavour hadrons, are identified as tracks reconstructed in the tracker combined with matching hits or tracks in the muon system, and matching energy deposits in the calorimeters. Charged hadrons are identified as charged particles not identified as electrons or muons. Finally, neutral hadrons are identified as HCAL energy clusters not matching any charged-particle track, or as ECAL and HCAL energy excesses with respect to the expected charged-hadron energy deposit.

For each event, particles originating from the same interaction vertex are clustered into jets with the infrared and collinear safe anti-\kt algorithm~\cite{Cacciari:2008gp, Cacciari:2011ma}, using a distance parameter $R=0.4$ (AK4 jets). Compared to the $R=0.5$ jets that were used in Run 1 physics analyses, jets reconstructed with $R=0.4$ are found to still contain most of the particles from the hadronization process, while at the same time being less sensitive to particles from additional {\Pp\Pp} interactions (known as pileup) appearing in the same or adjacent bunch crossings. For studies involving boosted topologies, jets are clustered with a larger distance parameter $R = 0.8$ (AK8 jets). The jet momentum is determined as the vectorial sum of all particle momenta in the jet. Jet energy corrections are derived from the simulation and are confirmed with in situ measurements using the energy balance in dijet, multijet, photon + jet, and leptonically decaying {\Zj} events~\cite{Khachatryan:2198719}. The jet energy resolution amounts typically to 15\% at 10\GeV, 8\% at 100\GeV, and 4\% at 1\TeV~\cite{Khachatryan:2198719}. For the studies presented here, jets are required to lie within the tracker acceptance ($\abs{\eta} < 2.4$) and have $\pt > 20\GeV$. The missing transverse momentum vector is defined as the projection of the negative vector sum of the momenta of all reconstructed particles in an event on the plane perpendicular to the beams. Its magnitude is referred to as {\ptmiss}.

The reconstructed vertex with the largest value of summed physics-object $\pt^2$ is taken to be the primary $\Pp\Pp$ interaction vertex (PV). The physics objects are the jets, clustered using the jet finding algorithm with the tracks assigned to the vertex as inputs, and the associated missing transverse momentum, taken as the negative vector sum of the \pt of those jets.

The energy of electrons is determined from a combination of the track momentum at the main interaction vertex, the corresponding ECAL cluster energies, and the energies of all bremsstrahlung photons associated with the track. The momentum resolution for electrons with $\pt \approx 45\GeV$ from $\PZ \to \Pe \Pe$ decays ranges from 1.7\% for nonshowering electrons, \ie not producing additional photons and electrons, in the barrel region ($\abs{\eta}<1.48$), to 4.5\% for showering electrons in the endcaps ($1.48 < \abs{\eta} < 3.0$)~\cite{Khachatryan:2015hwa}. Muons with $20 < \pt < 100\GeV$ have a relative \pt resolution of 1.3--2.0\% in the barrel and less than 6\% in the endcaps. The \pt resolution in the barrel is better than 10\% for muons with \pt up to 1\TeV~\cite{Chatrchyan:2012xi}. The energy of charged hadrons is determined from a combination of the track momentum and the corresponding ECAL and HCAL energy deposits, corrected for zero-suppression effects and for the response function of the calorimeters to hadronic showers. Finally, the energy of neutral hadrons is obtained from the corresponding corrected ECAL and HCAL energy deposits.

Events of interest are selected using a two-tiered trigger system~\cite{Khachatryan:2016bia}. The level-1 trigger (L1), composed of custom hardware processors, uses information from the calorimeters and muon detectors to select events at a rate of around 100\unit{kHz}. The second level, known as the high-level trigger (HLT), consists of a farm of processors running a version of the full event reconstruction software optimized for fast processing, and reduces the event rate to less than 1\unit{kHz} before data storage.

A more detailed description of the CMS detector, together with a definition of the coordinate system used and the relevant kinematic variables, can be found in Ref.~\cite{Chatrchyan:2008zzk}.

\section{Data and simulated samples}
\label{sec:simulation}
The results presented in this paper are based on the {\Pp\Pp} collision data set recorded at a centre-of-mass energy of 13\TeV by the CMS detector in 2016, corresponding to an integrated luminosity of $35.9\fbinv$. Various event generators are used to model the relevant physics processes. The interactions between particles and the material of the CMS detector are simulated using \GEANT~4~\cite{Agostinelli:2002hh,ALLISON2016186,2006ITNS...53..270A}. The data and simulated samples are used to determine the heavy-flavour jet identification efficiency in various event topologies. When measuring the heavy-flavour jet identification efficiency or when comparing the data to the simulation, the number of simulated events is large enough to neglect the statistical uncertainty in the simulation unless mentioned otherwise.

The pair production of top quarks and electroweak single top quark production is performed with the \POWHEG~2.0 generator at next-to-leading order (NLO) accuracy~\cite{Nason:2004rx,Frixione:2007vw,Alioli:2010xd,Campbell:2014kua,Re:2010bp,Alioli:2009je}. The value of the top quark mass used for the generation of the simulated samples is 172.5\GeV. The systematic uncertainty related to the value of the top quark mass $m_t$ is evaluated by varying it by $\pm$1\GeV. Alternative samples are used to assess parton shower uncertainties, as well as factorization and normalization scale uncertainties at the matrix element and parton shower levels.
Diboson {\PW\PW}, {\PW\PZ}, and {\ZZ} events, referred to collectively as ``VV'' events, are generated at NLO accuracy with the \MGvATNLO~2.2.2 generator~\cite{Alwall:2014hca}, including \textsc{MadSpin}~\cite{Artoisenet:2012st} and the FxFx merging scheme~\cite{Frederix:2012ps} between jets from matrix element calculations and the parton shower description, or with the \POWHEG~2.0 generator~\cite{Melia:2011tj,Nason:2013ydw}. The {\Zj} and {\Wj} events are generated with \MGvATNLO~2.2.2 at leading order (LO), using the MLM matching scheme~\cite{MLM}. Samples of events with a Kaluza--Klein graviton~\cite{Randall} decaying to two Higgs bosons are also simulated with \MGvATNLO~2.2.2 at LO for graviton masses ranging between 1 and 3.5\TeV.
Background events comprised uniquely of jets produced through the strong interaction (multijet events) are generated with \PYTHIA~8.205~\cite{Sjostrand:2014zea} in different $\hat{p}_{\text{T}}$ bins, where $\hat{p}_{\text{T}}$ is defined as the average \pt of the final-state partons. Muon-enriched multijet samples are produced by forcing the decay of charged pions and kaons into muons and by requiring a generated muon with $\pt > 5\GeV$.

\PYTHIA~8.205 is also used for the parton showering and hadronization of all the simulated samples with the CMS underlying event tunes CUETP8M1~\cite{Khachatryan:2015pea} using the NNPDF~2.3~\cite{NNPDF2} parton distribution functions. In the case of top quark pair production a modification of this tune is used, CUETP8M2T4~\cite{CMS:2016kle} using the NNPDF~3.0~\cite{NNPDF3} parton distribution functions.

Pileup interactions are modelled by overlaying the simulated events with additional minimum bias collisions generated with \PYTHIA~8.205. These additional simulated events are then reweighted to match the observed number of pileup interactions or the primary vertex multiplicity in data.

\section{Heavy-flavour jet discriminating variables}
\label{sec:vars}
\subsection{Properties of heavy-flavour jets}
\begin{figure}[hbtp]
  \centering
    \includegraphics[width=0.49\textwidth]{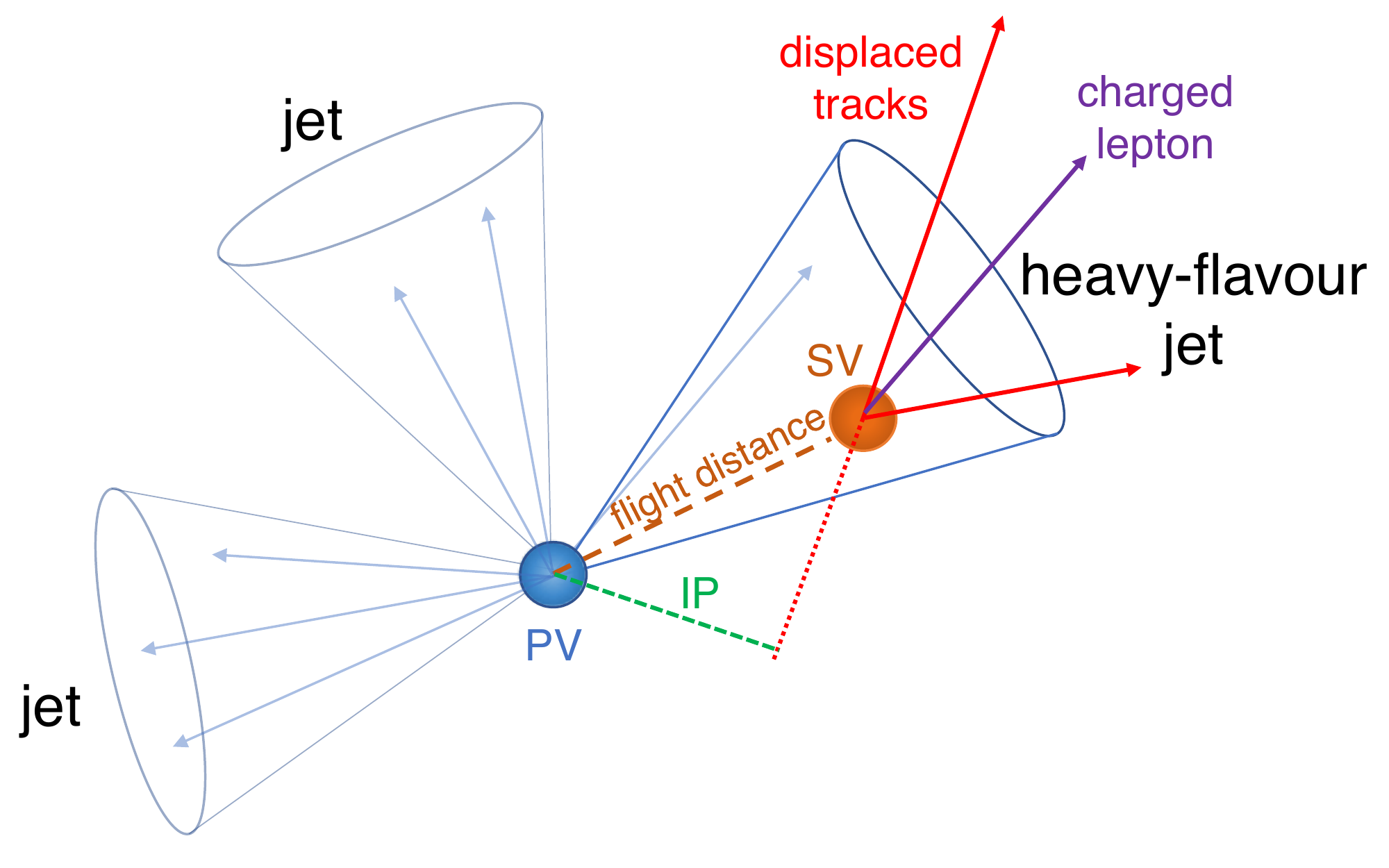}
    \caption{Illustration of a heavy-flavour jet with a secondary vertex (SV) from the decay of a {\cPqb} or {\cPqc} hadron resulting in charged-particle tracks (including possibly a soft lepton) that are displaced with respect to the primary interaction vertex (PV), and hence with a large impact parameter (IP) value.}
    \label{fig:schemebtagging}
\end{figure}

Algorithms for heavy-flavour jet identification use variables connected to the properties of heavy-flavour hadrons present in jets resulting from the radiation and hadronization of {\cPqb} or {\cPqc} quarks. For instance, the lifetime of hadrons containing {\cPqb} quarks is of the order of 1.5\,ps, while the lifetime of {\cPqc} hadrons is 1\,ps or less. This leads to typical displacements of a few \unit{mm} to one \unit{cm} for {\cPqb} hadrons, depending on their momentum, thus giving rise to displaced tracks from which a secondary vertex (SV) may be reconstructed, as illustrated in Fig.~\ref{fig:schemebtagging}.
The displacement of tracks with respect to the primary vertex is characterized by their impact parameter, which is defined as the distance between the primary vertex and the tracks at their points of closest approach. The vector pointing from the primary vertex to the point of closest approach is referred to as the impact parameter vector. The impact parameter value can be defined in three spatial dimensions (3D) or in the plane transverse to the beam line (2D). The longitudinal impact parameter is defined in one dimension, along the beam line. The impact parameter is defined to be positive or negative, with a positive sign indicating that the track is produced ``upstream''. This means that the angle between the impact parameter vector and the jet axis is smaller than $\pi/2$, where the jet axis is defined by the primary vertex and the direction of the jet momentum.
In addition, {\cPqb} and {\cPqc} quarks have a larger mass and harder fragmentation compared to the light quarks and massless gluons. As a result, the decay products of the heavy-flavour hadron have, on average, a larger \pt relative to the jet axis than the other jet constituents. In approximately 20\% (10\%) of the cases, a muon or electron is present in the decay chain of a heavy {\cPqb} (\cPqc) hadron. Hence, apart from the properties of the reconstructed secondary vertex or displaced tracks, the presence of charged leptons is also exploited for heavy-flavour jet identification techniques and for measuring their performance in data.

In order to design and optimize heavy-flavour identification techniques, a reliable method is required for assigning a flavour to jets in simulated events. The jet flavour is determined by clustering not only the reconstructed final-state particles into jets, but also the generated {\cPqb} and {\cPqc} hadrons that do not have {\cPqb} and {\cPqc} hadrons as daughters respectively. To prevent these generated hadrons from affecting the reconstructed jet momentum, the modulus of the hadron four-momentum is set to a small number, retaining only the directional information. This procedure is known as ghost association~\cite{Cacciari:2007fd}. Jets containing at least one {\cPqb} hadron are defined as {\cPqb} jets; the ones containing at least one {\cPqc} hadron and no {\cPqb} hadron are defined as {\cPqc} jets. The remaining jets are considered to be light-flavour (or ``udsg'') jets.
Since pileup interactions are not included during the hard-scattering event generation, jets from pileup interactions (``pileup jets'') in the simulation are tentatively identified as jets without a matched generated jet. The generated jets are reconstructed with the jet clustering algorithm mentioned in Section~\ref{sec:cms} applied to the generated final-state particles (excluding neutrinos). The matching between the reconstructed PF jets and the generated jets with $\pt > 8\GeV$ is performed by requiring the angular distance between them to be $\Delta R=\sqrt{(\Delta\eta)^2+(\Delta\phi)^2} < 0.25$.
Using this flavour definition, jets arising from gluon splitting to {\cPqb}{\cPaqb} are considered as {\cPqb} jets. In Sections~\ref{sec:boostedalgos},~\ref{sec:ak4eff} and~\ref{sec:boostedeff}, these ${\Pg\to{\cPqb}{\cPaqb}}$ jets are often shown as a separate category. In this case, two {\cPqb} hadrons without {\cPqb} hadron daughters should be clustered in the jet.
The studies presented in Sections~\ref{sec:vars} and~\ref{sec:ak4algos} are based on simulated events. For these studies, jets are removed if they are closer than $\Delta R=0.4$ to a generated charged lepton from a direct V boson decay. In addition, electrons or muons originating from gauge boson decays that are reconstructed as jets are removed if they carry more than 60\% of the jet \pt, \ie $\pt^{\ell}/\pt^{\text{jet}}<0.6$ is required, where $\pt^{\ell}$ ($\pt^{\text{jet}}$) is the \pt of the lepton (jet). No additional identification or isolation requirements are applied for muons or electrons.

\subsection{Track selection and variables}
\label{sec:tracks}
\begin{figure}[hbtp]
  \centering
    \includegraphics[width=0.49\textwidth]{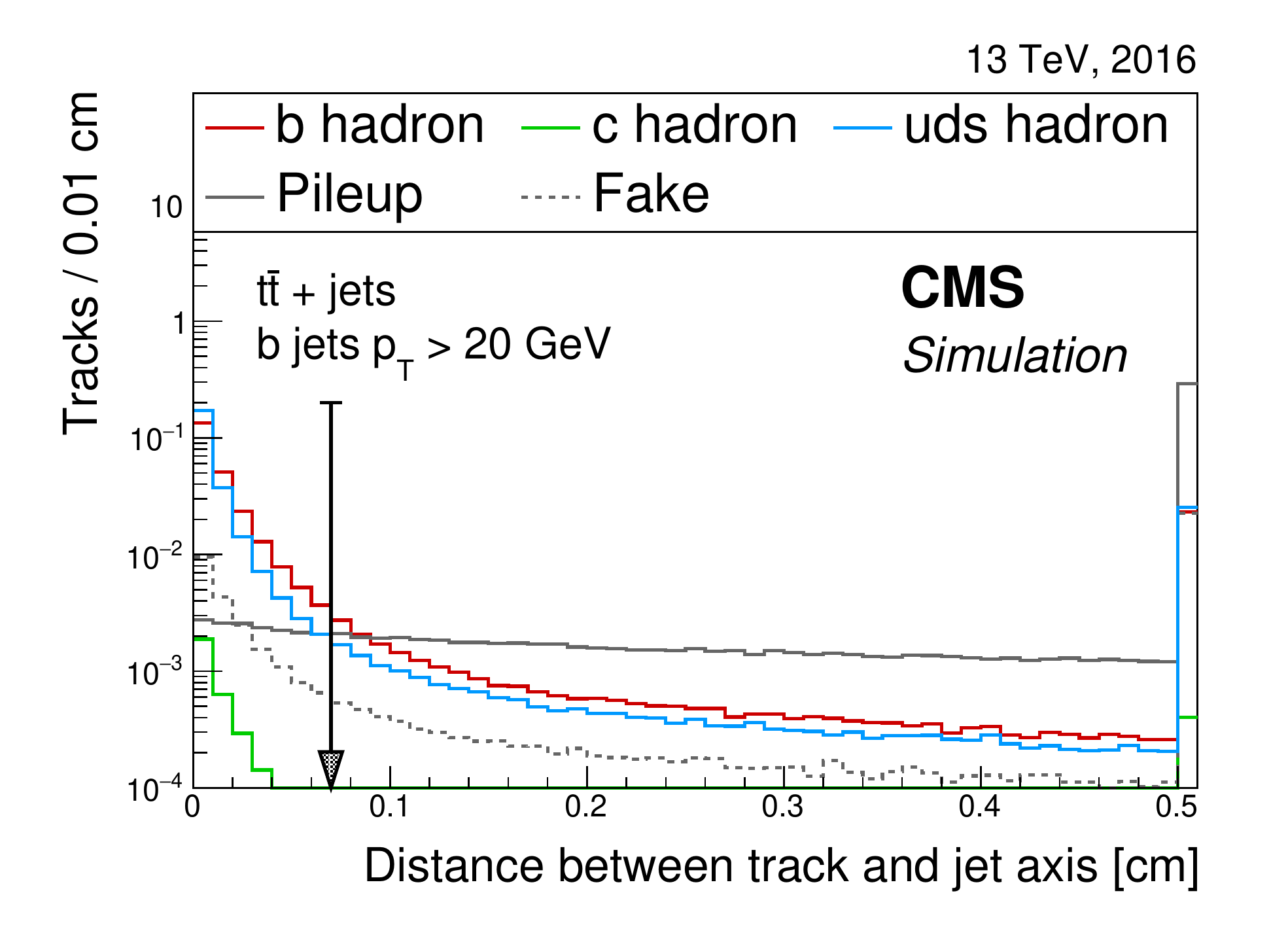}
    \includegraphics[width=0.49\textwidth]{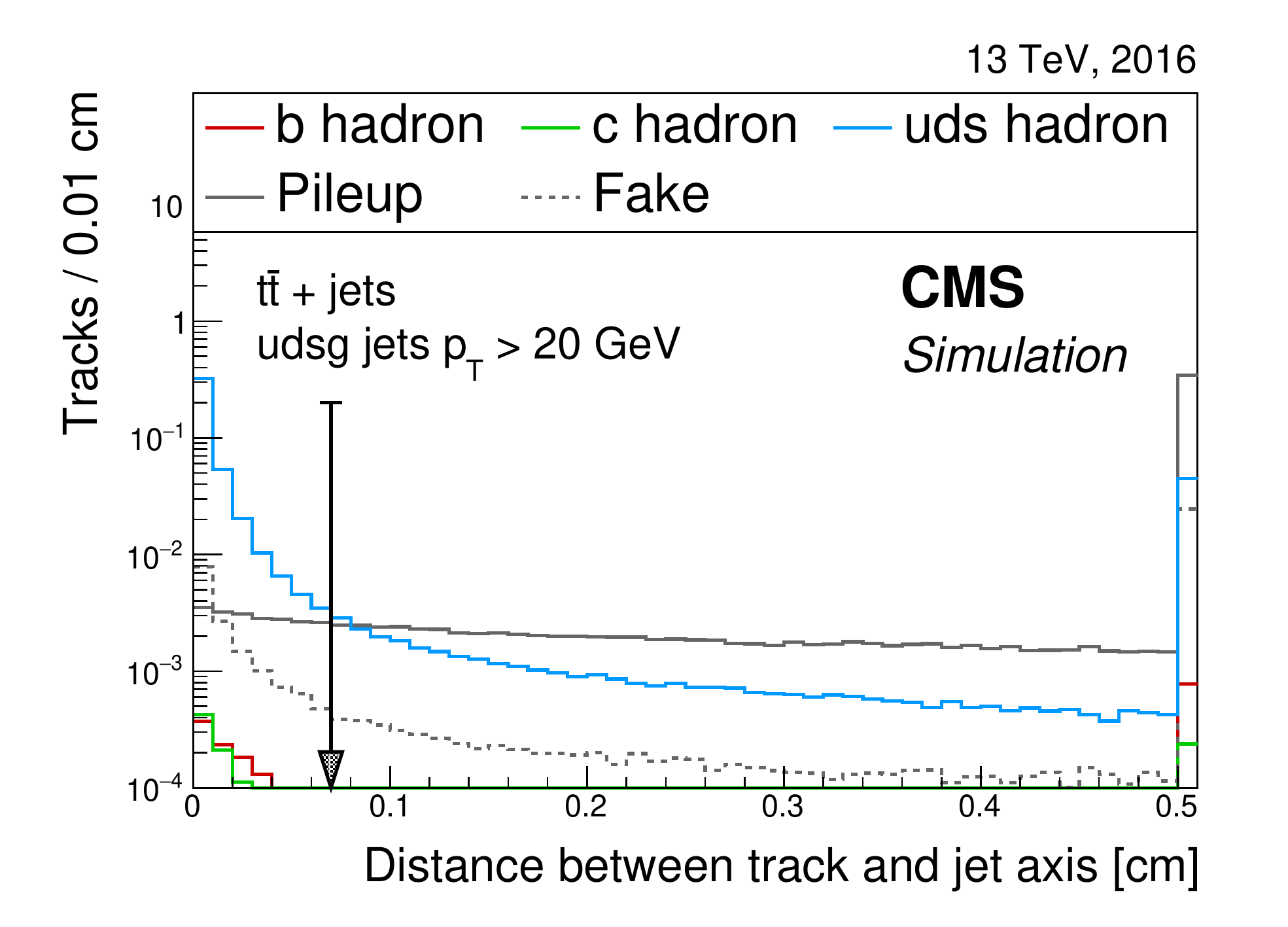}
    \caption{Distribution of the distance between a track and the jet axis at their point of closest approach for tracks associated with {\cPqb} (left) and light-flavour (right) jets in \ttbar events. This distance is required to be smaller than 0.07\unit{cm}, as indicated by the arrow. The tracks are divided into categories according to their origin as defined in the text. The distributions are normalized such that their sum has unit area. The last bin includes the overflow entries.}
    \label{fig:jettrackdistance}
\end{figure}

\begin{figure}[hbtp]
  \centering
    \includegraphics[width=\textwidth]{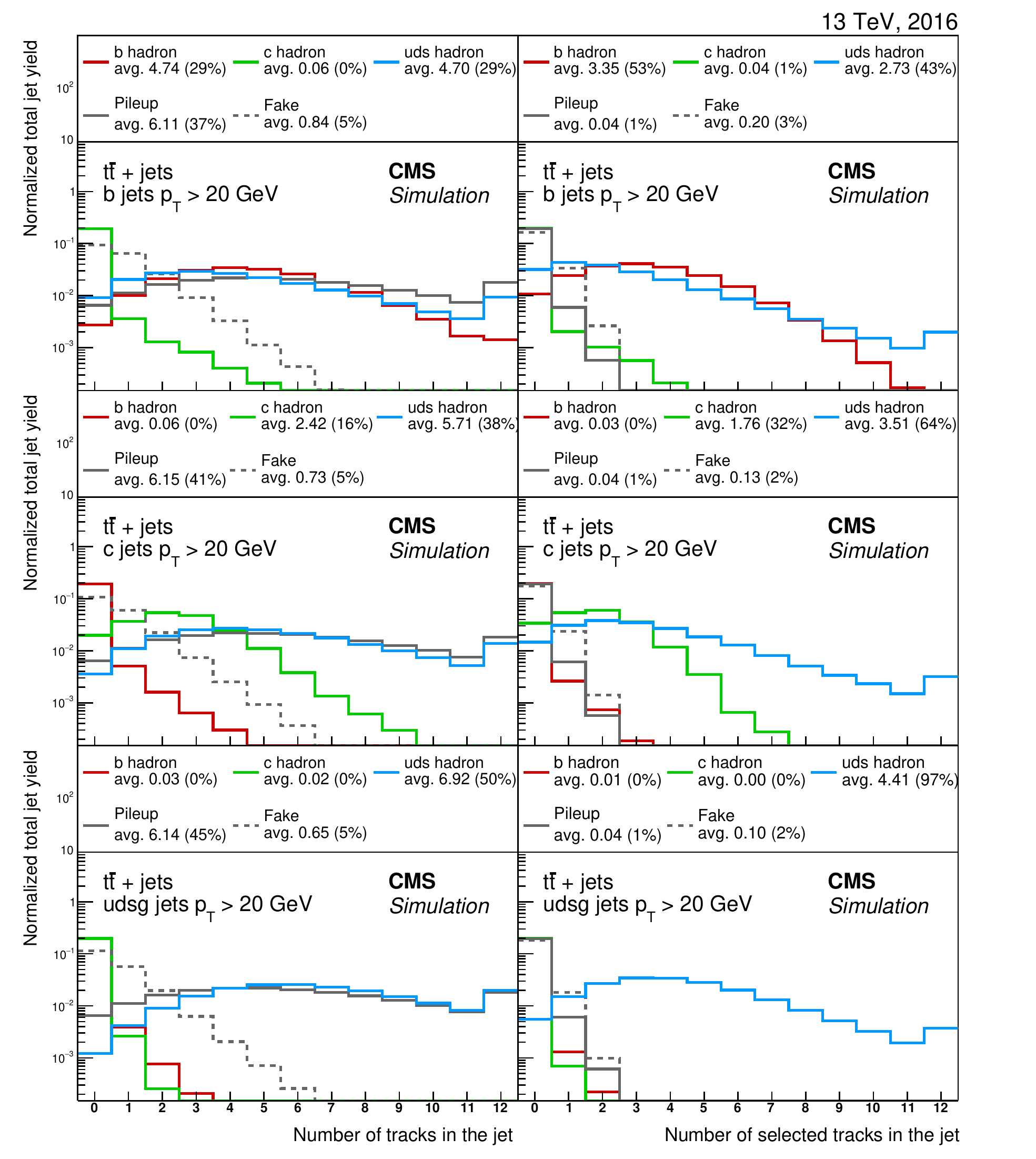}
    \caption{Fraction of tracks from different origins before (left) and after (right) applying the track selection requirements on {\cPqb} (upper), {\cPqc} (middle), and light-flavour (lower) jets in \ttbar events. The average number of tracks of each origin is given in the legend as well as the average fraction of tracks of a certain origin with respect to the total number of tracks in the jet, indicated in per cent. The number of tracks corresponding to pileup vertices or mismeasured tracks is strongly reduced after applying the track selection requirements. The distributions are normalized such that their sum has unit area. The last bin includes the overflow entries.}
    \label{fig:numberoftracksorigin}
\end{figure}

\begin{figure}[hbtp]
  \centering
    \includegraphics[width=0.49\textwidth]{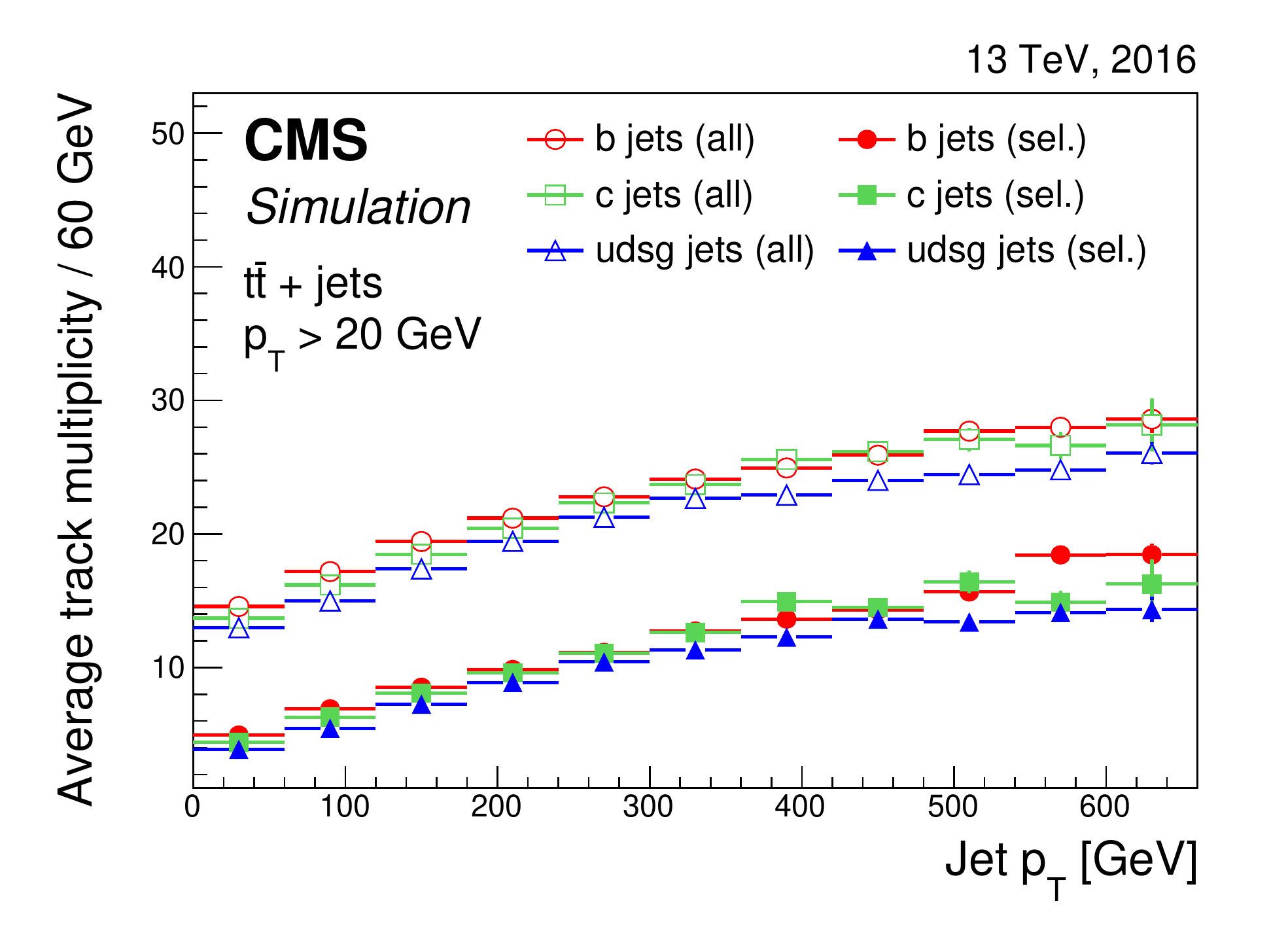}
    \includegraphics[width=0.49\textwidth]{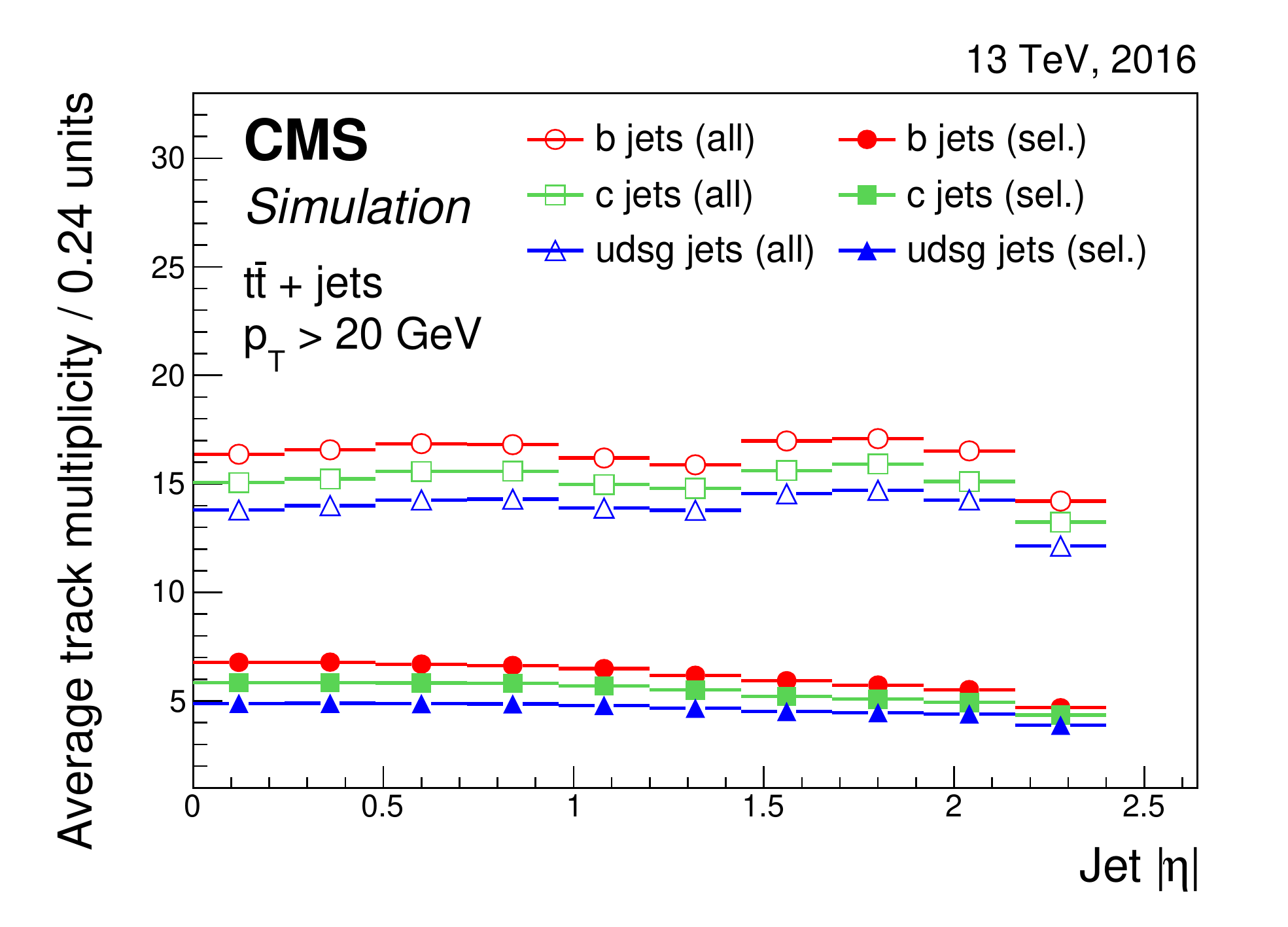}
    \caption{Average track multiplicity as a function of the jet \pt (left) and $\abs{\eta}$ (right) for jets of different flavours in \ttbar events before (open symbols) and after (filled symbols) applying the track selection requirements.}
    \label{fig:trackmult}
\end{figure}

The properties of the tracks clustered within the jet represent the basic inputs of all heavy-flavour jet identification (tagging) algorithms. Input variables for the tagging algorithms are constructed from the tracks after applying appropriate selection criteria. In particular, to ensure a good momentum and impact parameter resolution, tracks are required to have $\pt > 1\GeV$, a $\chi^2$ value of the trajectory fit normalized to the number of degrees of freedom below 5, and at least one hit in the pixel layers of the tracker detector. The last of these requirements is less stringent than the requirement used for {\cPqb} jet identification in Run 1, where at least eight hits were required in the pixel and strip tracker combined, of which at least two were pixel detector hits. The requirement on the number of hits was relaxed to cope with saturation effects that were observed at high occupancy in the readout electronics of the strip tracker during the first part of the 2016 data taking, leading to a reduced tracking and {\cPqb} tagging performance. The issues with the readout electronics have been fully resolved, with no side effects on the tracking performance, but the relaxed requirement on the number of hits was kept since there was no impact on the final {\cPqb} tagging performance.
Apart from the requirements on the quality of the tracks, the presence of tracks from long-lived {\PKzS} or {\PgL} hadrons as well as from material interactions is reduced by requiring the track decay length, defined as the distance from the primary vertex to the point of closest approach between the track and the jet axis, to be less than 5\unit{cm}. The contribution from tracks originating from pileup vertices is reduced with the following set of requirements: the absolute value of the transverse (longitudinal) impact parameter of the track is required to be smaller than 0.2 (17)\unit{cm} and the distance between the track and the jet axis at their point of closest approach is required to be less than 0.07\unit{cm}. Figure~\ref{fig:jettrackdistance} presents typical distributions of the latter variable for jets in \ttbar events after applying the rest of the track selection requirements, showing the origin of each track separately. The origin of a track is labelled with ``{\cPqb} hadron'' if the track corresponds to a particle originating from a {\cPqb} hadron decay. A track corresponding to a particle from the decay of a {\cPqc} hadron that itself originates from the decay of a {\cPqb} hadron is also labelled as ``{\cPqb} hadron''. The category with the ``{\cPqc} hadron'' label contains only tracks corresponding to a particle from the decay of a {\cPqc} hadron without a {\cPqb} hadron ancestor. The label ``uds hadron'' indicates tracks corresponding to particles without heavy-flavour hadron ancestors. The label ``pileup'' refers to tracks from charged particles originating from a different primary vertex. A category with mismeasured tracks is defined containing tracks that are more likely to have been misreconstructed, \eg by wrongly combining hits created by different particles. A track belongs to this category if the number of hits from the simulated charged particle closest to the track over the number of hits associated with the track, is less than 75\%. This category is labelled as ``fake''.
In Fig.~\ref{fig:numberoftracksorigin}, the impact of the track selection requirements on the number of tracks in a given category is shown for various jet flavours in \ttbar events. The track selection requirements clearly enhance the fraction of tracks originating from heavy-flavour hadron decays in bottom and charm jets. The track selection requirements reduce the number of tracks in the fake and pileup categories to a few per cent for all jet flavours.
Figure~\ref{fig:trackmult} shows the track multiplicity dependence on the jet \pt and $\abs{\eta}$ for various jet flavours in \ttbar events before and after applying the track selection requirements. For {\cPqb} jets, the average track multiplicity is higher than for light-flavour jets, before and after applying the track selection requirements, and the ratio of the average track multiplicity for {\cPqb} jets to other jet flavours is roughly constant.
The average track multiplicity increases with increasing jet \pt for all jet flavours.
Before the track selection, the average track multiplicity is almost constant with respect to the jet $\abs{\eta}$. The small variations seen are due to the tracker geometry that has an impact on the track reconstruction efficiency. In addition, since the $\eta$ of the jet is defined as the $\eta$ of the jet axis, some of the charged particles in the jet are outside the tracker acceptance for high jet $\abs{\eta}$ values, resulting in a lower track multiplicity in the highest bin. When the track selection requirements are applied, the average track multiplicity decreases with respect to the jet $\abs{\eta}$, because of the relatively larger impact of the track selection requirements near the edge of the acceptance window for the tracker.

\begin{figure}[hbtp]
  \centering
    \includegraphics[width=0.49\textwidth]{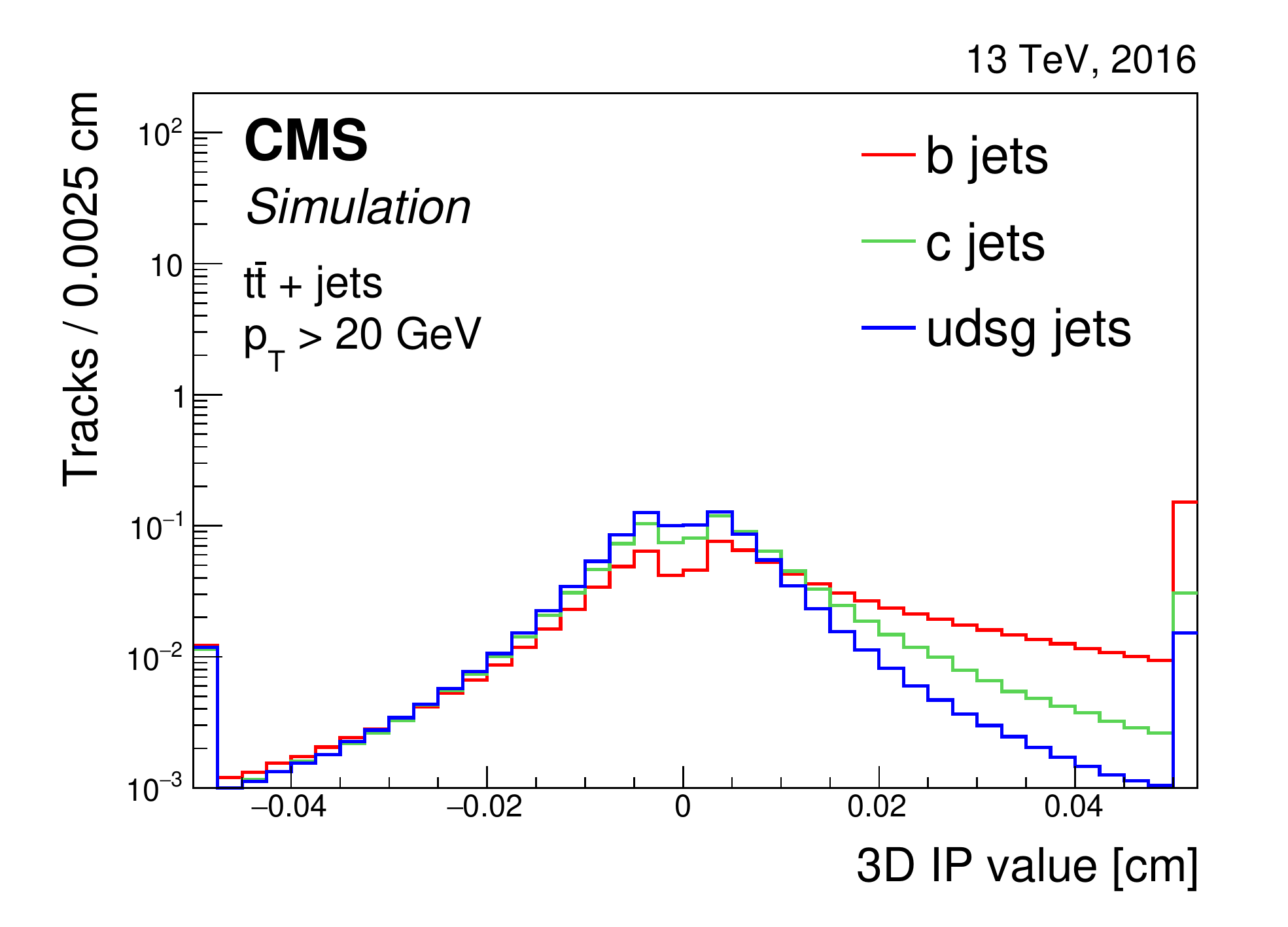}
    \includegraphics[width=0.49\textwidth]{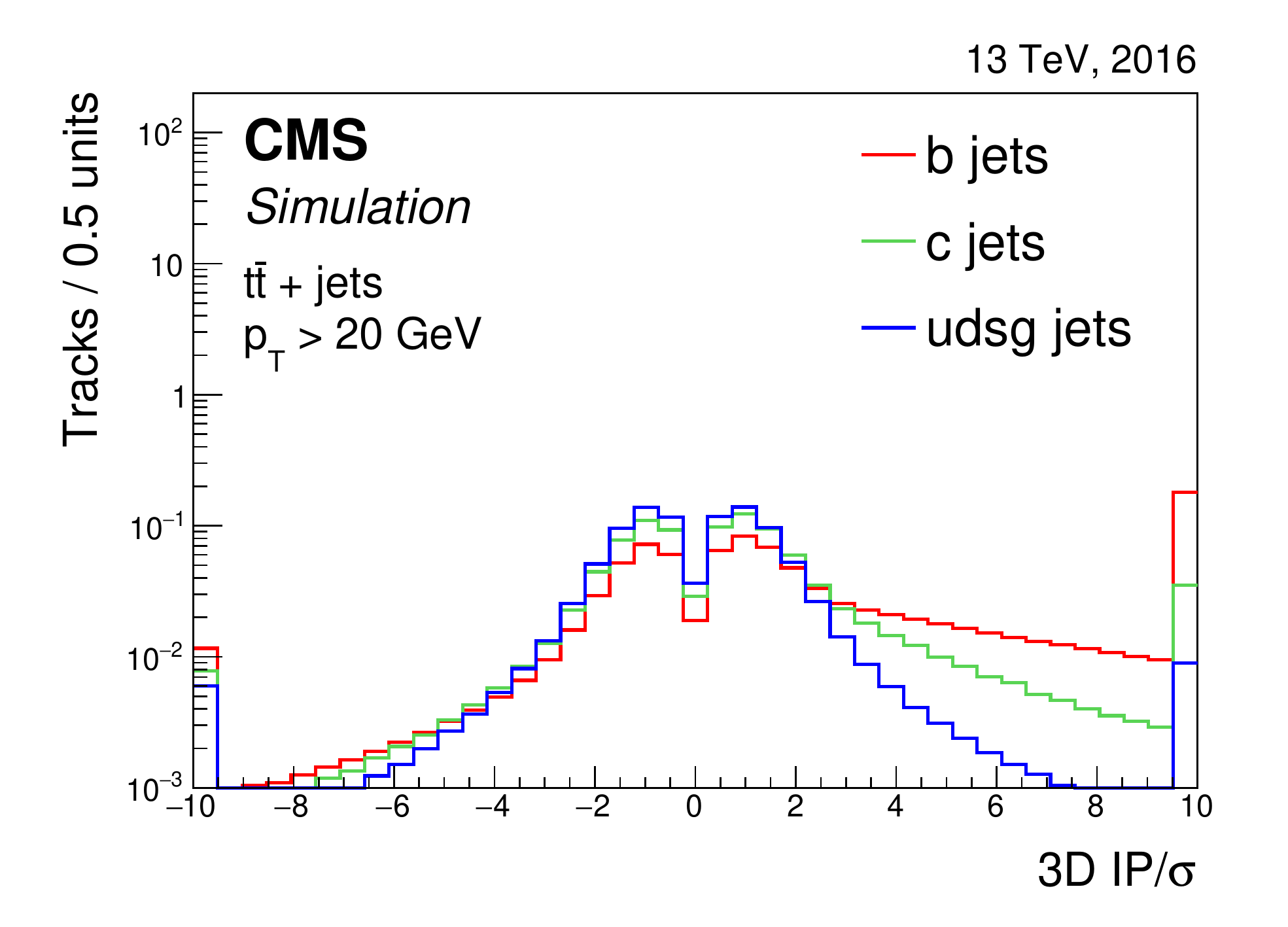}
    \includegraphics[width=0.49\textwidth]{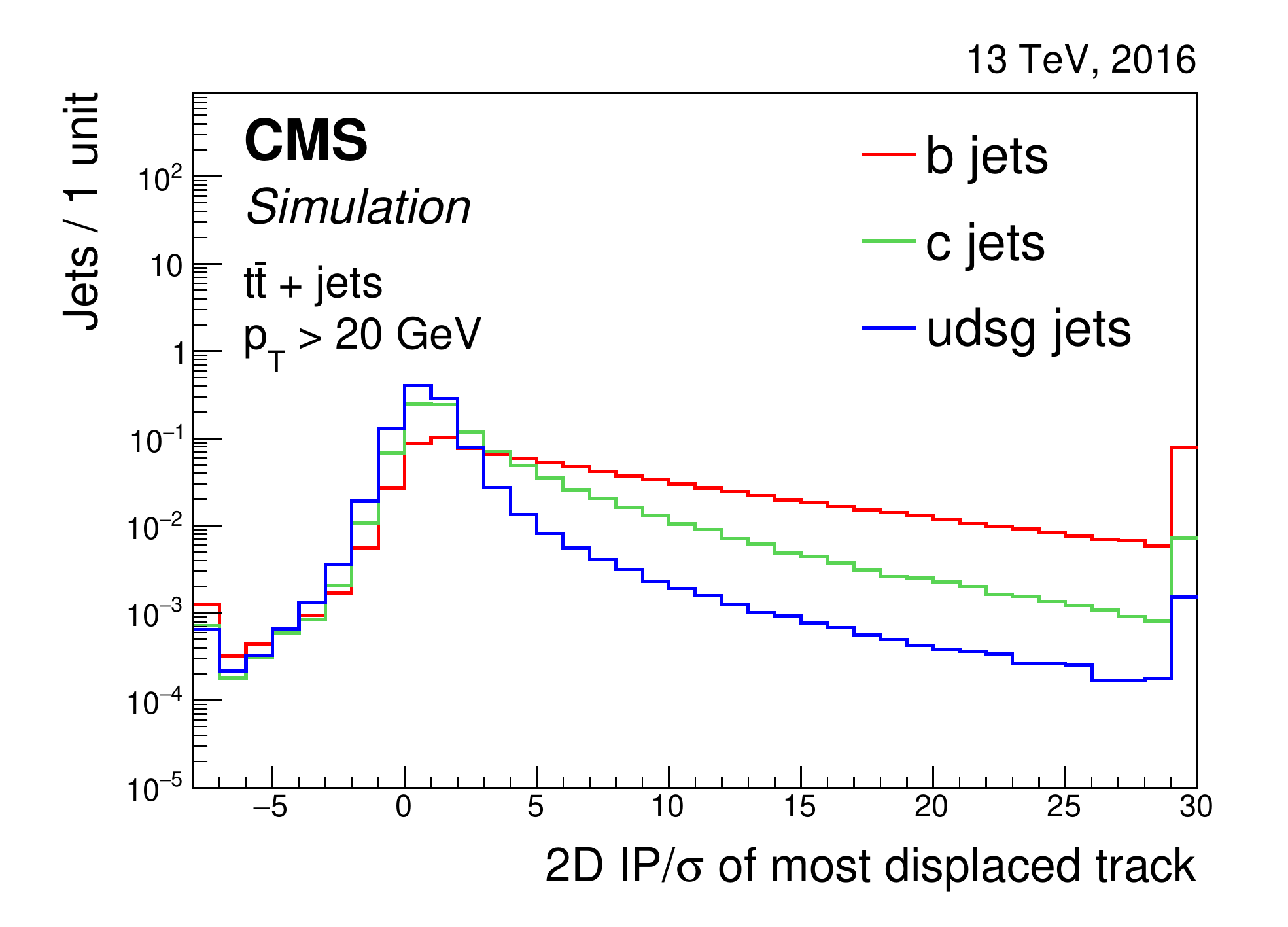}
    \includegraphics[width=0.49\textwidth]{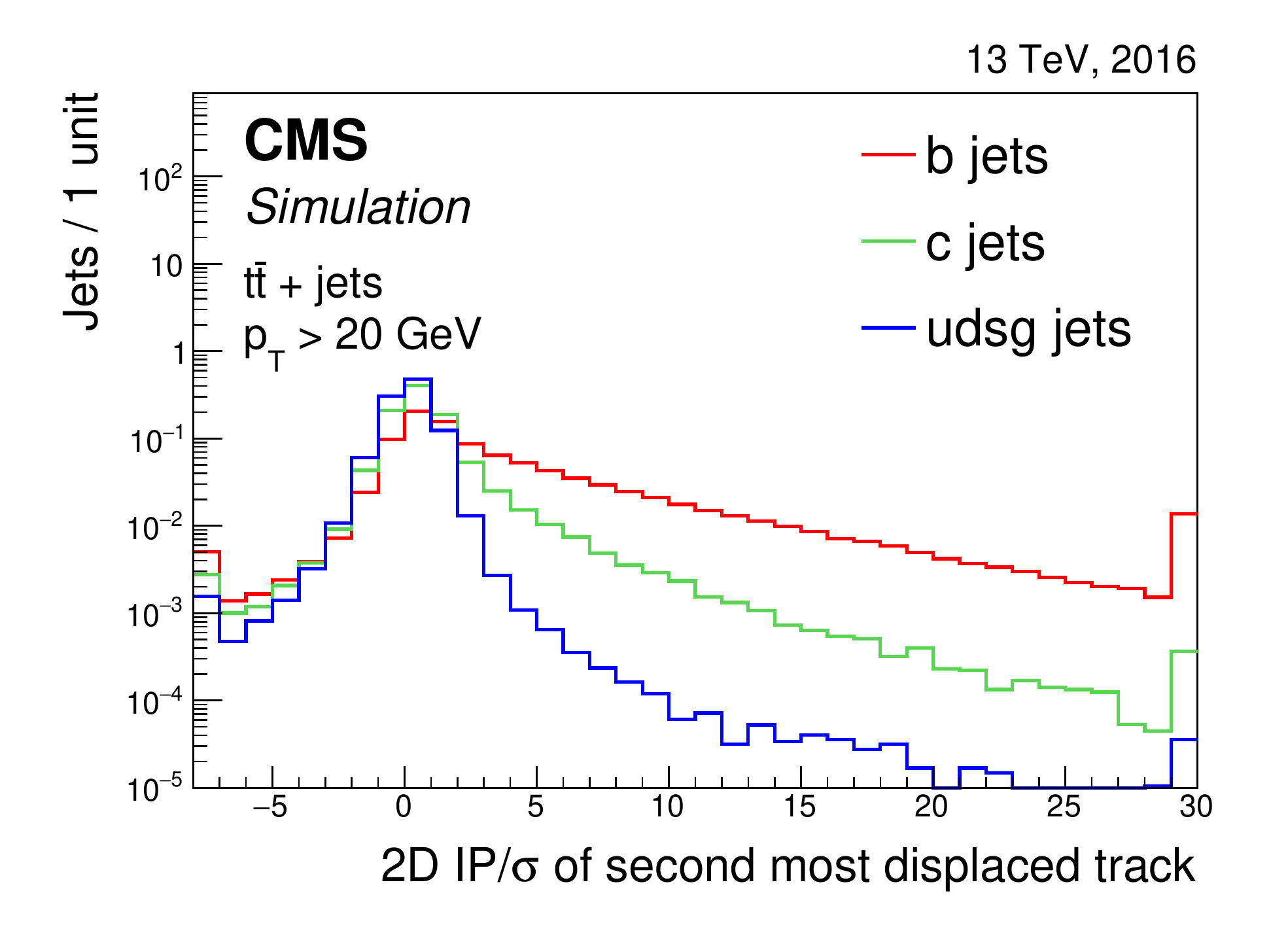}
    \caption{Distribution of the 3D impact parameter value (upper left) and significance (upper right) for tracks associated with jets of different flavours in \ttbar events. Distribution of the 2D impact parameter significance for the track with the highest (lower left) and second-highest (lower right) 2D impact parameter significance for jets of different flavours in \ttbar events. The distributions are normalized to unit area. The first and last bin include the underflow and overflow entries, respectively.}
    \label{fig:IP}
\end{figure}

The aforementioned track selection requirements are always applied when reconstructing the variables used in the tagging algorithms. An exception is given by the variables relying on the inclusive vertex finding algorithm, as discussed in Section~\ref{sec:vertexing}.
Figure~\ref{fig:IP} shows the distribution of the 3D impact parameter and its significance for the different jet flavours. The impact parameter significance is defined as the impact parameter value divided by its uncertainty, IP$/\sigma$. In addition, the lower panels in Fig.~\ref{fig:IP} also show the distribution of the 2D impact parameter significance for the track with the highest and second-highest 2D impact parameter significance for different jet flavours. From Fig.~\ref{fig:IP} it is clear that tracks in heavy-flavour jets have larger impact parameter and impact parameter significance compared to tracks in light-flavour jets.
The lower left panel in Fig.~\ref{fig:IP} shows that tracks with a large impact parameter significance are also present in light-flavour jets. These originate from the decays of relatively long-lived hadrons, for example {\PKzS} or {\PgL}, or from heavy-flavour hadrons where the tracks have been incorrectly clustered into a light-flavour jet. For the track with the second-highest impact parameter significance in light-flavour jets, the distribution is much more symmetric as expected for hadrons with a short lifetime.

\subsection{Secondary vertex reconstruction and variables}
\label{sec:vertexing}
If the secondary vertex from the decay of a heavy-flavour hadron is reconstructed, powerful discriminating variables can be derived from it. An example is the (corrected) secondary vertex mass, which is directly related to the mass of the heavy-flavour hadron. The corrected secondary vertex mass is defined as $\sqrt{M_{\text{SV}}^2 + p^2{\text{sin}}^2\theta} + p\text{sin} \theta$, where $M_{\text{SV}}$ is the invariant mass of the tracks associated with the secondary vertex, $p$ is the secondary vertex momentum obtained from the tracks associated with it, and $\theta$ the angle between the secondary vertex momentum and the vector pointing from the primary vertex to the secondary vertex, which is referred to as the secondary vertex flight direction. Using this definition, the secondary vertex mass is corrected for the observed difference between its flight direction and its momentum, taking into account particles that were not reconstructed or which failed to be associated with the secondary vertex. It should be noted that the energy of a track is obtained using its momentum and assuming the {\Pgppm} mass~\cite{Patrignani:2016xqp}.
Another example of a discriminating secondary vertex variable is its flight distance (significance), defined as the 2D or 3D distance between the primary and secondary vertex positions (divided by the uncertainty on the secondary vertex flight distance). Reconstructing the secondary vertex from the heavy-flavour hadron decay is not always possible for two main reasons: the heavy-flavour hadron decays too close to the primary vertex, or there are less than two selected tracks. The latter may be due to having less than two charged particles in the decay, less than two reconstructed tracks, or less than two tracks passing the selection requirements.

Two algorithms for reconstructing secondary vertices are used. The first one is the adaptive vertex reconstruction (AVR) algorithm~\cite{Waltenberger:1166320}. This secondary vertex reconstruction algorithm was used for {\cPqb} jet identification by the CMS Collaboration during the LHC Run 1~\cite{BTV12001}. The algorithm uses the tracks clustered within jets and passing the selection requirements discussed in Section~\ref{sec:tracks}. In addition, the tracks are required to be within $\Delta R<0.3$ of the jet axis and to have a track distance below 0.2\unit{cm}. The vertex pattern recognition iteratively fits all tracks with an outlier-resistant adaptive vertex fitter~\cite{Fruhwirth:2007hz}. At each iteration, tracks close enough to the fitted vertex are removed and a new iteration is made with the remaining tracks. Given that the first iteration often finds a vertex close to the primary vertex, the first iteration is explicitly run with a constraint on the primary vertex. Vertices are rejected if it is found that they share more than 65\% of their tracks with the primary vertex, or if their 2D secondary vertex flight distance is more than 2.5\unit{cm} or less than 0.01\unit{cm}. In addition, the 2D secondary vertex flight distance significance is required to be larger than 3. To reduce the impact of long-lived hadron decays and material interactions, only secondary vertices with $M_{\text{SV}}<6.5\GeV$ are considered. Pairs of tracks are rejected if they are compatible with the mass of the relatively long-lived {\PKzS} hadron within 50\MeV. Additionally, the angular distance between the jet axis and the secondary vertex flight direction should satisfy $\Delta R < 0.4$. When all these requirements are fulfilled, the reconstructed AVR secondary vertex is associated with the jet.

At the start of LHC Run 2, the inclusive vertex finding (IVF) algorithm was adopted as the standard secondary vertex reconstruction algorithm used to define variables for heavy-flavour jet tagging. In contrast with AVR, which uses as input the selected tracks clustered in the reconstructed jets, IVF uses as input all reconstructed tracks in the event with $\pt >0.8\GeV$ and a longitudinal $\text{IP} <0.3$\unit{cm}. The algorithm was initially developed to perform a measurement of the angular correlations between the {\cPqb} jets in {\cPqb\cPaqb} pair production~\cite{Khachatryan:2011wq}. It is well suited for {\cPqb} hadron decays at small relative angle giving rise to overlapping, or completely merged, jets.
The IVF procedure starts by identifying seed tracks with a 3D impact parameter value of at least 50\micron and a 2D impact parameter significance of at least 1.2. After identifying the seed tracks, the procedure includes the following steps:
\begin{itemize}
\item \textbf{Track clustering:} The compatibility between a seed track and any other track is evaluated using requirements on the distance at the point of closest approach of the two tracks and the angle between them. In addition, the distance between the seed track and any other track at their points of closest approach is required to be smaller than the distance between the track and the primary vertex at their points of closest approach.
\item \textbf{Secondary vertex fitting and cleaning:} In order to determine the position of the secondary vertices, the sets of clustered tracks are fitted with the adaptive vertex fitter also used in the AVR algorithm. After the fit, secondary vertices with a 2D (3D) flight distance significance smaller than 2.5 (0.5) are removed. For IVF vertices used in the {\cPqc} tagging algorithm presented in Section~\ref{sec:ctagger}, the threshold is relaxed to 1.25 (0.25). In addition, if two secondary vertices share 70\% or more of their tracks, or if the significance of the flight distance between the two secondary vertices is less than 2, one of the two secondary vertices is dropped from the collection of secondary vertices.
\item \textbf{Track arbitration:} At this stage, a track could be assigned to both the primary vertex and secondary vertex. To resolve this ambiguity, a track is discarded from the secondary vertex if it is more compatible with the primary vertex. This is the case if the angular distance between the track and the secondary vertex flight direction is $\Delta R > 0.4$, and if the distance between the secondary vertex and the track is larger than the absolute impact parameter value of the track.
\item \textbf{Secondary vertex refitting and cleaning:} The secondary vertex position is refitted after track arbitration and if there are still two or more tracks associated with the secondary vertex. After refitting the secondary vertex positions, a second check for duplicate vertices is performed. This time, a secondary vertex is removed from the collection of secondary vertices when it shares at least 20\% of its tracks with another secondary vertex and the significance of the flight distance between the two secondary vertices is less than 10.
 \end{itemize}
The selection criteria applied to the remaining IVF secondary vertices are mostly the same as in the case of the AVR vertices. However, to maximize the secondary vertex reconstruction efficiency, some requirements are relaxed. In particular, secondary vertices are rejected when they share 80\% or more of their tracks, and when the 2D flight distance significance is less than 2 (1.5) for secondary vertices used in {\cPqb} ({\cPqc}) tagging algorithms. The remaining secondary vertices are then associated with the jets by requiring the angular distance between the jet axis and the secondary vertex flight direction to satisfy $\Delta R < 0.3$.

\begin{figure}[hbtp]
  \centering
    \includegraphics[width=0.49\textwidth]{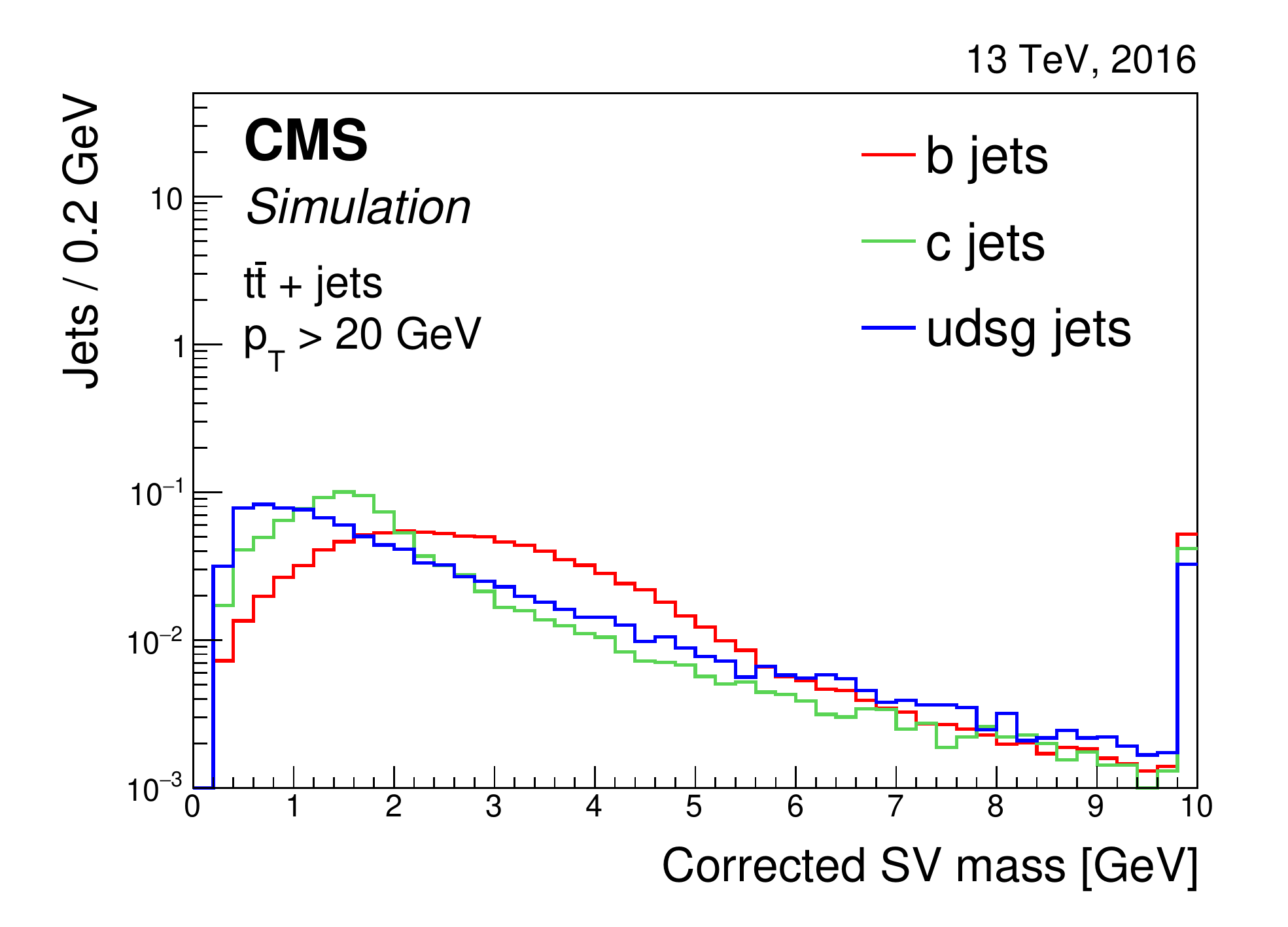}
    \includegraphics[width=0.49\textwidth]{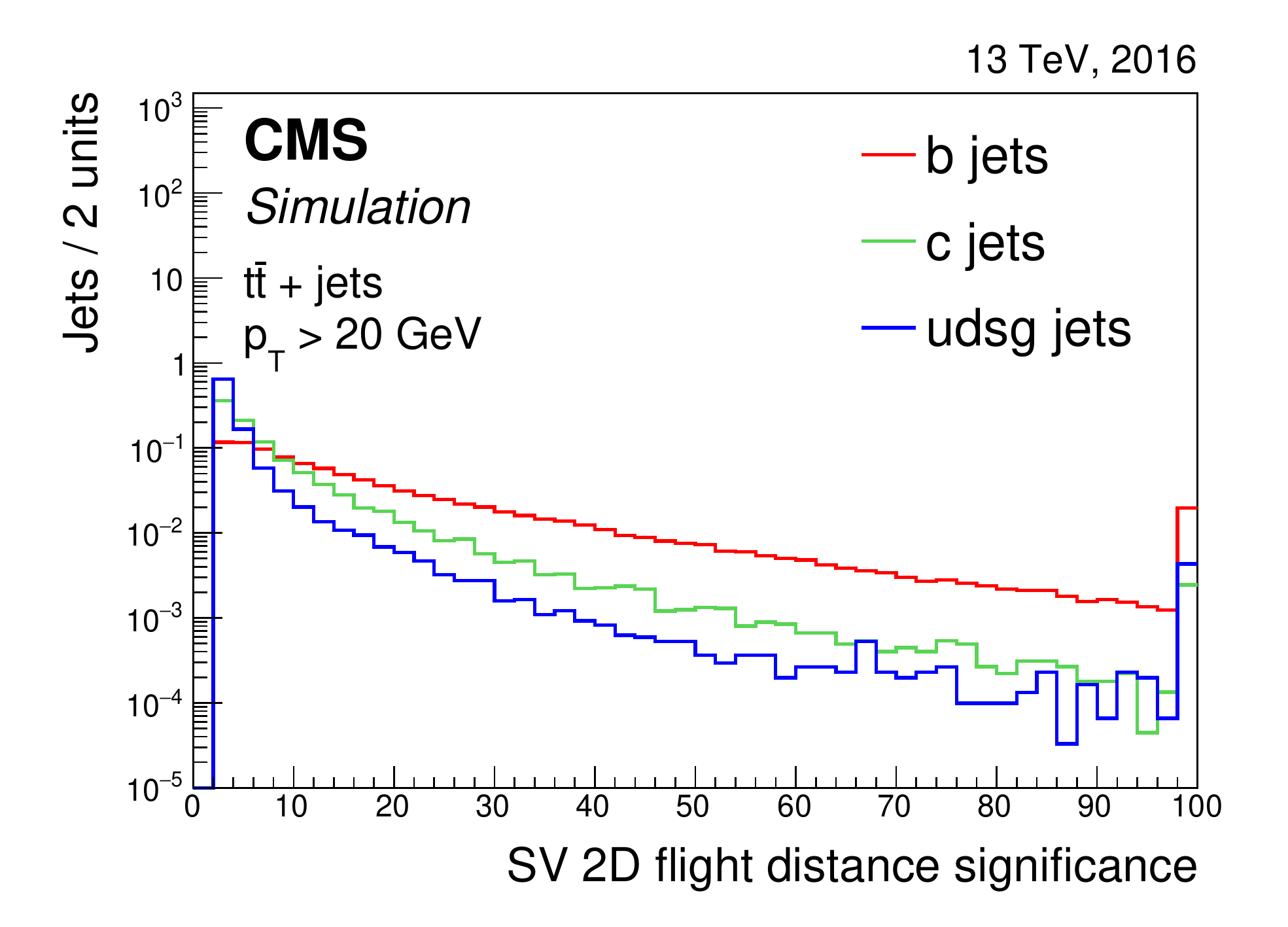}
    \caption{Distribution of the corrected secondary vertex mass (left) and of the secondary vertex 2D flight distance significance (right) for jets containing an IVF secondary vertex. The distributions are shown for jets of different flavours in \ttbar events and are normalized to unit area. The last bin includes the overflow entries.}
    \label{fig:SVinfo}
\end{figure}

Figure~\ref{fig:SVinfo} shows the discriminating power between the various jet flavours for the IVF secondary vertex mass (left) and 2D flight distance significance (right). The secondary vertex mass for {\cPqb} jets peaks at higher values compared to that of the other jet flavours. For {\cPqc} jets, a peak is observed around 1.5\GeV, as expected from the lower mass of {\cPqc} compared to {\cPqb} hadrons.

\begin{figure}[hbtp]
  \centering
    \includegraphics[width=0.49\textwidth]{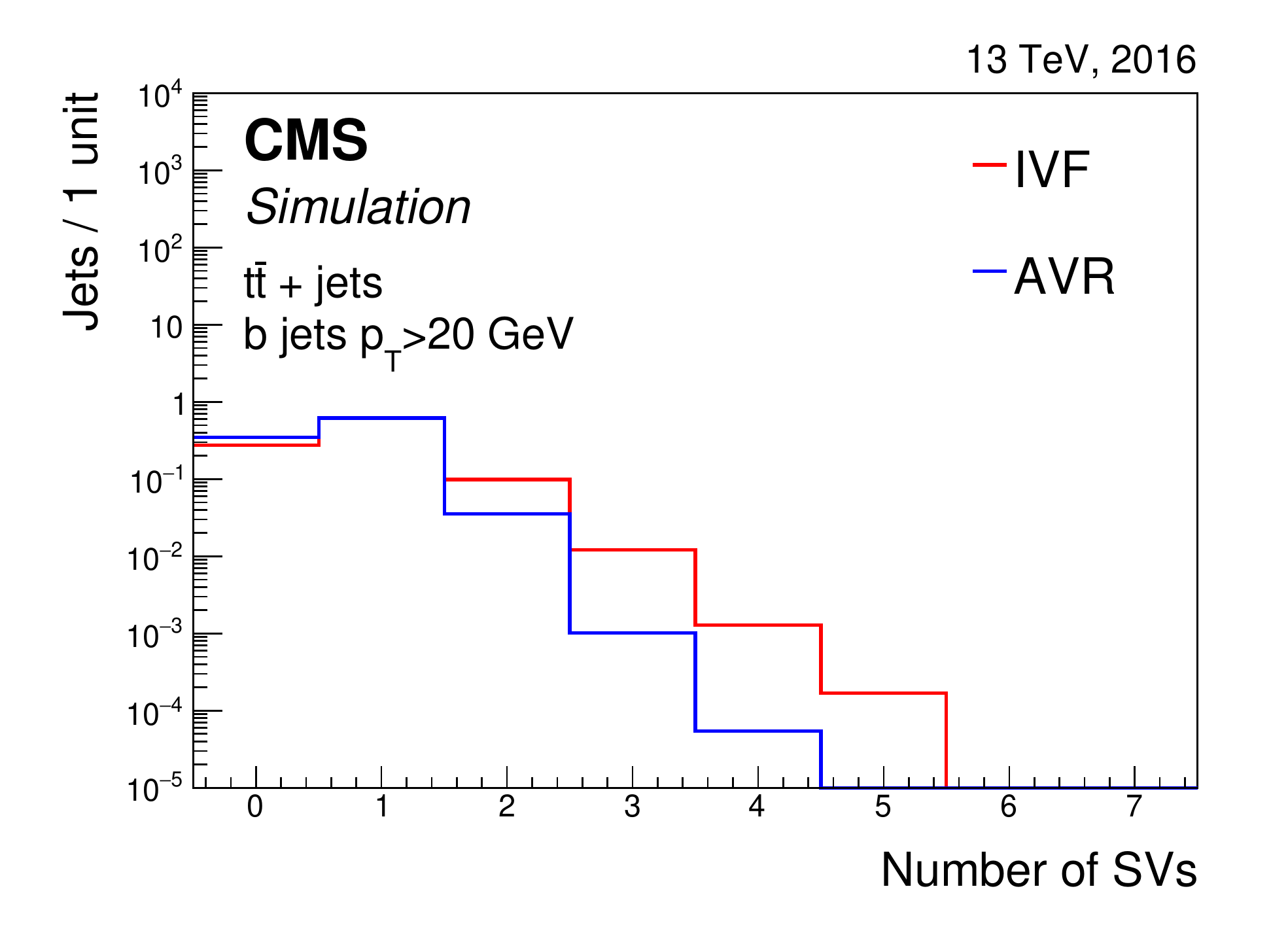}
    \includegraphics[width=0.49\textwidth]{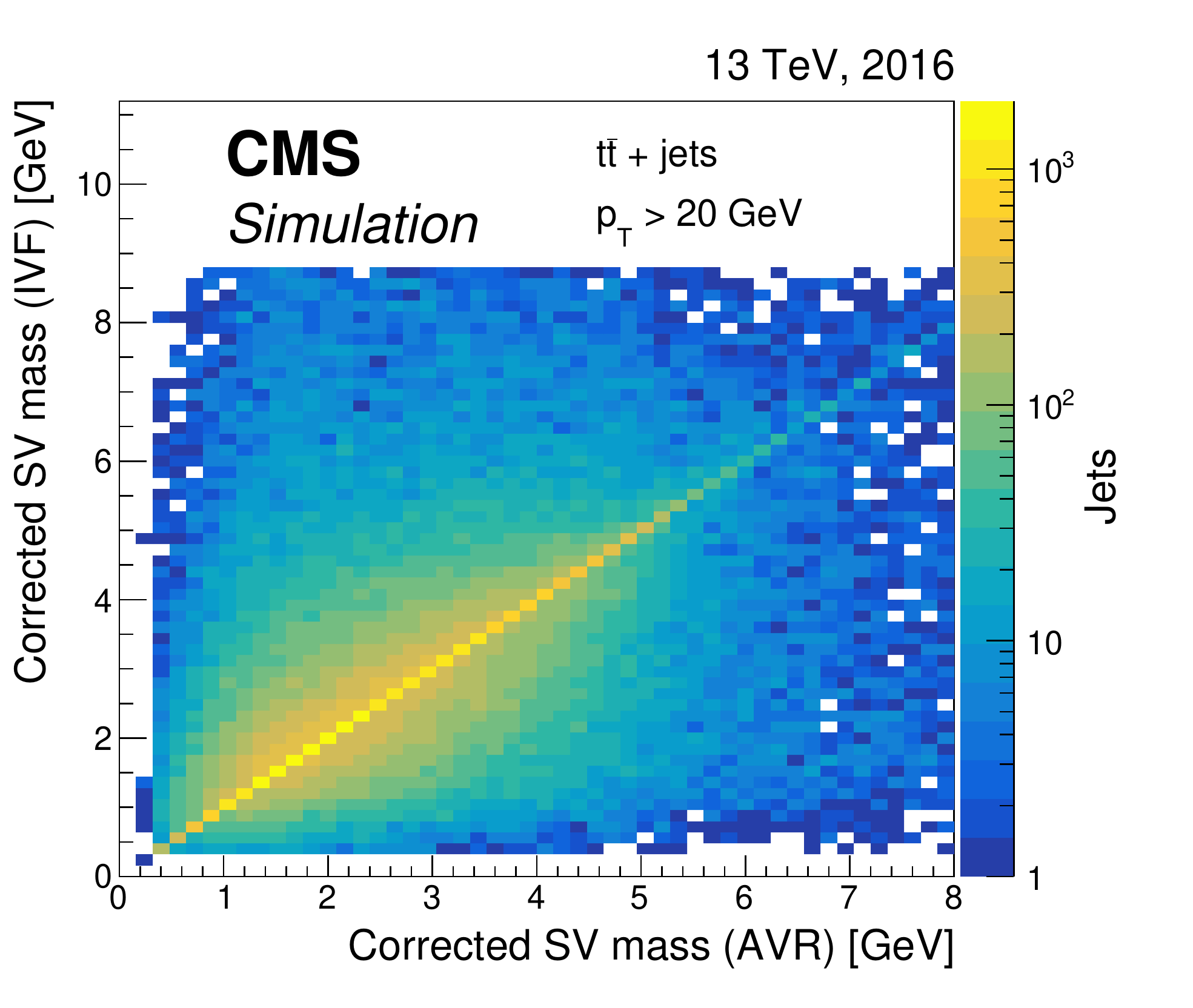}
    \caption{Distribution of the number of secondary vertices in {\cPqb} jets for the two vertex finding algorithms described in the text (left). The distributions are normalized to unit area. Correlation between the corrected secondary vertex mass for the vertices obtained with the two vertex finding algorithms (right). Both panels show jets in \ttbar events.}
    \label{fig:IVFvsAVR}
\end{figure}

The secondary vertex reconstruction efficiency for jets is defined as the number of jets containing a reconstructed secondary vertex divided by the total number of jets. For jets with $\pt>20$\GeV in \ttbar events, the efficiency for reconstructing a secondary vertex for {\cPqb} (udsg) jets using the IVF algorithm is about 75\% (12\%), compared to 65\% (4\%) for reconstructing a secondary vertex with the AVR algorithm. However, the efficiency gain is largest for {\cPqc} jets with an IVF secondary vertex reconstruction efficiency of about 37\%, compared to 23\% for the efficiency of the AVR algorithm. Averaged over all jet flavours, 66\% of the IVF secondary vertices in jets are also found by the AVR algorithm. The other way around, 86\% of the AVR secondary vertices are also found by the IVF algorithm.
Figure~\ref{fig:IVFvsAVR} (left) compares the number of secondary vertices in {\cPqb} jets for the IVF and AVR algorithms. As expected, more secondary vertices are reconstructed with the IVF algorithm because of the inclusive approach of using all tracks instead of only those associated with the jet and passing the selection requirements. The right panel in Fig.~\ref{fig:IVFvsAVR} shows the correlation between the corrected mass of the secondary vertices obtained with the two approaches. From the correlation it is clear that the same secondary vertex is found in most cases. Since the efficiency of the IVF algorithm is higher, IVF secondary vertices are used to compute the secondary vertex variables for the heavy-flavour jet identification algorithms. AVR secondary vertices are only used in one of the b jet identification algorithms discussed in Section~\ref{sec:ak4algos}.

\subsection{Soft-lepton variables}
\label{sec:SLvars}
Although an electron or muon is present in only 20\% (10\%) of the {\cPqb} ({\cPqc}) jets, the properties of this low-energy nonisolated ``soft lepton'' (SL) permit the selection of a pure sample of heavy-flavour jets. Therefore, some of the heavy-flavour taggers use the properties of these soft leptons. Soft muons are defined as particles clustered in the jet passing the loose muon identification criteria and with a \pt of at least 2\GeV~\cite{Chatrchyan:2012xi}. Electrons are associated with a jet by requiring $\Delta R<0.4$. Soft electrons should pass the loose electron identification criteria, have an associated track with at least three hits in the pixel layers, and be identified as not originating from a photon conversion~\cite{Khachatryan:2015hwa}.

\begin{figure}[hbtp]
  \centering
    \includegraphics[width=0.49\textwidth]{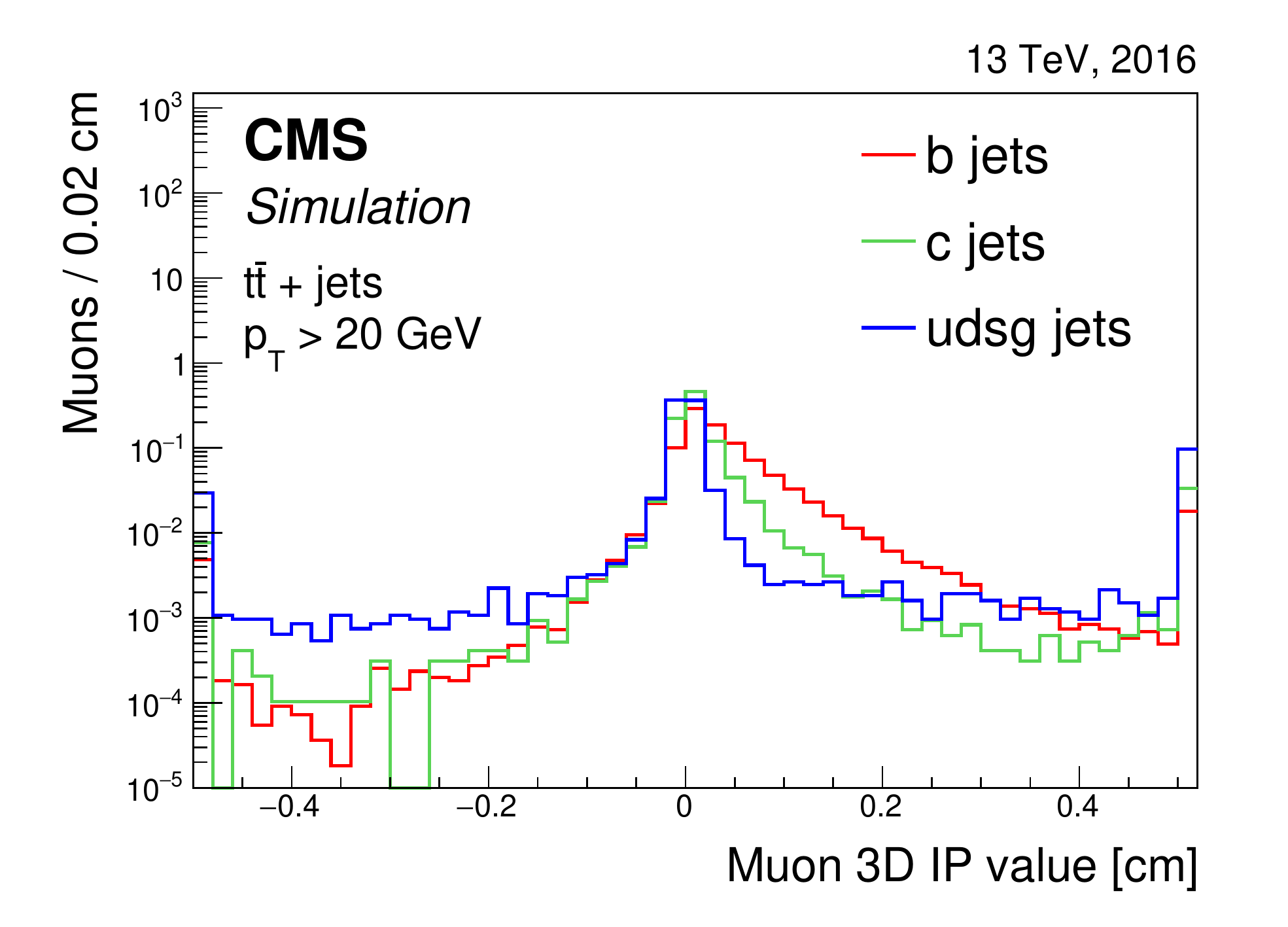}
    \includegraphics[width=0.49\textwidth]{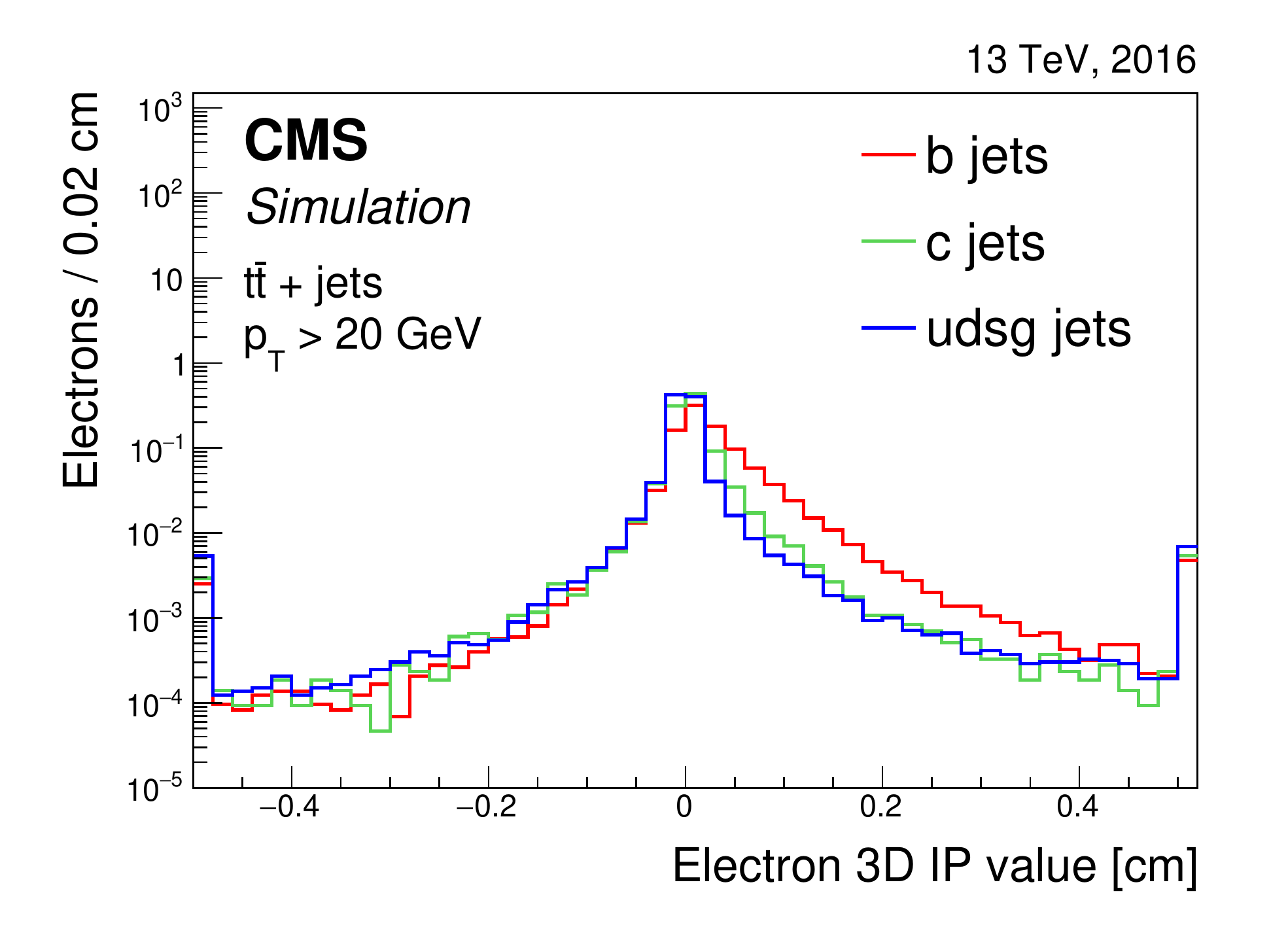}
    \caption{Distribution of the 3D impact parameter value for soft muons (left) and soft electrons (right) for jets of different flavours in \ttbar events. The distributions are normalized to unit area. The first and last bins include the underflow and overflow entries, respectively.}
    \label{fig:leptonIPsig}
\end{figure}

Discriminating variables using soft lepton information are typically similar to the variables based on track information alone. As an example, Fig.~\ref{fig:leptonIPsig} shows the distribution of the 3D impact parameter value of soft leptons associated with jets.
The 3D impact parameter value of the soft lepton discriminates between the various jet flavours. For the low-\pt muons expected from the heavy-flavour hadron decays, it should be noted that the impact parameter resolution is worse than at high \pt~\cite{Chatrchyan:2012xi}, which is reflected in the relatively large spread of the impact parameter values. The soft lepton variables are used in the soft lepton algorithms discussed in Section~\ref{sec:SLalgo} and in the {\cPqc} tagger discussed in Section~\ref{sec:ctagger}.

\section{Heavy-flavour jet identification algorithms}
\label{sec:ak4algos}
\subsection{The \texorpdfstring{{\cPqb}}{b} jet identification}
The jet probability (JP) and combined secondary vertex (CSV) taggers used during Run 1~\cite{BTV12001} are also used for the Run 2 analyses. Likewise, the combined multivariate analysis (cMVA) tagger, which combines the discriminator values of various taggers, was retrained. Apart from the retraining, the CSV algorithm was also optimized and the new version is referred to as CSVv2. In addition, another version of the CSV algorithm was developed that uses deep machine learning~\cite{PhysRevD.94.112002} (DeepCSV). These taggers are presented in more detail in the Sections~\ref{sec:JP} to~\ref{sec:SLalgo}. The new developments result in a performance that is significantly better than that of the Run 1 taggers, as discussed in Section~\ref{sec:perfak4b}.

\subsubsection{Jet probability taggers}
\label{sec:JP}
\begin{figure}[hbtp]
  \centering
    \includegraphics[width=0.49\textwidth]{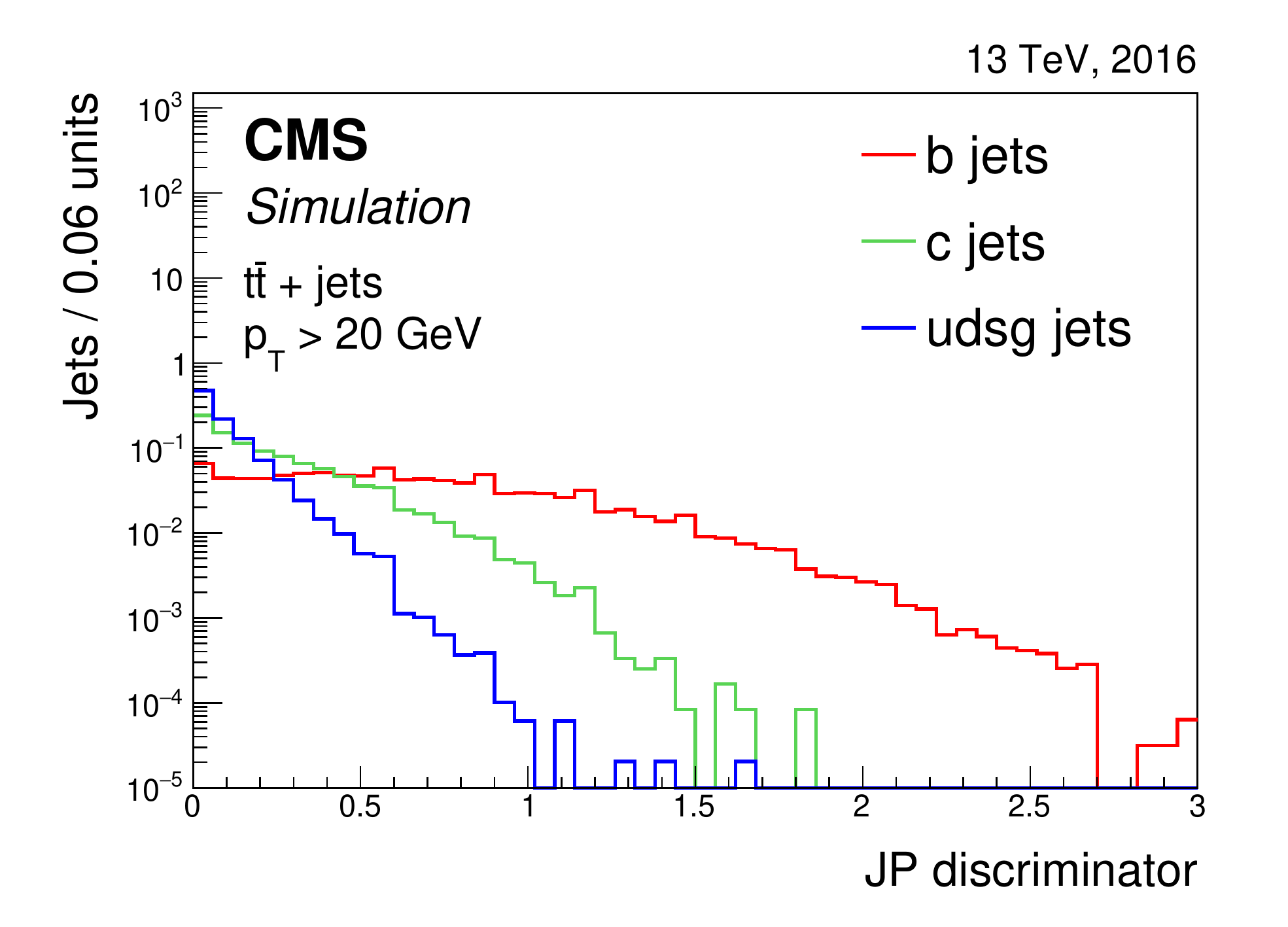}
    \includegraphics[width=0.49\textwidth]{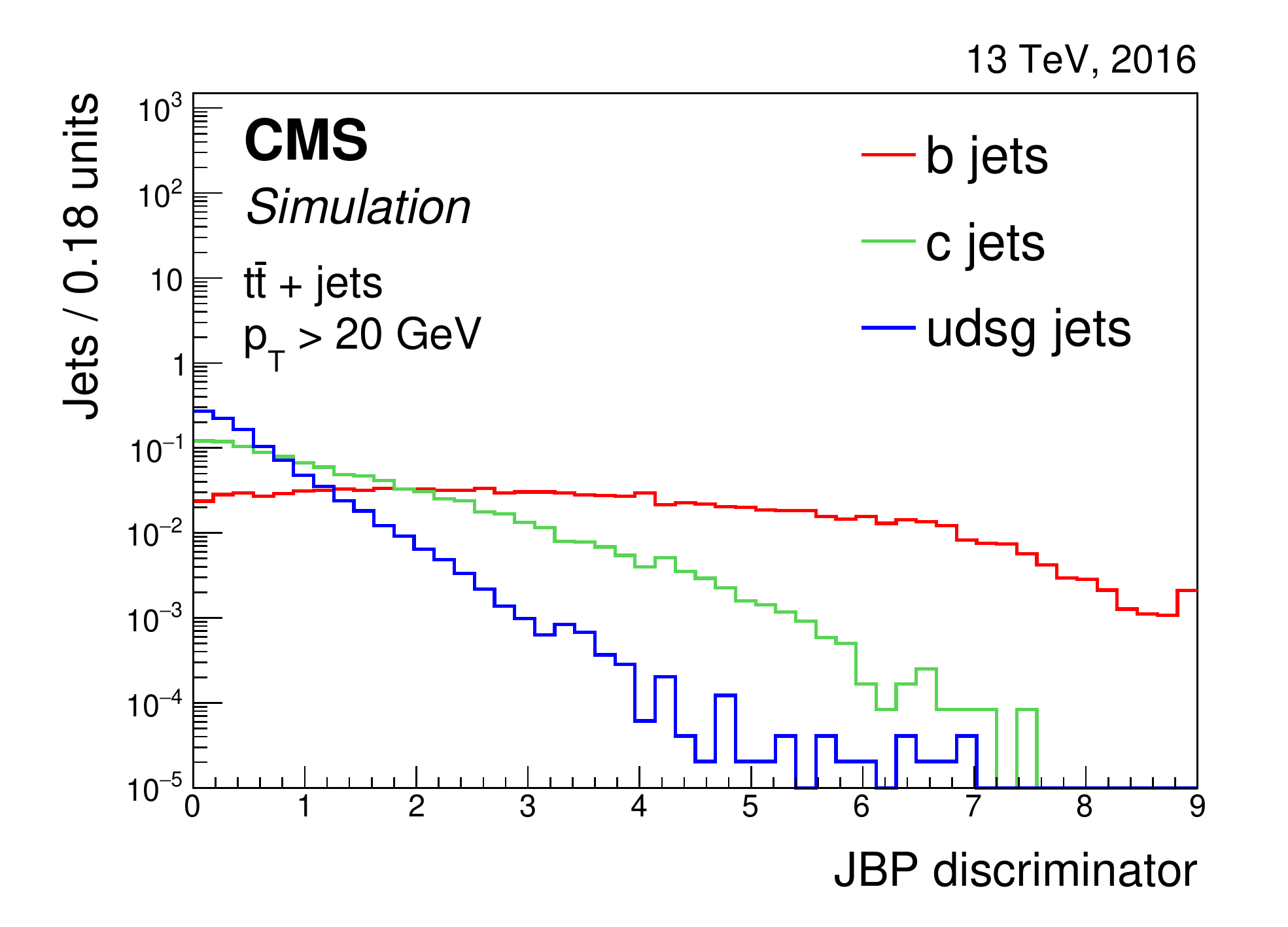}
    \caption{Distribution of the JP (left) and JBP (right) discriminator values for jets of different flavours in \ttbar events. Jets without selected tracks are assigned a negative value. The distributions are normalized to unit area. The first and last bin include the underflow and overflow entries, respectively.}
    \label{fig:JPdiscr}
\end{figure}

There are two jet probability taggers, the JP and JBP algorithms. The JP algorithm is described in Ref.~\cite{BTV12001} and uses the signed impact parameter significance of the tracks associated with the jet to obtain a likelihood for the jet to originate from the primary vertex. This likelihood, or jet probability, is obtained as follows. The negative impact parameter significance of tracks from light-flavour jets reflects the resolution of the measured track impact parameter values. Hence, the distribution of the negative impact parameter significance is used as a resolution function. The probability for a track to originate from the primary vertex, $P_{\text{tr}}$, is obtained by integrating the resolution function ${\mathcal{R}}(s)$ from $-\infty$ to the negative of the absolute track impact parameter significance, $-|\text{IP}|/\sigma$:
\begin{linenomath}
\begin{equation}
\label{eq:JP}
P_{\text{tr}} = \int_{-\infty}^{-|\text{IP}|/\sigma}{\mathcal{R}}(s)\rd{}s.
\end{equation}
\end{linenomath}
The resolution function depends strongly on the quality of the reconstructed track, e.g. the number of hits in the pixel and strip layers of the tracker. Moreover, the probability for a given track to originate from the primary vertex will be smaller for tracks with a large number of missing hits. Therefore, different resolution functions are defined for various track quality classes. In addition, the track quality may be different in data and simulated events. To calibrate the JP algorithm, the resolution functions are determined separately for data and simulation. Using Eq.~(\ref{eq:JP}), tracks corresponding to particles from the decay of a displaced particle will have a low track probability, indicating that the track is not compatible with the primary vertex.
The individual track probabilities are combined to obtain a jet probability $P_{\text{j}}$ as follows:
\begin{linenomath}
\begin{equation}
P_{\text{j}} = \Pi \sum_{\text{tr}=0}^{N-1} \frac{({-\ln \Pi})^{\text{tr}}}{\text{tr}!},
\end{equation}
\end{linenomath}
where $\Pi$ is the product over the track probabilities, $P_{\text{tr}}$, and the sum runs over the selected tracks index ${\text{tr}}$, with $N$ the number of selected tracks associated with the jet. To avoid instabilities due to the multiplication of small track probabilities, the probability is set to 0.5\% for track probabilities below 0.5\%. Only tracks with a positive impact parameter and for which the angular distance between the track and the jet axis satisfies $\Delta R < 0.3$ are used.  A variant of the JP algorithm also exists for which the four tracks with the highest impact parameter significance get a higher weight in the jet probability calculation. This algorithm is referred to as jet {\cPqb} probability (JBP) and uses tracks with $\Delta R < 0.4$. For a light-flavour jet misidentification probability of around 10\%, the JBP algorithm has a {\cPqb} jet identification efficiency of 80\% compared to 78\% for the JP algorithm. The discriminators for the jet probability algorithms were constructed to be proportional to $-\ln P_{\text{j}}$. Figure~\ref{fig:JPdiscr} shows the distributions of the discriminator values for the JP and JBP algorithms. The discontinuities in the discriminator distributions are due to the minimum track probability threshold of 0.5\%.

The jet probability algorithms are interesting for two reasons. First, the fact that the calibration of the resolution function is performed independently for data and simulation results in a robust reference tagger. Second, these algorithms rely only on the impact parameter information of the tracks. Therefore, they are used by some methods when measuring the efficiency of other {\cPqb} jet identification algorithms that rely on secondary vertex or soft lepton information, as discussed in Sections~\ref{sec:ak4eff} and~\ref{sec:boostedeff}.

\subsubsection{Combined secondary vertex taggers}
\paragraph{The CSVv2 tagger}
\label{sec:CSVv2}

The CSVv2 algorithm is based on the CSV algorithm described in Ref.~\cite{BTV12001} and combines the information of displaced tracks with the information on secondary vertices associated with the jet using a multivariate technique. Two variants of the CSVv2 algorithm exist according to whether IVF or AVR vertices are used. As baseline, IVF vertices are used in the CSVv2 algorithm, otherwise we refer to it as CSVv2 (AVR). At least two tracks per jet are required. When calculating the values of the track variables, the tracks are required to have an angular distance with respect to the jet axis of $\Delta R < 0.3$. Moreover, any combination of two tracks compatible with the mass of the {\PKzS} meson in a window of 30\MeV is rejected. Jets that have neither a selected track nor a secondary vertex are assigned a default output discriminator value of $-1$.

\begin{figure}[hbtp]
  \centering
    \includegraphics[width=0.49\textwidth]{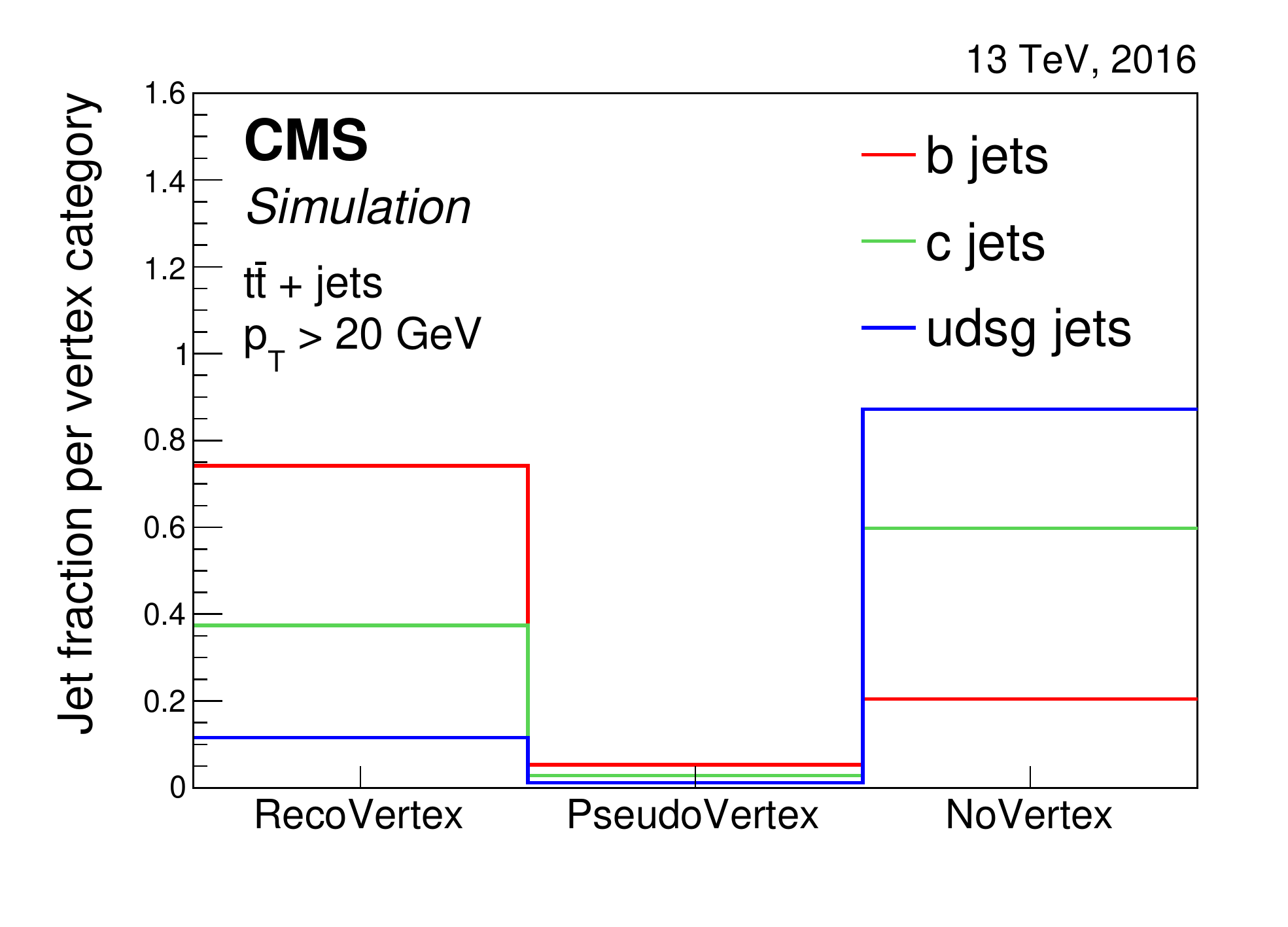}
    \includegraphics[width=0.49\textwidth]{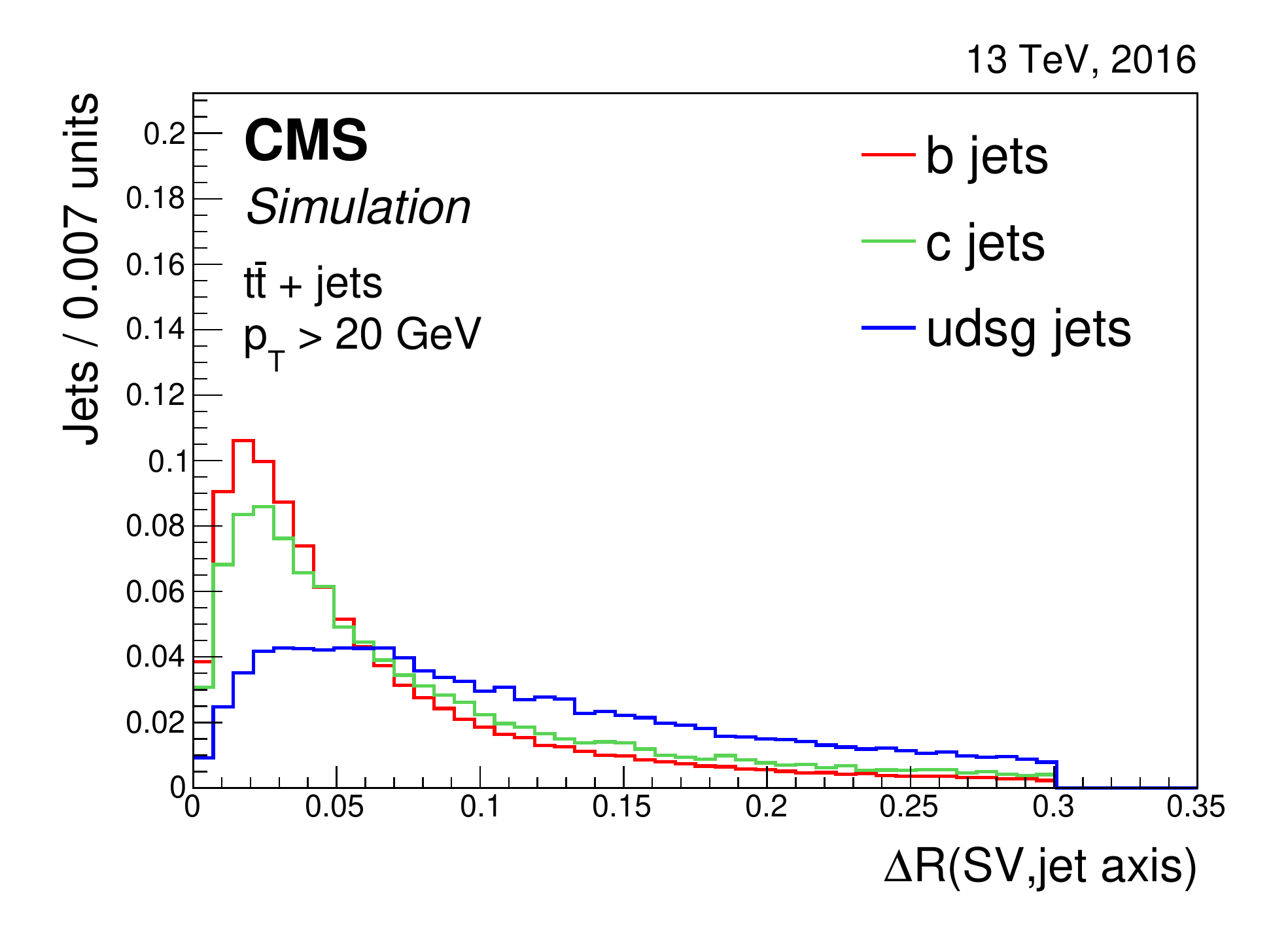}
    \caption{Vertex category for secondary vertices reconstructed with the IVF algorithm (left), and the distribution of the angular distance between the IVF secondary vertex flight direction and the jet axis (right) for jets of different flavours in \ttbar events. The distributions are normalized to unit area. }
    \label{fig:vertexcat}
\end{figure}

In a first step, the algorithm has to learn the features, \eg input variable distributions corresponding to the various jet flavours, and combine them into a single discriminator output value. This step is the so-called ``training'' of the algorithm. During this step, it is important to ensure that the algorithm does not learn any unwanted behaviour, such as {\cPqb} jets having a higher jet \pt, on average, compared to other jets in a sample of \ttbar events. To avoid discrimination between jet flavours caused by different jet \pt and $\eta$ distributions, these distributions are reweighted to obtain the same spectrum for all jet flavours in the training sample. The training is performed on inclusive multijet events in three independent vertex categories:
\begin{itemize}
\item \textbf{RecoVertex}: The jet contains one or more secondary vertices.
\item \textbf{PseudoVertex}: No secondary vertex is found in the jet but a set of at least two tracks with a 2D impact parameter significance above two and a combined invariant mass at least 50\MeV away from the {\PKzS} mass are found. Since there is no real secondary vertex reconstruction, no fit is performed, resulting in a reduced number of variables.
\item \textbf{NoVertex}: Containing jets not assigned to one of the previous two categories. Only the information of the selected tracks is used.
\end{itemize}
Figure~\ref{fig:vertexcat} shows the fraction of jets of each flavour in the various vertex categories of the CSVv2 algorithm using jets in \ttbar events with \pt above 20\GeV, where the secondary vertices in the RecoVertex category are obtained with the IVF algorithm.
The following discriminating variables are combined in the algorithm.
\begin{itemize}
\item The ``SV 2D flight distance significance'', defined as the 2D flight distance significance of the secondary vertex with the smallest uncertainty on its flight distance for jets in the RecoVertex category.
\item The ``number of SV'', defined as the number of secondary vertices for jets in the RecoVertex category.
\item The ``track $\eta_{\text{rel}}$'', defined as the pseudorapidity of the track relative to the jet axis for the track with the highest 2D impact parameter significance for jets in the RecoVertex and PseudoVertex categories.
\item The ``corrected SV mass'', defined as the corrected mass of the secondary vertex with the smallest uncertainty on its flight distance for jets in the RecoVertex category or the invariant mass obtained from the total summed four-momentum vector of the selected tracks for jets in the PseudoVertex category.
\item The ``number of tracks from SV'', defined as the number of tracks associated with the secondary vertex for jets in the RecoVertex category or the number of selected tracks for jets in the PseudoVertex category.
\item The ``SV energy ratio'', defined as the energy of the secondary vertex with the smallest uncertainty on its flight distance divided by the energy of the total summed four-momentum vector of the selected tracks.
\item The ``$\Delta R({\text{SV, jet}})$'', defined as the $\Delta R$ between the flight direction of the secondary vertex with the smallest uncertainty on its flight distance and the jet axis for jets in the RecoVertex category, or the $\Delta R$ between the total summed four-momentum vector of the selected tracks for jets in the PseudoVertex category.
\item The ``3D IP significance of the first four tracks'', defined as the signed 3D impact parameter significances of the four tracks with the highest 2D impact parameter significance.
\item The ``track  $p_{\text{T,rel}}$'', defined as the track \pt relative to the jet axis, \ie the track momentum perpendicular to the jet axis, for the track with the highest 2D impact parameter significance.
\item The ``$\Delta R({\text{track, jet}})$'', defined as the $\Delta R$ between the track and the jet axis for the track with the highest 2D impact parameter significance.
\item The ``track $p_{\text{T,rel}}$ ratio'', defined as the track \pt relative to the jet axis divided by the magnitude of the track momentum vector for the track with the highest 2D impact parameter significance.
\item The ``track distance'', defined as the distance between the track and the jet axis at their point of closest approach for the track with the highest 2D impact parameter significance.
\item The ``track decay length'', defined as the distance between the primary vertex and the track at the point of closest approach between the track and the jet axis for the track with the highest 2D impact parameter significance.
\item The ``summed tracks $E_{\text{T}}$ ratio'', defined as the transverse energy of the total summed four-momentum vector of the selected tracks divided by the transverse energy of the jet.
\item The ``$\Delta R({\text{summed tracks, jet}})$'', defined as the $\Delta R$ between the total summed four-momentum vector of the tracks and the jet axis.
\item The ``first track 2D IP significance above {\cPqc} threshold'', defined as the 2D impact parameter significance of the first track that raises the combined invariant mass of the tracks above 1.5\GeV. This track is obtained by summing the four-momenta of the tracks adding one track at the time. Every time a track is added, the total four-momentum vector is computed. The 2D impact parameter significance of the first track that is added resulting in a mass of the total four-momentum vector above the aforemention threshold is used as a variable. The threshold of 1.5\GeV is related to the {\cPqc} quark mass.
\item The number of selected tracks.
\item The jet \pt and $\eta$.
\end{itemize}

The discriminating variables in each vertex category are combined into a neural network, specifically a feed-forward multilayer perceptron with one hidden layer~\cite{Sarle94neuralnetworks}. The number of nodes in the hidden layer is different for the three different vertex categories and is set to twice the number of input variables. The discriminator values of the three vertex categories are combined with a likelihood ratio taking into account the fraction of jets of each flavour expected in \ttbar events. The fraction of jets of each flavour is obtained as a function of the jet \pt and $\abs{\eta}$, using 19 exclusive bins in total. Two dedicated trainings are performed, one with {\cPqc} jets, and one with light-flavour jets as background. The final discriminator value is a linear combination of the output of these two trainings with relative weights of $1:3$ for the output of the network trained against {\cPqc} and light-flavour jets, respectively. The value of these relative weights is inspired by \ttbar events where one of the two {\PW} bosons decays into quarks and the other into leptons, and provides the best performance for a wide variety of physics topologies compared to alternative relative weights.

\begin{figure}[hbtp]
  \centering
    \includegraphics[width=0.49\textwidth]{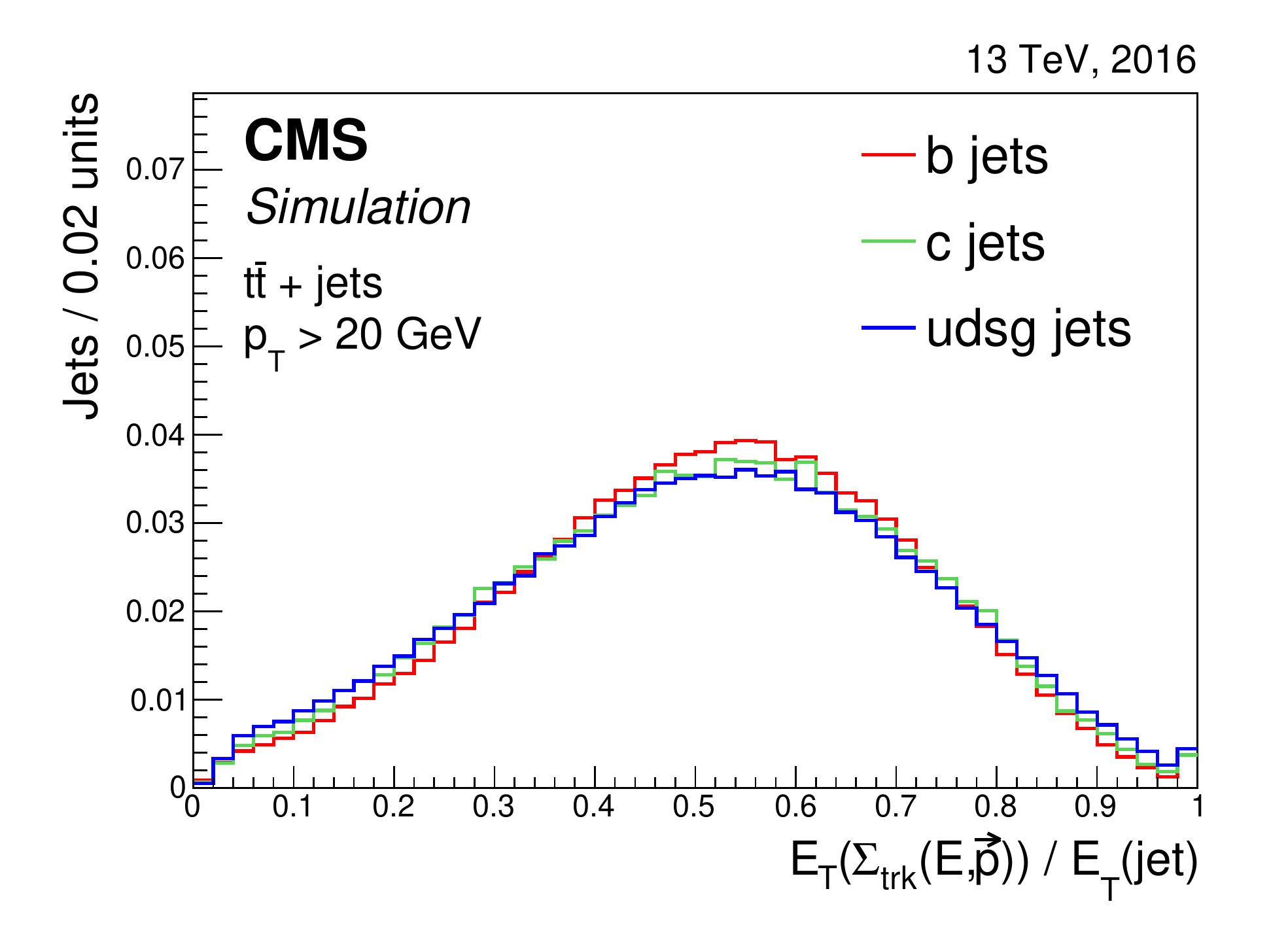}
    \includegraphics[width=0.49\textwidth]{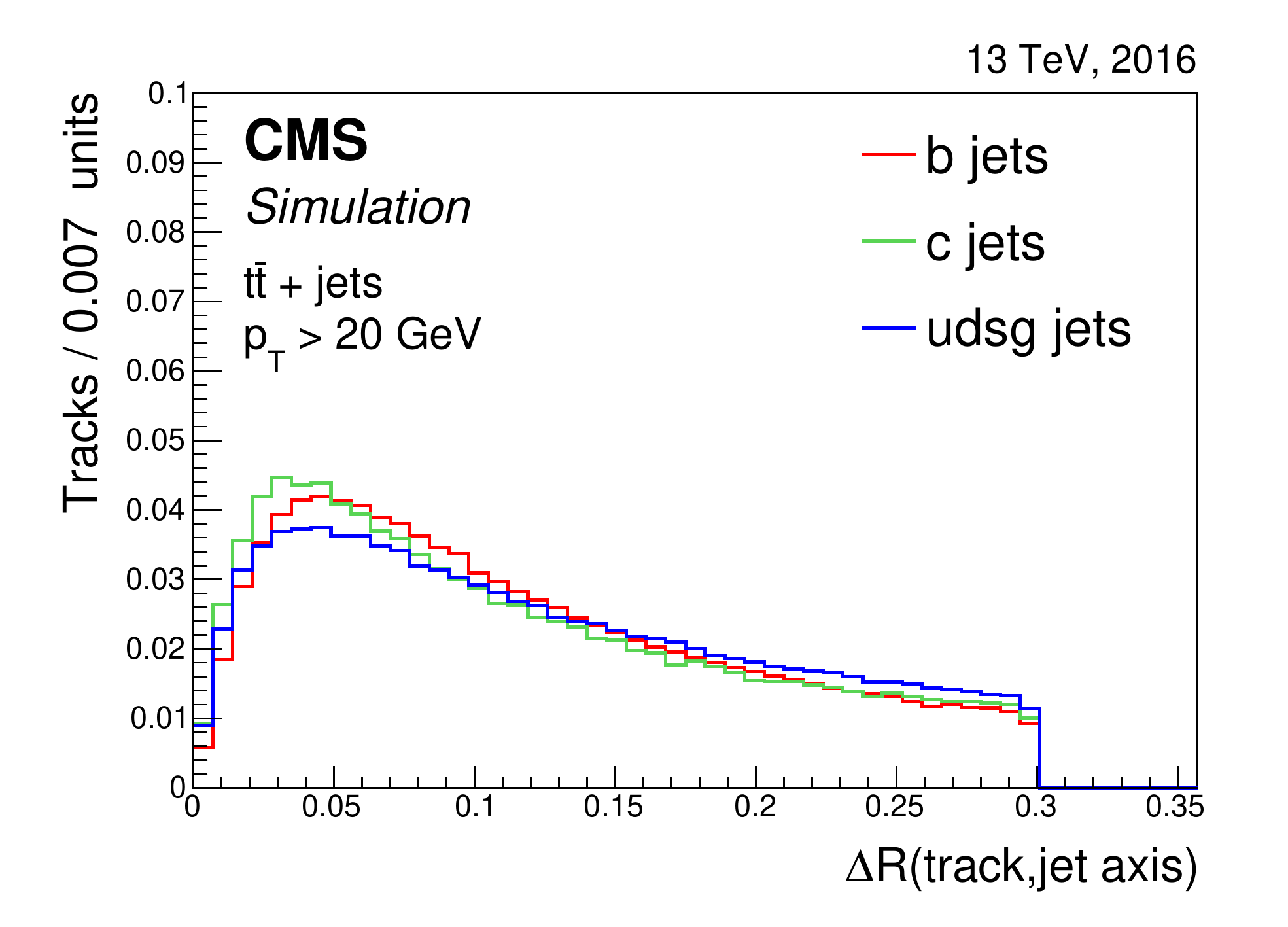}
    \caption{Distribution of the transverse energy of the total summed four-momentum vector of the selected tracks divided by the jet transverse energy (left), and angular distance between the track and the jet axis (right) for jets of different flavours in \ttbar events. The distributions are normalized to unit area. The last bin in the left panel includes the overflow entries.}
    \label{fig:moreVars}
\end{figure}

The main differences from the Run 1 version of the CSV algorithm are the following:
\begin{itemize}
\item \textbf{The secondary vertex reconstruction algorithm}: The secondary vertices are reconstructed with the IVF algorithm.
\item \textbf{Input variables}: Table~\ref{tab:CSVvsCSVv2Vars} lists the variables used for the Run 1 version of the CSV algorithm and for the CSVv2 algorithm. Figure~\ref{fig:moreVars} shows two of the variables used for the CSVv2 algorithm and not for the CSV algorithm.
\item \textbf{Multilayer perceptron}: In the previous version of the algorithm the input variables in a certain vertex category were combined with a likelihood ratio. Depending on the type of correlations present between the input variables, the likelihood ratio performs at a comparable level to the other multivariate methods. The likelihood ratio is particularly useful because of its simplicity and when a small number of variables are used. However, to increase the performance of the algorithm, more input variables were added and combined into an artificial neural network.
\item \textbf{Jet \pt and $\eta$ dependence}: The correlation of some of the input variables with the jet \pt and $\eta$ is taken into account by including the jet kinematics as input variables, after reweighting the distributions to be the same for all jet flavours. In the past, the training was performed in bins of the jet kinematics. In the current procedure, the bins of jet kinematics are only used to combine the vertex categories after the training.
\end{itemize}

\begin{table*}[htbp]
\centering
\topcaption{Input variables used for the Run 1 version of the CSV algorithm and for the CSVv2 algorithm. The symbol ``x'' (``\NA{}'') means that the variable is (not) used in the algorithm
\label{tab:CSVvsCSVv2Vars}}
\begin{tabular}{lcc}
Input variable & Run 1 CSV & CSVv2 \\
\hline
SV 2D flight distance significance & x  & x  \\
Number of SV & \NA  & x  \\
Track $\eta_{\text{rel}}$ & x  & x  \\
Corrected SV mass & x  & x  \\
Number of tracks from SV & x  & x  \\
SV energy ratio & x  & x  \\
$\Delta R({\text{SV, jet}})$ & \NA  & x  \\
3D IP significance of the first four tracks & x  & x  \\
Track $p_{\text{T,rel}}$ & \NA  & x  \\
$\Delta R({\text{track, jet}})$ & \NA  & x  \\
Track $p_{\text{T,rel}}$ ratio  & \NA  & x  \\
Track distance & \NA  & x  \\
Track decay length & \NA  & x  \\
Summed tracks $E_{\text{T}}$ ratio & \NA  & x  \\
$\Delta R({\text{summed tracks, jet}})$ & \NA  & x  \\
First track 2D IP significance above {\cPqc} threshold & \NA  & x  \\
Number of selected tracks & \NA  & x  \\
Jet \pt & \NA  & x  \\
Jet $\eta$ & \NA  & x  \\
\end{tabular}
\end{table*}

\begin{figure}[hbtp]
  \centering
    \includegraphics[width=0.49\textwidth]{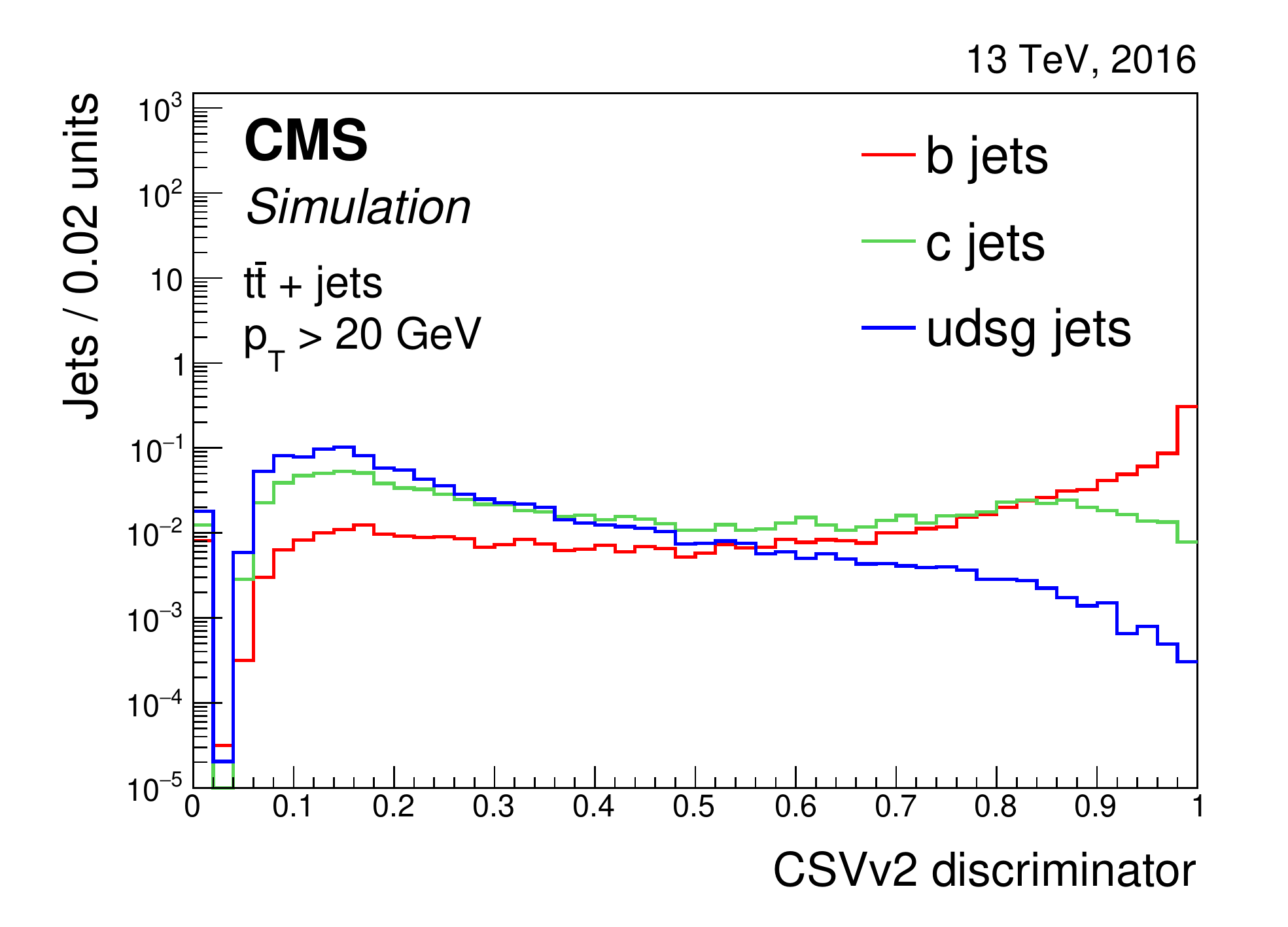}
    \includegraphics[width=0.49\textwidth]{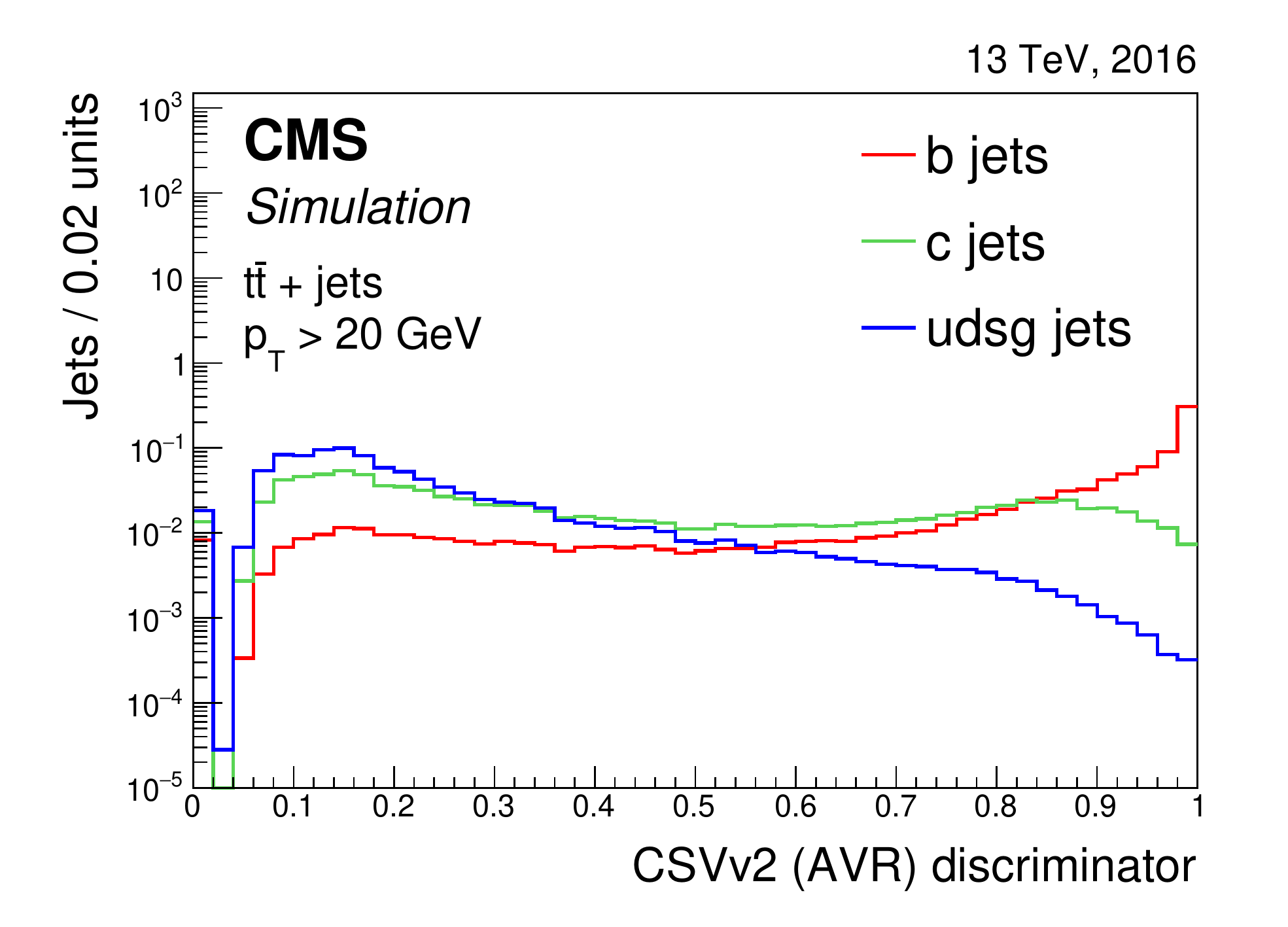}
    \caption{Distribution of the CSVv2 (left) and CSVv2(AVR) (right) discriminator values for jets of different flavours in \ttbar events. The distributions are normalized to unit area. Jets without a selected track and secondary vertex are assigned a negative discriminator value. The first bin includes the underflow entries.}
    \label{fig:CSVv2discr}
\end{figure}

Figure~\ref{fig:CSVv2discr} shows the distribution of the discriminator values for the various jet flavours for both versions of the CSVv2 algorithm.

\paragraph{The DeepCSV tagger}
\label{sec:DeepCSV}

The identification of jets from heavy-flavour hadrons can be improved by using the advances in the field of deep machine learning~\cite{PhysRevD.94.112002}. A new version of the CSVv2 tagger, ``DeepCSV'', was developed using a deep neural network with more hidden layers, more nodes per layer, and a simultaneous training in all vertex categories and for all jet flavours.

The same tracks and IVF secondary vertices are used in this approach as for the CSVv2 tagger. The same input variables are also used, with only one difference, namely that for the track-based variables up to six tracks are used in the training of the DeepCSV. Jets are randomly selected in such a way that similar jet \pt and $\eta$ distributions are obtained for all jet flavours. These jet \pt and $\eta$ distributions are also used as input variables in the training to take into account the correlation between the jet kinematics and the other variables. The distribution of all input variables is preprocessed to centre the mean of each distribution around zero and to obtain a root-mean-square value of unity. All of the variables are presented to the multivariate analysis (MVA) in the same way because of the preprocessing. This speeds up the training. In case a variable cannot be reconstructed, \eg because there are less than six selected tracks (or no secondary vertex), the variable values associated with the missing track or vertex are set to zero after the preprocessing.

The training is performed using jets with \pt between 20\GeV and 1\TeV, and within the tracker acceptance. The relative ratio of the number of jets of each flavour is set to $2:1:4$ for ${\cPqb}:{\cPqc}:{\text{udsg}}$ jets. A mixture of \ttbar and multijet events is used to reduce the possible dependency of the training on the heavy-flavour quark production process.

The training of the deep neural network is performed using the \textsc{Keras}~\cite{chollet2015keras} deep learning library, interfaced with the \textsc{TensorFlow}~\cite{tensorflow2015-whitepaper} library that is used for low-level operations such as convolutions. The neural network uses four hidden layers that are fully connected, each with 100 nodes. Increasing the number of hidden layers and the number of nodes per layer had negligible effects on the performance. Each node in one of the hidden layers uses a rectified linear unit as its activation function to define the output of the node given the input values. For the nodes in the last layer, a normalized exponential function is used for the activation to be able to interpret the output value as a probability for a certain jet flavour category, $P(f)$. The output layer contains five nodes corresponding to five jet flavour categories used in the training. These categories are defined according to whether the jet contains exactly one {\cPqb} hadron, at least two {\cPqb} hadrons, exactly one {\cPqc} hadron and no {\cPqb} hadrons, at least two {\cPqc} hadrons and no {\cPqb} hadrons, or none of the aforementioned categories. Each of these categories is completely independent of the others. The reason for defining five flavour categories in the training is to provide analyses with the possibility to identify jets containing two {\cPqb} or {\cPqc} hadrons.

\begin{figure}[hptb]
  \centering
    \includegraphics[width=0.49\textwidth]{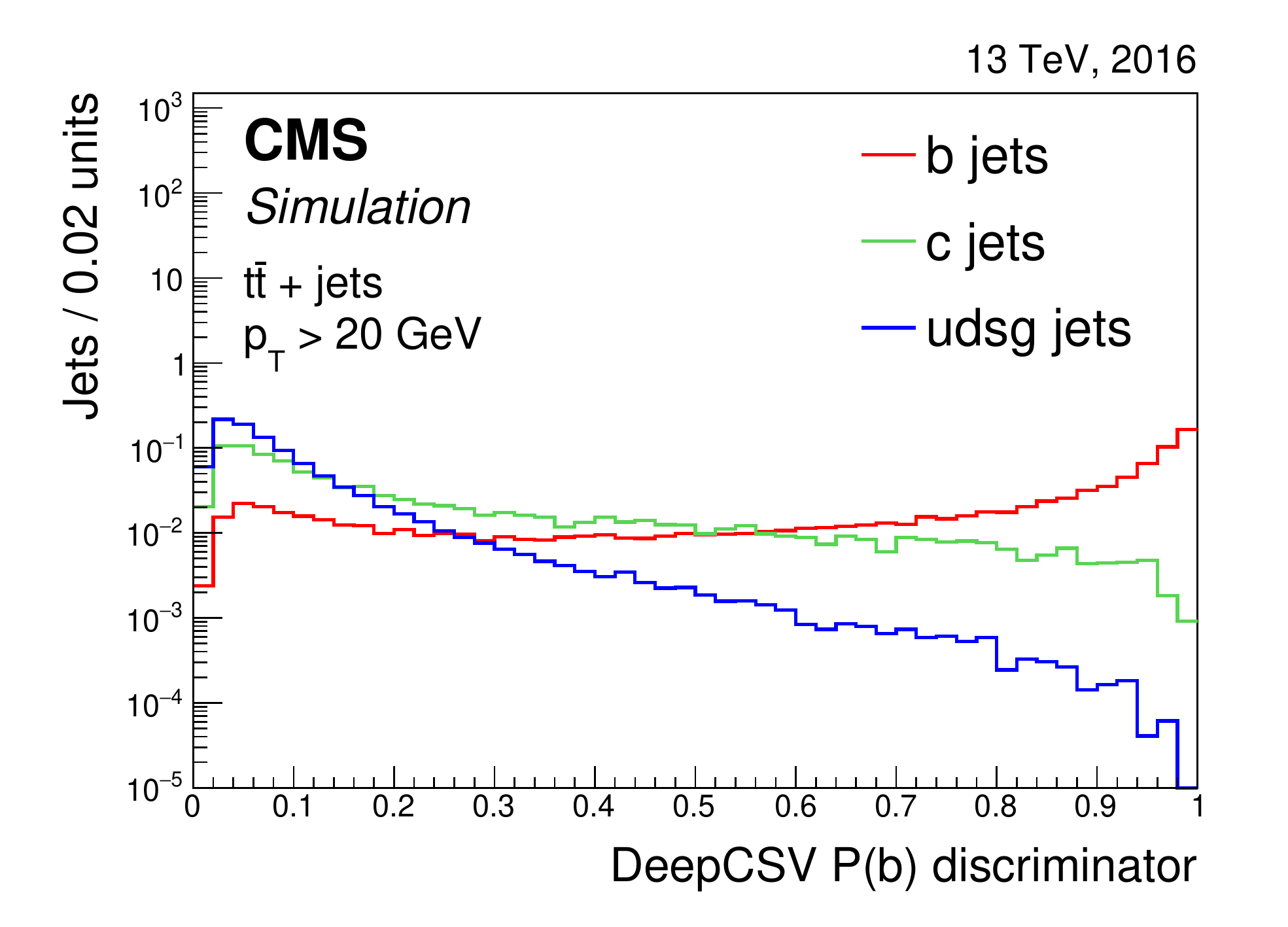}
    \includegraphics[width=0.49\textwidth]{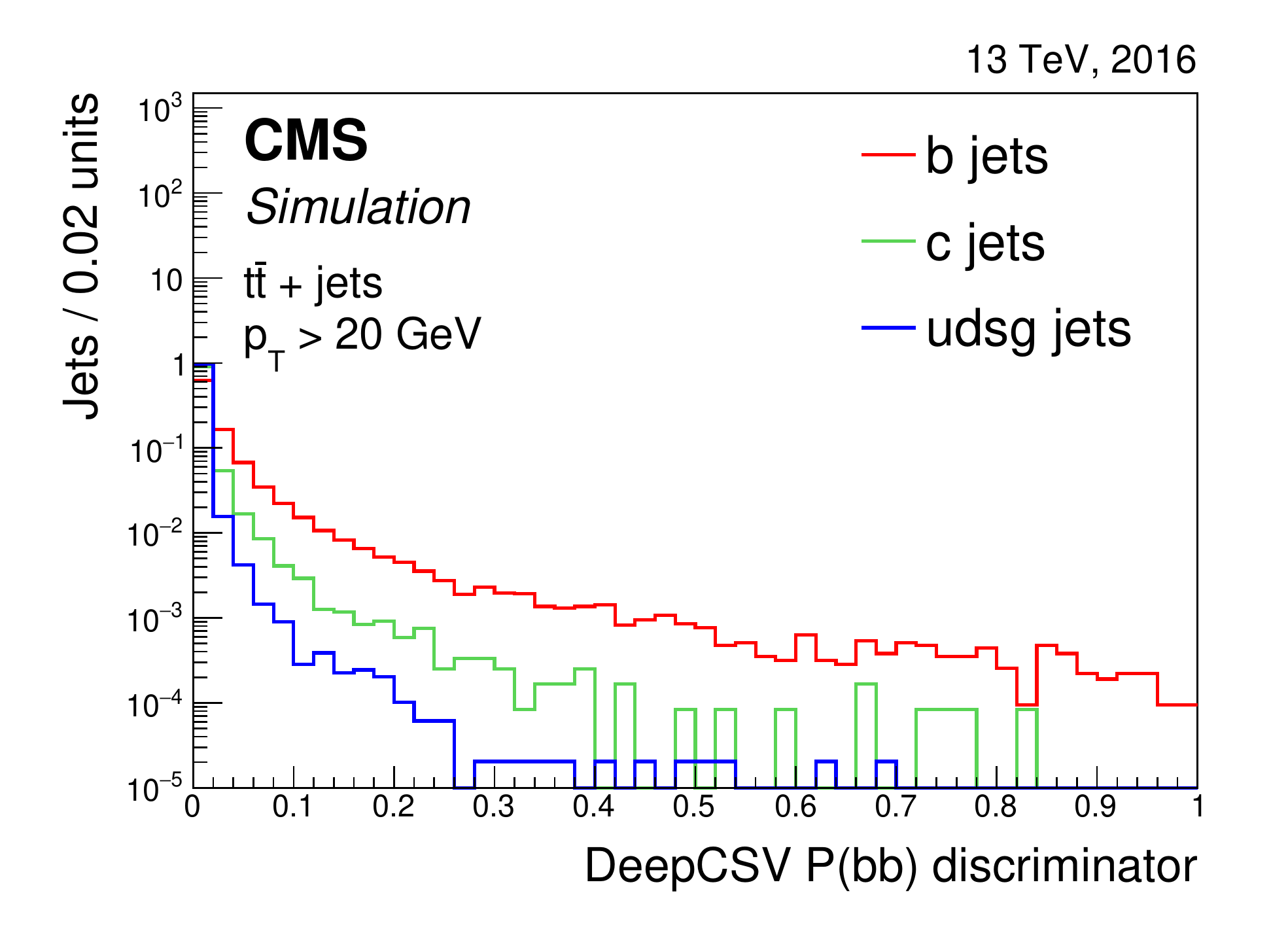}\\
    \includegraphics[width=0.49\textwidth]{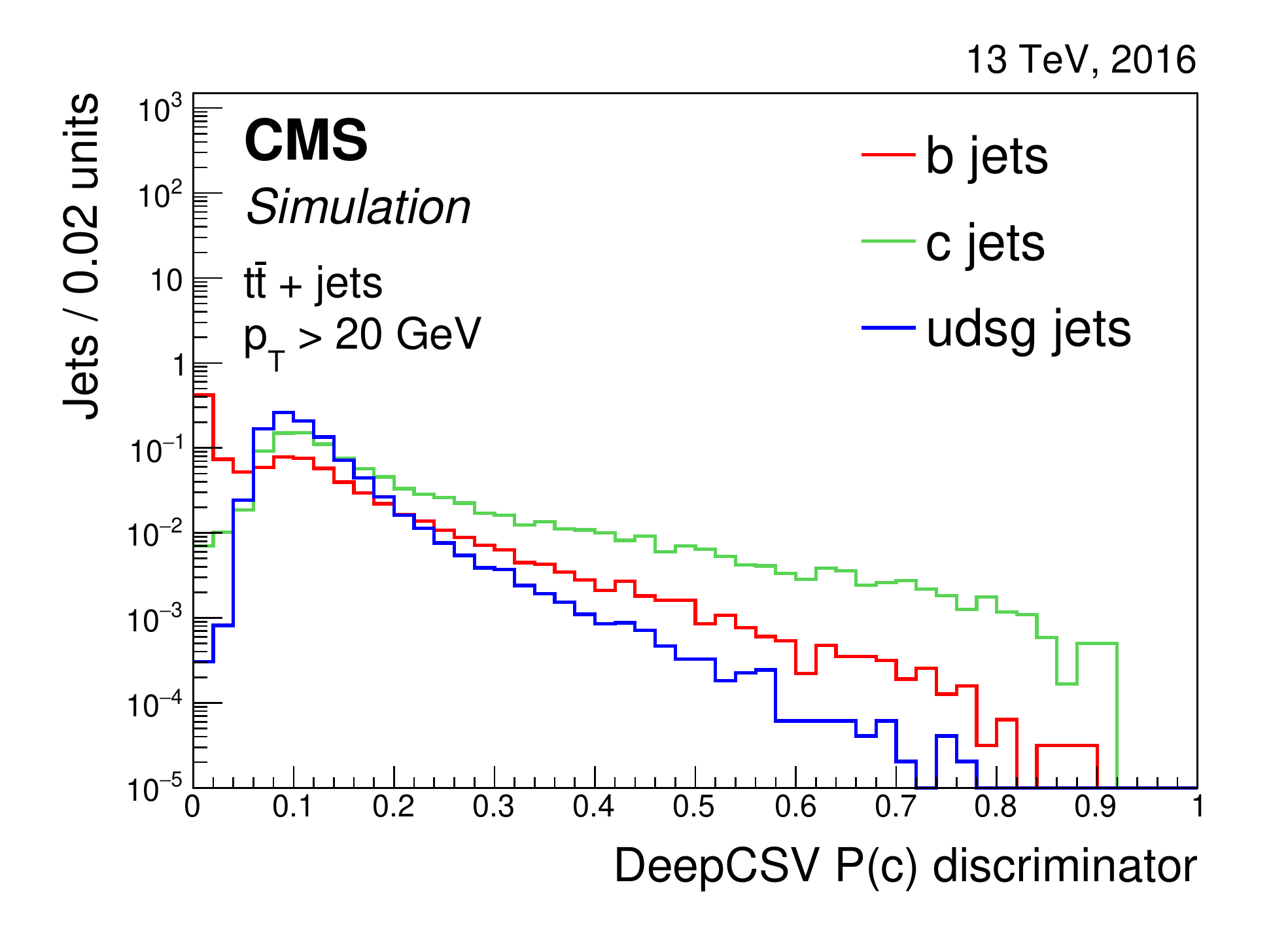}
    \includegraphics[width=0.49\textwidth]{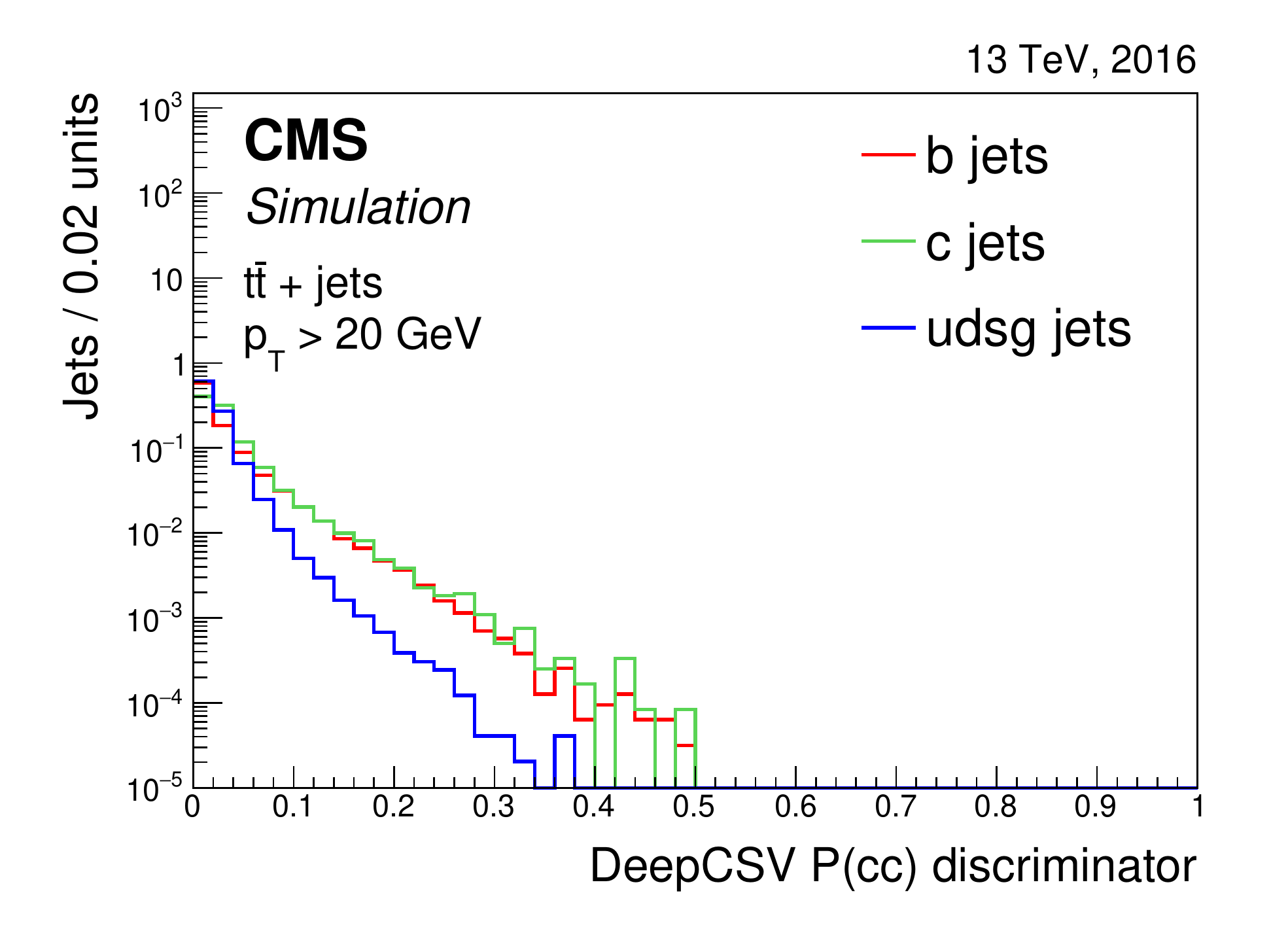}\\
    \includegraphics[width=0.49\textwidth]{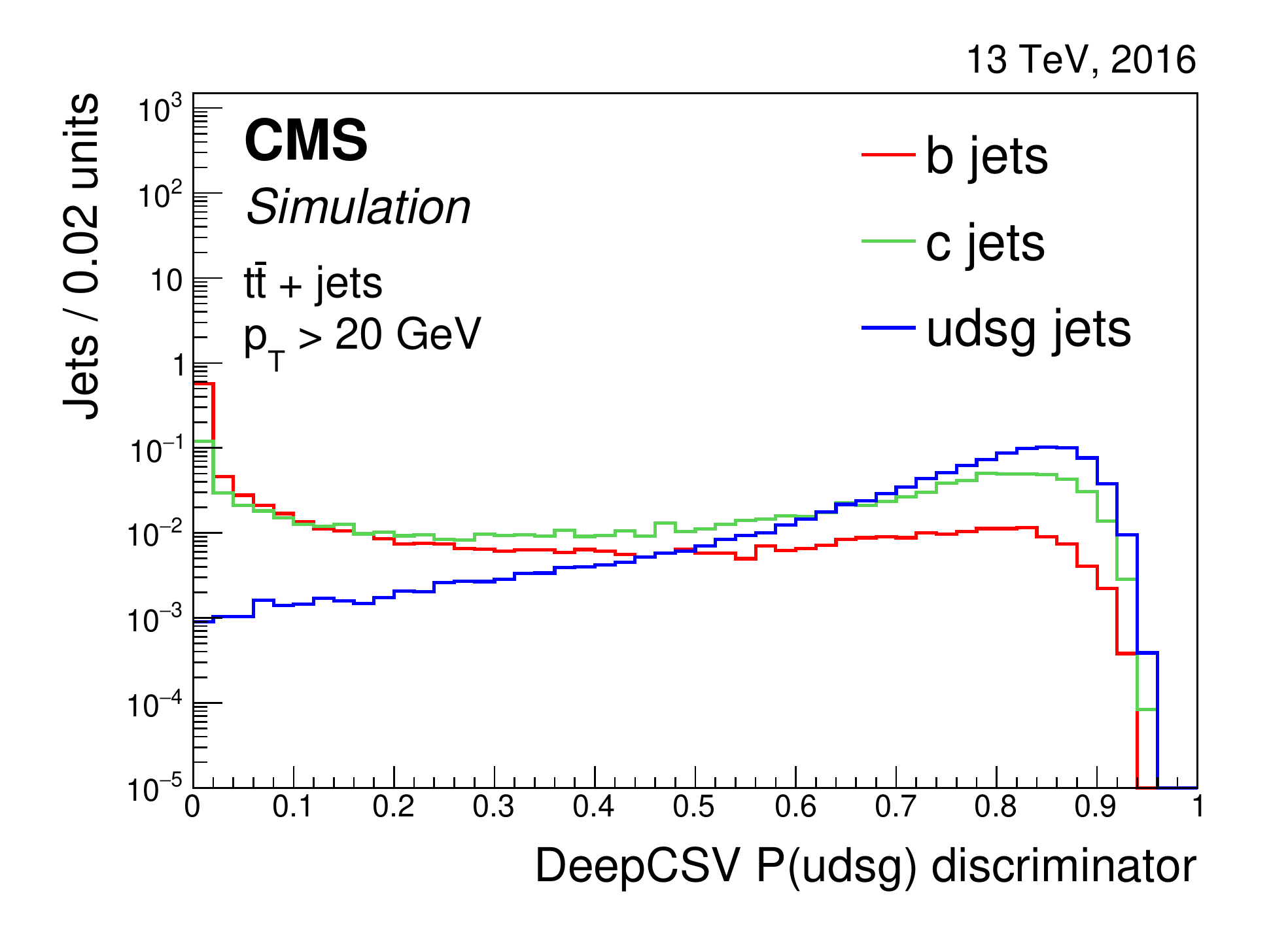}
    \includegraphics[width=0.49\textwidth]{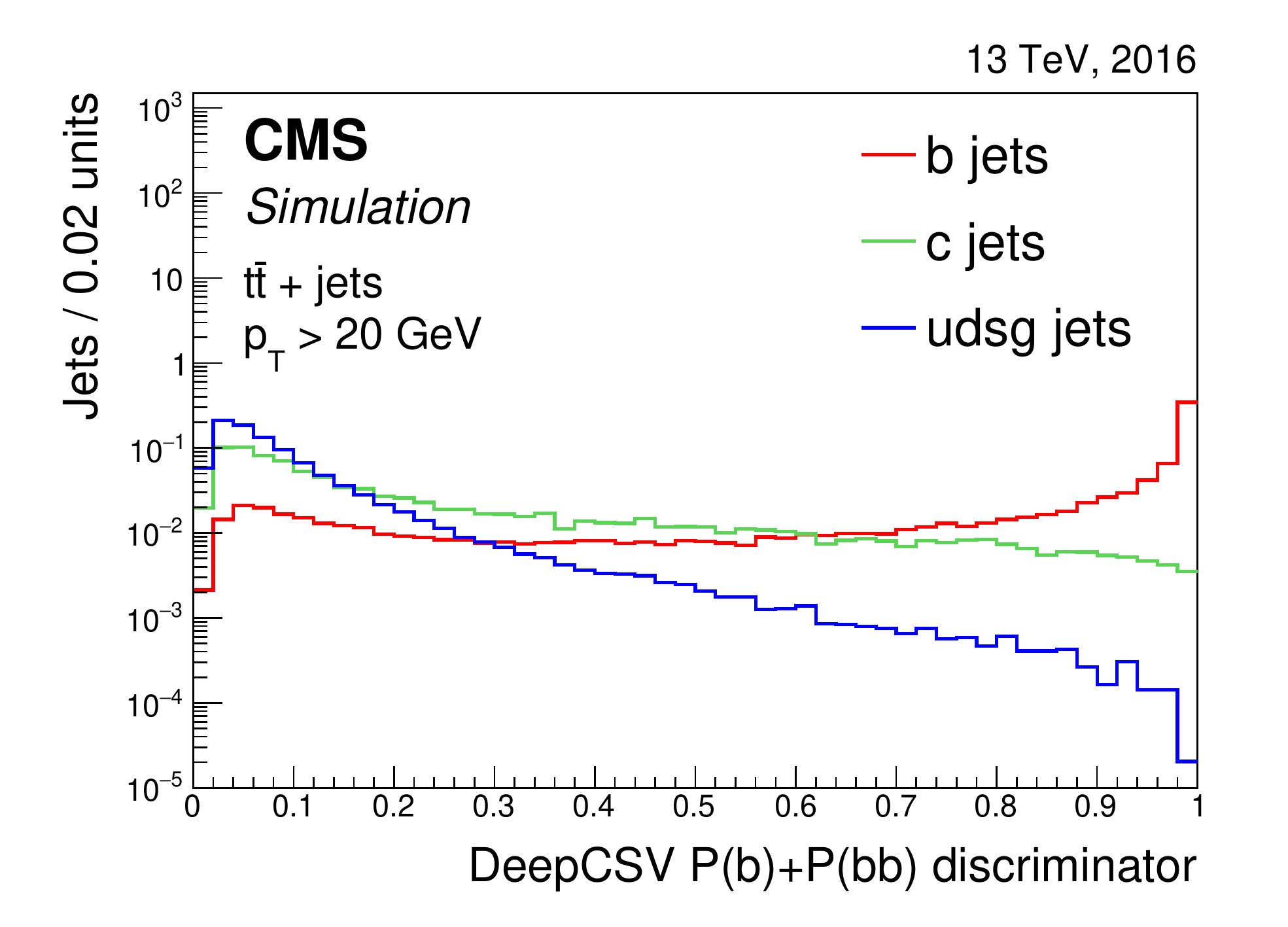}
    \caption{Distribution of the DeepCSV $P(\cPqb)$ (upper left), $P(\cPqb\cPqb)$ (upper right), $P(\cPqc)$ (middle left), $P(\cPqc\cPqc)$ (middle right), $P({\text{udsg}})$ (lower left), and $P(\cPqb)+P(\cPqb\cPqb)$ (lower right) discriminator values for jets of different flavours in \ttbar events. Jets without a selected track and without a secondary vertex are assigned a discriminator value of 0. The distributions are normalized to unit area. }
    \label{fig:DeepCSV}
\end{figure}

Figure~\ref{fig:DeepCSV} shows the discriminator distribution for each of the DeepCSV probabilities $P(f)$.
The lower right panel in Fig.~\ref{fig:DeepCSV} also shows the $P(\cPqb)+P(\cPqb\cPqb)$ discriminator used to tag {\cPqb} jets in physics analyses. It has been checked that summing the probabilities for these two categories is equivalent to using a combined training for these categories.

\subsubsection{Soft-lepton and combined taggers}
\label{sec:SLalgo}
\begin{figure}[hbtp]
  \centering
    \includegraphics[width=0.49\textwidth]{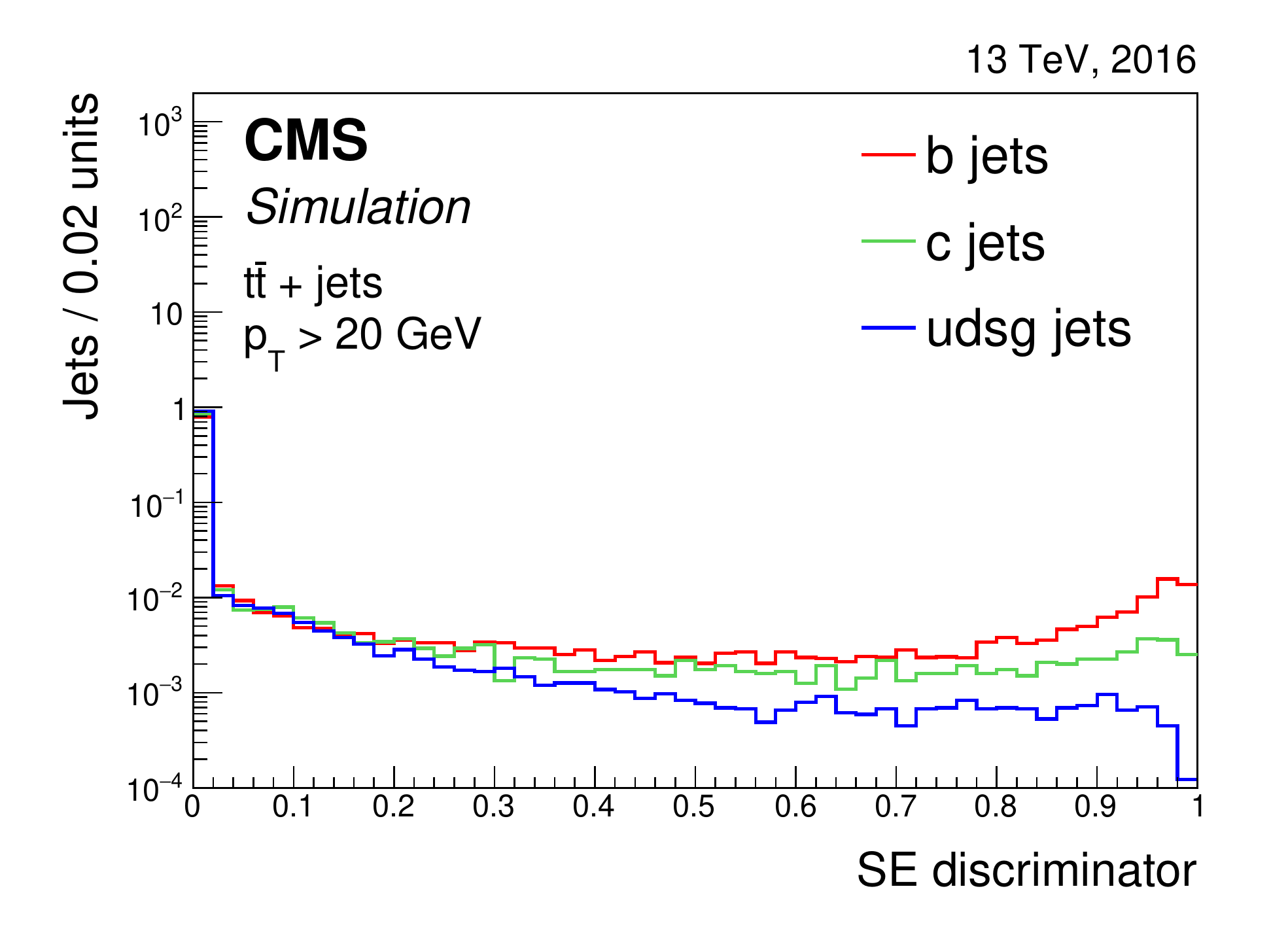}
    \includegraphics[width=0.49\textwidth]{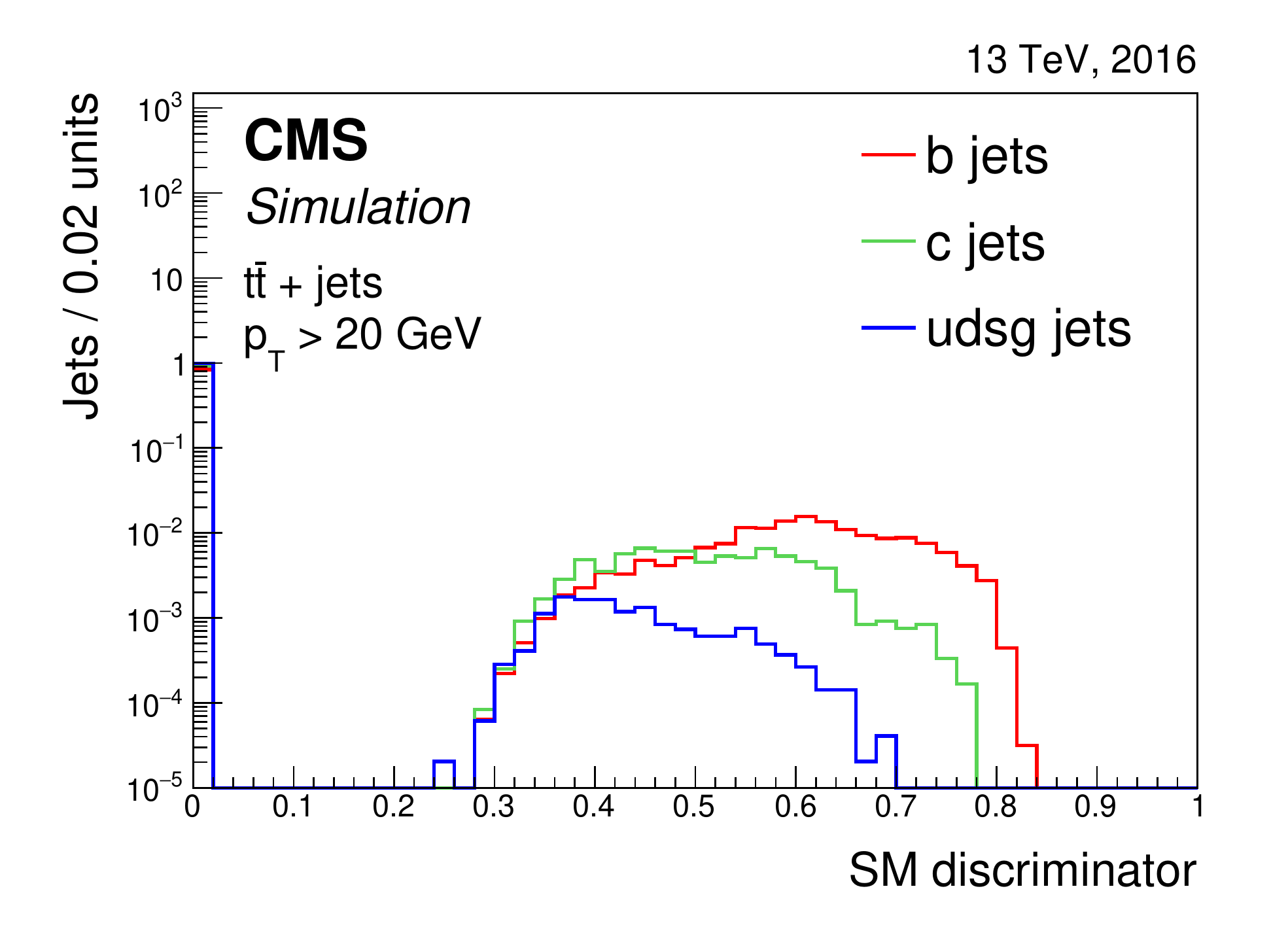}
    \caption{Distribution of the soft-electron (left) and soft-muon (right) discriminator values for jets of different flavours in \ttbar events. Jets without a soft lepton are assigned a discriminator value of 0. The distributions are normalized to unit area.}
    \label{fig:SLdiscr}
\end{figure}

Soft leptons, \ie electrons or muons reconstructed as described in Section~\ref{sec:SLvars} are sometimes present in a jet. When they are, the information related to the charged lepton is used to construct a soft-electron (SE) and soft-muon (SM) tagger.
The discriminating variables that are used as input for the boosted decision tree (BDT) are the 2D and 3D impact parameter significance of the lepton, the angular distance between the jet axis and the lepton, $\Delta R$, the ratio of the \pt of the lepton to that of the jet, and the \pt of the lepton relative to the jet axis, $\pt^{\text{rel}}$. In the case of the SE algorithm an MVA-based electron identification variable is also used as input.
The distributions of the SE and SM discriminator values are shown in Fig.~\ref{fig:SLdiscr}. The different range for the algorithm output values is related to different settings in the training when combining the input variables with a BDT.

\begin{figure}[hbtp]
  \centering
    \includegraphics[width=0.49\textwidth]{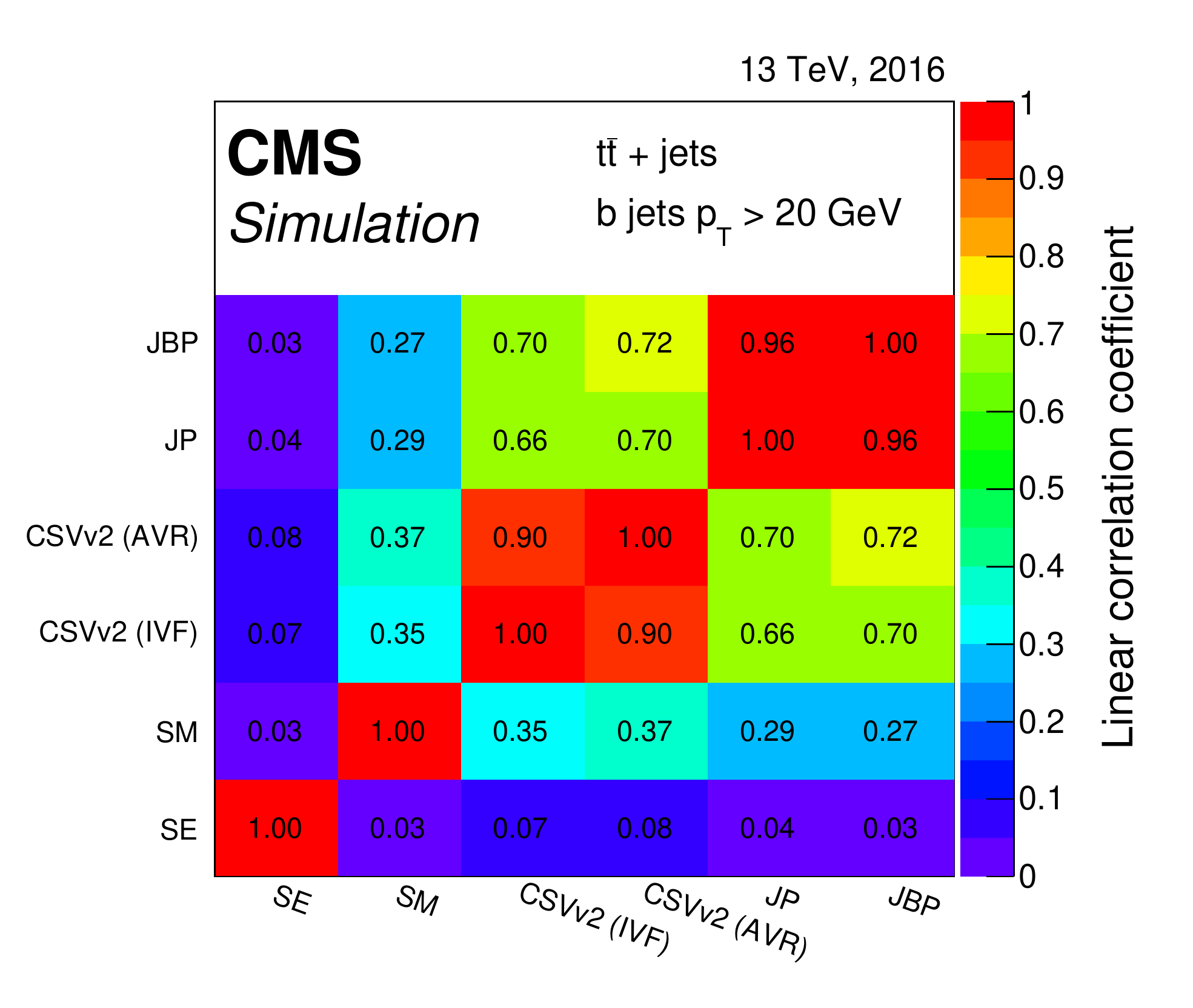}
    \includegraphics[width=0.49\textwidth]{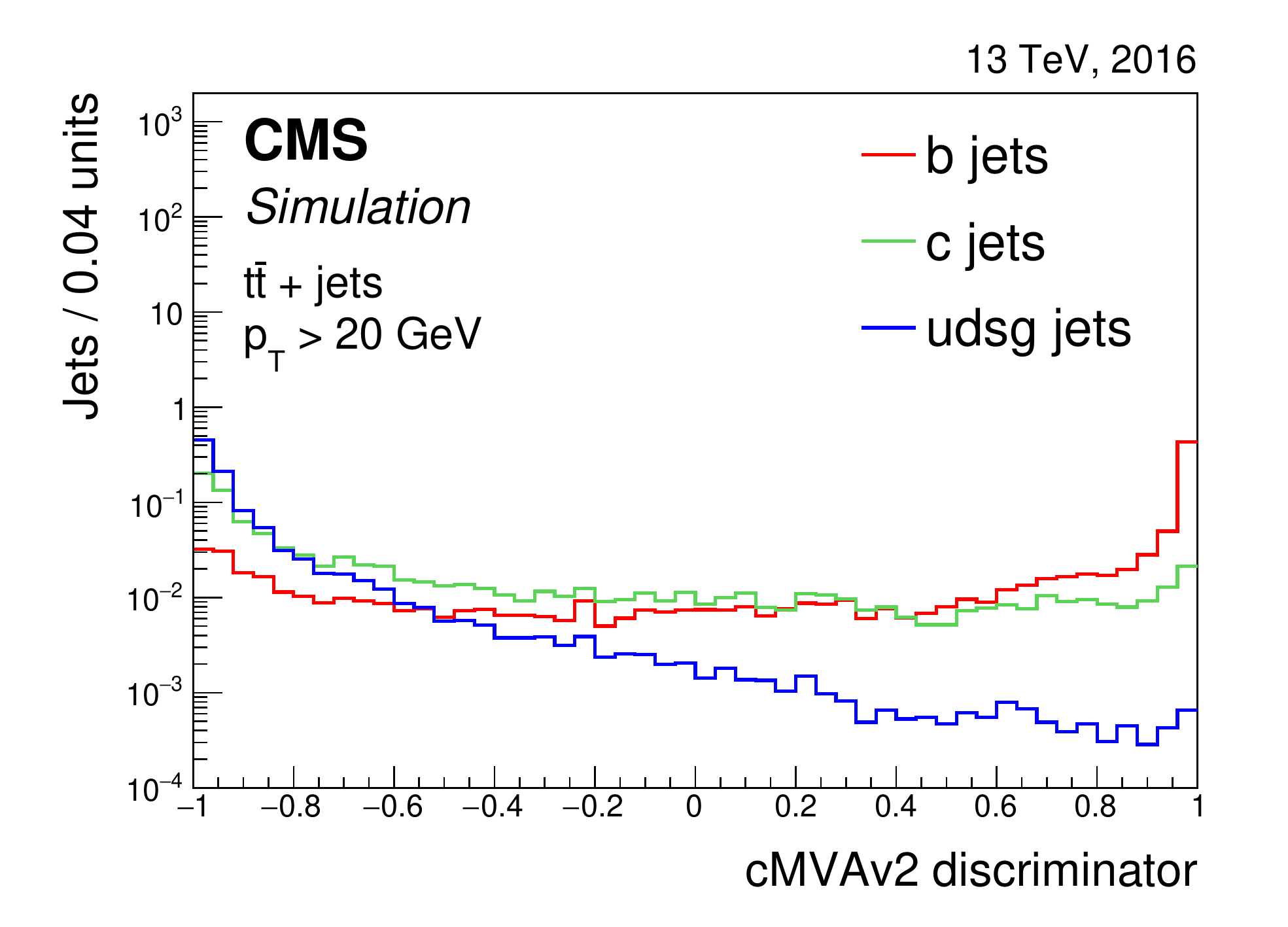}
    \caption{Correlation between the different input variables for the cMVAv2 tagger for {\cPqb} jets in \ttbar events (left), and distribution of the cMVAv2 discriminator values (right), normalized to unit area, for jets of different flavours in \ttbar events.}
    \label{fig:cMVAv2discr}
\end{figure}

As a soft lepton is only present in a relatively small fraction of heavy-flavour jets, the soft lepton taggers are not always able to discriminate heavy-flavour jets from other jets. Therefore they are not used standalone, but rather as input for a combined tagger. The combined tagger, cMVAv2, uses six {\cPqb} jet identification discriminators as input variables, namely the two variants of the JP algorithm, the SE and SM algorithms, and the two variants of the CSVv2 algorithm. The training is performed using the open source \textsc{scikit-learn} package~\cite{scikit-learn} and the variables are combined using a gradient boosting classifier (GBC) as BDT. Prior to the training, the jet \pt and $\eta$ distributions are reweighted to obtain a similar distribution for all jet flavours. Although the correlation between the two CSVv2 discriminator values is close to 100\%, a small improvement is seen in the case where the vertex finding algorithms reconstruct different secondary vertices. Figure~\ref{fig:cMVAv2discr} shows the correlation between the input variables of the cMVAv2 algorithm for {\cPqb} jets as well as the distribution of the cMVAv2 discriminator values for various jet flavours obtained in a \ttbar sample. The correlation between the input variables is similar for other jet flavours. Adding the SL taggers or one of the JP taggers as input variables for the cMVAv2 algorithm results in a similar large performance gain with respect to the CSVv2 algorithm. Adding the other JP tagger and CSVv2 (AVR) algorithm results only in a modest performance gain. The performance of the cMVAv2 tagger for discriminating {\cPqb} jets against other jet flavours is discussed more extensively in Section~\ref{sec:perfak4b}.

It is relevant to note that the DeepCSV discriminator output was not included as an input variable, as this algorithm was developed after the cMVAv2 tagger. Further optimizations are ongoing, in particular in the context of the new pixel tracker installed in 2017~\cite{Dominguez:1481838}.

\subsubsection{Performance in simulation}
\label{sec:perfak4b}
\begin{figure}[hbtp]
  \centering
    \includegraphics[width=0.9\textwidth]{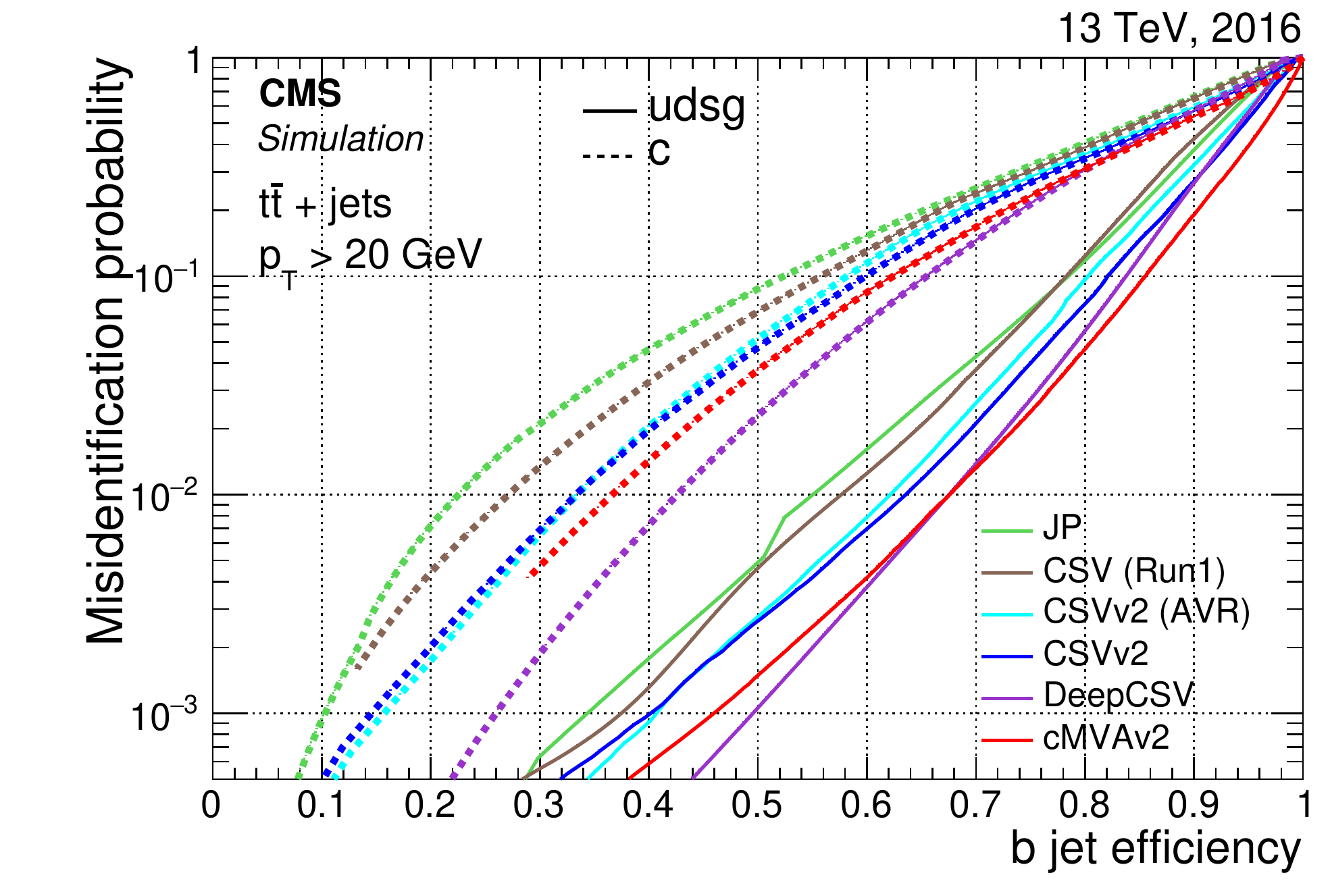}
    \caption{Misidentification probability for {\cPqc} and light-flavour jets versus {\cPqb} jet identification efficiency for various {\cPqb} tagging algorithms applied to jets in \ttbar events.}
    \label{fig:perfMC}
\end{figure}

The tagging efficiency of the JP, CSVv2, cMVAv2, and DeepCSV taggers is determined using simulated {\Pp}{\Pp} collision events. The efficiency (misidentification probability) to correctly (wrongly) tag a jet with flavour $f$ is defined as the number of jets of flavour $f$ passing the tagging requirement divided by the total number of jets of flavour $f$. Figure~\ref{fig:perfMC} shows the {\cPqb} jet identification efficiency versus the misidentification probability for either {\cPqc} or light-flavour jets in simulated \ttbar events requiring jets with $\pt > 20\GeV$ and $\abs{\eta}<2.4$ for various {\cPqb} taggers. In this figure, the tagging efficiency is integrated over the \pt and $\eta$ distributions of the jets in the \ttbar sample. The tagging efficiency is also shown for the Run 1 version of the CSV algorithm. It should be noted that the CSV algorithm was trained on simulated multijet events at centre-of-mass energy of 7\TeV using anti-\kt jets clustered with a distance parameter $R=0.5$. Therefore, the comparison is not completely fair. The performance improvement expected from a retraining is typically of the order of 1\%. The absolute improvement in the {\cPqb} jet identification efficiency for the CSVv2 (AVR) algorithm with respect to the CSV algorithm is of the order of 2--4\% when the comparison is made at the same misidentification probability value for light-flavour jets. An additional improvement of the order of 1--2\% is seen when using IVF vertices instead of AVR vertices in the CSVv2 algorithm. The cMVAv2 tagger performs around 3--4\% better than the CSVv2 algorithm for the same misidentification probability for light-flavour jets.
The DeepCSV $P(\cPqb)+P(\cPqb\cPqb)$ tagger outperforms all the other {\cPqb} jet identification algorithms, when discriminating against {\cPqc} jets or light-flavour jets, except for {\cPqb} jet identification efficiencies above 70\% where the cMVAv2 tagger performs better when discriminating against light-flavour jets. The absolute {\cPqb} identification efficiency improves by about 4\% with respect to the CSVv2 algorithm for a misidentification probability for light-flavour jets of 1\%.
Three standard working points are defined for each {\cPqb} tagging algorithm using jets with $\pt>30\GeV$ in simulated multijet events with $80 < \hat{p}_{\text{T}} <120\GeV$. The average jet \pt in this sample of events is about 75\GeV. These working points, ``loose'' (L), ``medium'' (M), and ``tight'' (T), correspond to thresholds on the discriminator after which the misidentification probability is around 10\%, 1\%, and 0.1\%, respectively, for light-flavour jets. The efficiency for correctly identifying {\cPqb} jets in simulated \ttbar events for each of the three working points of the various taggers is summarized in Table~\ref{tab:OP}.
\begin{table*}[htbp]
\centering
\topcaption{Taggers, working points, and corresponding efficiency for {\cPqb} jets with $\pt>20$\GeV in simulated \ttbar events. The numbers in this table are for illustrative purposes since the {\cPqb} jet identification efficiency is integrated over the \pt and $\eta$ distributions of jets.
\label{tab:OP}}
\begin{tabular}{lcccc}
Tagger & Working point & $\varepsilon_{{\cPqb}}$ (\%)& $\varepsilon_{{\cPqc}}$ (\%)& $\varepsilon_{\text{udsg}}$ (\%) \\
\hline
 & JP L &  78 & 37  &  9.6 \\
Jet probability (JP) & JP M &  56 & 12  &  1.1 \\
 & JP T & 36 & 3.3  &  0.1 \\
\hline
 & CSVv2 L & 81 & 37  &  8.9 \\
Combined secondary vertex (CSVv2) & CSVv2 M & 63 & 12  & 0.9  \\
 & CSVv2 T &41 &  2.2 &  0.1 \\
\hline
 & cMVAv2 L & 84 & 39  &  8.3 \\
Combined MVA (cMVAv2) & cMVAv2 M &  66 & 13  & 0.8 \\
 & cMVAv2 T  & 46 & 2.6  & 0.1 \\
 \hline
  & DeepCSV L &  84 & 41  &  11  \\
Deep combined secondary vertex  & DeepCSV M &  68 &  12 & 1.1  \\
(DeepCSV) $P(\cPqb)+P(\cPqb\cPqb)$ & DeepCSV T & 50 & 2.4  & 0.1  \\
\end{tabular}
\end{table*}

\begin{figure}[hbtp]
  \centering
    \includegraphics[width=\textwidth]{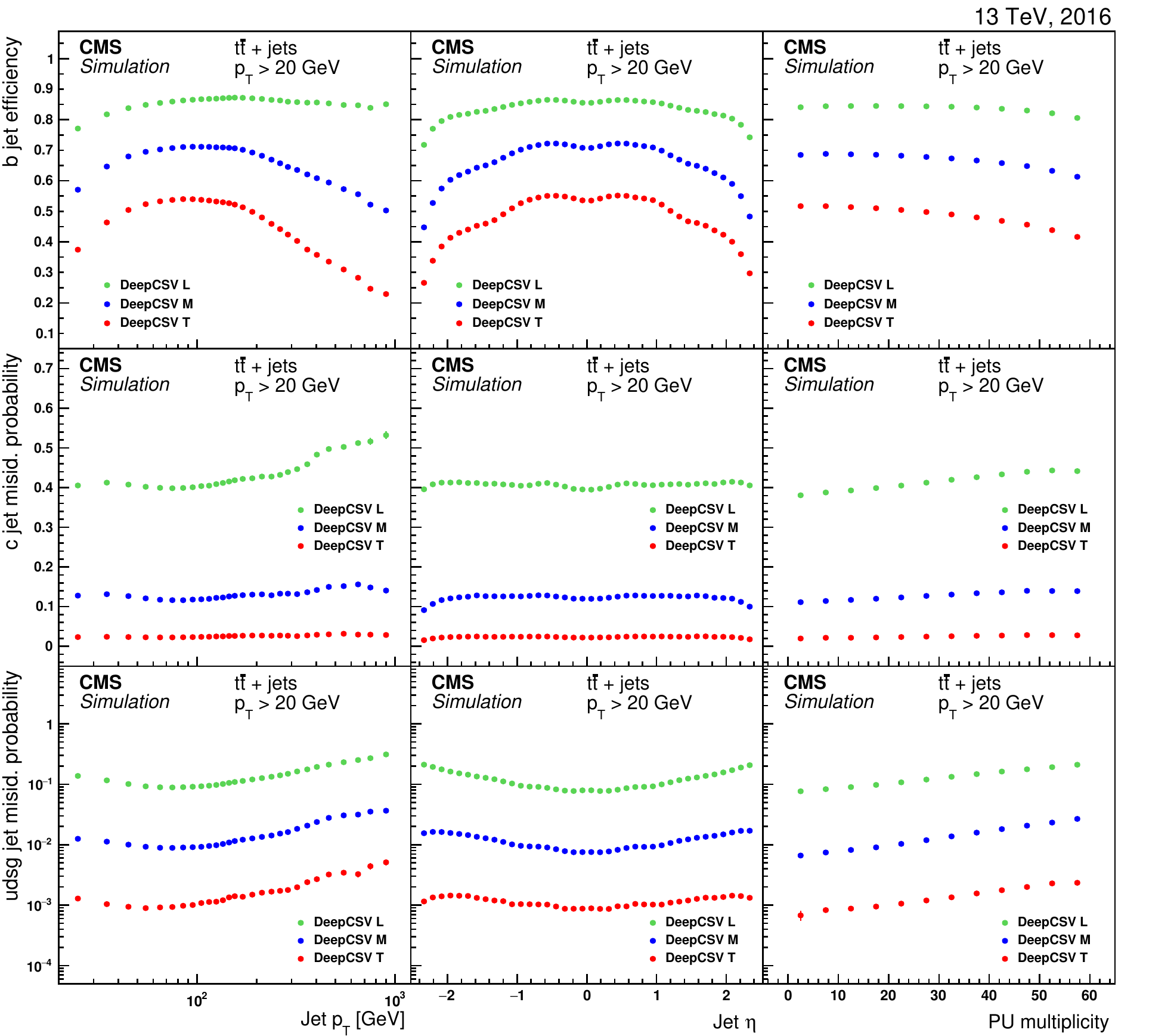}
    \caption{Efficiencies and misidentification probabilities for the DeepCSV $P(\cPqb)+P(\cPqb\cPqb)$ tagger as a function of the jet \pt (left), jet $\eta$ (middle), and PU multiplicity, \ie the number of inelastic {\Pp\Pp} collisions in the event (right), for {\cPqb} (upper),  {\cPqc} (middle), and light-flavour (lower) jets in \ttbar events. Each panel shows the efficiency for the three different working points with different colours. }
    \label{fig:DeepCSVeff}
\end{figure}

The tagging efficiency depends on the jet \pt, $\eta$, and the number of pileup interactions in the event. This dependency is illustrated for the DeepCSV $P(\cPqb)+P(\cPqb\cPqb)$ tagger in Fig.~\ref{fig:DeepCSVeff} using jets with $\pt > 20\GeV$ in \ttbar events. A parameterization of the efficiency as a function of the jet \pt is provided in Appendix~\ref{sec:app}. The efficiency for correctly identifying {\cPqb} jets is maximal for jets with $\pt \approx 100\GeV$ and decreases at low- and high-\pt values. The lower efficiency at low jet \pt is due to the larger uncertainty on the track impact parameter resolution. At high jet \pt, there are two main effects. First, the misidentification probability for light-flavour jets increases because of the larger number of tracks present in the jet, as can be seen from Fig.~\ref{fig:trackmult}. Second, at higher jet transverse momenta, jets are more collimated and their charged particles are closer together, resulting in merged hits in the innermost layers of the tracking system. This effect impacts the track reconstruction efficiency and hence also the {\cPqb} jet identification efficiency.
Due to the higher track reconstruction efficiency and the better resolution of the track parameters at small $\abs{\eta}$ values~\cite{TRK-11-001}, the algorithms are more efficient in identifying {\cPqb} jets in the barrel region of the CMS silicon tracker ($\abs{\eta}<1$). The efficiency for misidentifying light-flavour jets increases with an increasing number of pileup interactions. This is explained as follows. First, the increasing number of pileup interactions results in a higher probability to choose the wrong primary vertex resulting in light-flavour jets that are displaced, and {\cPqb} jets for which the displacement is wrong. Second, the increasing number of pileup interactions results in a higher occupancy in the tracker, leading to a larger number of wrongly reconstructed tracks as well as more tracks from a different interaction vertex that are clustered in the jets associated with the primary vertex. It was checked that all taggers presented in Table~\ref{tab:OP} show a similar dependence with respect to the number of pileup interactions, and jet \pt and $\abs{\eta}$.

\subsection{The \texorpdfstring{{\cPqc}}{c} jet identification}
As can be seen from Figs.~\ref{fig:IP},~\ref{fig:SVinfo}, and~\ref{fig:leptonIPsig} in Section~\ref{sec:vars}, the distributions of the tagging variables for {\cPqc} jets lie in between the distributions for {\cPqb} and light-flavour jets. This is due to the lifetime of the {\cPqc} hadrons being shorter than that of the {\cPqb} hadrons. In addition, the secondary vertex multiplicity is also lower and the smaller {\cPqc} quark mass results in a smaller track \pt relative to the jet axis. Therefore, it is particularly challenging to efficiently identify jets originating from {\cPqc} quarks.

\subsubsection{Algorithm description}
\label{sec:ctagger}
The {\cPqc} jet identification algorithm uses properties related to displaced tracks, secondary vertices, and soft leptons inside the jets. The secondary vertices are obtained using the IVF algorithm with modified parameters for {\cPqc} jets as described in Section~\ref{sec:vertexing}. Based on the presence or absence of a secondary vertex associated with a jet, three secondary vertex categories are defined in the same way as for the CSVv2 algorithm. The presence or absence of a soft lepton, as discussed in the previous paragraph, leads to the definition of three soft lepton categories, independent of the secondary vertex categories:
\begin{itemize}
\item \textbf{NoSoftLepton}: Including jets without soft leptons found inside the jet;
\item \textbf{SoftMuon}: At least one soft muon was found inside the jet;
\item \textbf{SoftElectron}: No soft muon, but at least one soft electron was found inside the jet.
\end{itemize}
With this categorization, jets containing a muon and an electron will be assigned to the SoftMuon category.
Like for the {\cPqb} tagging algorithms, the displaced tracks are ordered by decreasing 2D impact parameter significance, and the secondary vertices are ordered by increasing uncertainty on their 3D flight distance. Some variables are only defined if a secondary vertex was reconstructed or if a soft lepton was found inside the jet. Whenever a variable is not available, a default value is assigned to it. The variables used are similar to the ones used in the CSVv2 algorithm (Section~\ref{sec:CSVv2}) and in the SM or SE algorithms (Section~\ref{sec:SLalgo}). For track- and lepton-based variables, up to two tracks or leptons are used (if available), while for the secondary vertex variables only the first secondary vertex is used (if available). The list of variables used is the following:
\begin{itemize}
\item The vertex-lepton category.
\item The 2D and 3D impact parameter significance of the first two tracks, and the 3D impact parameter significance of the first two leptons.
\item The pseudorapidity of the track (lepton) relative to the jet axis for the first two tracks (leptons).
\item The track (lepton) \pt relative to the jet axis, \ie the track momentum perpendicular to the jet axis, for the first two tracks (leptons).
\item The track \pt relative to the jet axis divided by the magnitude of the track momentum vector, for the first two tracks.
\item The track momentum parallel to the jet direction, for the first two tracks.
\item The track momentum parallel to the jet direction divided by the magnitude of the track momentum vector, for the first two tracks.
\item The $\Delta R$ between the track (lepton) and the jet axis for the first two tracks (leptons).
\item The distance between the track and the jet axis at their point of closest approach, for the first two tracks.
\item The track decay length, \ie the distance between the primary vertex and the track at the point of closest approach between the track and the jet axis, for the first two tracks.
\item The transverse energy of the total summed four-momentum vector of the selected tracks divided by the transverse energy of the jet.
\item The $\Delta R$ between the total summed four-momentum vector of the tracks and the jet axis.
\item The 2D and 3D impact parameter significance of the first track that raises the combined invariant mass of the tracks above 1.5\GeV. This track is obtained by summing the four-momenta of the tracks adding one track at the time. Every time a track is added, the total four-momentum vector is computed. The 2D impact parameter significance of the first track that is added resulting in a mass of the total four-momentum vector above the aforemention threshold is used as a variable. The threshold of 1.5\GeV is related to the {\cPqc} quark mass.
\item The lepton \pt divided by the jet \pt, for the first two leptons.
\item The lepton momentum parallel to the jet direction divided by the magnitude of the jet momentum, for the first two leptons.
\item The 2D and 3D flight distance significance of the first secondary vertex.
\item The secondary vertex energy ratio, defined as the energy of the secondary vertex with the smallest uncertainty on its flight distance divided by the energy of the total summed four-momentum vector of the selected tracks.
\item The corrected secondary vertex mass.
\item The ``massVertexEnergyFraction'' variable, which is defined as $X/(X+0.04)$, where $X$ is the corrected secondary vertex mass divided by the average {\cPqb} meson mass~\cite{Patrignani:2016xqp} multiplied by the scalar sum of the track energies (assuming the pion mass) for tracks associated with the secondary vertex divided by the scalar sum of the track energies for track associated with the jet:
\begin{equation}
X = \frac{m_{\,\mathrm{SV}}\mathrm{[\GeVns{}]}}{5.2794} \, \frac{\sum\limits_{\mathrm{SV}\ \mathrm{tracks}}E_i}{\sum\limits_{\mathrm{jet}\ \mathrm{tracks}}E_i}.
\end{equation}
This variable is first defined in Section~7 of Ref.~\cite{Lehmacher:1555437}.
\item The ``vertexBoost'' variable, defined as $Y^2/(Y^2+10)$, where $Y$ is the square root of the average {\cPqb} meson mass~\cite{Patrignani:2016xqp} multiplied with the scalar sum of the track \pt for tracks associated with the vertex, divided by the product of the corrected secondary vertex mass and the square root of the jet \pt. This variable is related to the boost of the secondary vertex. This variable is first defined in Section~7 of Ref.~\cite{Lehmacher:1555437}.
\item The number of tracks associated with the first secondary vertex.
\item The number of secondary vertices.
\item The number of tracks associated with the jet.
\end{itemize}
								
The training of the algorithm was performed on simulated multijet events. As in the case of the DeepCSV tagger, the variables are first preprocessed to centre their mean at zero and obtain a root-mean-square of unity. Two weights are applied for each jet in the training. To avoid introducing any unwanted dependence on the jet kinematics in the tagger, a first weight is applied to flatten the jet \pt and $\eta$ distributions in the whole training sample for all jet flavours. Simultaneously, a second weight skews the relative contribution of the different secondary vertex categories in the multijet sample to fit the observed ones in the \ttbar sample. Two trainings are performed: one for discriminating {\cPqc} jets from light-flavour jets (CvsL) and another one for discriminating {\cPqc} jets from {\cPqb} jets (CvsB). The training of the two discriminators was performed with the \textsc{scikit-learn} package~\cite{scikit-learn} using a GBC as implementation of the BDT.

The GBC settings were optimized by varying them over a wide range of values, to ensure the optimal setting was contained within the scanned range. Both the CvsL and CvsB trainings were optimized by scanning a range of the parameters and comparing the final performance curves. The best performance was achieved with the number of boosting stages set to 500, the learning rate to 0.05, the minimum number of samples required to split an internal node to 0.6\% and a maximum depth of the individual regression estimators of 15\,(8) for the CvsL (CvsB) training. Some of the optimized values did not change the performance visibly when being varied, but they were chosen to reduce the computation time without a loss in performance.

\begin{figure}[hbtp]
  \centering
    \includegraphics[width=0.49\textwidth]{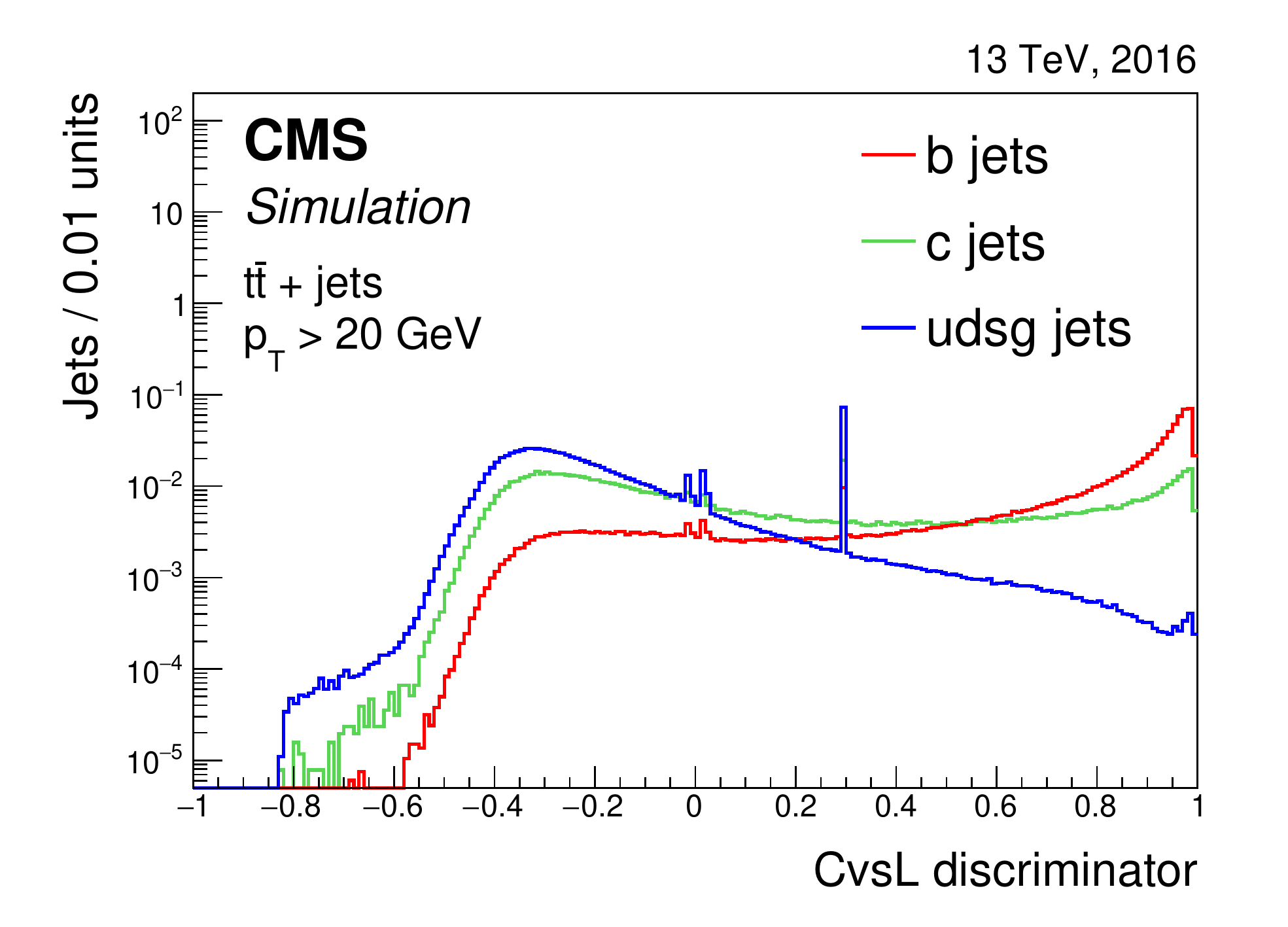}
    \includegraphics[width=0.49\textwidth]{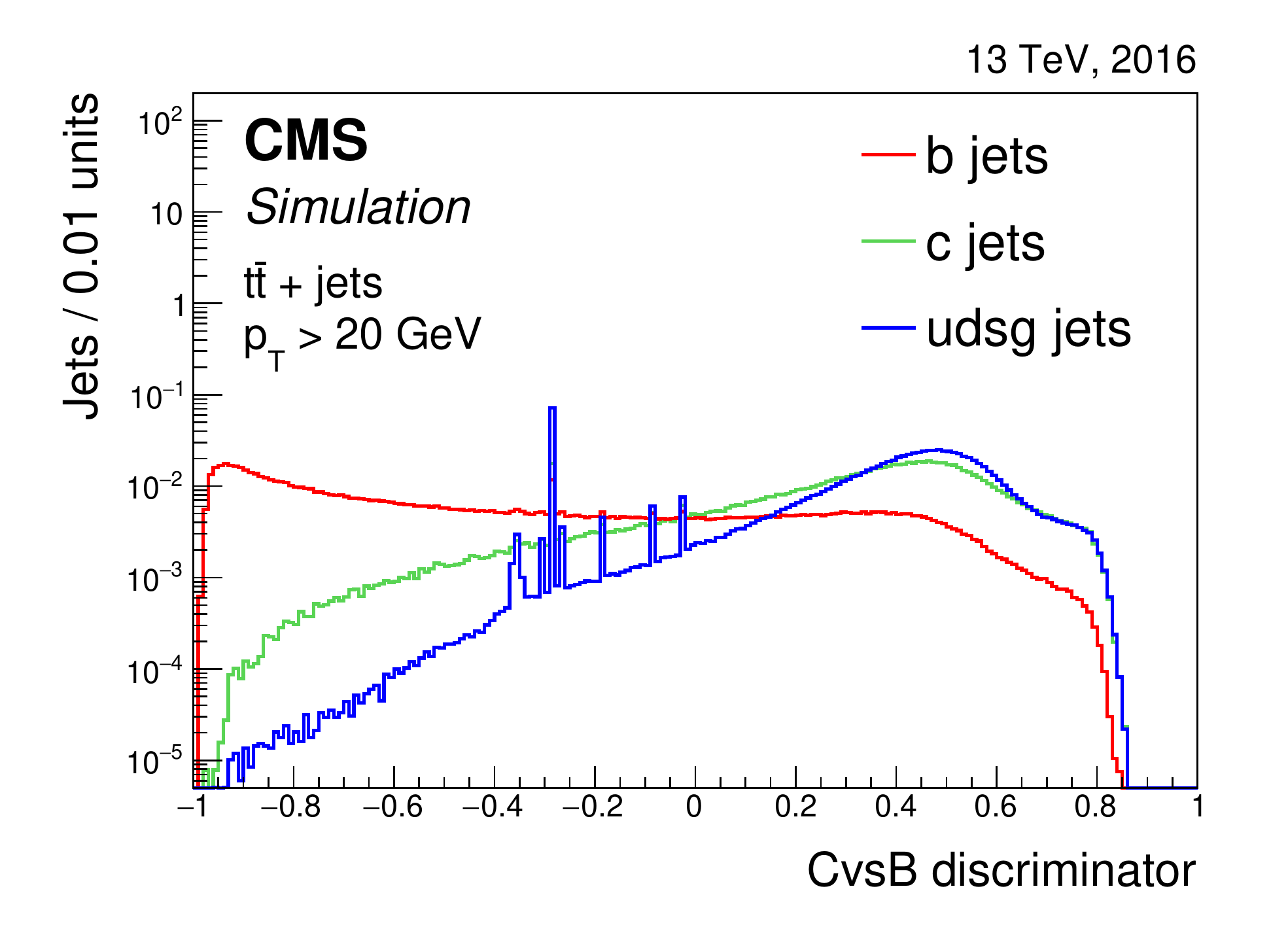}
    \caption{Distribution of the CvsL (left) and CvsB (right) discriminator values for jets of different flavours in \ttbar events. The spikes originate from jets without a track passing the track selection criteria, as discussed in the text. The distributions are normalized to unit area. }
    \label{fig:ctaggerdistribution}
\end{figure}

Figure~\ref{fig:ctaggerdistribution} shows the output discriminator distributions for the CvsL and CvsB taggers.
The discriminator distributions exhibit spikes, which originate from the default values for most input variables if a jet has no track passing the selection criteria. These spikes do not affect any physics analyses, as the discriminator thresholds defining the working points are not just before or after a spike.

\subsubsection{Performance in simulation}
\label{sec:perfak4c}
\begin{figure}[hbtp]
  \centering
    \includegraphics[width=0.49\textwidth]{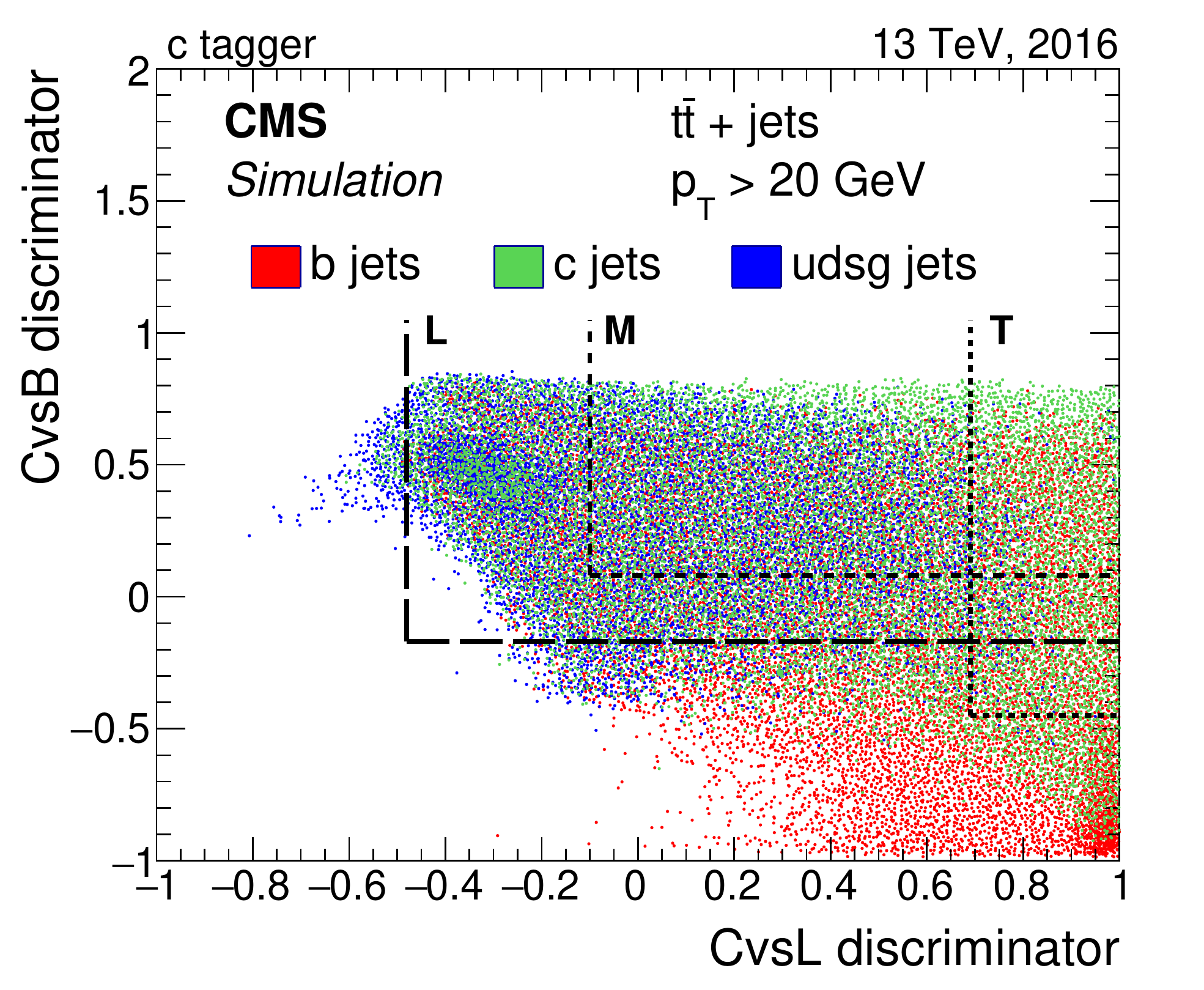}
    \includegraphics[width=0.49\textwidth]{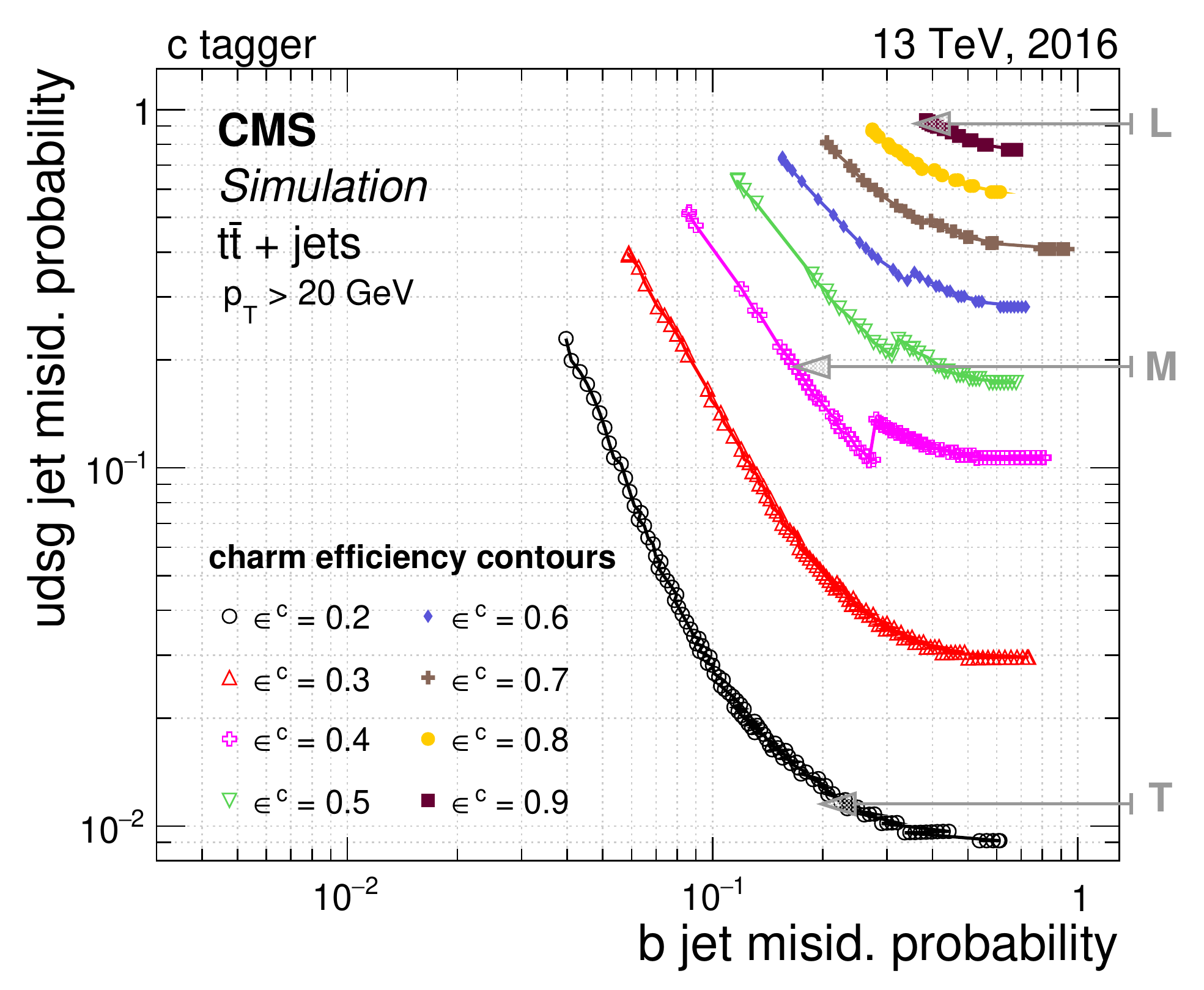}
    \caption{Correlation between CvsL and CvsB taggers for the various jet flavours (left), and misidentification probability for light-flavour jets versus misidentification probability for {\cPqb} jets for various constant {\cPqc} jet efficiencies (right) in \ttbar events. The L, M, and T working points discussed in the text are indicated by the dashed lines (left) or arrows (right). The discontinuity in the curves corresponding to {\cPqc} tagging efficiencies between 0.4 and 0.7 are due to the spike in the CvsL distribution of Figure~\ref{fig:ctaggerdistribution}.}
    \label{fig:corrctagger}
\end{figure}

The performance is evaluated using jets with $\pt>20$\GeV and $\abs{\eta}<2.4$ in a sample of simulated \ttbar events. The left panel in Fig.~\ref{fig:corrctagger} shows the correlation between the CvsL and CvsB discriminators for various jet flavours. Discriminator values close to one correspond to signal-like {\cPqc} jets. Therefore, the {\cPqc} jets populate the upper right corner of this figure, whereas {\cPqb} jets and light-flavour jets populate the region near the bottom right and the upper left corners, respectively. In the upper left corner there is a relatively large fraction of {\cPqc} jets because of the similarity of {\cPqc} jets and light-flavour jets at CvsL discriminator values below $-0.3$ and CvsB discriminator values above $+0.5$, as can be seen in Fig.~\ref{fig:ctaggerdistribution}. In order to discriminate {\cPqc} jets from other jet flavours and to evaluate the performance of the {\cPqc} tagger, thresholds are applied on both CvsL and CvsB to select the upper right corner of this phase space. Three working points have been defined corresponding to the efficiency for correctly identifying {\cPqc} jets. These are indicated by the dashed lines. The loose working point has a high efficiency for {\cPqc} jets and rejects primarily {\cPqb} jets, whereas the tight working point rejects primarily light-flavour jets. Table~\ref{tab:OPc} summarizes the efficiencies for the three working points.
\begin{table*}[htb]
\centering
\topcaption{Efficiency for the working points of the {\cPqc} tagger and corresponding efficiency for the different jet flavours obtained using jets with $\pt > 20\GeV$ in simulated \ttbar events. The numbers quoted are for illustrative purposes since the efficiency is integrated over the \pt and $\eta$ distributions of the jets.
\label{tab:OPc}}
\begin{tabular}{lccccc}
Working point & $\varepsilon_{\cPqc}$ (\%) & $\varepsilon_{\cPqb}$ (\%) & $\varepsilon_{\text{udsg}}$ (\%) \\
\hline
{\cPqc} tagger L  &  88 &  36 &  91  \\
{\cPqc} tagger M &  40 &  17  &  19  \\
{\cPqc} tagger T  &  19 &  20  &  1.2  \\
\end{tabular}
\end{table*}

The right panel in Fig.~\ref{fig:corrctagger} shows the light-flavour and {\cPqb} jet misidentification probabilities for constant {\cPqc} tagging efficiencies. The arrows indicate the {\cPqc} jet identification efficiency and misidentification probability for {\cPqb} and light-flavour jets corresponding to the three working points. The discontinuous transition in each of the curves for {\cPqc} tagging efficiencies between 0.35 and 0.7 are due to the largest spike in the CvsL distribution in the left panel in Fig.~\ref{fig:ctaggerdistribution}.

\begin{figure}[hbtp]
  \centering
    \includegraphics[width=0.49\textwidth]{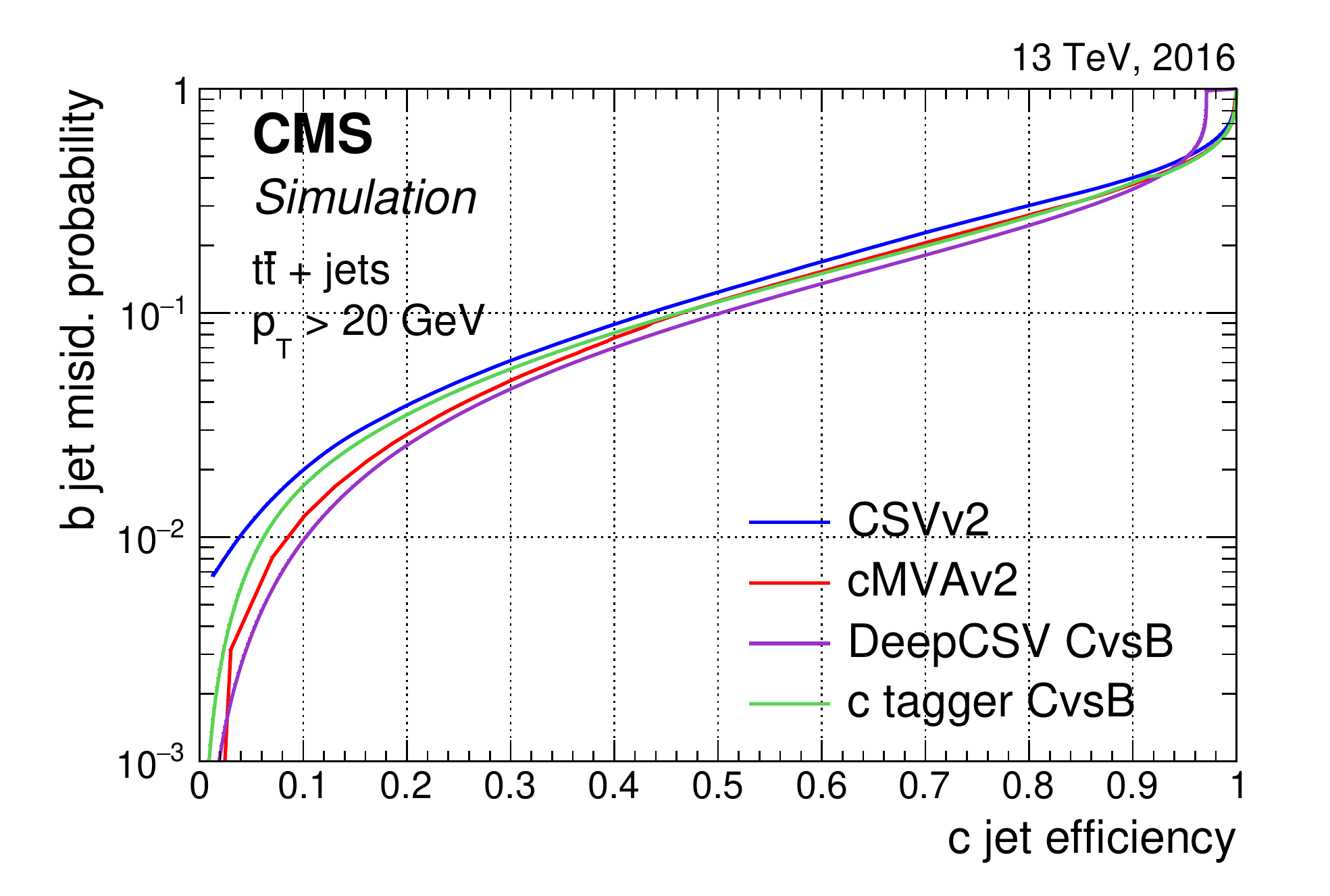}
    \includegraphics[width=0.49\textwidth]{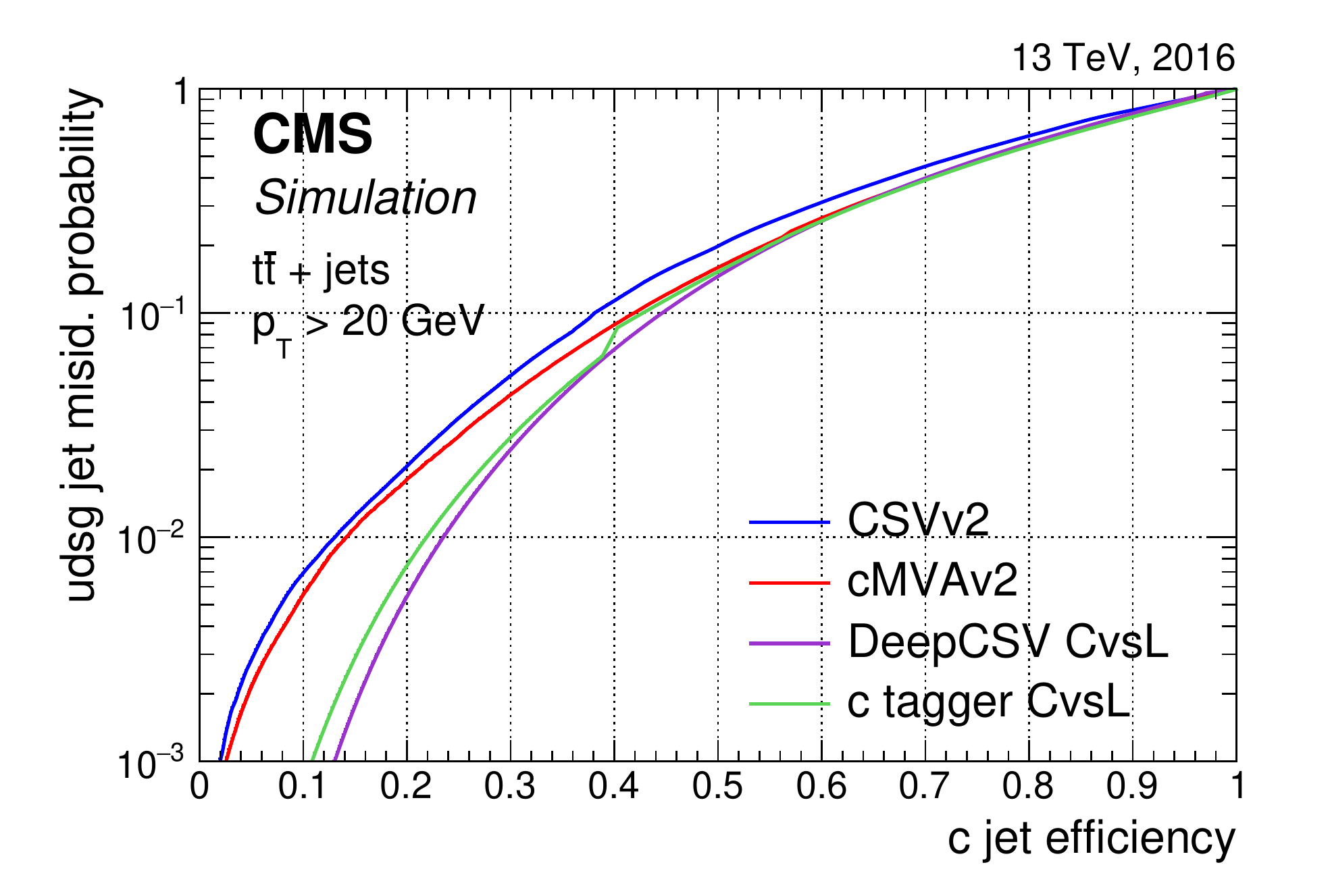}
    \caption{Misidentification probability for {\cPqb} jets (left) or light-flavour jets (right) versus {\cPqc} jet identification efficiency for various {\cPqc} tagging algorithms applied to jets in \ttbar events.}
    \label{fig:ctaggervsbtaggers}
\end{figure}

In Fig.~\ref{fig:ctaggervsbtaggers} the performance of the CvsL and CvsB taggers is compared with the cMVAv2 and CSVv2 {\cPqb} tagging algorithms. In the right panel of this figure, the transition in the performance of the curve for a {\cPqc} jet identification efficiency around 0.4 is due to the largest spike in the CvsL discriminator distribution. The performance of the CvsB tagger is similar to the performance of both {\cPqb} taggers, except at small {\cPqb} jet misidentification probabilities where the CvsB tagger is performing slightly worse than the cMVAv2 tagger. The CvsL tagger outperforms the cMVAv2 and CSVv2 tagger for small light-flavour jet misidentification probabilities. The DeepCSV tagger described in Section~\ref{sec:DeepCSV} is outperforming the dedicated {\cPqc} tagger. For the discrimination between {\cPqc} and {\cPqb} jets, the DeepCSV probabilities corresponding to the five flavour categories defined in Section~\ref{sec:DeepCSV}, are combined in the following way:
\begin{linenomath}
\begin{equation}
\mathrm{DeepCSV}\,\mathrm{CvsB} = \frac{P({\cPqc})+P({\cPqc}{\cPqc})}{1-P({\text{udsg}})},
\end{equation}
\end{linenomath}
where the numerator corresponds to the probability to identify {\cPqc} jets and the denominator to the probability to identify {\cPqb} or {\cPqc} jets.
Similarly, for the discrimination between {\cPqc} and light-flavour jets, the discriminator is constructed:
\begin{linenomath}
\begin{equation}
\mathrm{DeepCSV}\,\mathrm{CvsL} = \frac{P({\cPqc})+P({\cPqc}{\cPqc})}{1-(P({\cPqb})+P({\cPqb}{\cPqb}))},
\end{equation}
\end{linenomath}
with the numerator giving the probability to identify {\cPqc} jets and the denominator the probability to identify light-flavour or {\cPqc} jets.
The comparison with the DeepCSV algorithm used for {\cPqc} tagging should be considered as an illustration for the performance of future {\cPqc} taggers since the working points are not yet defined and the efficiency in data is not yet measured.

\section{Identification of \texorpdfstring{\cPqb}{b} jets in boosted topologies}
\label{sec:boostedalgos}
\subsection{Boosted \texorpdfstring{\cPqb}{b} jet identification with the CSVv2 algorithm}
At the high centre-of-mass energy of the LHC, particles decaying to {\cPqb} quarks can be produced with a large Lorentz boost. Examples are boosted top quarks decaying to ${\cPqb}{\PW}\to{\cPqb}{\cPq}{\cPaq}$, or boosted Higgs or {\cPZ} bosons decaying to ${\cPqb}{\cPaqb}$. As a result of the large boost of the parent particle the decay products often give rise to overlapping jets. In order to capture all the decay products, the jets are reconstructed with a distance parameter of $R=0.8$ (AK8). Jet substructure techniques can then be applied to resolve the subjets corresponding to the decay products in the AK8 jet~\cite{Larkoski:2014wba,Dasgupta:2013ihk,Thaler:2010tr,JME16003}. In this paper, the soft-drop algorithm~\cite{Larkoski:2014wba,Dasgupta:2013ihk}, which recursively removes soft wide-angle radiation from a jet, is used to resolve the substructure of the AK8 jets. The subjet axes are obtained by reclustering the jet constituents using the anti-\kt algorithm and undoing the last step of the clustering procedure.

\begin{figure}[hptb]
  \centering
    \includegraphics[width=0.9\textwidth]{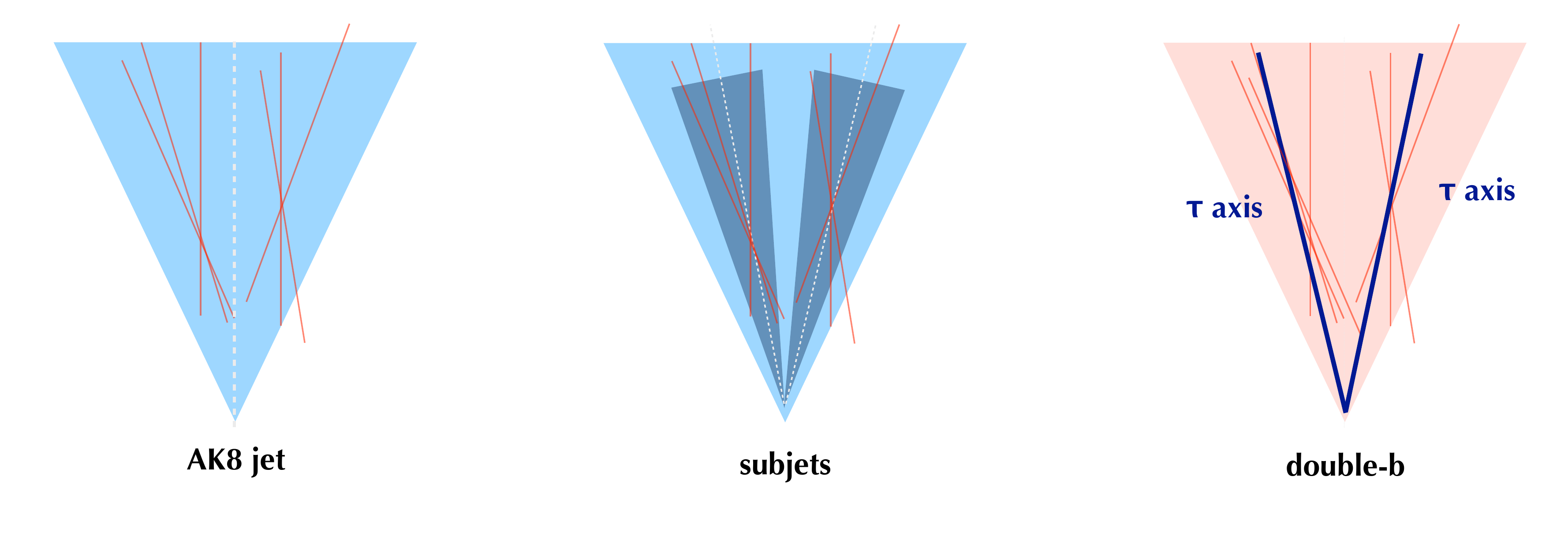}
    \caption{Schematic representation of the AK8 jet (left) and subjet (middle) {\cPqb} tagging approaches, and of the double-{\cPqb} tagger approach (right).}
    \label{fig:boostedHbbtagging}
\end{figure}

When the decay of the boosted particle contains a {\cPqb} quark, {\cPqb} tagging can be applied either on the AK8 jet or on its subjets. In both cases the CSVv2 algorithm is used. In the first approach the CSVv2 algorithm is applied to the AK8 jet but using looser requirements for the track-to-jet and vertex-to-jet association criteria, consistent with the $R=0.8$ parameter. In the second approach the CSVv2 algorithm is applied to the subjets. The two approaches are illustrated by the scheme in Fig.~\ref{fig:boostedHbbtagging} (left and middle).

\begin{figure}[hbtp]
  \centering
    \includegraphics[width=0.49\textwidth]{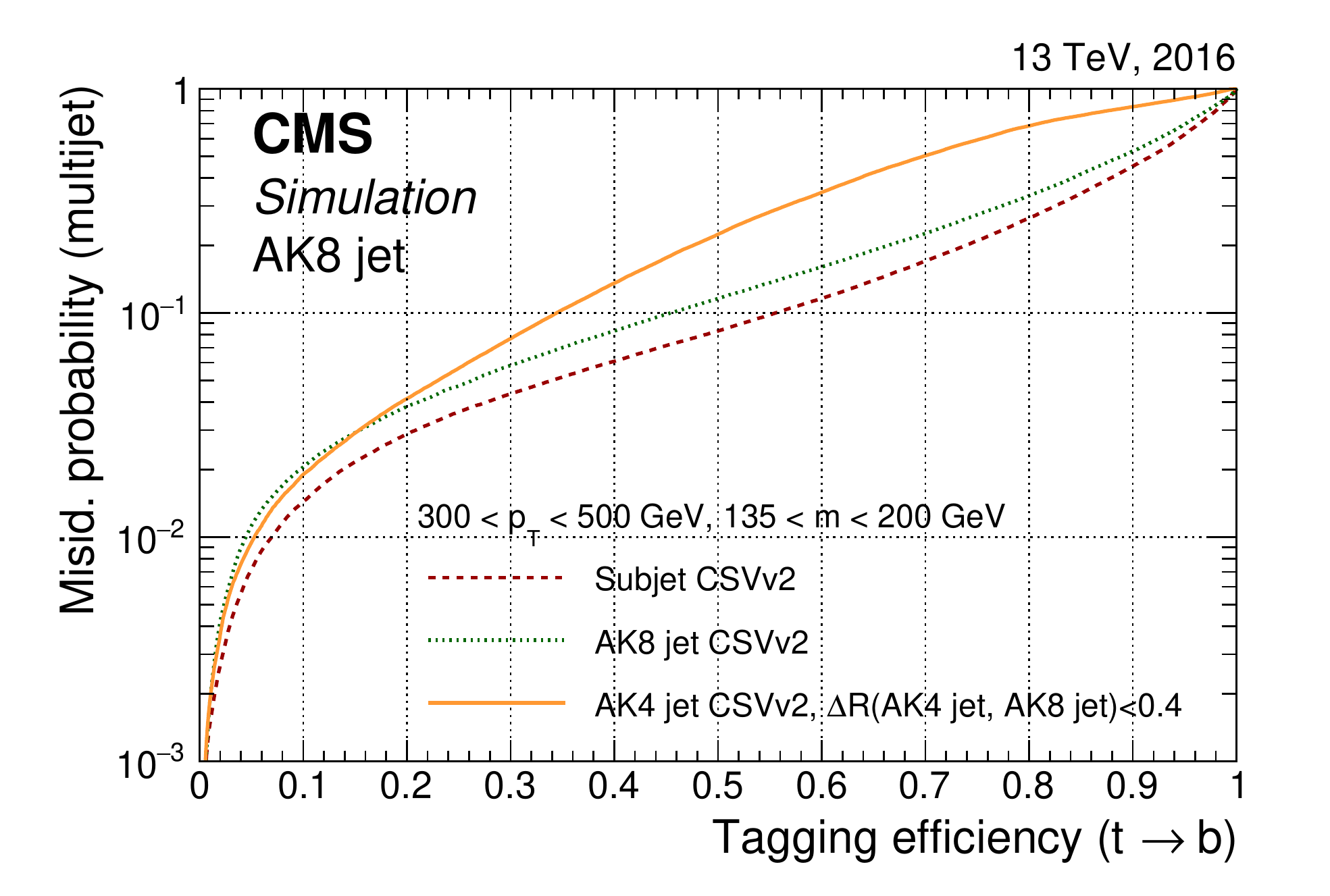}
    \includegraphics[width=0.49\textwidth]{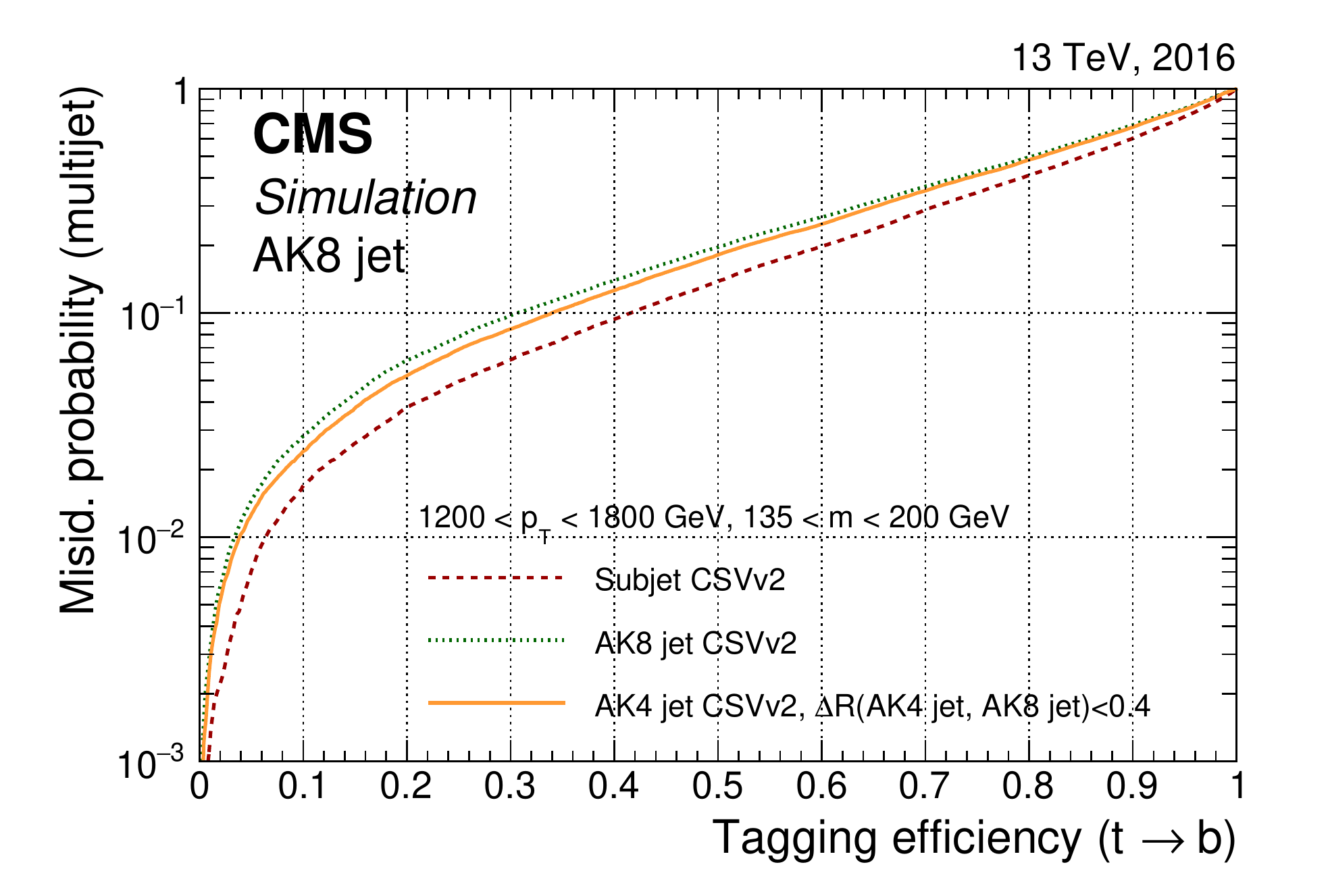}
    \caption{Misidentification probability for jets in an inclusive multijet sample versus the efficiency to correctly tag boosted top quark jets. The CSVv2 algorithm is applied to three different types of jets: AK8 jets, their subjets, and AK4 jets matched to AK8 jets. The AK8 jets are selected to have a pruned jet mass between 135 and 200\GeV, and $300<\pt<500$\GeV (left), or $1.2<\pt<1.8$\TeV (right).}
    \label{fig:ROCboostedtop}
\end{figure}

\begin{figure}[hbtp]
  \centering
    \includegraphics[width=\textwidth]{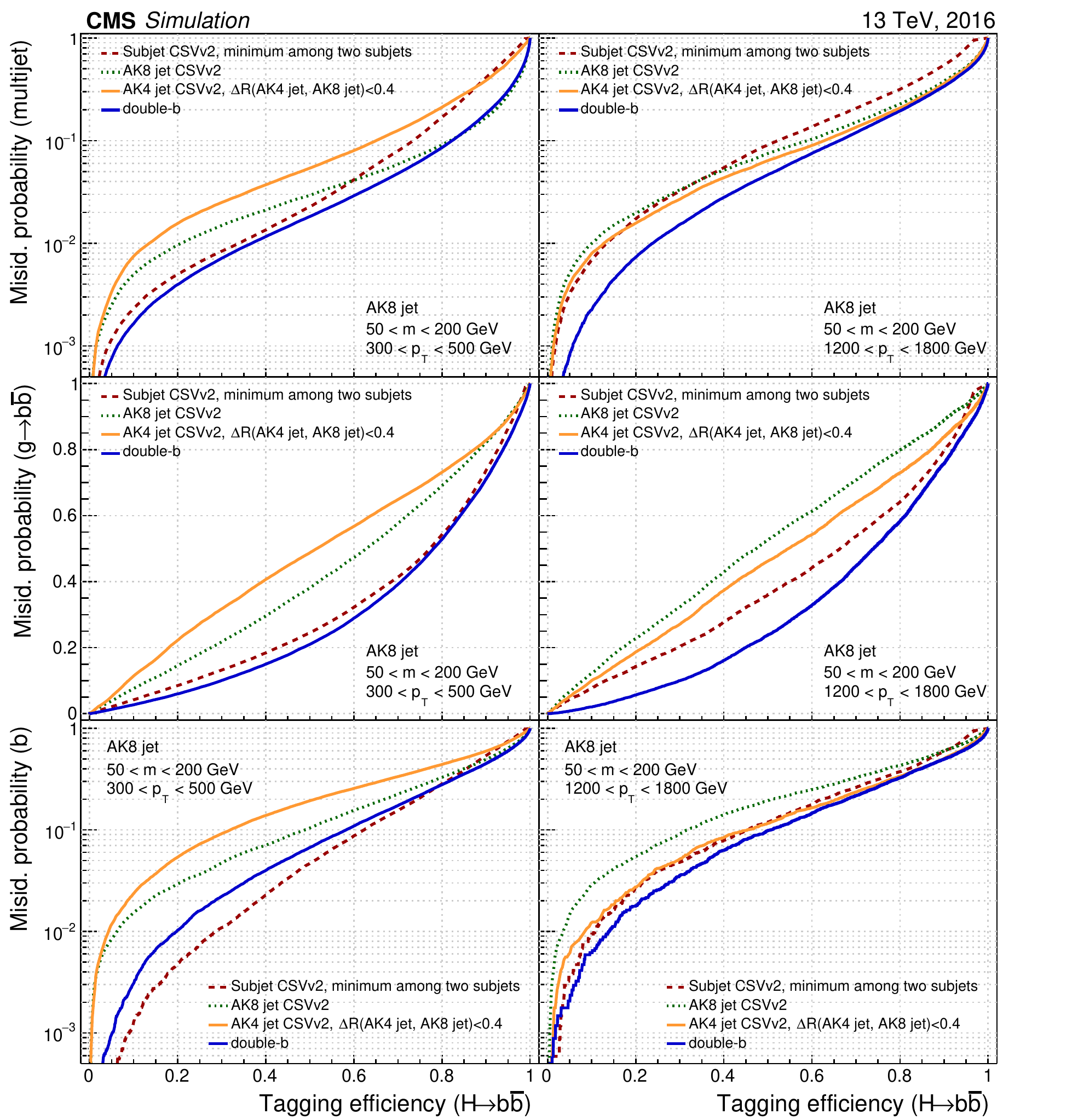}
    \caption{Misidentification probability using jets in a multijet sample (upper), for $\Pg\to{\cPqb}{\cPaqb}$ jets (middle), and for single {\cPqb} jets (lower), versus the efficiency to correctly tag $\PH\to {\cPqb}{\cPaqb}$ jets. The CSVv2 algorithm is applied to three different types of jets: AK8 jets, their subjets, and AK4 jets matched to AK8 jets. For the subjet {\cPqb} tagging curves, both subjets are required to be tagged. The double-{\cPqb} tagger, described in Section~\ref{sec:doubleb}, is applied to AK8 jets. The AK8 jets are selected to have a pruned jet mass between 50 and 200\GeV, and $300<\pt<500$\GeV (left), or $1.2<\pt<1.8$\TeV (right).}
    \label{fig:ROCHbbQCD}
\end{figure}

To illustrate the performance of {\cPqb} tagging in various boosted topologies, AK8 and subjet {\cPqb} tagging are compared in Figs.~\ref{fig:ROCboostedtop} and~\ref{fig:ROCHbbQCD}. When studying the performance of {\cPqb} tagging in various boosted topologies, jets originating from the decay of boosted top quarks (boosted top quark jets) are obtained from a $Z'$ sample, where the $Z'$ decays to \ttbar, with ${\cPqt}\to{\cPqb}{\PW}\to{\cPqb}{\cPq}{\cPaq}$. The boosted top quark jets are then defined as jets containing at least one {\cPqb} hadron. Jets originating from the decay of boosted Higgs bosons ($\PH\to{\cPqb}{\cPaqb}$ jets) are obtained from a Kaluza--Klein graviton sample, where the graviton decays to two Higgs bosons, with $\PH\to{\cPqb}{\cPaqb}$. The $\PH\to{\cPqb}{\cPaqb}$ jets are then defined as jets containing at least two {\cPqb} hadrons. Jets from a sample of inclusive multijet events are used to determine the misidentification probability.

To obtain a performance similar to what is expected in physics analyses, the jet mass is used to select jets consistent with the top quark or Higgs boson mass. While the jet mass for these particles arises from the kinematics of the decay products present in the jet, the single-parton jet mass arises mostly from soft-gluon radiation.
This soft radiation can be removed by applying jet grooming methods~\cite{Butterworth:2008tr,Krohn2010,PhysRevD.81.094023}, shifting the single-parton jet mass to smaller values. In this paper, jet pruning~\cite{PhysRevD.81.094023} is applied to the AK8 jets. The jet mass obtained from the jet four-momentum after pruning is referred to as the pruned jet mass. Jets are then selected when they have a pruned jet mass between 50 (135) and 200\GeV for {\cPqb} tagging boosted $\PH\to{\cPqb}{\cPaqb}$ (top quark) jets.

Figure~\ref{fig:ROCboostedtop} shows the {\cPqb} tagging efficiency for boosted top quark jets versus the misidentification probability using jets from a background sample of multijet events. The performance of AK8 and subjet {\cPqb} tagging is compared. When {\cPqb} tagging is applied to the subjets of boosted top quark jets, at least one of the subjets is required to be tagged. In addition, the performance of {\cPqb} tagging applied to AK4 jets matched to AK8 jets within $\Delta R\text{(AK4,AK8)}<0.4$ is also shown. When {\cPqb} tagging is applied to AK4 jets matched to the AK8 jet, at least one of the AK4 jets is required to be tagged. In Fig.~\ref{fig:ROCboostedtop} (left), for jets with $300<\pt<500$\GeV, the AK8 jet {\cPqb} tagging is more efficient than AK4 jet {\cPqb} tagging. In contrast, in Fig.~\ref{fig:ROCboostedtop} (right), for jets with $\pt>1200$\GeV, AK8 and AK4 jet {\cPqb} tagging perform similarly. This can be understood as due to the fact that at large jet \pt most of the tracks and the secondary vertex are also present in the AK4 jet because of the larger boost. In both cases, subjet {\cPqb} tagging is more efficient than AK8 jet {\cPqb} tagging when identifying the {\cPqb} jet from the boosted top quark decay.

Figure~\ref{fig:ROCHbbQCD} shows the efficiency for identifying $\PH\to{\cPqb}{\cPaqb}$ jets versus the misidentification probability using jets from a background sample of inclusive multijet events, $\Pg\to{\cPqb}{\cPaqb}$ jets or single {\cPqb} jets. When {\cPqb} tagging is applied to the subjets of the $\PH\to{\cPqb}{\cPaqb}$ jet, both subjets are required to be tagged. Similarly, both AK4 jets matched with the AK8 jet are required to be tagged.

When the misidentification probability is determined using inclusive multijet events, as illustrated in the upper panels of Fig.~\ref{fig:ROCHbbQCD}, AK8 jet {\cPqb} tagging performs well at the highest $\PH\to{\cPqb}{\cPaqb}$ jet tagging efficiencies, while subjet {\cPqb} tagging performs better at lower $\PH\to{\cPqb}{\cPaqb}$ jet tagging efficiencies. This can be understood as follows. Some of the input variables used in the CSVv2 tagger rely on the jet axis, as mentioned in Section~\ref{sec:CSVv2}. An example is the $\Delta R$ between the secondary vertex flight direction and the jet axis. This variable is expected to have, on average, a smaller value for {\cPqb} jets compared to other jets, as can be seen in the right panel of Fig.~\ref{fig:vertexcat}. When {\cPqb} tagging is applied to the AK8 jet, the AK8 jet axis is used to calculate some of the variables. However, when two {\cPqb} hadrons are present in the jet, the $\Delta R$ between the secondary vertex flight direction and the AK8 jet axis or between the track and the AK8 jet axis may be quite large. Therefore, it is better to calculate these variables with respect to their respective subjet axes. On the other hand, at the highest $\PH\to{\cPqb}{\cPaqb}$ jet tagging efficiencies, subjet {\cPqb} tagging does not fully use variables that rely on the information of the full AK8 jet, such as the number of secondary vertices. This results in a worse performance of subjet {\cPqb} tagging compared to AK8 jet {\cPqb} tagging at the highest $\PH\to{\cPqb}{\cPaqb}$ jet tagging efficiencies.

The middle panels of Fig.~\ref{fig:ROCHbbQCD} show the efficiency for $\PH\to {\cPqb}{\cPaqb}$ jets versus the misidentification probability for $\Pg\to{\cPqb}{\cPaqb}$ in multijet events. Both for jets with $300<\pt<500$\GeV and $1.2<\pt<1.8$\TeV, subjet {\cPqb} tagging performs better than AK8 jet {\cPqb} tagging. This is understood as due to the fact that the information from both {\cPqb} hadrons is better used by the subjet {\cPqb} tagging approach.

As can be seen in the bottom panels of Fig.~\ref{fig:ROCHbbQCD}, also in the case where the background is composed of single {\cPqb} jets, subjet {\cPqb} tagging performs better. The lower misidentification probability at the same efficiency is explained by the fact that for the subjet {\cPqb} tagging, the two subjets are required to be tagged. Requiring both subjets to be tagged while there is only one {\cPqb} hadron present in the background jets results in a lower misidentification probability. It is worth noting that these performance curves look very similar to the performance curves obtained when {\cPqb} jets from boosted top quarks are considered as background instead of single {\cPqb} jets from multijet events.

The left panels in Fig.~\ref{fig:ROCHbbQCD} demonstrate that AK8 jet {\cPqb} tagging is more efficient than AK4 jet {\cPqb} tagging using jets with $300<\pt<500$\GeV. The reason is that at low jet \pt not all the tracks and secondary vertex are associated with the two AK4 jets, while they are associated with the AK8 jet. In contrast, using jets with $\pt>1200$\GeV, requiring the two AK4 jets to be tagged results in a similar performance or better than when the AK8 jet is required to be tagged. This can be explained by the fact that the high jet \pt results in tracks and secondary vertices that are more collimated and fully contained in the AK4 jets.

Figs.~\ref{fig:ROCboostedtop} and~\ref{fig:ROCHbbQCD} demonstrate that the performance of subjet and AK8 jet {\cPqb} tagging depends not only on the signal jets to be {\cPqb} tagged and on the background jets under consideration, but also on the jet \pt.

\subsection{The double-\texorpdfstring{\cPqb}{b} tagger}
\label{sec:doubleb}
As mentioned in the previous section, the approaches of {\cPqb} tagging AK8 jets, as well as applying subjet {\cPqb} tagging, have limitations when identifying $\PH\to {\cPqb}{\cPaqb}$ jets. In this section, a novel approach is presented to discriminate $\PH\to{\cPqb}{\cPaqb}$ candidates from single-parton jets in multijet events. The strategy followed when developing the new ``double-b'' tagging algorithm is to fully use not only the presence of two {\cPqb} hadrons inside the AK8 jet but also the correlation between the directions of the momenta of the two {\cPqb} hadrons. Although the algorithm is developed using simulated $\PH\to{\cPqb}{\cPaqb}$ events, any dependence of the algorithm performance on the mass or \pt of the {\cPqb}{\cPaqb} pair is avoided. This strategy allows the usage of the tagger in physics analyses with a large range of jet \pt. The dependence on the jet mass is avoided as this variable is often used to define a region for the estimation of the background. In addition, this strategy also permits the use of the double-{\cPqb} tagger for the identification of boosted ${\PZ}\to{\cPqb}{\cPaqb}$ jets or any other boosted {\cPqb}{\cPaqb} resonance where the kinematics of the decay products are similar.

A variable sensitive to the substructure is the N-subjettiness, $\tau_{\text{N}}$~\cite{Thaler:2010tr}, which is a jet shape variable, computed under the assumption that the jet has N subjets, and it is defined as the \pt-weighted distance between each jet constituent and its nearest subjet axis ($\Delta R$):
\begin{linenomath}
\begin{equation}
\tau_{\text{N}}=\frac{1}{d_0}\sum_{k} \pt^{k}\min(\Delta R_{1,k},\ldots, \Delta R_{\text{N},k}),
\end{equation}
\end{linenomath}
where $k$ runs over all jet constituents. The normalization factor is $d_0 = \sum_k \pt^{k} R_0$ and $R_0$ is the original jet distance parameter, \ie $R_0=0.8$. The $\tau_{\text{N}}$ variable has a small value if the jet is consistent with having N or fewer subjets. The subjet axes are used as a starting point for the $\tau_{\text{N}}$ minimization. After the minimization, the $\tau_{\text{N}}$ axes, also called $\tau$ axes, are obtained. These are then used to estimate the directions of the partons giving rise to the subjets, as schematically illustrated in Fig.~\ref{fig:boostedHbbtagging} (right).

Many of the CSVv2 variables are also used in the double-{\cPqb} tagger algorithm. The variables rely on reconstructed tracks, secondary vertices obtained using the IVF algorithm, as well as the system of two secondary vertices. Tracks with $\pt > 1\GeV$ are associated with jets in a cone of $\Delta R < 0.8$ around the jet axis. Each track is then associated with the closest $\tau$ axis, where the distance of a track to the $\tau$ axis is defined as the distance at their point of closest approach. The selection requirements applied to tracks in the CSVv2 algorithm are also applied here, using the $\tau$ axis instead of the jet axis. The reconstructed secondary vertices are associated first with jets in a cone $\Delta R < 0.7$ and then to the closest $\tau$ axis within that jet. For each $\tau$ axis, the track four-momenta of the constituent tracks from all the secondary vertices associated with a given $\tau$ axis are added to compute the secondary vertex mass and \pt for that $\tau$ axis.

Input variables are selected that discriminate between $\PH\to{\cPqb}{\cPaqb}$ jets and other jet flavours, and that improve the discrimination against the background from inclusive multijet production by at least 5\% compared to the performance of the tagger without the variable. In addition, as mentioned earlier, variables are chosen that do not have a strong dependence on the jet \pt or jet mass. This procedure resulted in the following list of variables:
\begin{itemize}
\item The four tracks with the highest impact parameter significance.
\item The impact parameter significance of the first two tracks ordered in decreasing impact parameter significance, for each $\tau$ axis.
\item The 2D impact parameter significance, of the first two tracks (first track) that raise the total mass above 5.2 (1.5)\GeV. These tracks are obtained as explained in Section~\ref{sec:CSVv2} in the context of the CSVv2 algorithm. In the case of the highest threshold, also the second track above the threshold mass is used. The thresholds of 5.2\GeV and 1.5\GeV are related to the {\cPqb} and {\cPqc} hadron masses, respectively.
\item The secondary vertex energy ratio, defined as the total energy of all secondary vertices associated with a given $\tau$ axis divided by the total energy of all the tracks associated with the AK8 jet that are consistent with the primary vertex, for each of the two $\tau$ axes.
\item The number of secondary vertices associated with the jet.
\item The 2D secondary vertex flight distance significance, for the secondary vertex with the smallest uncertainty on the 3D flight distance, for each of the two $\tau$ axes.
\item The $\Delta R$ between the secondary vertex with the smallest 3D flight distance uncertainty and its $\tau$ axis, for each of the two $\tau$ axes.
\item The relative pseudorapidity, $\eta_{\text{rel}}$, of the tracks from all secondary vertices with respect to their $\tau$ axis for the three leading tracks ordered in increasing $\eta_{\text{rel}}$, for each of the two $\tau$ axes.
\item The total secondary vertex mass, defined as the invariant mass of all tracks from secondary vertices associated with the same $\tau$ axis, for each of the two $\tau$ axes.
\item The information related to the system of two secondary vertices, the $z$ variable, defined as:
\begin{linenomath}
\begin{equation}
z=\Delta R(\text{SV}_0,\text{SV}_1)\frac{\pt(\text{SV}_1)}{m(\text{SV}_0,\text{SV}_1)}
\label{eq:zvar}
\end{equation}
\end{linenomath}
where $\text{SV}_0$ and $\text{SV}_1$ are the secondary vertices with the smallest 3D flight distance uncertainty associated with the two $\tau$ axes, $\pt(\text{SV}_1)$ is the \pt of the secondary vertex associated with the second $\tau$ axis, and $\Delta R(\text{SV}_0,\text{SV}_1)$ is the distance between the two secondary vertices, and $m(\text{SV}_0,\text{SV}_1)$ is the invariant mass corresponding to the summed four-momenta of the two secondary vertices.
\end{itemize}
The most discriminating variables are the impact parameter significance for the most displaced tracks, the 2D impact parameter significance for the first track above the (5.2\GeV) {\cPqb}-hadron mass threshold, and the secondary vertex energy ratio for the secondary vertex with the smallest 3D flight distance uncertainty ($\text{SV}_0$).

\begin{figure}[hptb]
  \centering
    \includegraphics[width=0.49\textwidth]{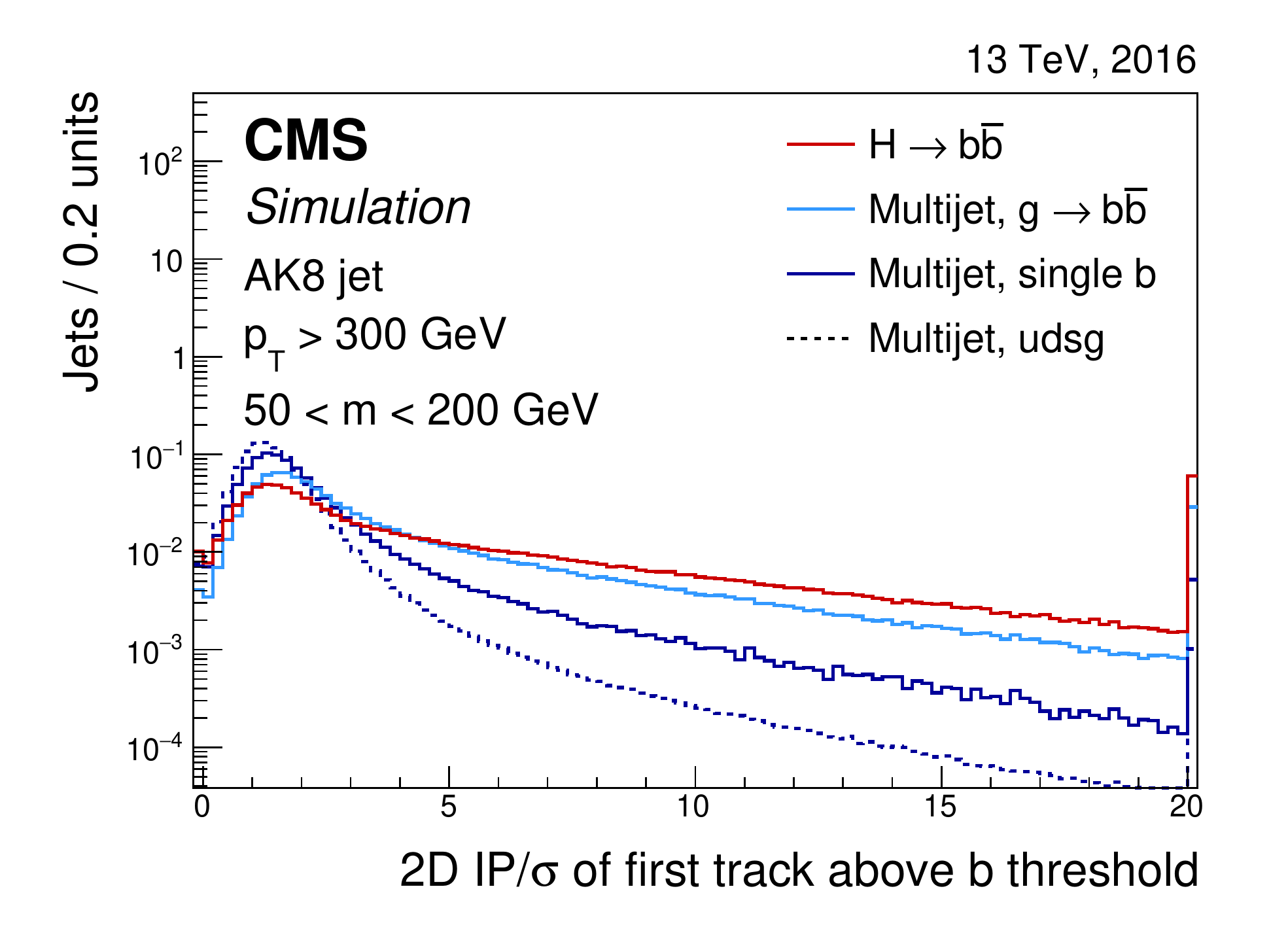}
    \includegraphics[width=0.49\textwidth]{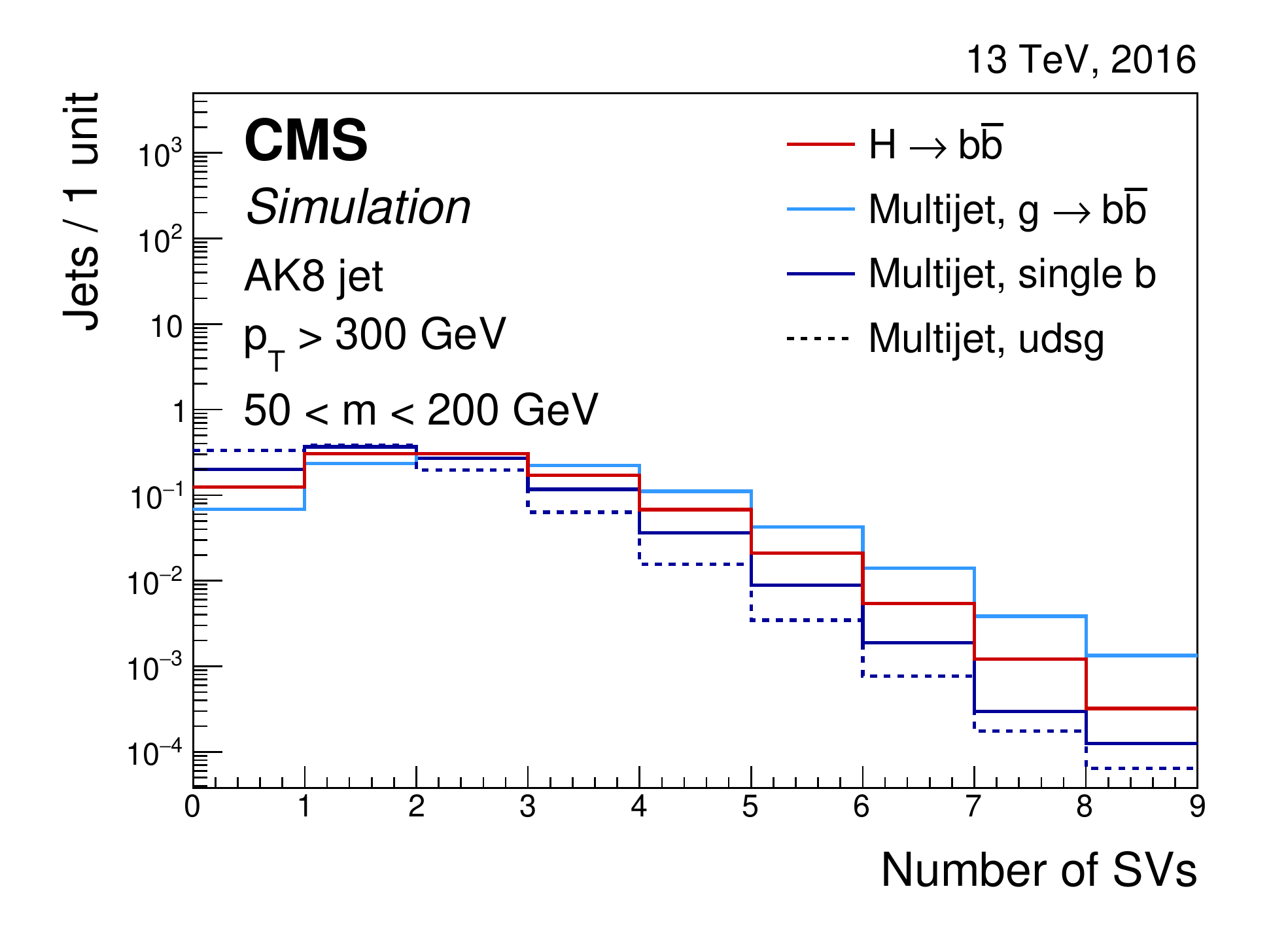}\\
    \includegraphics[width=0.49\textwidth]{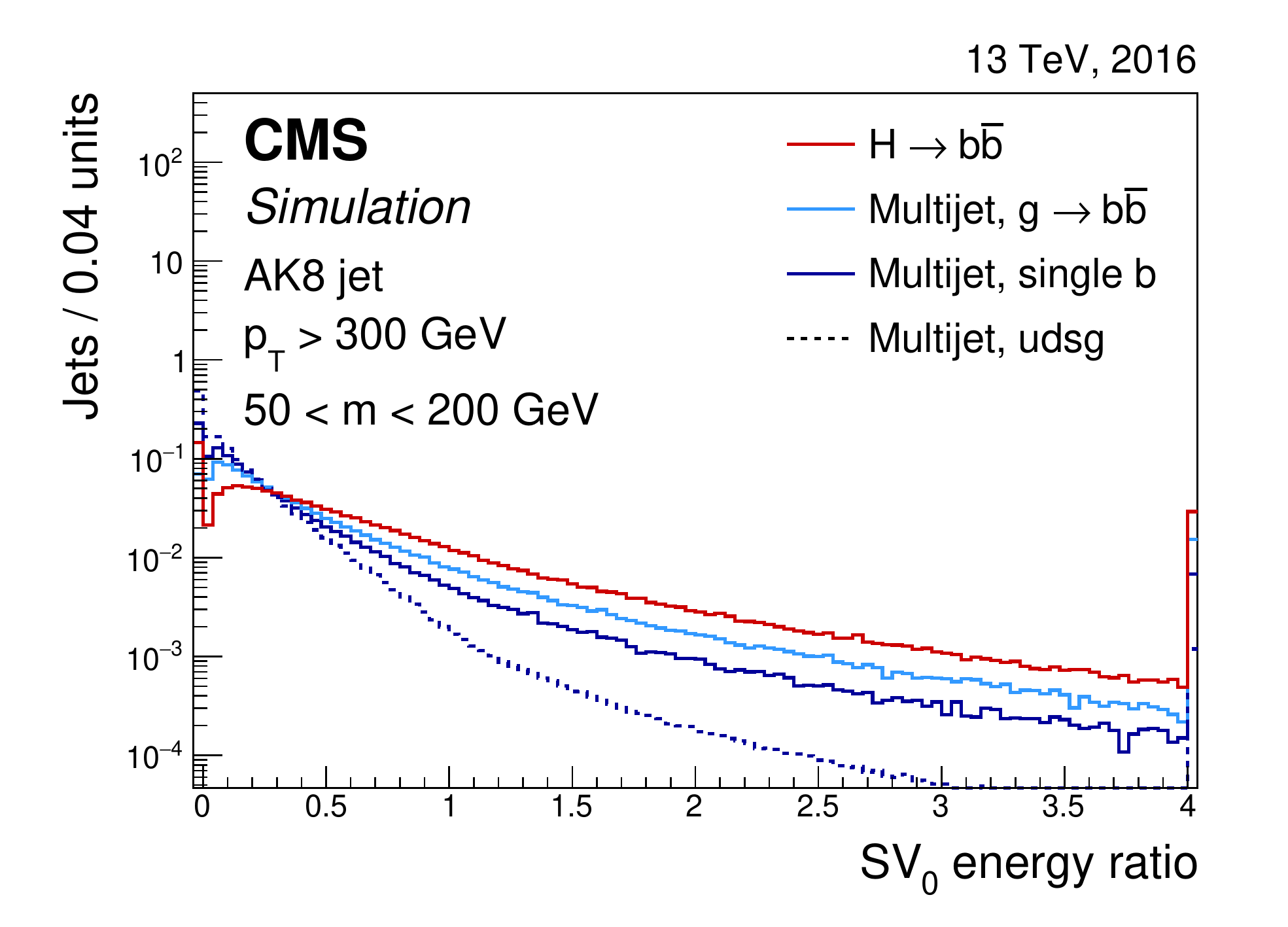}
    \includegraphics[width=0.49\textwidth]{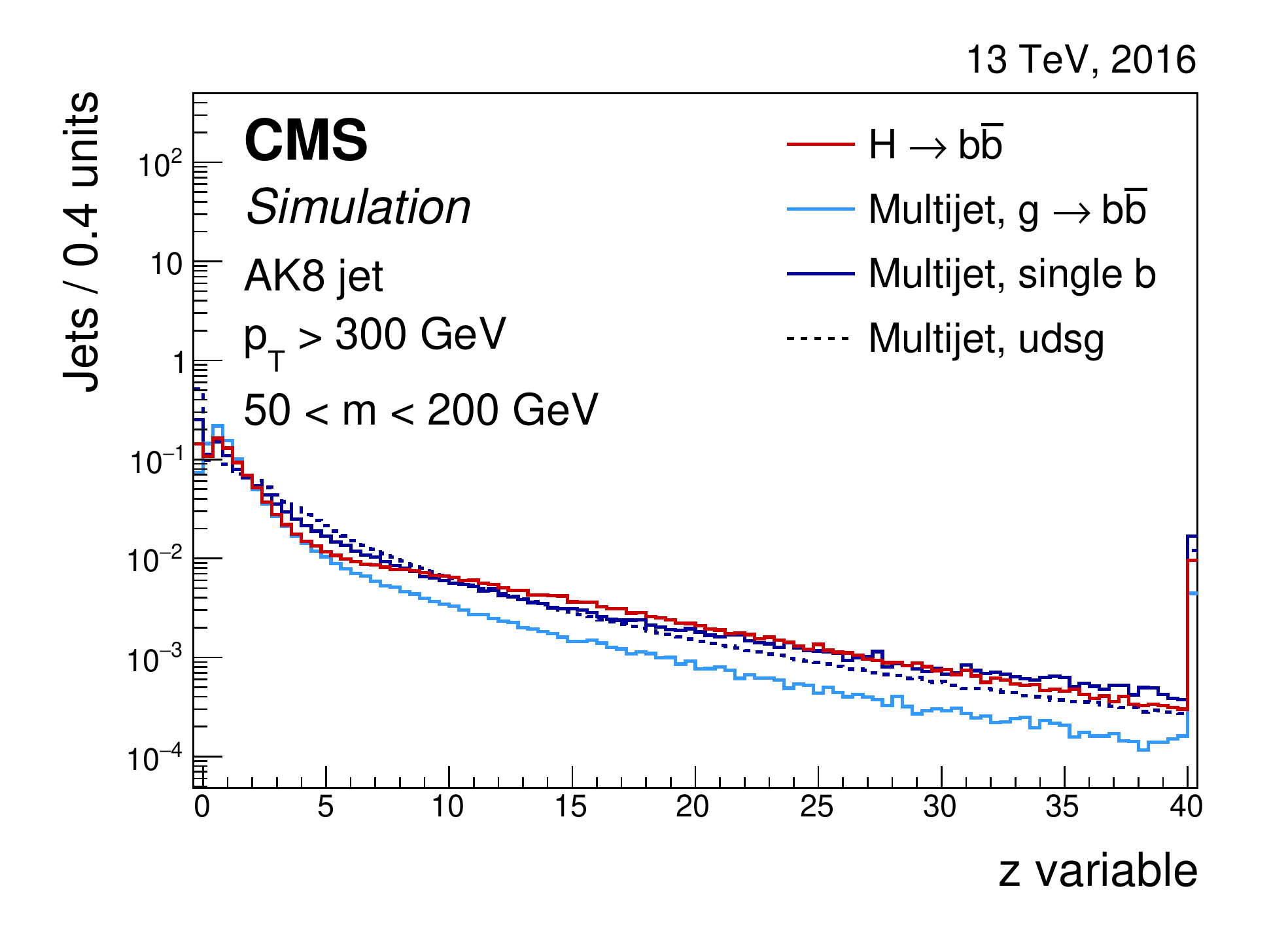}
    \caption{Distribution of 2D impact parameter significance for the most displaced track raising the mass above the {\cPqb} hadron mass threshold as described in the text (upper left), number of secondary vertices associated with the AK8 jet (upper right), vertex energy ratio for the secondary vertex with the smallest 3D flight distance uncertainty (lower left), and $z$ variable described in the text (lower right). Comparison between $\PH\to{\cPqb}{\cPaqb}$ jets from simulated samples of a Kaluza--Klein graviton decaying to two Higgs bosons, and jets in an inclusive multijet sample containing zero, one, or two b quarks. The AK8 jets are selected with $\pt>300$\GeV and pruned jet mass between 50 and 200\GeV. The distributions are normalized to unit area. The last bin includes the overflow entries.}
    \label{fig:varsDoubleB}
\end{figure}

Figure~\ref{fig:varsDoubleB} shows the distributions for some of the input variables for the signal $\PH\to{\cPqb}{\cPaqb}$ jets and using jets from inclusive multijet production containing zero, one, or two {\cPqb} quarks. Distributions are shown separately for $\Pg\to{\cPqb}{\cPaqb}$, single {\cPqb} quark, and light-flavour jets production. The secondary vertex multiplicity and the vertex energy ratio for $SV_{0}$, along with the impact parameter significance of the first track raising the total invariant mass of all tracks above the {\cPqb} hadron mass threshold show a good separation between the $\PH\to {\cPqb}{\cPaqb}$ jets and the different background contributions. The $z$ variable, Eq.~(\ref{eq:zvar}), shows good discrimination against $\Pg\to{\cPqb}{\cPaqb}$ jets since it uses the different kinematic properties of the $\PH\to{\cPqb}{\cPaqb}$ and $\Pg\to{\cPqb}{\cPaqb}$ decays.

Several variables related to the properties of soft leptons arising from the {\cPqb} hadron decay were also investigated. Despite a small gain in performance, these variables were excluded as input variables since they could introduce a bias in the efficiency measurement from data. The bias could arise when using muon information both to define input variables and to select a sample of jets containing a muon for the efficiency measurement in data, presented in Section~\ref{sec:boostedeff}.

The discriminating variables are combined using a BDT and the \textsc{tmva} package~\cite{TMVA}. The training is performed using $\PH\to{\cPqb}{\cPaqb}$ jets from simulated events with a Kaluza--Klein graviton decaying to two Higgs bosons as signal, and jets from inclusive multijet production as background. Jets are selected when they have a pruned mass between 50 and 200\GeV and \pt between 300 and 2500\GeV. The jet \pt distributions for the simulated signal and background jet samples are similar, therefore no dedicated reweighting of the samples was performed.

\begin{figure}[hbtp]
  \centering
    \includegraphics[width=0.49\textwidth]{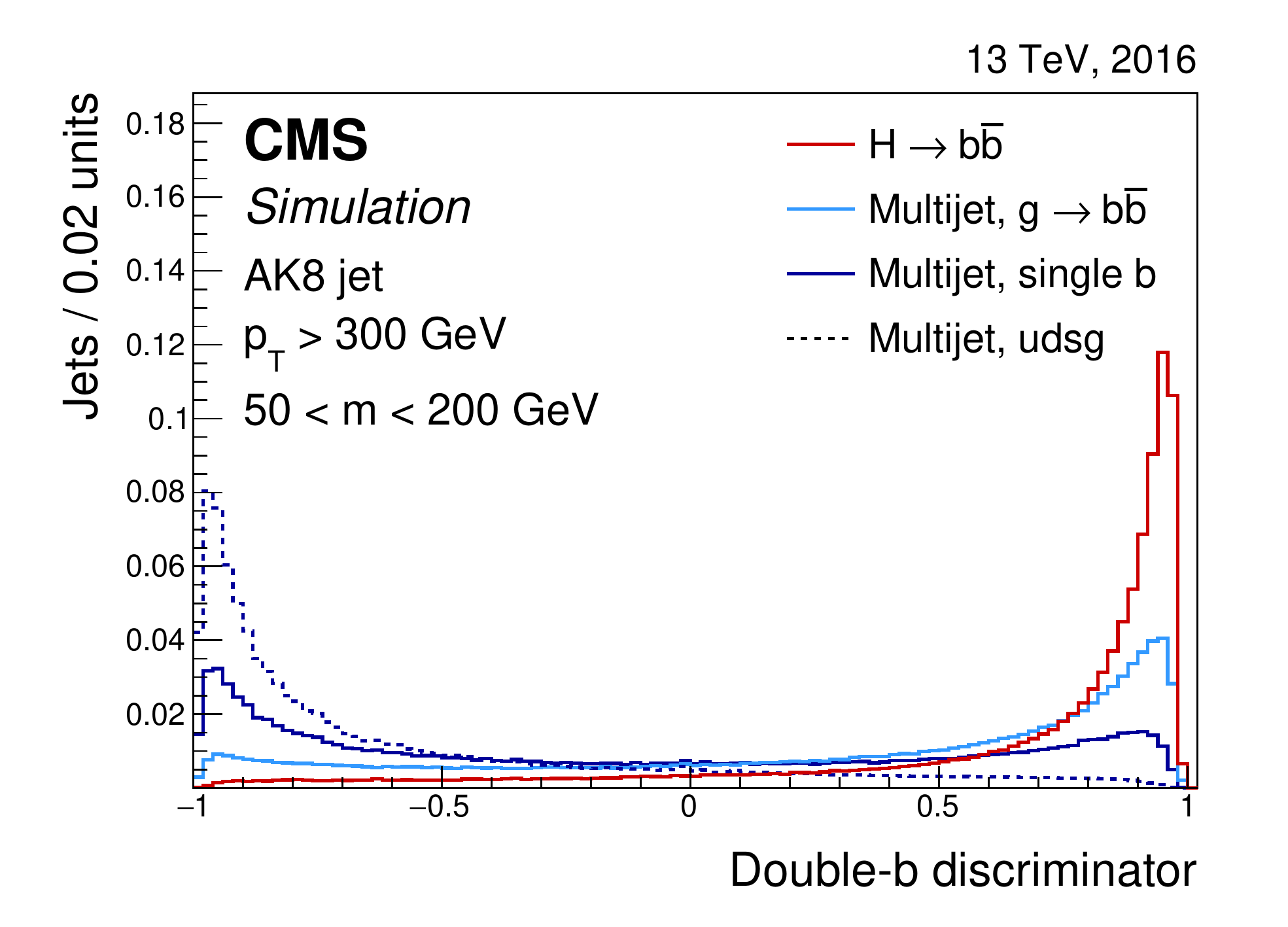}\\
    \includegraphics[width=0.49\textwidth]{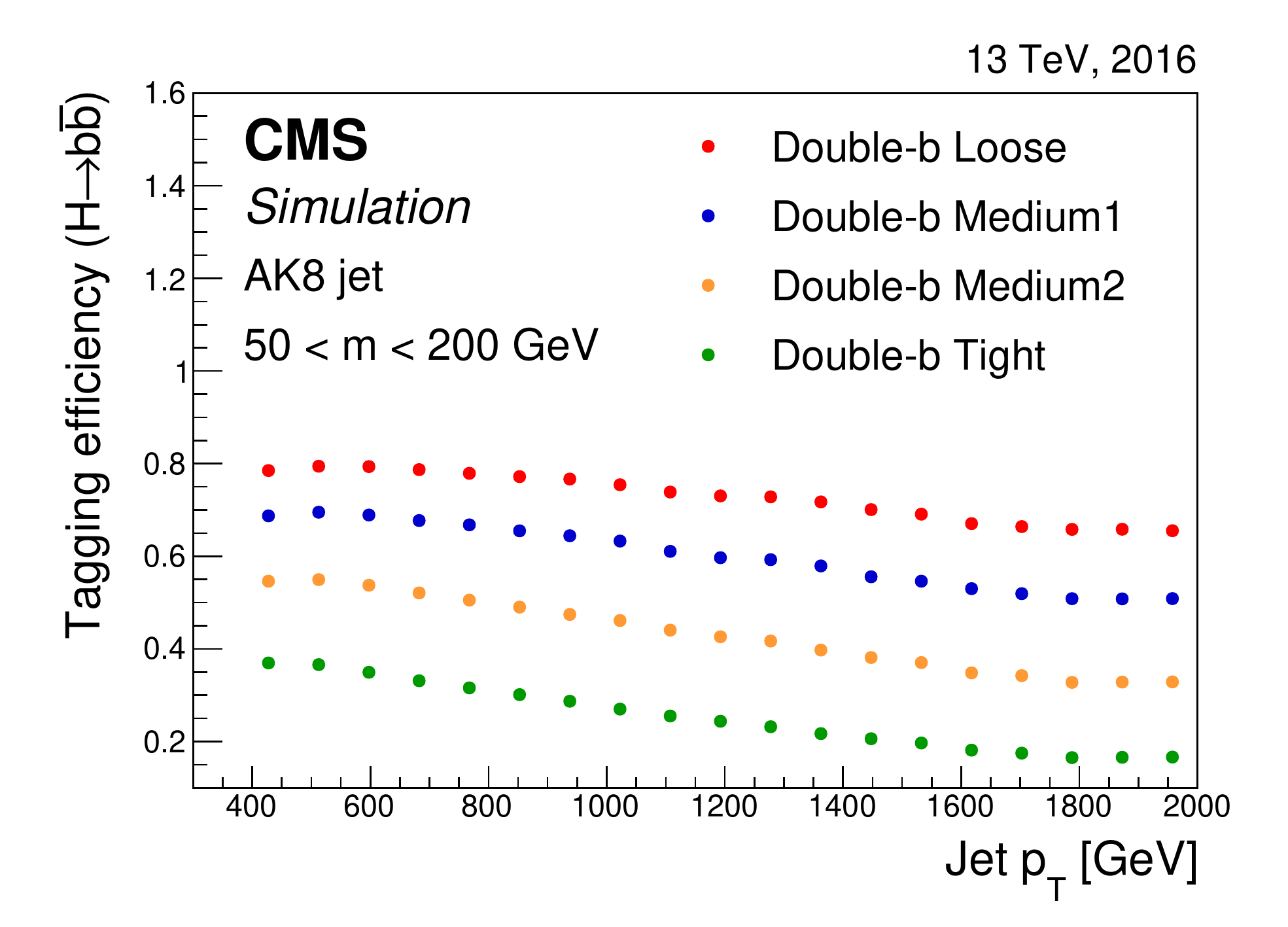}
    \includegraphics[width=0.49\textwidth]{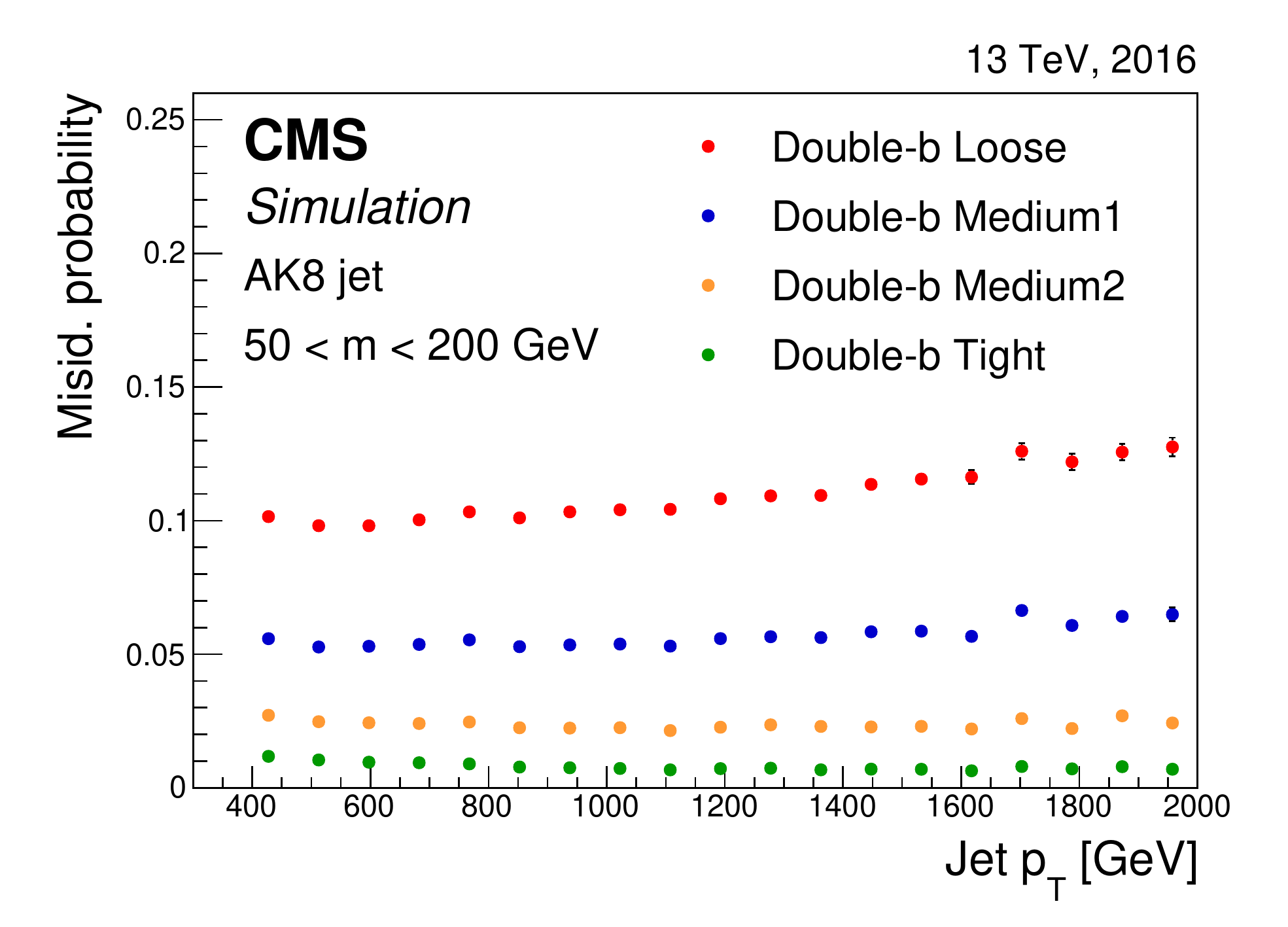}
    \caption{Distribution of the double-{\cPqb} tagger discriminator values normalized to unit area for $\PH\to{\cPqb}{\cPaqb}$ jets in simulated samples of a Kaluza--Klein graviton decaying to two Higgs bosons, and for jets in an inclusive multijet sample containing zero, one, or two b quarks (upper). Efficiency to correctly tag $\PH\to{\cPqb}{\cPaqb}$ jets (lower left) and misidentification probability using jets in an inclusive multijet sample (lower right) for four working points of the double-{\cPqb} tagger as a function of the jet \pt. The AK8 jets are selected with $\pt>300$\GeV and pruned jet mass between 50 and 200\GeV.}
    \label{fig:EffDoubleBHbbQCD}
\end{figure}

The distribution of the double-{\cPqb} discriminator values is shown in the upper panel of Fig.~\ref{fig:EffDoubleBHbbQCD}. Four working points are defined corresponding to about 75, 65, 45 and 25\% signal efficiency for a jet \pt of about 1\TeV. The signal efficiencies and misidentification probabilities as functions of the jet \pt for these four working points are shown in the lower panels of Fig.~\ref{fig:EffDoubleBHbbQCD}.
The decreasing signal efficiency at high jet \pt originates from the larger collimation of particles, which results in a lower track reconstruction efficiency due to close by hits. The reduced track reconstruction efficiency for high jet \pt results in a lower tagging efficiency for high jet \pt.

The performance of the double-{\cPqb} tagger is compared with that of the CSVv2 tagger applied to AK8 jets or their subjets. The top and middle panels in Fig.~\ref{fig:ROCHbbQCD} show the performance when the background consists of jets from inclusive multijet production or $\Pg\to{\cPqb}{\cPaqb}$ jets. In these cases, the double-{\cPqb} tagger outperforms the AK8 jet and subjet {\cPqb} tagging approaches for all jet \pt ranges. At high jet \pt the improvement is larger compared to low jet \pt, thereby providing an important gain in the searches for heavy resonances where mostly high-\pt jets are expected. When the background is composed of single {\cPqb} jets, as shown in the bottom panels of Fig.~\ref{fig:ROCHbbQCD}, subjet {\cPqb} tagging outperforms the double-{\cPqb} tagger at low jet \pt, while the two approaches are similar at high jet \pt. The lower misidentification probability for single {\cPqb} jets at the same $\PH\to{\cPqb}{\cPaqb}$ jet tagging efficiency for subjet {\cPqb} tagging at low jet \pt is explained by the fact that the two subjets are very well separated at low jet \pt and the variables related to the AK8 jet used in the double-{\cPqb} tagger are less efficient. In contrast, at high jet \pt the subjets are much closer together, resulting in shared tracks and secondary vertices and thereby leading to a more similar performance.

Whether it is better to use subjet {\cPqb} tagging or the double-{\cPqb} tagger in a physics analysis depends strongly on the flavour composition and \pt distribution of the jets from the signal and background processes under consideration.

\section{Performance of \texorpdfstring{\cPqb}{b} jet identification at the trigger level}
\label{sec:trigger}
The identification of {\cPqb} jets at the trigger level is essential to collect events that do not pass standard lepton, jet, or missing \pt triggers, and to increase the purity of the recorded sample for analyses requiring {\cPqb} jets in the final state. The L1 trigger uses information from the calorimeters and muon detectors to reconstruct objects such as charged leptons and jets. Identification of {\cPqb} jets is not possible at that stage as it relies on the reconstructed tracks from charged particles available only at the HLT. In this section, we describe {\cPqb} jet identification at the HLT. A detailed description of the CMS trigger system can be found in Ref.~\cite{Khachatryan:2016bia}.

\begin{figure}[htbp]
  \centering
    \includegraphics[width=0.9\textwidth]{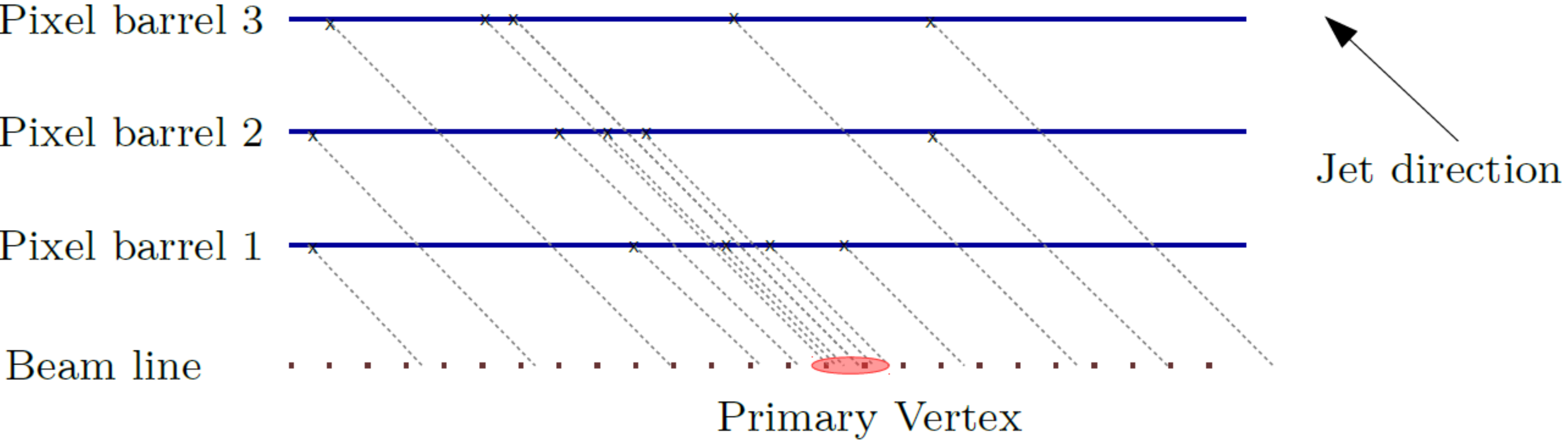}
    \caption{Scheme of the fast primary vertex finding algorithm used to determine the position of the vertex along the beam line. The pixel detector hits from the tracks in a jet are projected along the calorimeter jet direction onto the beam line.}
    \label{fig:fastPVfinding}
\end{figure}

Because of latency constraints at the HLT, it is not feasible to reconstruct the tracks and primary vertex with the algorithms used for offline reconstruction. The time needed for track finding can be significantly reduced if the position of the primary vertex is known. While the position in the transverse plane is defined with a precision of 20\micron, its position along the beam line is not known~\cite{TRK-11-001}. However, it is possible to obtain a rough estimate of the primary vertex position along the beam line by projecting onto the $z$ direction the position of the silicon pixel tracker hits (pixel detector hits) compatible with the jets. A pixel tracker hit in the barrel (endcap) is compatible with a jet when the difference in azimuthal angle between the hit and the jet is less than 0.21 (0.14). The region along the beam line with the highest number of projected pixel detector hits is most likely to correspond to the position of the primary vertex. This concept is illustrated in Fig.~\ref{fig:fastPVfinding}: the direction of the tracks in a jet is assumed to be approximately the same as the direction of the jet obtained using the calorimeter information.

This fast primary vertex (FPV) finding algorithm is sensitive to pixel detector hits from pileup interactions. Therefore, a number of selection requirements based on the shape of the charge deposition clusters associated with the pixel detector hits are applied to select those that most likely correspond to a particle with a large \pt. In addition, only pixel detector hits compatible with up to four leading jets with $\pt>30$\GeV and $\abs{\eta}<2.4$ are used. Finally, each pixel detector hit is assigned a weight reflecting the probability that it corresponds to a track in one of the considered jets. The weight is obtained by using information related to the shape of the charge deposition cluster, the azimuthal angle between the jet and the cluster, and the jet \pt. Since the spread of projected hits from the primary vertex is proportional to the distance from the beam line, a larger weight is assigned to pixel detector hits closer to the beam line.

\begin{figure}[hptb]
  \centering
    \includegraphics[width=0.49\textwidth]{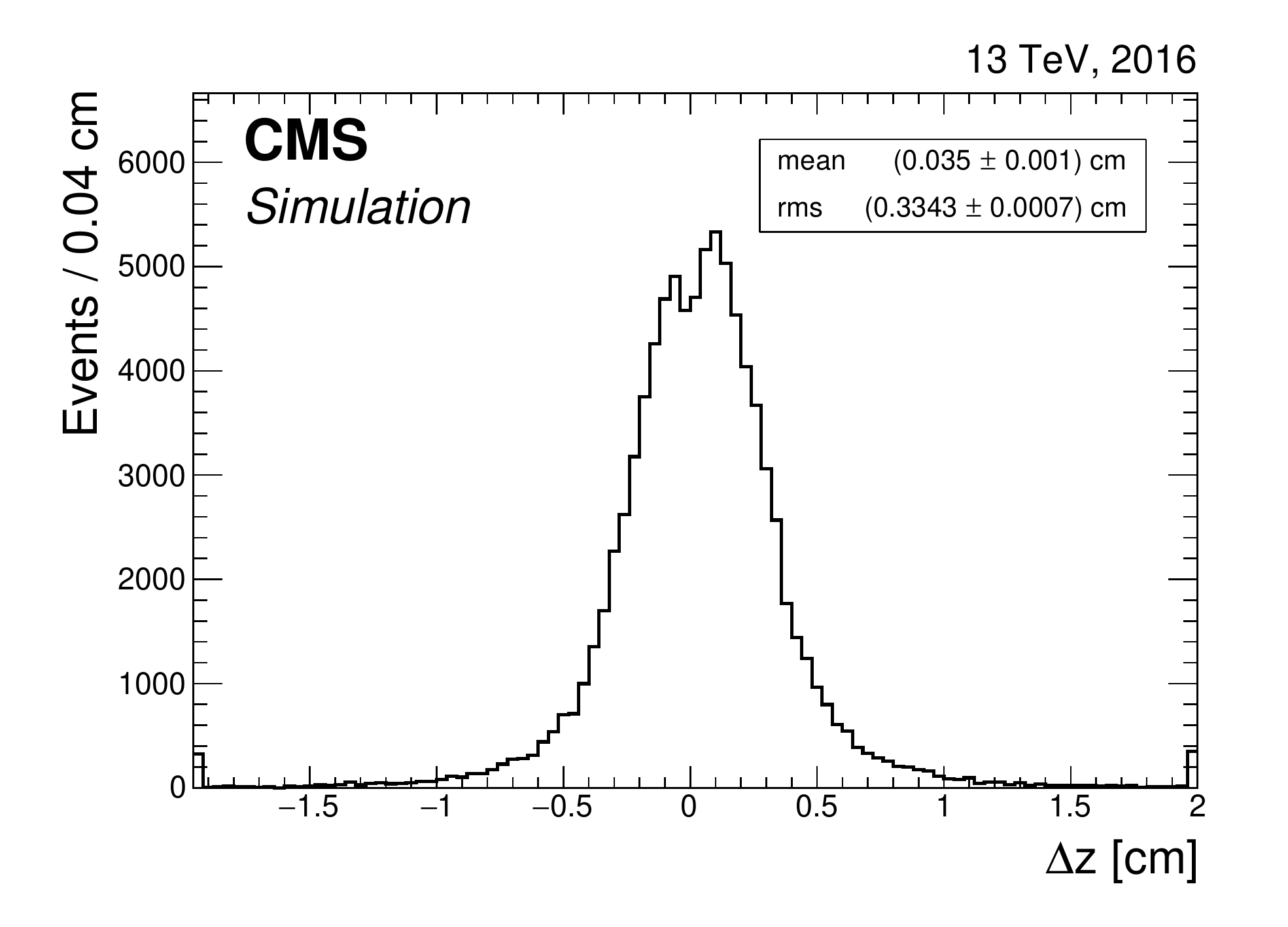}
    \includegraphics[width=0.49\textwidth]{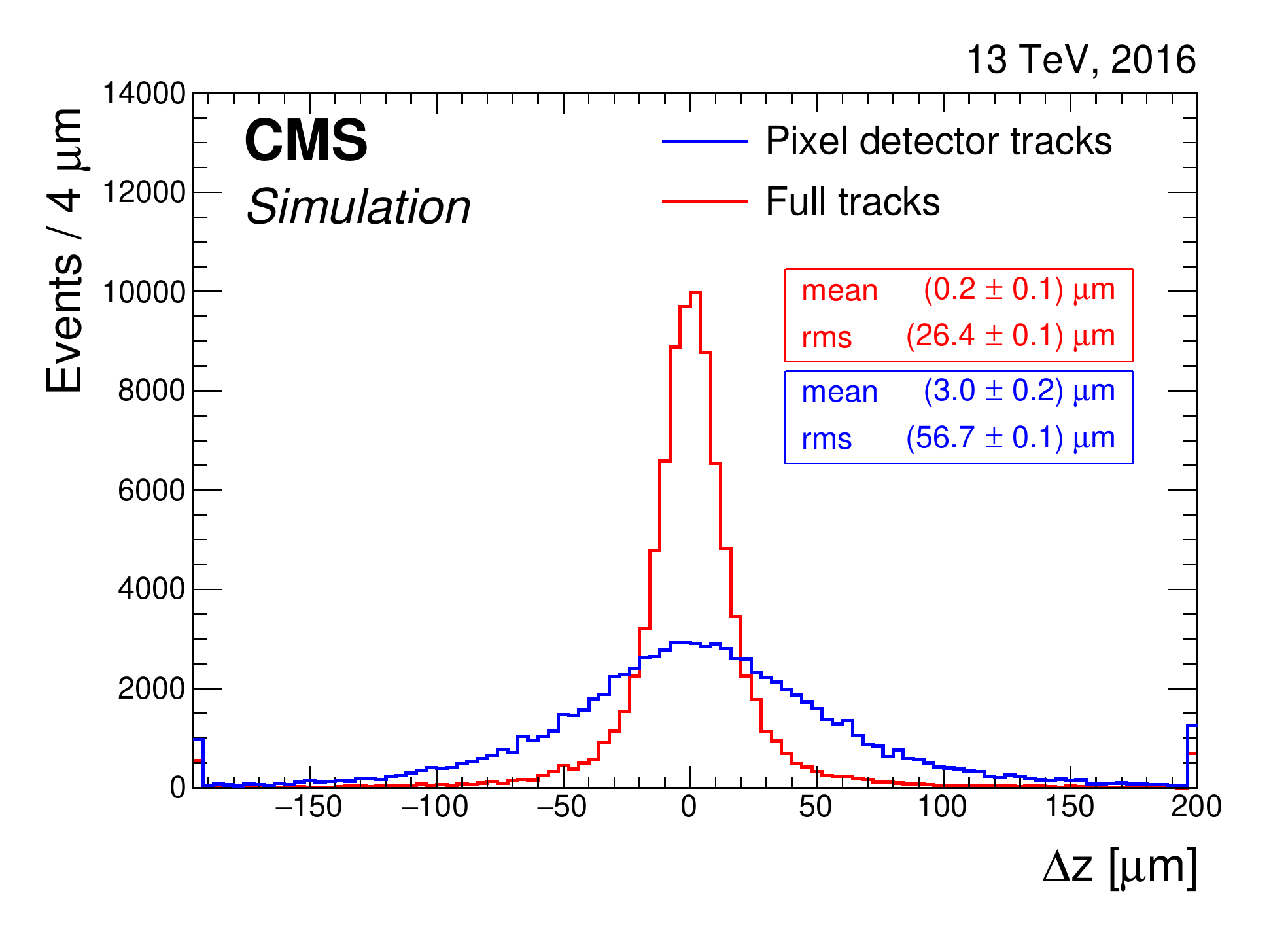}
    \caption{Distribution of residuals on the position of the primary vertex along the beam line using the fast primary vertex finding algorithm described in the text (left), and on the position of the primary vertex along the beam line after refitting with the tracks reconstructed at the HLT (right). The distributions are obtained using simulated multijet events with 35 pileup interactions on average and a flat $\hat{p}_{\text{T}}$ spectrum between 15 and 3000\GeV for the leading jet. Events are selected for which the scalar sum of the \pt of the jets is above 250\GeV. The first and last bin of each histogram contain the underflow and overflow entries, respectively.}
    \label{fig:resolutionFastPV}
\end{figure}

Figure~\ref{fig:resolutionFastPV} (left) shows that the resolution of the primary vertex along the beam line, $\Delta z$, is about 3\unit{mm} for simulated multijet events with 35 pileup interactions on average. Here, events are selected if the scalar sum of the calorimeter jet transverse momenta exceeds 250\GeV. The double-peak structure is caused by a bias in the FPV reconstruction that finds the primary vertex closer to the centre of the CMS detector than it is in reality in the simulation. This bias originates from the higher number of projected hits at the centre of the detector because of the detector geometry and pileup interactions. The efficiency of the FPV algorithm to reconstruct the primary vertex within 1.5\unit{cm} of its true position along the beam line is close to 99\%.

Since {\cPqb} tagging relies on the precise measurement of the displaced tracks with respect to the primary vertex, it is crucial to use tracks that use the information of both the pixel and the silicon strip tracker to improve the spatial and momentum resolutions. To reduce the HLT algorithm processing time, these tracks are only reconstructed when originating near the primary vertex and if they are close to the direction of the leading jets, sorted according to decreasing jet \pt. Up to eight jets with $\pt>30$\GeV and $\abs{\eta}<2.4$ are considered in an event. In the first step, the trajectories of charged particles are reconstructed from the pixel detector hits. To reduce the reconstruction time, tracks are only reconstructed when they have a longitudinal (transverse) impact parameter below 15 (2)\unit{mm} and are compatible with the direction of one of the jets. For simulated \ttbar events with 35 pileup interactions on average, this approach of regional pixel tracking reduces the track reconstruction time by a factor of almost 40 with respect to pixel tracking without constraints. Using the reconstructed pixel tracks, the efficiency to find the primary vertex within 0.2\unit{mm} of its true position along the beam line is around 97.5\%. To increase the efficiency even further, the variable
\begin{linenomath}
\begin{equation}
R = \frac{\sum_{j=1}^2\sum_{i}{\pt}^{i,j}}{\sum_{j=1}^2{\pt}^{j}}
\end{equation}
\end{linenomath}
is defined, where ${\pt}^{i,j}$ is the \pt of track $i$ associated with the leading or subleading jet ($j=1$ or 2) and ${\pt}^{j}$ is the \pt of jet $j$ obtained from the calorimeter deposits. To calculate $R$, tracks from the two leading jets are used if they have a $\chi^2$ of the track fit below 20, which reduces the effect of tracks reconstructed from a wrong combination of pixel hits. The impact of mismeasured tracks is reduced by setting the track \pt to 20\GeV if it is larger than this value. If the primary vertex position is not correctly reconstructed, the value of $R$ will be small. If $R<0.10$, the reconstruction of pixel detector tracks is run without the primary vertex position and using instead the direction of the two leading jets. The pixel detector tracks obtained in this way are then used to obtain a new position for the primary vertex, partially recovering the efficiency loss. The primary vertex position for all events is refitted using the reconstructed pixel detector tracks, resulting in a resolution that is much improved, as can be seen in Fig.~\ref{fig:resolutionFastPV} (right). Pairs of vertices that are closer than 70\micron to each other are merged into a single vertex. After the full procedure, the efficiency to find the primary vertex within 0.2\unit{mm} of its true position is larger than 98.5\%, and the resolution on the position of the primary vertex along the beam line is less than 60\micron, using simulated multijet events with 35 pileup interactions on average.

In the second step, the tracks are reconstructed using the information from the pixel and strip detectors. An iterative procedure is applied that is similar to the offline track reconstruction except for the number of iterations and the seeds used for track finding in each iteration. In the first iteration, the pixel tracks reconstructed as described above with $\pt>0.9$\GeV are used as seeds if they have a transverse (longitudinal) impact parameter below 1\,(3)\unit{mm}. For the second iteration, triplets of pixel hits are used with $\pt>0.5$\GeV and a transverse (longitudinal) impact parameter $<$0.5\,(1)\unit{mm}. The last iteration uses pairs of pixel hits with $\pt>1.2$\GeV and a transverse (longitudinal) impact parameter ${<}0.25$ (0.5)\unit{mm}. It is worth noting that the requirements on the impact parameter do not have a large impact on the reconstruction efficiency for displaced tracks. When refitting the primary vertex using the reconstructed tracks, the resolution on its position along the beam line further improves to less than 30\micron, as shown in Fig.~\ref{fig:resolutionFastPV}.

The reconstructed tracks and the refitted primary vertex are then used to reconstruct secondary vertices with the IVF vertex reconstruction algorithm. These vertices and tracks are then used as input for the CSVv2 algorithm described in Section~\ref{sec:ak4algos}. No dedicated training of the CSVv2 algorithm is used at the HLT, as studies have not shown any improvement in performance.
The processing time of regional tracking used for {\cPqb} tagging with up to eight leading jets with $\pt>30$\GeV is on average 87~ms, not including the jet reconstruction time. The processing time was evaluated using data with the highest number of pileup interactions observed in 2016 (49 pileup interactions on average) and selecting events using a trigger threshold of 250\GeV on the scalar sum of the calorimeter jet transverse momenta. As a comparison, the average global processing time of the HLT farm is limited to about 200\unit{ms} per event. The {\cPqb} tagging algorithm was run in about 6\% of the events accepted by the L1 trigger.

\begin{figure}[hptb]
  \centering
    \includegraphics[width=0.49\textwidth]{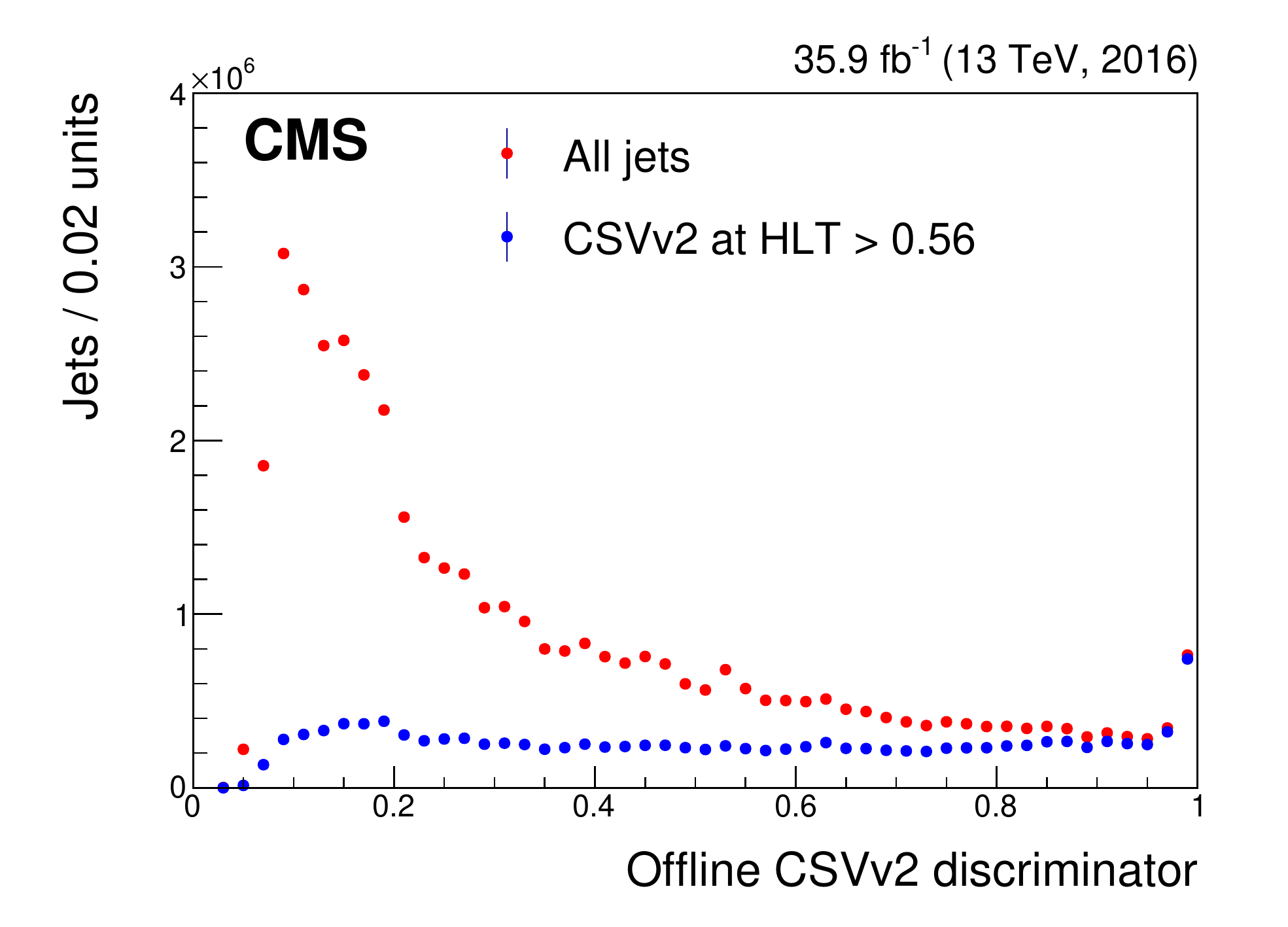}
    \includegraphics[width=0.49\textwidth]{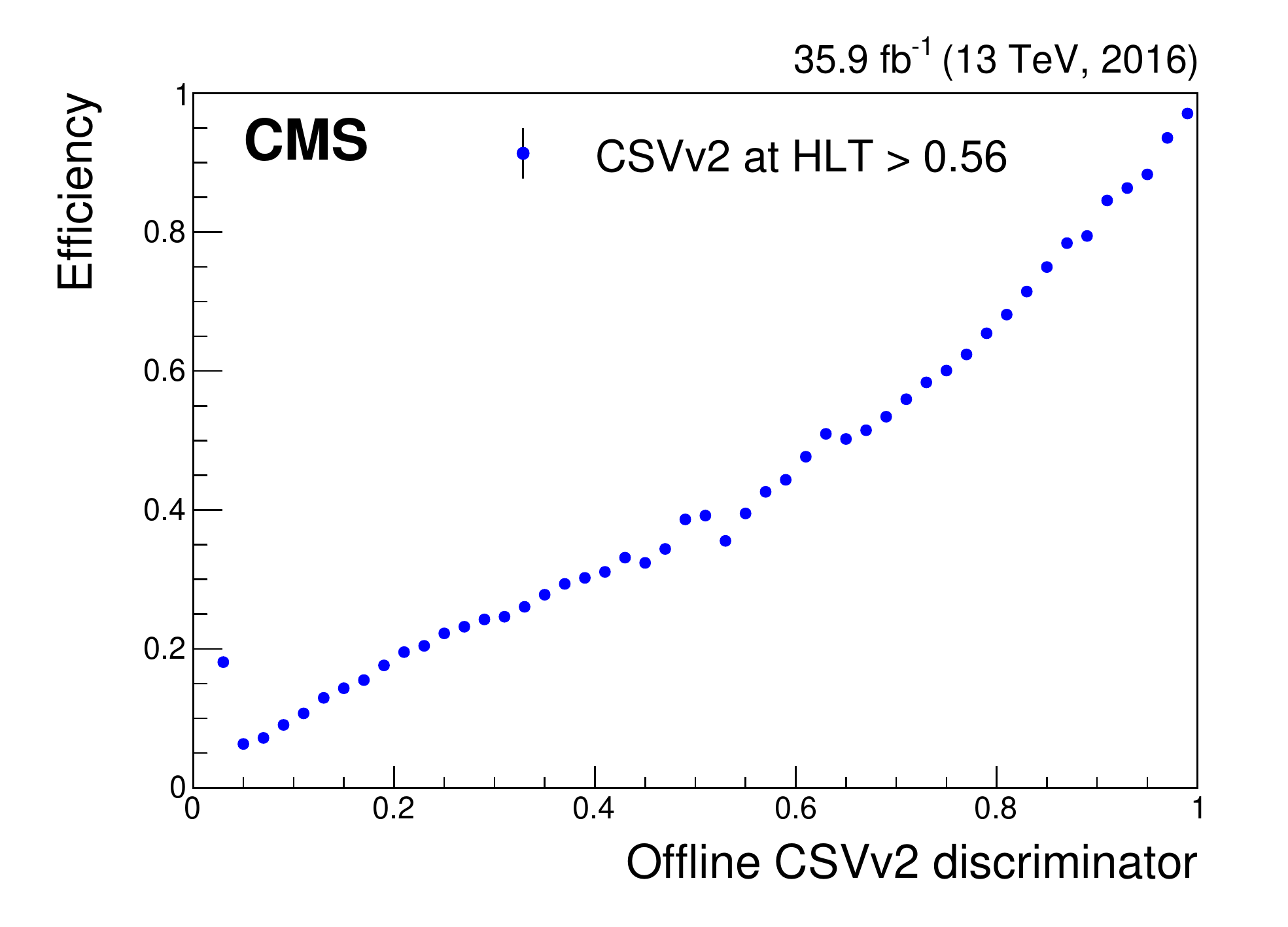}
    \caption{Offline CSVv2 discriminator distribution for all jets and for jets with a value of the CSVv2 discriminator at the HLT exceeding 0.56 (left), and {\cPqb} tagging efficiency at the HLT as a function of the offline CSVv2 discriminator value (right).}
    \label{fig:onlinebtagging}
\end{figure}

The performance of {\cPqb} tagging at the HLT is evaluated using data collected during 2016, selecting events with at least four calorimeter jets with $\pt > 45\GeV$ and $\abs{\eta}<2.4$ and with the sum of the \pt of the jets at the HLT above 800\GeV. Offline CSVv2 discriminator distributions are shown in Fig.~\ref{fig:onlinebtagging} using all jets (in red) as well as using jets with an HLT CSVv2 discriminator exceeding 0.56 (in blue). An estimate of the reduction factor for the trigger rate when requiring a single b tagged jet at HLT is determined as the number of jets passing the initial trigger, based on the sum of the \pt of the jets, divided by the number of jets passing the trigger and having an HLT CSVv2 discriminator above 0.56. The {\cPqb} tagging efficiency for a threshold of 0.56 on the HLT CSVv2 discriminator is shown as a function of the offline CSVv2 discriminator value in Fig.~\ref{fig:onlinebtagging} (right). In both panels, the structure at a discriminator value of $\approx 0.5$ is caused by jets from pileup interactions. In the right panel, the discontinuity indicates that these jets do not behave exactly in the same manner at the HLT and offline, due to their different track reconstruction. The larger efficiency for CSVv2 discriminator values below 0.05 is due to jets for which the chosen primary vertex at the HLT and offline is different. In particular, the primary vertex position is wrongly reconstructed at the HLT, resulting in an apparent displaced jet with a high CSVv2 discriminator value at the HLT and a small offline CSVv2 discriminator value. The impact of this effect is relatively small since there are only a few jets with an offline CSVv2 discriminator value below 0.05, as can be seen in the left panel.

\begin{figure}[hptb]
  \centering
    \includegraphics[width=0.49\textwidth]{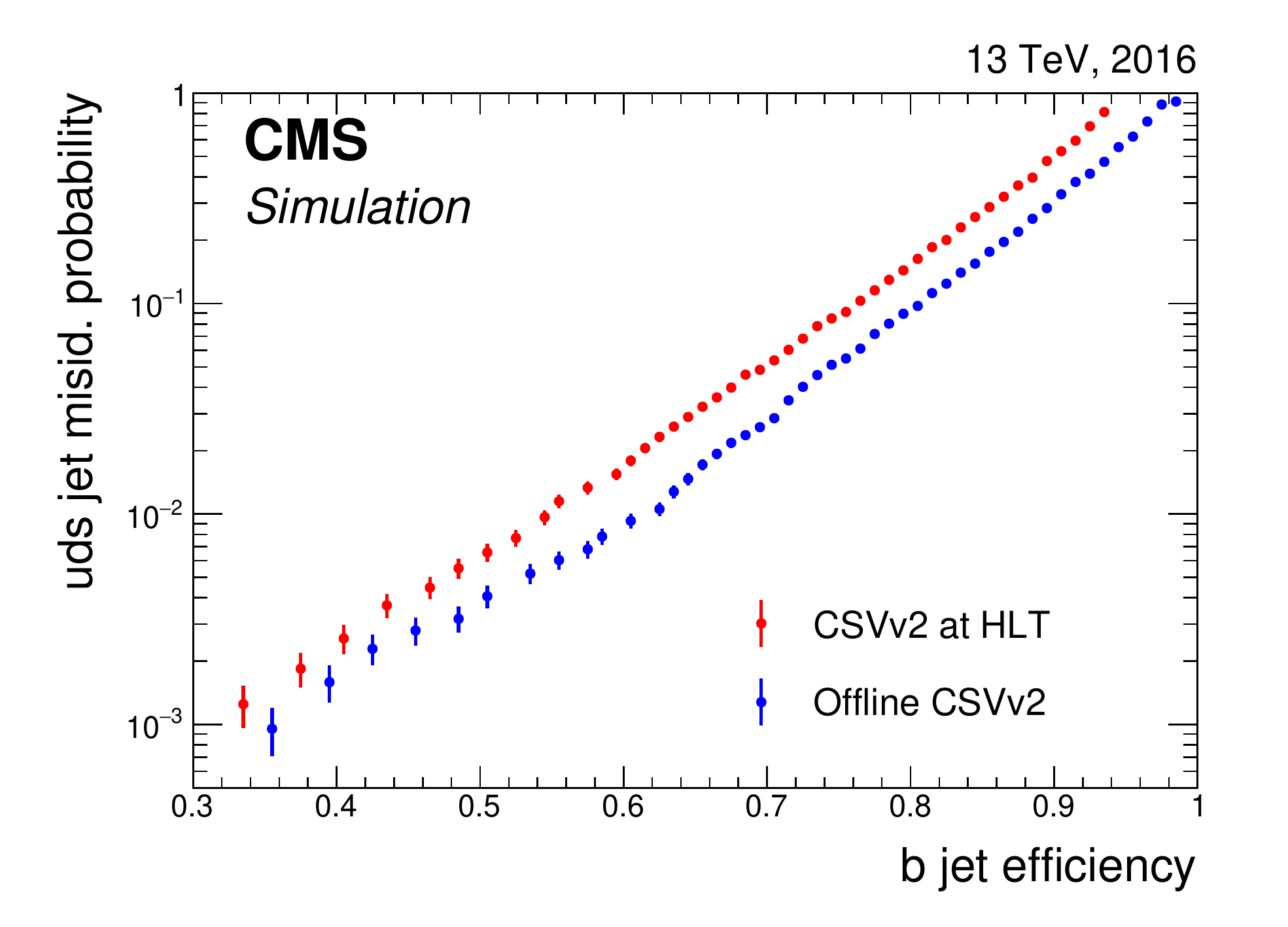}
    \includegraphics[width=0.49\textwidth]{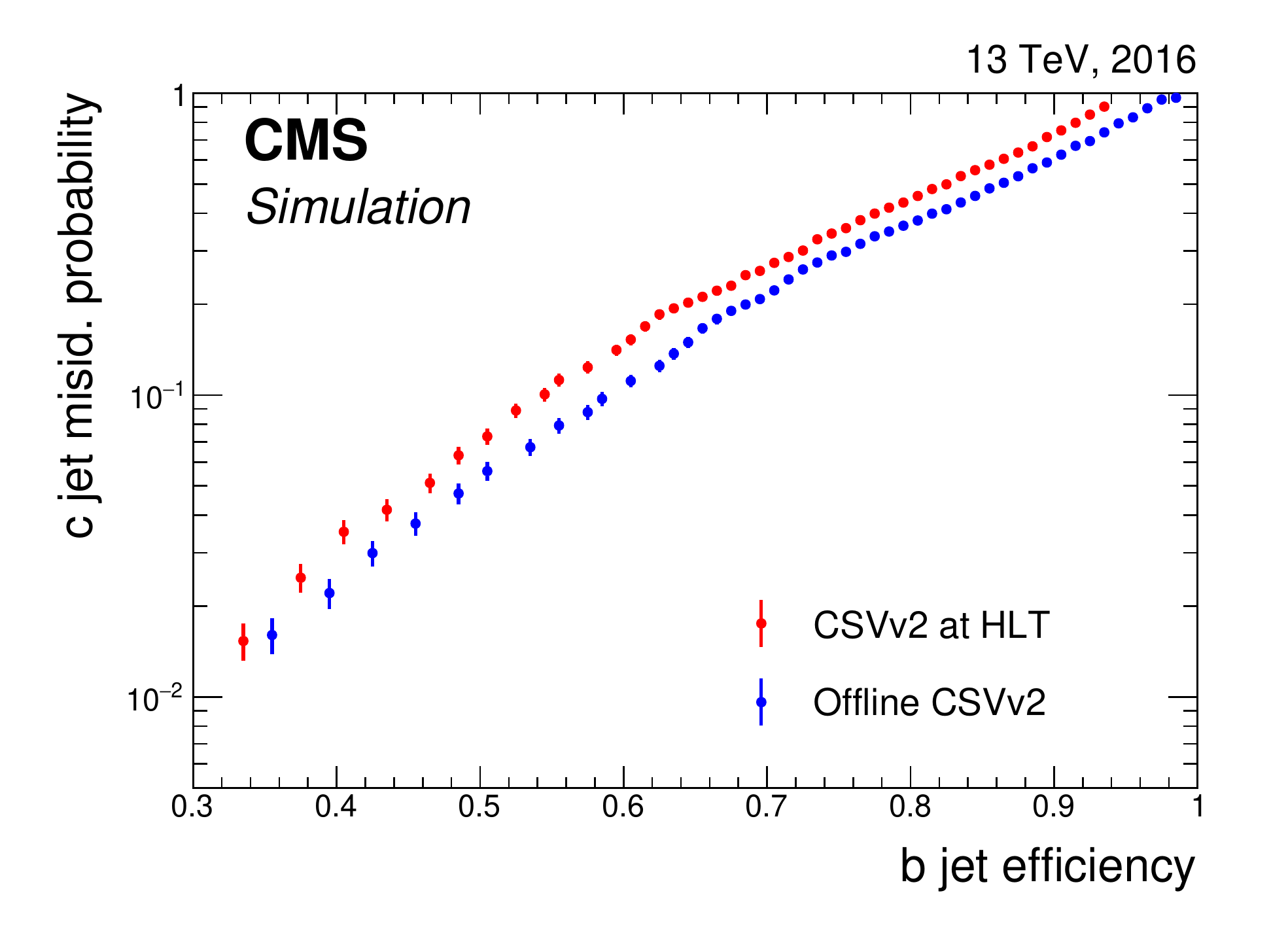}
    \caption{Comparison of the misidentification probability for light-flavour jets (left) and {\cPqc} jets (right) versus the {\cPqb} tagging efficiency at the HLT and offline for the CSVv2 algorithm applied on simulated \ttbar events for which the scalar sum of the jet \pt for all jets in the event exceeds 250\GeV. }
    \label{fig:onlinevsofflinebtagging}
\end{figure}

Figure~\ref{fig:onlinevsofflinebtagging} compares the HLT and offline {\cPqb} tagging performance using jets in simulated \ttbar events with 35 pileup interactions on average. Events are selected if the scalar sum of the jet transverse momenta exceeds 250\GeV. Up to eight leading jets are used with $\pt>30$\GeV and $\abs{\eta}<2.4$.
As expected, the {\cPqb} tagging performance of the offline reconstruction is better than at the HLT. The maximum {\cPqb} jet identification efficiency at the HLT is $\approx 95$\% because of three effects that occur more frequently at the HLT:
\begin{itemize}
\item The primary vertex is not reconstructed or not identified as the vertex corresponding to the jets on which the {\cPqb} tagging algorithm is applied.
\item Since the track reconstruction efficiency at the HLT is lower, it happens more often that less than two tracks are associated with the jet, resulting in no valid discriminator value being assigned to the jet.
\item There are at least two reconstructed tracks, but they do not pass the track selection requirements applied in the CSVv2 algorithm.
\end{itemize}

In the future, the {\cPqb} tagging performance at the HLT will be further improved by replacing the CSVv2 tagger with the DeepCSV tagger.

\section{Measurement of the tagging efficiency using data}
\label{sec:ak4eff}
In the previous sections, the performance of the taggers was studied on simulated samples. In this section, we present the methods used to measure the efficiency of the heavy-flavour tagging algorithms applied on the data. In Section~\ref{sec:ak4comm}, the data are compared to the simulation for a few input variables as well as for the output discriminator distributions. The measurement of the misidentification probability in the data is presented in Section~\ref{sec:negtag}. The tagging efficiency for {\cPqc} and {\cPqb} jets is presented in Sections~\ref{sec:SFc} and~\ref{sec:SFb}, respectively. Section~\ref{sec:SFiter} summarizes a method to measure data-to-simulation scale factors as a function of the discriminator value for the various jet flavours. The results of the various measurements are compared and discussed in Section~\ref{sec:SFconcl}.

\subsection{Comparison of data with simulation}
\label{sec:ak4comm}
The data are compared to simulation in different event topologies, chosen for their different jet flavour composition, and selected according to the following criteria:
\begin{itemize}
\item \textbf{Inclusive multijet sample}: Events are selected if they satisfy a trigger selection requiring the presence of at least one AK4 jet with $\pt>40$\GeV. Because of the high event rates only a fraction of the events that fulfill the trigger requirement are selected (prescaled trigger). The fraction of accepted events depends on the prescale value, which varies during the data-taking period according to the instantaneous luminosity. The data are compared to simulated multijet events using jets with $50 < \pt < 250\GeV$. This topology is dominated by light-flavour jets and contains also a contribution of jets from pileup interactions.
\item \textbf{Muon-enriched jet sample}: Events are considered if they satisfy an online selection requiring at least two AK4 jets with $\pt>40$\GeV of which at least one contains a muon with $\pt>5$\GeV. Also in this case, the trigger was prescaled. The data are compared to a sample of jets with $50 < \pt < 250\GeV$ and containing a muon selected from simulated muon-enriched multijet events. Because of the muon requirement this topology is dominated by jets containing heavy-flavour hadrons.
\item \textbf{Dilepton \ttbar sample}: At trigger level, events are selected by requiring the presence of at least one isolated electron and at least one isolated muon. Offline, the leading muon and electron are required to have $\pt>25$\GeV and be isolated, as expected for leptonic {\PW} boson decays~\cite{Khachatryan:2015hwa,Chatrchyan:2012xi}. Events are further considered if they contain at least two AK4 jets with $\pt>20$\GeV. In this event sample we expect an enrichment in {\cPqb} jets from top quark decays. There is also a small contribution from jets from pileup interactions due to the relatively low threshold on jet \pt.
\item \textbf{Single-lepton \ttbar sample}: Events are selected at trigger level by requiring the presence of at least one isolated electron or muon~\cite{Khachatryan:2015hwa,Chatrchyan:2012xi}. Offline, exactly one isolated electron or muon is required, satisfying tight identification criteria. The electron (muon) is required to have a $\pt>40~(30)$\GeV and $\abs{\eta}<2.4$. Events are further considered if they contain at least four jets with $\pt>25$\GeV. In this event sample a higher fraction of {\cPqc} jets is expected in comparison with the other samples. These {\cPqc} jets arise from the decay of the {\PW} boson to quarks.
\end{itemize}

\begin{figure}[hbtp]
  \centering
    \includegraphics[width=0.49\textwidth]{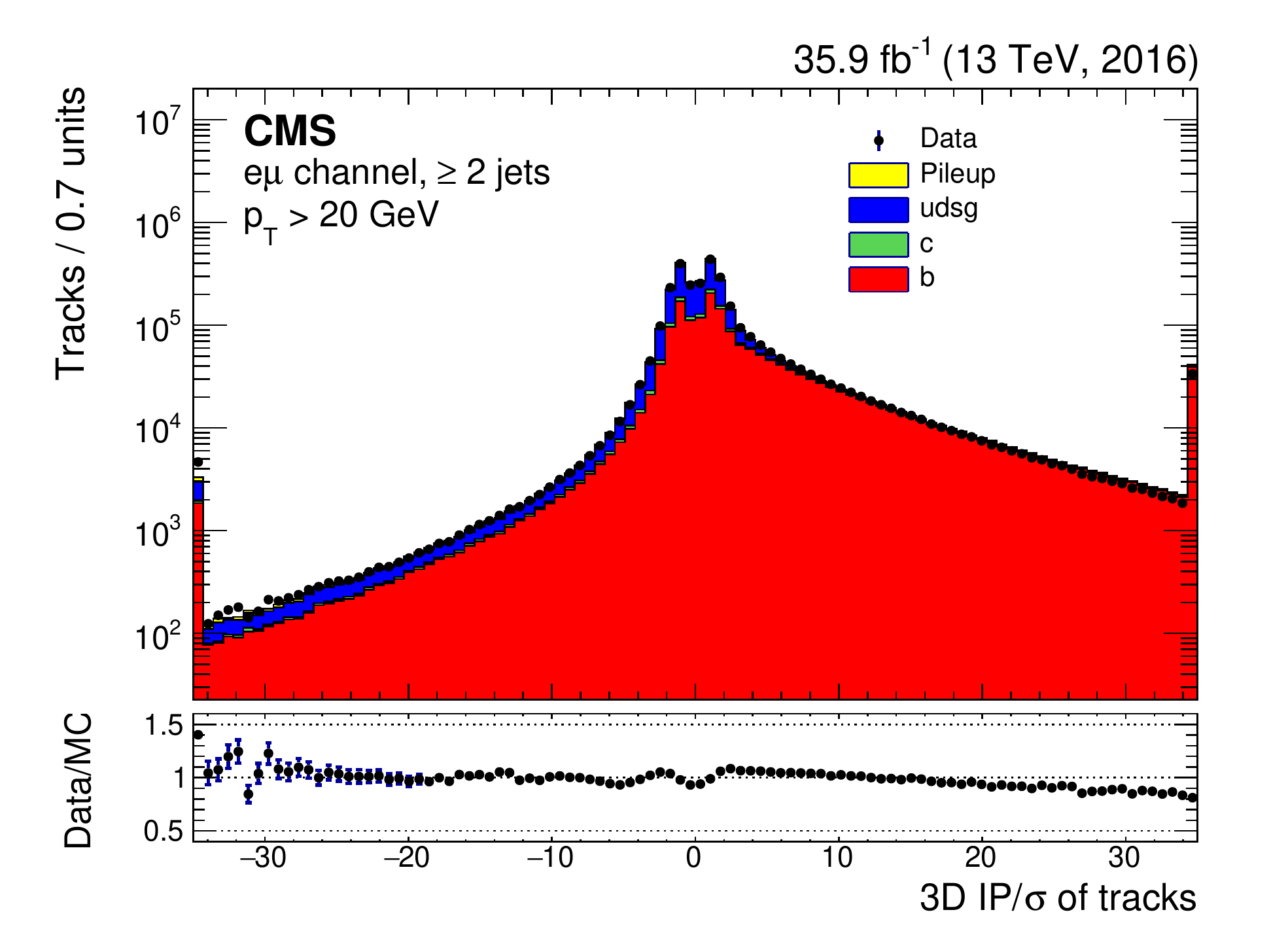}
    \includegraphics[width=0.49\textwidth]{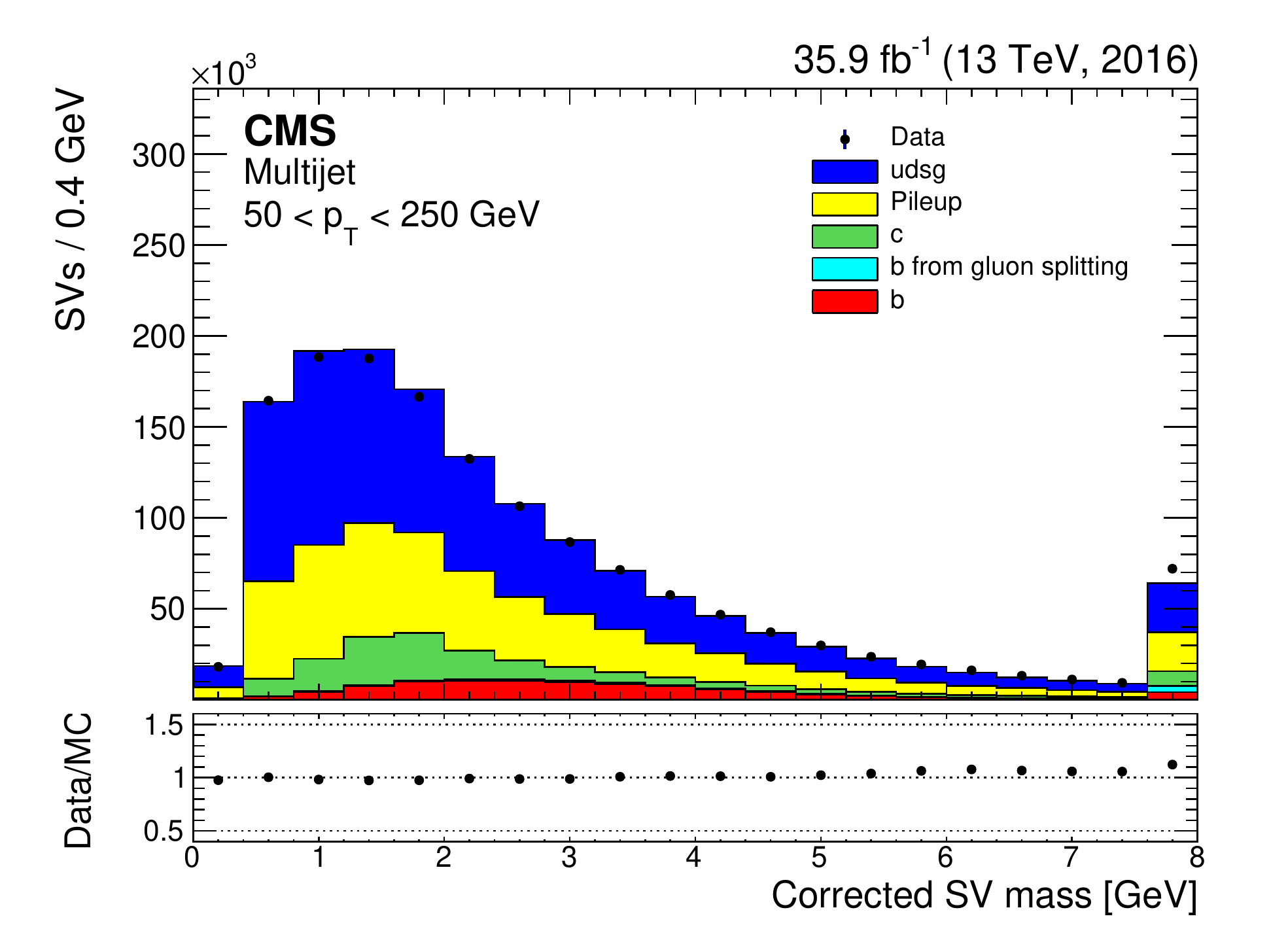}\\
    \includegraphics[width=0.49\textwidth]{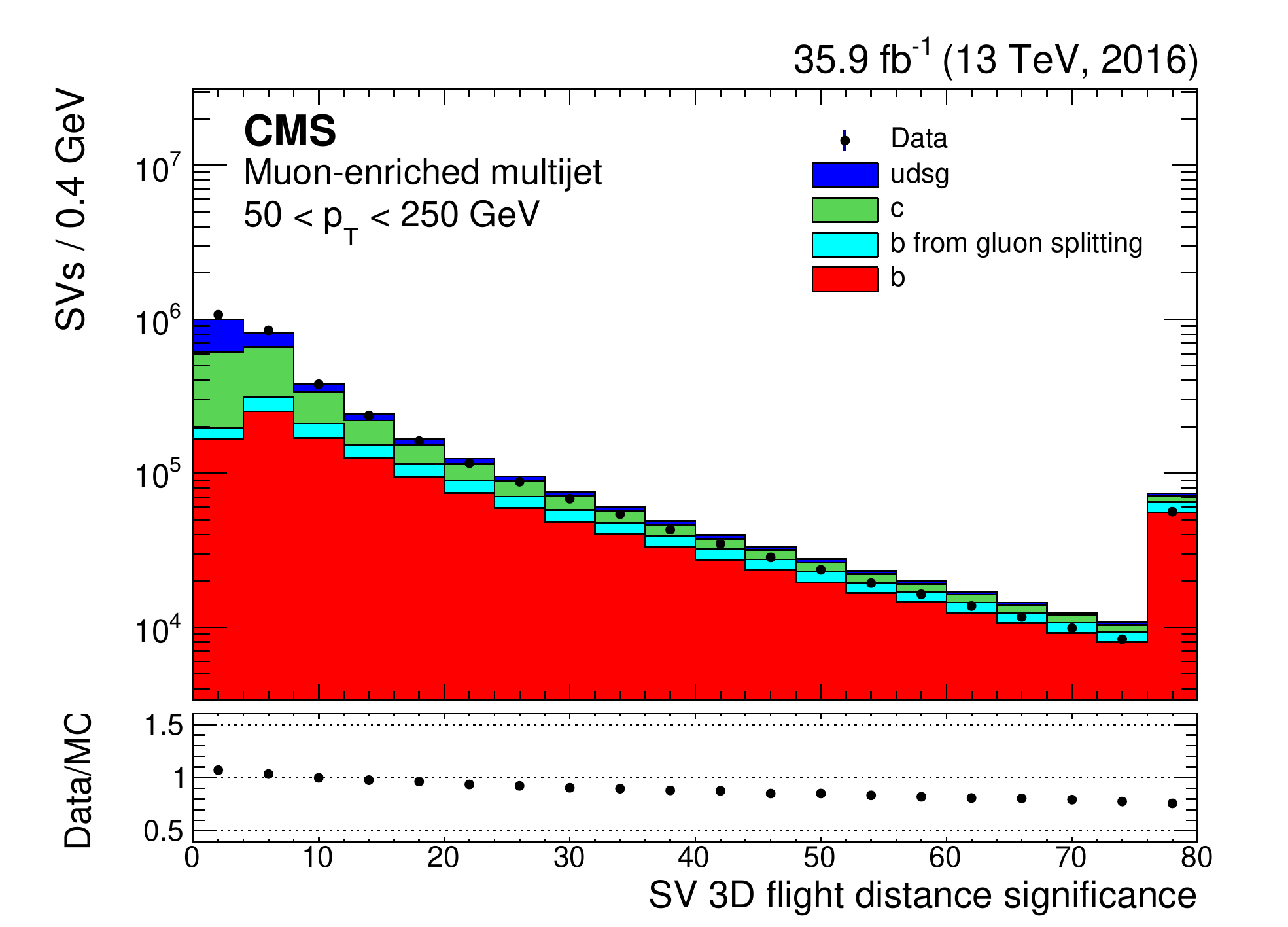}
    \includegraphics[width=0.49\textwidth]{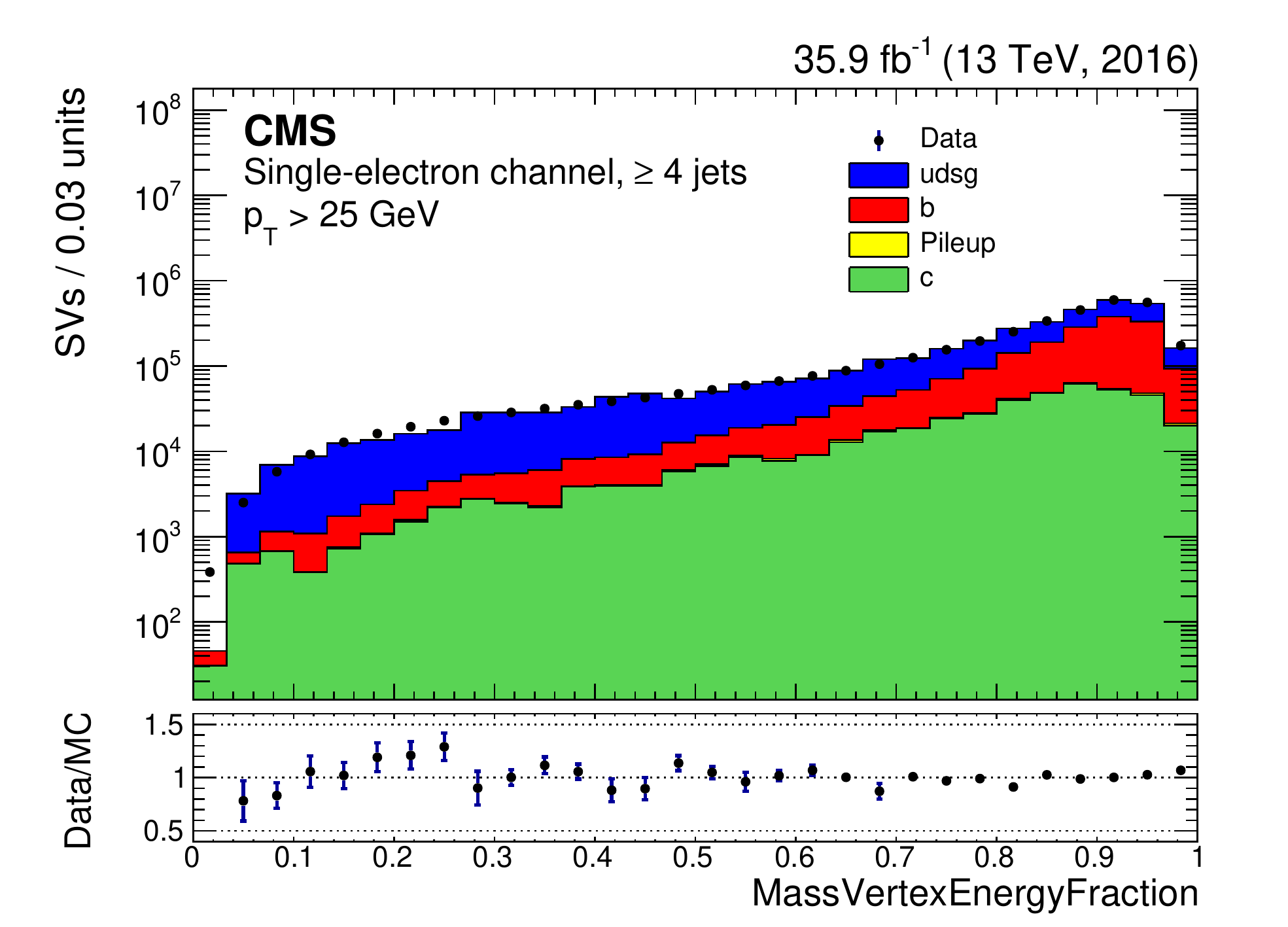}
    \caption{Examples of input variables used in heavy-flavour tagging algorithms in data compared to simulation. Impact parameter significance of the tracks in jets from the dilepton \ttbar sample (upper left), corrected secondary vertex mass for the secondary vertex with the smallest uncertainty in the 3D flight distance for jets in an inclusive multijet sample (upper right), secondary vertex flight distance significance for jets in a muon-enriched jet sample (lower left), and distribution of the massVertexEnergyFraction variable described in the text for jets in the single-lepton \ttbar sample (lower right). The simulated contributions of each flavour are shown with different colours. The total number of entries in the simulation is normalized to the number of observed entries in data. The first and last bin of each histogram contain the underflow and overflow entries, respectively.}
    \label{fig:ak4commvars}
\end{figure}

The distributions of all input variables and output discriminators in the four aforementioned event topologies are monitored to assess the agreement between data and simulation. Figure~\ref{fig:ak4commvars} shows a selection of four input variables. For the secondary vertex variables that are shown the secondary vertices are reconstructed with the IVF algorithm, discussed in Section~\ref{sec:vertexing}. In the top left panel, the 3D impact parameter significance of the tracks is shown for jets in the dilepton \ttbar sample. The observed discrepancy around zero is explained by the sensitivity of this variable to the tracker alignment and the uncertainty in the track parameters. The top right panel shows the corrected secondary vertex mass for the leading secondary vertex (sorted according to increasing uncertainty in the 3D flight distance), using jets in an inclusive multijet sample. The bottom left panel shows the 3D flight distance significance of the leading secondary vertex using jets in the muon-enriched jet sample. As was the case for the impact parameter significance, the disagreement between the data and the simulation is related to the sensitivity of this variable to the tracker alignment and the uncertainty in the track parameters and hence on the secondary vertex position. The bottom right panel shows the ``massVertexEnergyFraction'' variable, defined in Section~\ref{sec:ctagger}, using jets in the single-lepton \ttbar sample.

\begin{figure}[hbtp]
  \centering
    \includegraphics[width=0.49\textwidth]{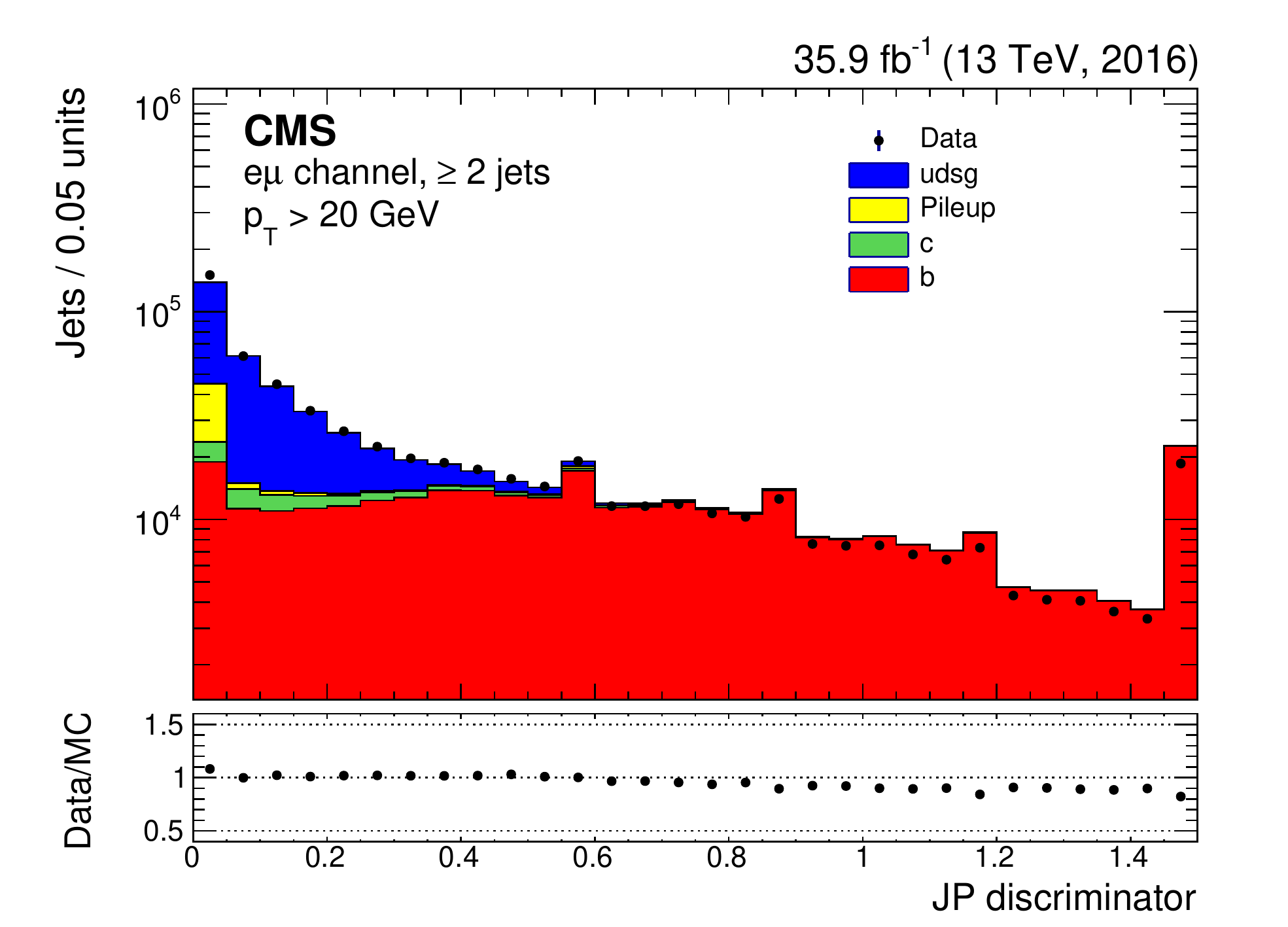}
    \includegraphics[width=0.49\textwidth]{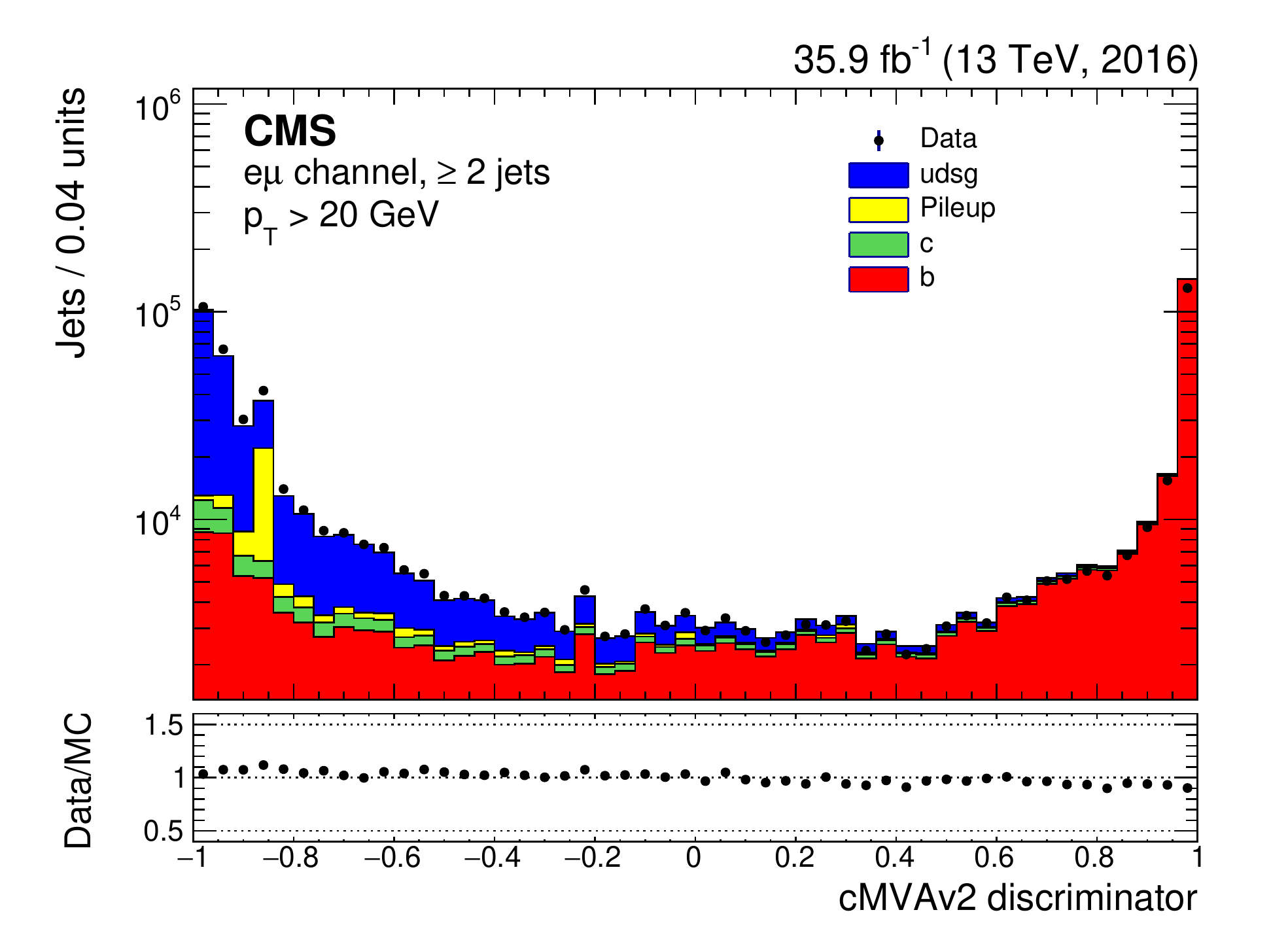}\\
    \includegraphics[width=0.49\textwidth]{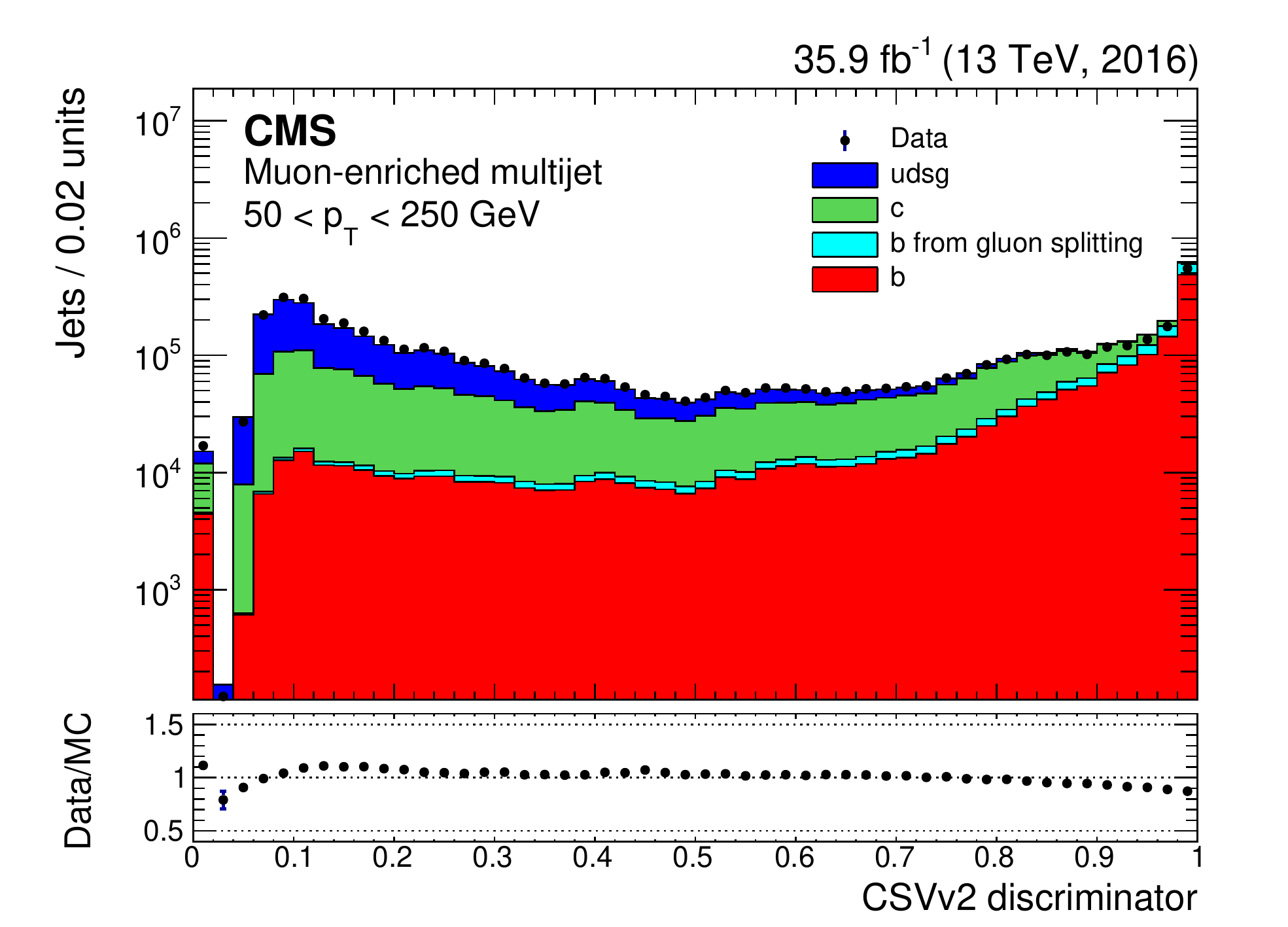}
    \includegraphics[width=0.49\textwidth]{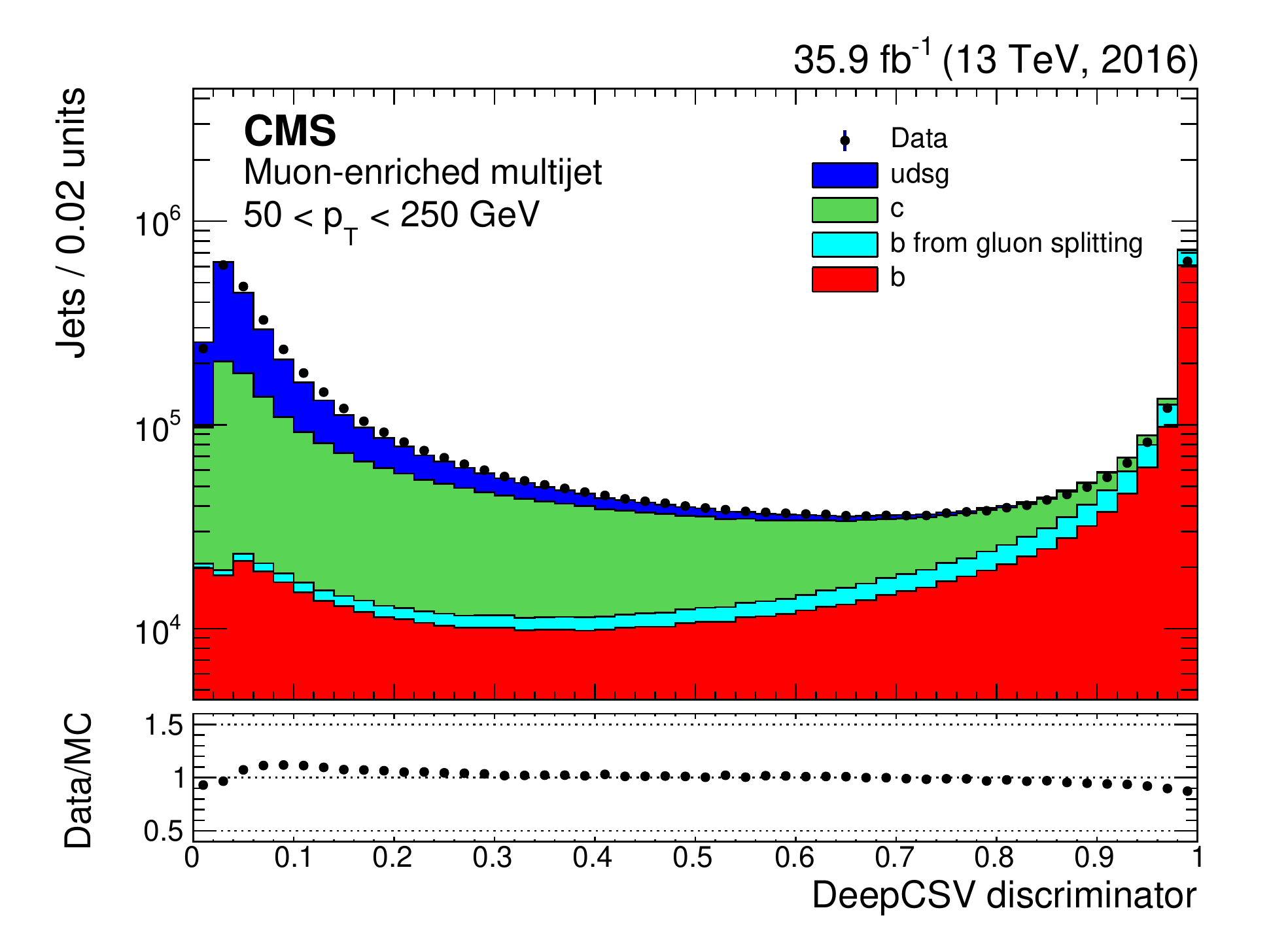}\\
    \includegraphics[width=0.49\textwidth]{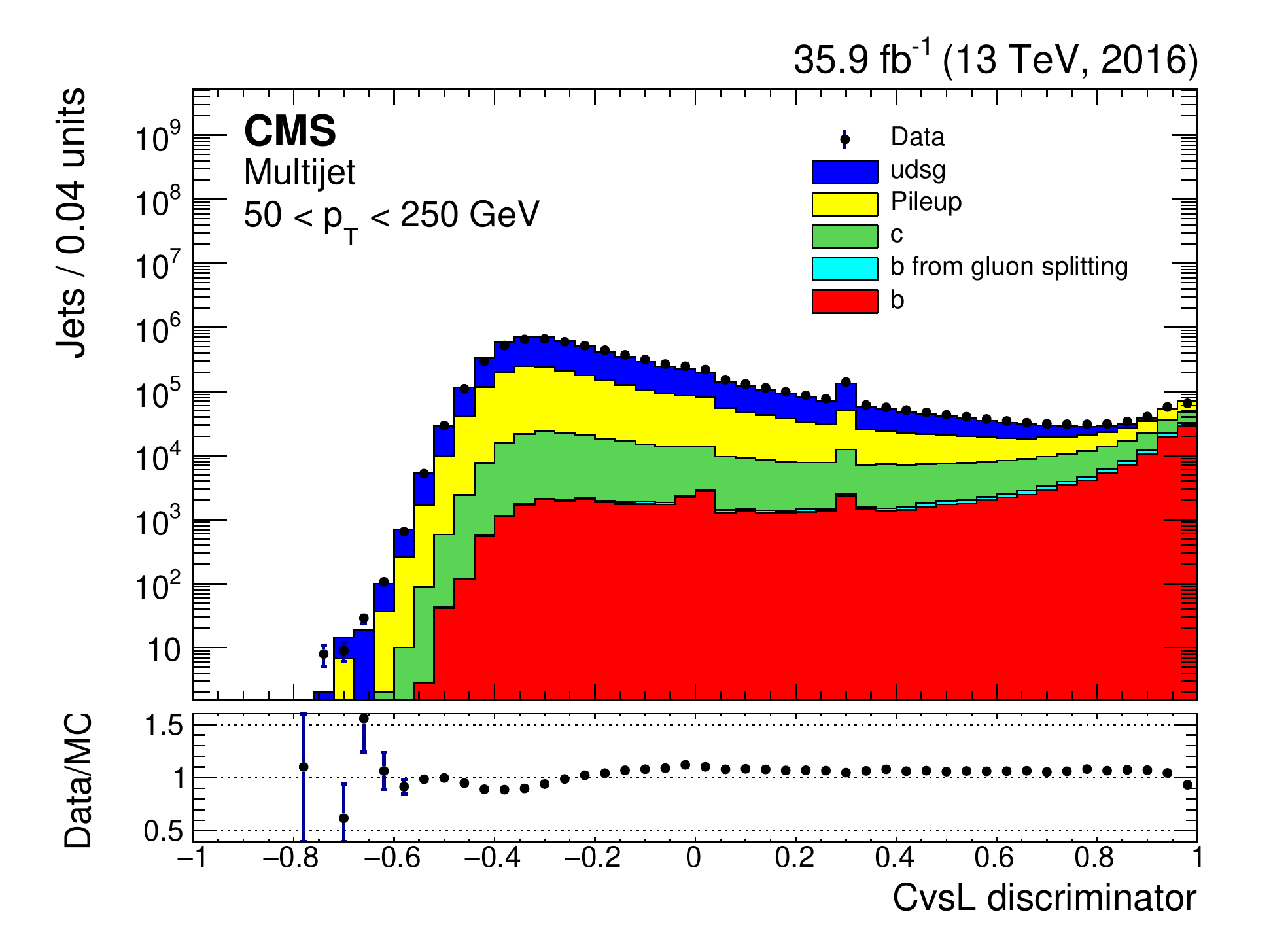}
    \includegraphics[width=0.49\textwidth]{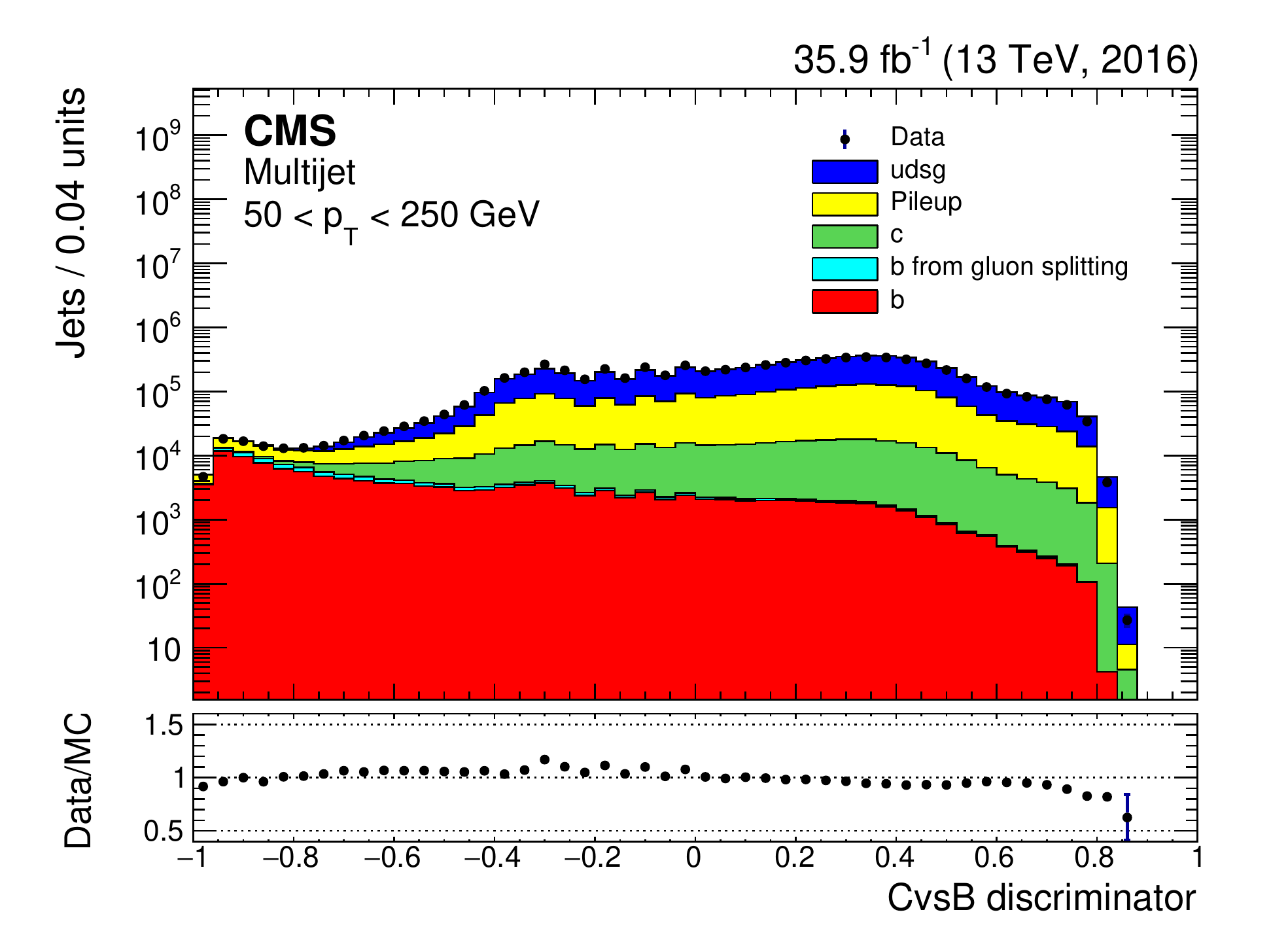}
    \caption{Examples of discriminator distributions in data compared to simulation. The JP (upper left) and cMVAv2 (upper right) discriminator values are shown for jets in the dilepton \ttbar sample, the CSVv2 (middle left) and DeepCSV (middle right) discriminators for jets in the muon-enriched multijet sample, and the CvsL (lower left) and CvsB (lower right) discriminators for jets in the inclusive multijet sample. The simulated contributions of each jet flavour are shown with different colours. The total number of entries in the simulation is normalized to the number of observed entries in data. The first and last bin of each histogram contain the underflow and overflow entries, respectively.}
    \label{fig:ak4commdiscr}
\end{figure}

While the simulation models the secondary vertex mass reasonably well, some discrepancies are observed for the impact parameter significance of the tracks and the secondary vertex flight distance. The imperfect modelling of the input variables will also have an impact on the modelling of the output discriminator distributions, which are shown in Fig.~\ref{fig:ak4commdiscr}. The upper panels show the JP and cMVAv2 discriminators using jets in the dilepton \ttbar sample. The discontinuities in the distribution of the JP discriminator values are due to the minimum track probability requirement of 0.5\%, as explained in Section~\ref{sec:JP}. The middle panels show the CSVv2 and DeepCSV discriminators using jets in the muon-enriched sample. The lower panels show the CvsL and the CvsB discriminators, using jets in the inclusive multijet sample. The discontinuities in both distributions arise from jets for which no tracks pass the track selection criteria, as discussed in Section~\ref{sec:ctagger}. Deviations of up to 20\% are observed at the highest discriminator values. These deviations may be related to the modelling of the detector in the simulation and to the accuracy of the generators in their modelling of the parton shower and hadronization. It is therefore important to measure the efficiencies directly from the data.
In physics analyses, the difference between the tagging efficiency in the data and simulation is then corrected for by taking into account a per jet data-to-simulation scale factor
\begin{linenomath}
\begin{equation}
SF_f=\varepsilon_f^{\text{data}}(\pt,\eta)/\varepsilon_f^{\text{MC}}(\pt,\eta),
\end{equation}
\end{linenomath}
where $\varepsilon_f^{\text{data}}(\pt,\eta)$ and $\varepsilon_f^{\text{MC}}(\pt,\eta)$ are the tagging efficiencies for a jet with flavour $f$ in data and simulation, respectively. For most of the efficiency measurements, the number of jets in the data is too limited to provide a dependence on the jet $\abs{\eta}$. For those methods, only the dependence on the jet \pt is measured. In simulation, the {\cPqb}/{\cPqc} tagging efficiency (misidentification probability) is defined as the number of {\cPqb}/{\cPqc} (light-flavour) jets that are tagged, according to the working point of a given algorithm (Section~\ref{sec:ak4algos}), with respect to the total number of {\cPqb}/{\cPqc} (light-flavour) jets. Using simulated events, the number of jets with flavour $f$ is determined by matching the jets with the generated hadrons.
In data, the tagging efficiency is measured with a pure sample of jets with a certain flavour $f$, using selection requirements that do not bias the jets with respect to the variables used in the tagging algorithm.

\subsection{The misidentification probability}
\label{sec:negtag}
The misidentification probability for light-flavour jets is measured with a sample of inclusive multijet events. The inclusive multijet data are collected using triggers requiring at least one jet above a certain \pt threshold, with $\pt>40$\GeV being its lowest value. Because of the high trigger rates for the lowest trigger thresholds, the triggers are prescaled. The selected events are reweighted to take into account the different prescales for each trigger threshold in order to obtain the same jet \pt distribution as if unprescaled triggers were used. The simulated events are reweighted to match the distribution of the number of pileup interactions in the data.

\begin{figure}[hbtp]
  \centering
  \includegraphics[width=.49\textwidth]{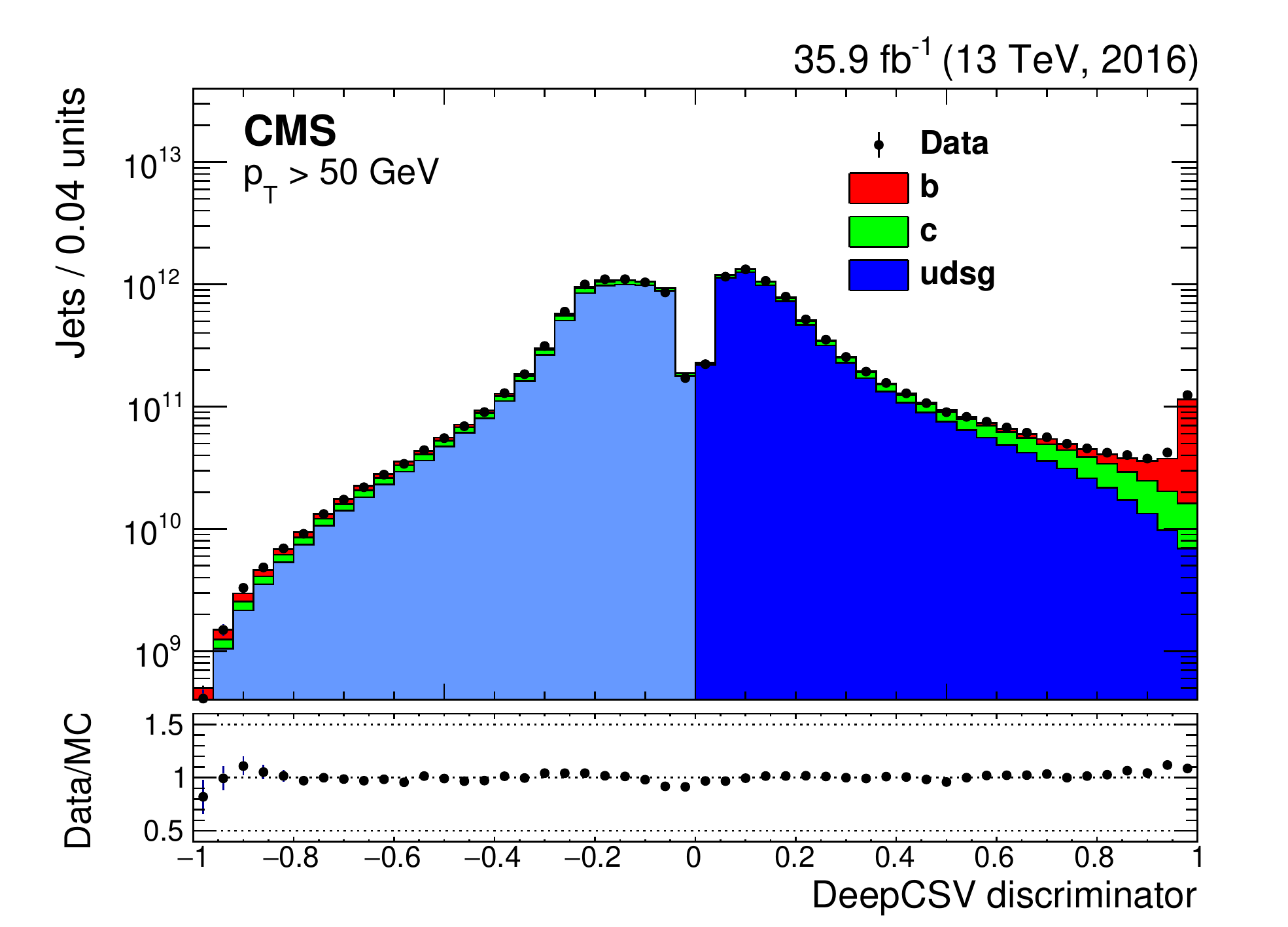}
  \includegraphics[width=.49\textwidth]{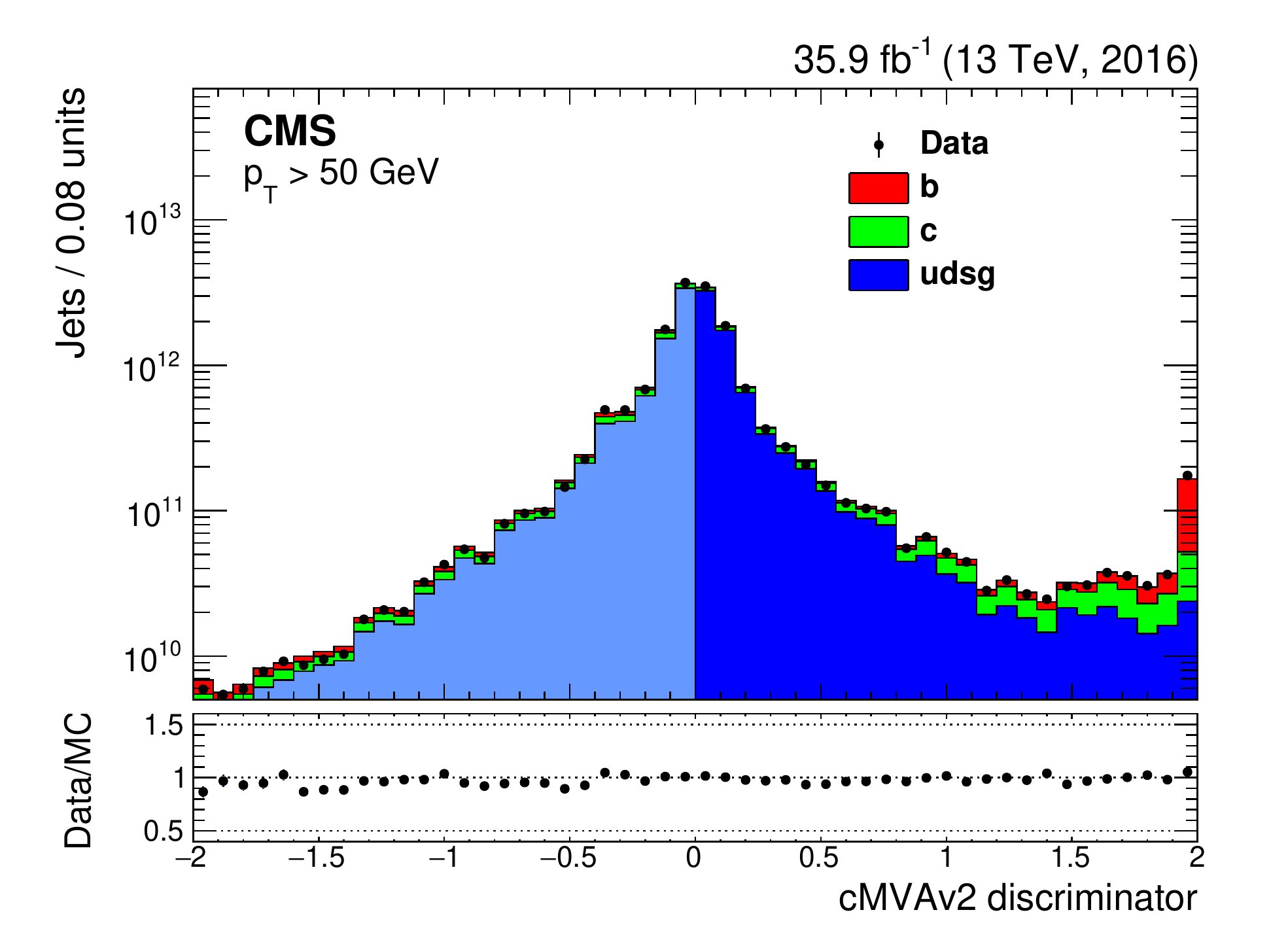}
  \caption{Distributions of the DeepCSV (left) and the cMVAv2 (right) discriminators for jets in an inclusive multijet sample. For visualization purposes the discriminator output of the negative DeepCSV tagger is shown with a negative sign. For the cMVAv2 tagger, the discriminator output of the positive tagger is shifted from $[-1,1]$ to $[0,2]$ and the discriminator values of the negative tagger are shown with a negative sign. The simulation is normalized to the number of entries in the data.}
  \label{fig:NegTag}
\end{figure}

The negative-tag method~\cite{BTV12001} is used for the measurement of the misidentification probability and the data-to-simulation scale factor, $SF_{\text{l}}$. The method is based on the definition of positive and negative taggers, which are identical to the default algorithms, except that for each jet only tracks with either positive or negative impact parameter values and secondary vertices with either positive or negative flight distance are used. To first order, the discriminator values for negative and positive taggers are expected to be symmetric for light-flavour jets, with nonzero values of the impact parameter and flight distance arising because of resolution effects. Some asymmetries are present for light-flavour jets due to long-lived hadrons, such as {\PKzS} and {\PgL} hadrons. The positive and negative discriminator distributions are presented in Fig.~\ref{fig:NegTag} using jets with \pt $>50$\GeV. For convenience, the discriminator values of the negative taggers are shown with a negative sign. Note that since the cMVAv2 and {\cPqc} tagger discriminator values range between $-1$ and 1, a shift was introduced such that the positive cMVAv2 discriminator is defined between 0 and 2, while the negative discriminator is shown with a negative sign and obtains values between $-2$ and 0. Deviations of up to 10\% are observed between the data and simulation for some discriminator values.

We define negative-tagged (positive-tagged) jets as the jets with a discriminator value of the negative (positive) tagger passing the working point of the tagger. The misidentification probability, $\varepsilon_{\text{l}}$, is determined from the fraction of negative-tagged jets passing the working point, $\varepsilon^-$, in an inclusive multijet sample
\begin{linenomath}
\begin{equation}
 \varepsilon_{\text{l}} = \varepsilon^- \, R_{\text{LF}}
\end{equation}
\end{linenomath}
where the correction factor $R_{\text{LF}}=\varepsilon_{\text{l}}^\text{MC} / \varepsilon^{-,\text{MC}}$ is the ratio of the misidentification probability of light-flavour jets to the negative tagging probability of all jets in simulation. The correction factor $R_{\text{LF}}$ is typically between 0.3 and 1, with the exact value depending on the working point and tagger.

Systematic uncertainties in the misidentification probability are related to possible effects that may have an impact on $R_\text{LF}$. In particular, the following systematic uncertainties are evaluated:
\begin{itemize}
\item \textbf{Fraction of heavy-flavour jets}: If the fraction of jets from heavy-flavour quarks in the negative-tag sample increases, the value of $R_\text{LF}$ decreases. The fraction of {\cPqb} jets has been measured by the CMS collaboration to agree with the simulation within $\pm$20\%~\cite{Chatrchyan:2012dk}. To assess the effect of this systematic uncertainty, the fraction of heavy-flavour jets in the simulation is varied by $\pm$20\%.
\item \textbf{Gluon fraction}: The fraction of gluon jets affects the misidentification probability in the simulation as well as the negative tagging probability, because of the larger track multiplicity in gluon jets compared to jets originating from light-flavour quarks. In addition, the fraction of gluon jets depends on the parton density and parton showering in the simulation. The systematic effect due to the uncertainty in the fraction of gluon jets is evaluated by varying the gluon fraction by $\pm$20\%~\cite{qcd-10-011}.
\item \textbf{{\PKzS} and {\PgL} decays (V$^0$)}: The observed numbers of reconstructed {\PKzS} and {\PgL} hadrons are found to be a factor of $1.30 \pm 0.30$ and $1.50 \pm 0.50$ larger than expected~\cite{TRK2010,V0}, respectively. To determine the nominal value of the data-to-simulation scale factor, the amount of reconstructed {\PKzS} or {\PgL} hadrons is reweighted in the simulation to be consistent with the observed yields. To obtain the size of the systematic effect due to the reweighting, the fraction of {\PKzS} and {\PgL} hadrons is varied by the uncertainty in the measured fraction, \ie by $\pm$30 and $\pm$50\%, respectively.
\item \textbf{Secondary interactions}: The rate of secondary interactions from photon conversions or nuclear interactions in the pixel tracker layers has been measured with a precision of $\pm$5\%~\cite{TRK2010,V0}. The number of secondary interactions is varied by this amount to obtain the systematic uncertainty in the data-to-simulation scale factor.
\item \textbf{Mismeasured tracks}: According to the simulation, there are more positive- than negative-tagged jets containing a reconstructed track that cannot be associated with a genuine charged particle. This is expected because the positive-tagged light-flavour jets contain {\PKzS} or {\PgL} hadrons, resulting in more hits and hence a higher probability for a wrong combination of those hits leading to a mismeasured track. To correct for this residual effect of mismeasured tracks, a $\pm$50\% variation of this contribution is taken into account for the systematic uncertainty in $R_\text{LF}$.
\item \textbf{Sign flip}: The number of jets with a negative tag is sensitive to the angular resolution on the jet axis and 3D impact parameter since these may affect the impact parameter sign. In particular, a difference between data and simulation in the probability of sign flips will affect the ratio of the negative tagging probability in data to that in simulation. The difference between data and simulation on the fraction of negative-tagged jets with respect to all tagged jets is measured with a muon-enriched jet sample and used to estimate the size of this systematic effect.
\item \textbf{Sampling}: The dependence of the data-to-simulation scale factor on the event topology is estimated by the trigger dependence of the scale factor. The scale factor is computed separately for each of the trigger requirements used to select the inclusive multijet sample. The maximum variation of the scale factor for these different measurements with respect to the nominal value using the unbiased jet \pt spectrum is taken as the size of the systematic effect.
\item \textbf{Pileup}: The simulated events are reweighted according to the observed amount of pileup interactions in data. A 5\% uncertainty in the total inelastic cross section of {\Pp}{\Pp} collisions~\cite{Aaboud:2016mmw} is propagated to the distribution of the number of pileup interactions to assess the impact of the uncertainty in the pileup reweighting.
\item \textbf{Statistical uncertainty in the simulation}: The limited amount of simulated multijet events is taken into account as an additional systematic uncertainty.
\end{itemize}

\begin{figure}[!hbtp]
  \centering
  \includegraphics[width=0.99\textwidth]{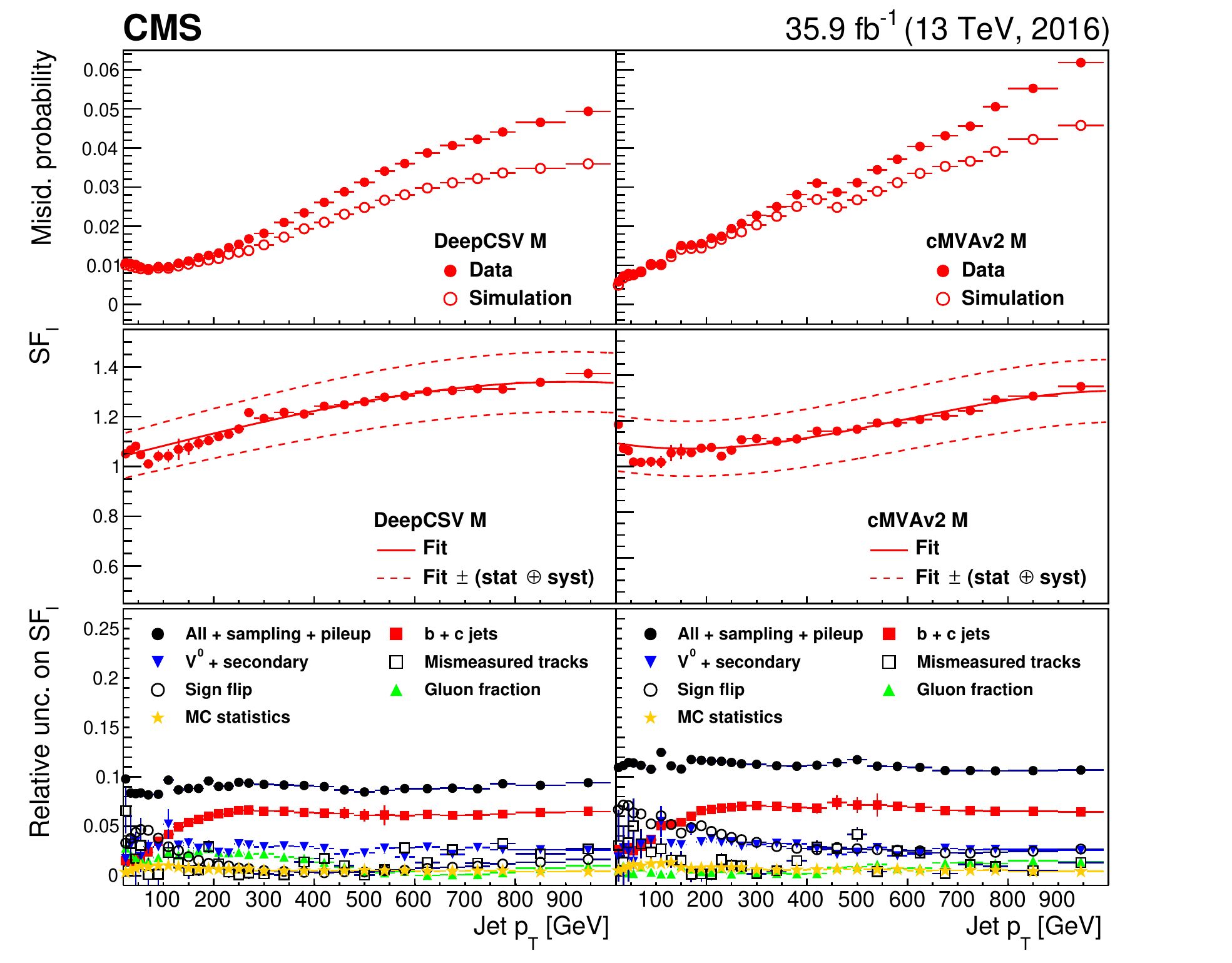}
    \caption{Misidentification probability, data-to-simulation scale factors, and relative uncertainty in the scale factors for light-flavour jets for the medium working point of the DeepCSV (left) and cMVAv2 (right) algorithm. The upper panels show the misidentification probability in data and simulation as a function of the jet \pt. The middle panels show the scale factors for light-flavour jets, where the solid curve is the result of a fit to the scale factors, and the dashed lines represent the overall statistical and systematic uncertainty in the measurement. The lower panels show the relative systematic uncertainties in the scale factors for light-flavour jets. The sampling and pileup uncertainties are not shown since they are below 1\%, but are included in the total systematic uncertainty covered by the black dots. }
    \label{Fig:negTag_WPM}
\end{figure}

Figure~\ref{Fig:negTag_WPM} shows an example of the measured misidentification probabilities, data-to-simulation scale factors, and relative systematic uncertainties for the medium working point of the DeepCSV and cMVAv2 taggers. In the top right panel of Fig.~\ref{Fig:negTag_WPM}, the ``step'' in the misidentification probability around 450\GeV is caused by the \pt- (and $\abs{\eta}$-) dependent weights for the jet flavours in the vertex categories in the training of the CSVv2 algorithm, discussed in Section~\ref{sec:CSVv2}.
The middle panels in Fig.~\ref{Fig:negTag_WPM} show the scale factors as a function of the jet \pt with the result of the fit superimposed. The fit functions are typically parameterized by a third degree polynomial with four free parameters. The dashed lines around the fit function represent the overall statistical and systematic uncertainty in the measurement. For jets with $\pt>1000\GeV$ the uncertainty in the scale factor is doubled. The scale factors are typically larger than one in a broad jet \pt range. The relative precision that is achieved on the scale factors for light-flavour jets when using {\cPqb} tagging algorithms is 5--10\% for the loose working point and rises to 20--30\% for the tight working point using jets with $20<\pt<1000\GeV$. The statistical uncertainty is typically a factor of 10 times smaller than the systematic uncertainty. For the {\cPqc} tagger, the relative precision varies between 3 and 7\% for the loose and tight working points, respectively. The reason for the smaller uncertainty for the {\cPqc} tagger compared to the {\cPqb} taggers is the different definition of the working points. The working points for the {\cPqc} tagger have a much higher misidentification probability for light-flavour jets, ranging from over 90\% for the loose working point to about a per cent for the tight working point, compared to 10\% and 0.1\%, respectively, for the {\cPqb} tagging algorithms (Section~\ref{sec:ak4algos}). The tight working point of the {\cPqc} tagger corresponds to a misidentification probability that is in between the loose and medium working points of the {\cPqb} taggers. Taking this into account, the corresponding systematic uncertainties are of a similar size.

\subsection{The \texorpdfstring{\cPqc}{c} jet identification efficiency}
\label{sec:SFc}
In this section, the methods are presented to obtain a jet sample enriched in {\cPqc} quark content, which is subsequently employed to measure the efficiency for (mis)identifying {\cPqc} jets in data. The efficiency in data and simulation is then used to determine the data-to-simulation scale factor for {\cPqc} jets, $SF_{\cPqc}$, for each algorithm and working point. The first method relies on the {\Wc} topology. The second method uses {\cPqc} jets from the {\PW} boson decay to quarks in the single-lepton \ttbar topology, where one of the {\PW} bosons decays into quarks and the other one into leptons.

\subsubsection{Measurement relying on \texorpdfstring{{\Wc}}{Wc} events}
\label{sec:Wc}
\begin{figure}[!htbp]
  \centering
    \includegraphics[width=0.8\textwidth]{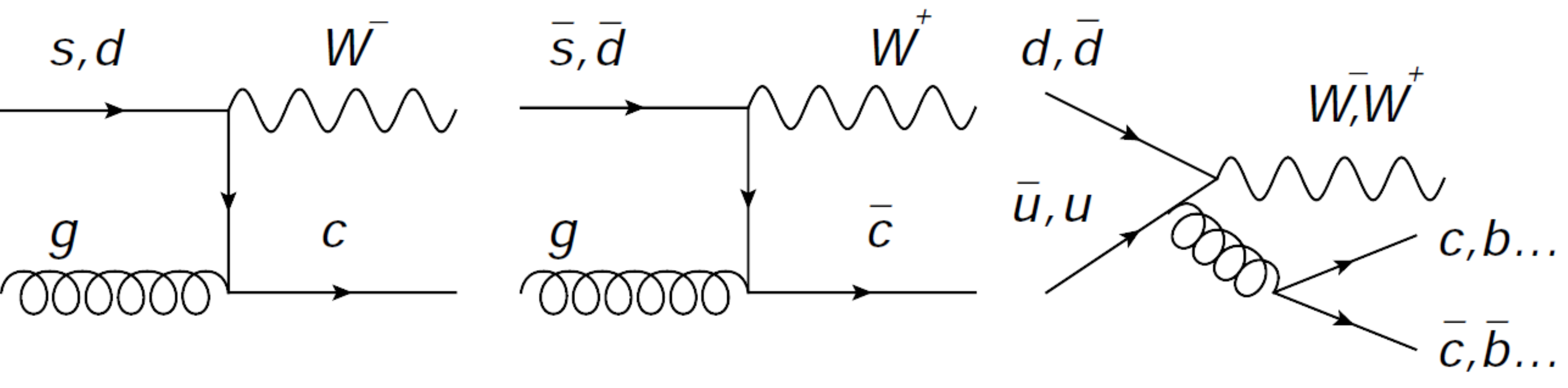}
  \caption{Leading order production of {\Wc} with opposite-sign electric charges (left and middle), and of ${\PW}+\text{\cPq\cPaq}$ through gluon splitting (right). In gluon splitting there is an additional {\cPqc} quark with the same sign as the {\PW} boson.}
  \label{fig:OSSS_diagram}
\end{figure}

The efficiency to identify {\cPqc} jets using heavy-flavour jet identification algorithms is measured with a sample enriched in {\cPqc} jets obtained from events with a {\PW} boson produced in association with a {\cPqc} quark. At leading order, the production of a {\PW} boson in association with a {\cPqc} quark proceeds mainly through $\cPqs + \Pg \to \PWmc$ and ${\cPaqs + \Pg}\to \PWmac$ as shown in Fig.~\ref{fig:OSSS_diagram} (left and middle). A key property of this production process is that the {\cPqc} quark and {\PW} boson have opposite-sign (OS) electric charge. The dominant background are ${\PW}+{\cPq\cPaq}$ events, which are produced with an equal amount of OS and same-sign (SS) events, as can be seen in Fig.~\ref{fig:OSSS_diagram} (right). After the event selection, a sample with a high purity of {\Wc} events is obtained by subtracting the SS distribution of a variable from the OS distribution for that variable. The remaining events are referred to as ``OS-SS''.

\begin{figure}[!htb]
  \centering
    \includegraphics[width=.49\textwidth]{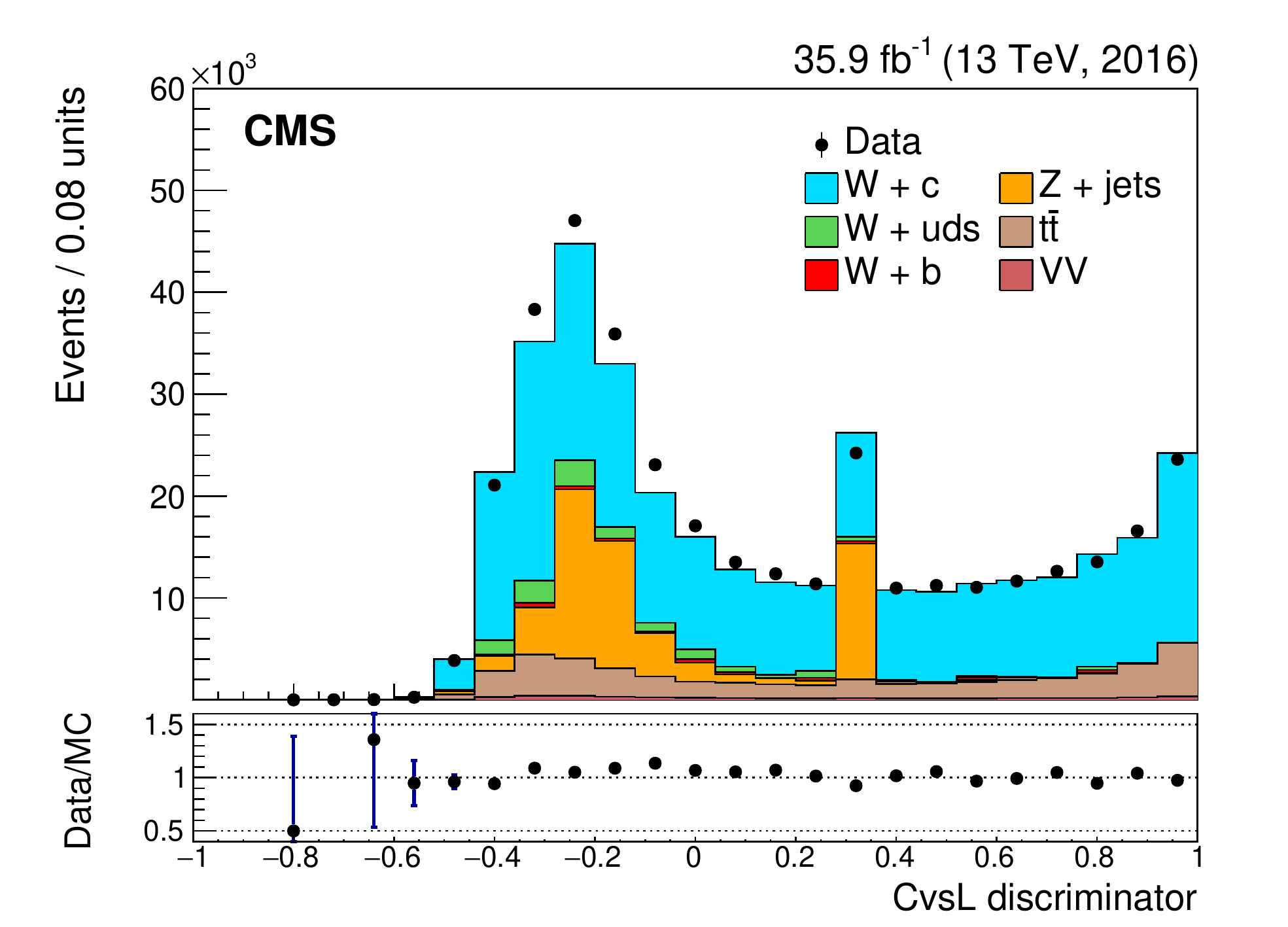}
    \includegraphics[width=.49\textwidth]{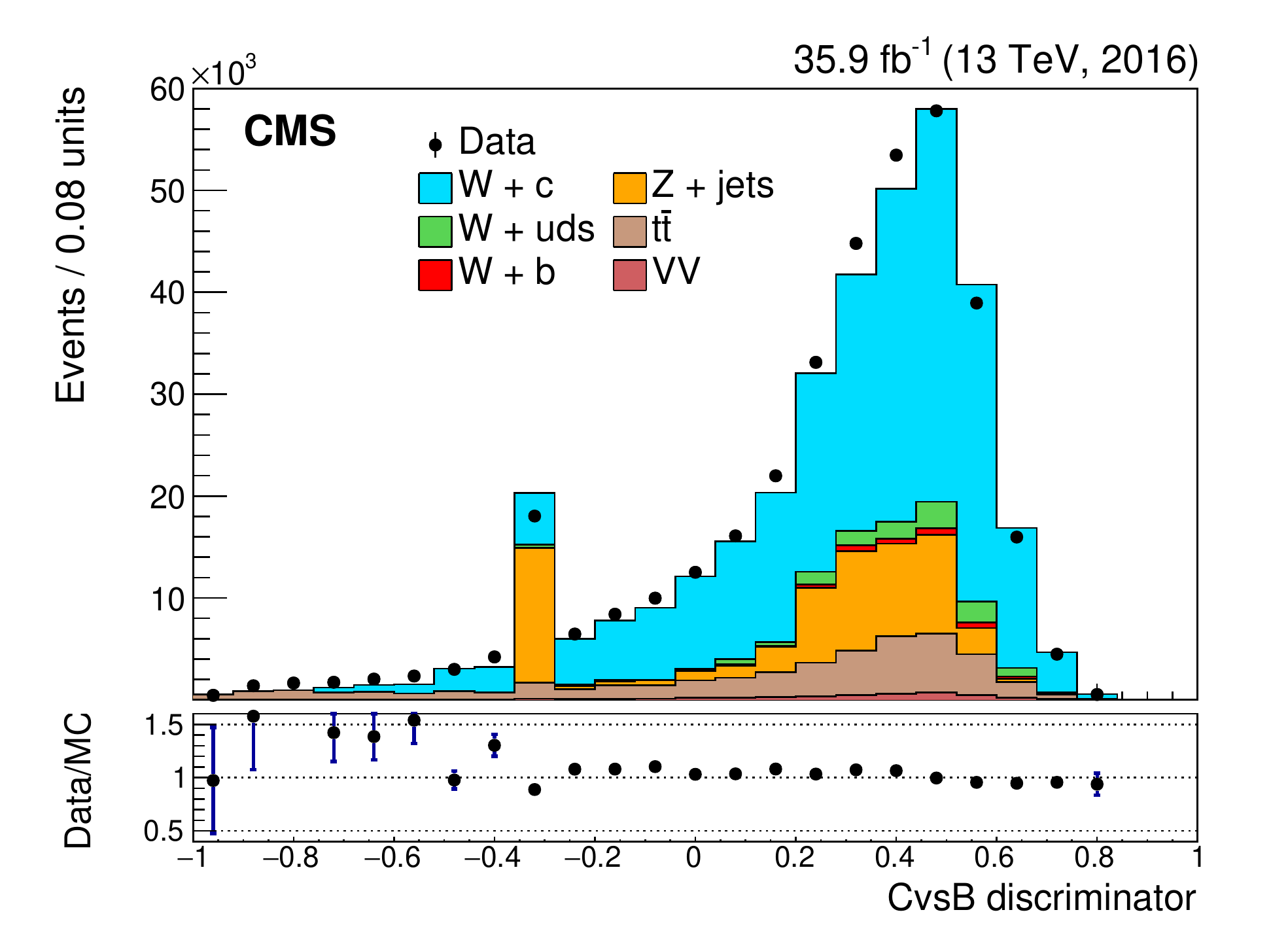}
    \caption{Distribution of the CvsL (left) and CvsB (right) discriminators in the $\PW \to \Pgm \Pgn$ and $\PW \to \Pe \Pgn$ channels after the OS-SS subtraction. The spikes originate from jets without a track passing the track selection criteria, as discussed in Section~\ref{sec:ctagger}. The last bin includes the overflow entries.}
    \label{fig:Wcharm_selection_ctagger}
\end{figure}

The {\Wc} events are selected according to the criteria of Ref.~\cite{WplusC}. Events are selected by requesting one isolated electron (muon) with a $\pt^{\ell}$ above 34 (26)\GeV and satisfying medium (tight) identification criteria~\cite{Khachatryan:2015hwa,Chatrchyan:2012xi}. When the event has more than one isolated electron or muon satisfying the selection criteria, the highest-\pt lepton is considered as the lepton from the {\PW} boson decay. The contribution from {\Zj} events is reduced by vetoing events with a same-flavour dilepton invariant mass between 70 and 110\GeV. To reduce the background from multijet events to a negligible level, the transverse mass $\MT = \sqrt{\smash[b]{\pt^{\ell} {\ptmiss} [1-\cos(\phi^{\ell}-\phi^{\ptmiss})]}}$ is required to be larger than 55\GeV. In this expression, $\phi^{\ell}$ and $\phi^{\ptmiss}$ ($\pt^{\ell}$ and \ptmiss) are the azimuthal angles (transverse momenta) of the isolated lepton and the \ptvecmiss vector, respectively. At least one jet is required in the tracker acceptance, with $\pt > 25\GeV$ and separated from the isolated lepton by $\Delta R>0.5$. In addition, the leading jet should contain a nonisolated soft muon among the jet constituents with $\pt < 25\GeV$. The charge of the {\cPqc} quark is determined from the charge of the soft muon inside the jet. The OS (SS) events are then defined as events for which the muon in the jet has the opposite (same) charge as the isolated lepton from the {\PW} decay.
After these requirements, the expected signal purity is about 60\% for $\PW \to \Pgm \Pgn$ events and 80\% for $\PW \to \Pe \Pgn$ events. Remaining {\Zj} and \ttbar events are the main sources of background for the $\PW \to \Pgm \Pgn$ channel, and \ttbar events for the $\PW \to \Pe \Pgn$ channel.
As an example, the distributions of the {\cPqc} tagger discriminators are shown in Fig.~\ref{fig:Wcharm_selection_ctagger} for the OS-SS sample, for the $\PW \to \Pgm \Pgn$ and $\PW \to \Pe \Pgn$ channels combined.

The efficiency to tag a {\cPqc} jet using a certain working point and tagger is obtained as the fraction of tagged {\cPqc} jets over the total number of {\cPqc} jets in {\Wc} events in the OS-SS sample:
\begin{linenomath}
\begin{equation}
\varepsilon_{\cPqc}=\frac{N(\Wc)^{\text{OS-SS}}_{\text{tagged}}}{N(\Wc)^{\text{OS-SS}}},
\end{equation}
\end{linenomath}
where the number of $\Wc$ events in data, ${N(\Wc)^{\text{OS-SS}}}$, is obtained as the observed number of OS-SS events times the fraction of {\Wc} events among these, derived from simulation as $f^{\text{MC}}_{\Wc} = 1-f^{\text{MC}}_{\text{bkg}}$. Analogously, $N(\Wc)^{\text{OS-SS}}_{\text{tagged}}$ corresponds to the number of {\Wc} events with a tagged {\cPqc} jet, obtained as the observed number of OS-SS events with a tagged {\cPqc} jet times the fraction of expected {\Wc} events with a tagged {\cPqc} jet, where the fraction is obtained as $f^{\text{tagged,MC}}_{\Wc} = 1-f^{\text{tagged,MC}}_{\text{bkg}}$. The simulated {\cPqc} jet tagging efficiency is obtained by repeating the procedure on simulated data.
Apart from the statistical uncertainty, the measurement may also be affected by several sources of systematic effects:
\begin{itemize}
\item \textbf{Background subtraction}: The number of {\Wc} events in data is obtained under the assumption that the fraction of (tagged) background events in data and simulation is the same. The effect of this assumption is quantified by varying $f^{\text{MC}}_{\text{bkg}}$ and $f^{\text{tagged,MC}}_{\text{bkg}}$ by 50\%. The impact on the measured efficiency for tagging {\cPqc} jets is of the order of 2\%, becoming one of the dominant uncertainties.
\item \textbf{Branching fraction of ${\PD} \to {\Pgm} X$ and fragmentation of ${\cPqc} \to {\PD}$}: The branching fractions for ${\PD} \to {\Pgm} X$ are varied to match the latest PDG data~\cite{Patrignani:2016xqp}. In particular, the branching fractions are shifted by $-2$\% for ${\PDp} \to {\Pgm} X$, $+13$\% for $\PDz \to {\Pgm} X$, and $+16$\% for ${\PD}_{\text{s}} \to {\Pgm} X$. In addition, also the fragmentation rate of a {\cPqc} quark to a {\PD} meson is varied to be consistent with the PDG data~\cite{Gladilin:2014tba}. This implies the following \PYTHIA~8 variations: $+37$\% for ${\cPqc}\to {\PDp}$, $-9$\% for ${\cPqc} \to {\PDz}$, and $-33$\% for ${\cPqc} \to {\PD}_{\text{s}}$. The difference in the measured {\cPqc} jet tagging efficiency after this simultaneous variation is less than 1\% and is taken as a systematic uncertainty.
\item \textbf{Number of tracks}: The uncertainty in the modelling of the number of selected tracks per jet in the simulation is taken into account by reweighting the distribution to match the data and remeasuring the data-to-simulation scale factor. The difference between the nominal scale factor value and the one after reweighting is less than 1\%.
\item \textbf{Soft-muon requirement}: Requiring a muon in a jet may introduce a potential bias in the efficiency measurement when the tagger also relies on muon variables, as is the case for the {\cPqc} tagger and the cMVAv2 tagger. The bias may arise if the tagger response is different for jets with and without a soft lepton. The potential bias is estimated by repeating the measurement using a modified version of the tagger, which treats the muon as a track and assigns a default value to the soft-muon input variables. The difference between the values measured with the modified tagger and the default one is taken as systematic uncertainty. The effect of this variation is less than 3\%. This is the dominant systematic uncertainty.
\item \textbf{Jet energy scale}: Since the measurements are performed in bins of jet \pt, the fraction of jets in each bin may vary depending on the jet energy corrections. The data-to-simulation scale factors are remeasured after varying the jet energies by $\pm$1 standard deviation of the nominal jet energy scale. The systematic effect due to this variation is less than 1\%.
\item \textbf{Electron and muon efficiency}: The uncertainties related to the lepton reconstruction and identification are taken into account by varying the corresponding correction factors within their uncertainty and reevaluating the efficiency for tagging {\cPqc} jets. The effect of this variation is smaller than 1\%.
\item \textbf{Pileup}: The effect of the uncertainty in the number of additional pileup interactions is evaluated as described in Section~\ref{sec:negtag}, having an impact on the {\cPqc} tagging efficiency below 1\%.
\item \textbf{Factorization and renormalization scales}: In Ref.~\cite{WplusC} the normalized cross section for {\Wc} events has been measured and the impact of the factorization and renormalization scales used at matrix element and parton shower levels was evaluated. The systematic uncertainty related to the variation of these scales was found to be well below 1\% because of the cross section normalization. When measuring data-to-simulation scale factors, this systematic uncertainty also cancels in the ratio.
\item \textbf{Parton distribution functions}: The NNPDF parton densities are varied within their uncertainties resulting in additional templates for the systematic uncertainty. The effect was found to be less than 1\%.
\end{itemize}
The total systematic uncertainty in the data-to-simulation scale factor measurement is obtained as the quadratic sum of the individual systematic uncertainties.

\begin{figure}[!htb]
  \centering
    \includegraphics[width=.49\textwidth]{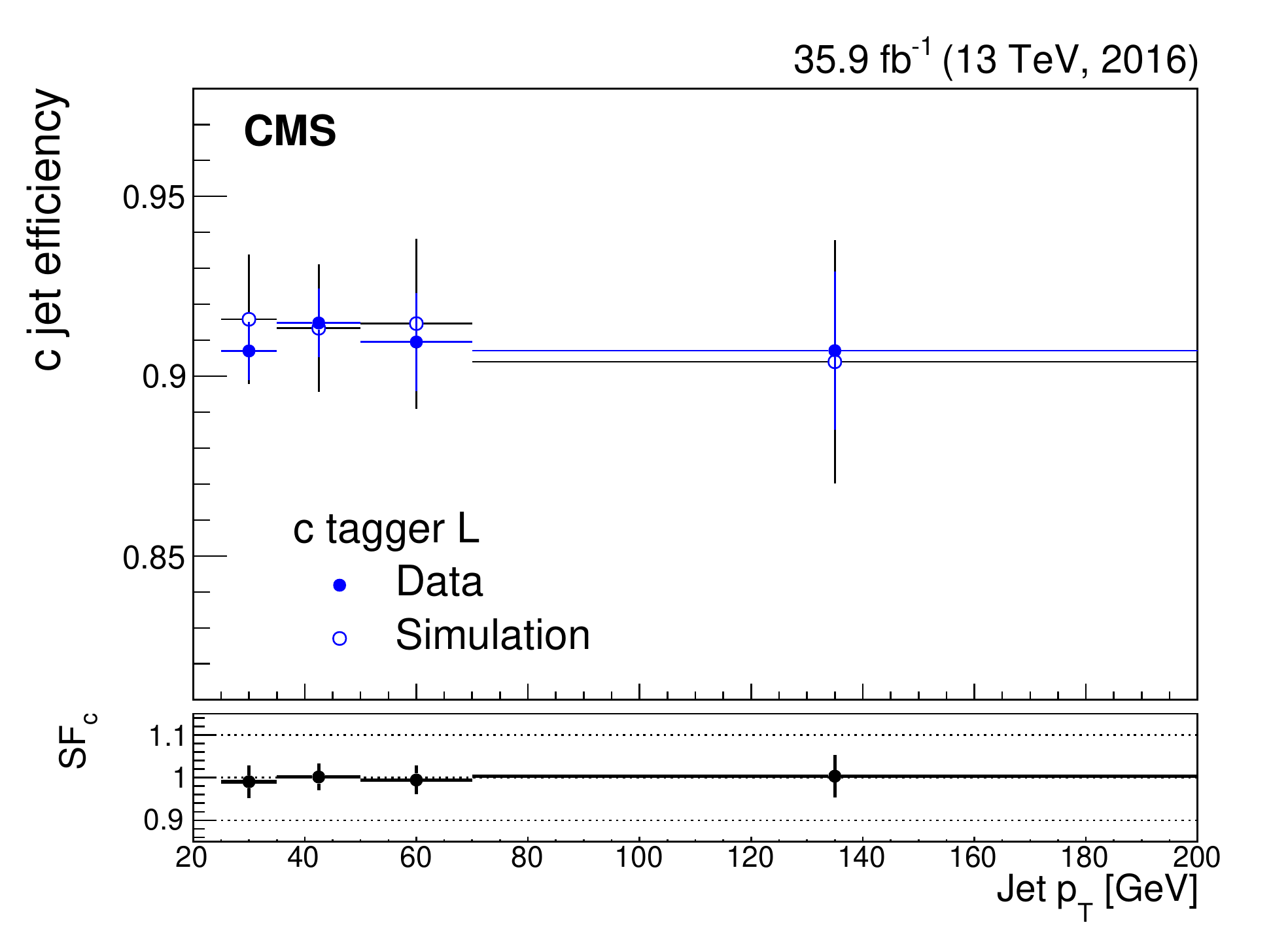}
    \includegraphics[width=.49\textwidth]{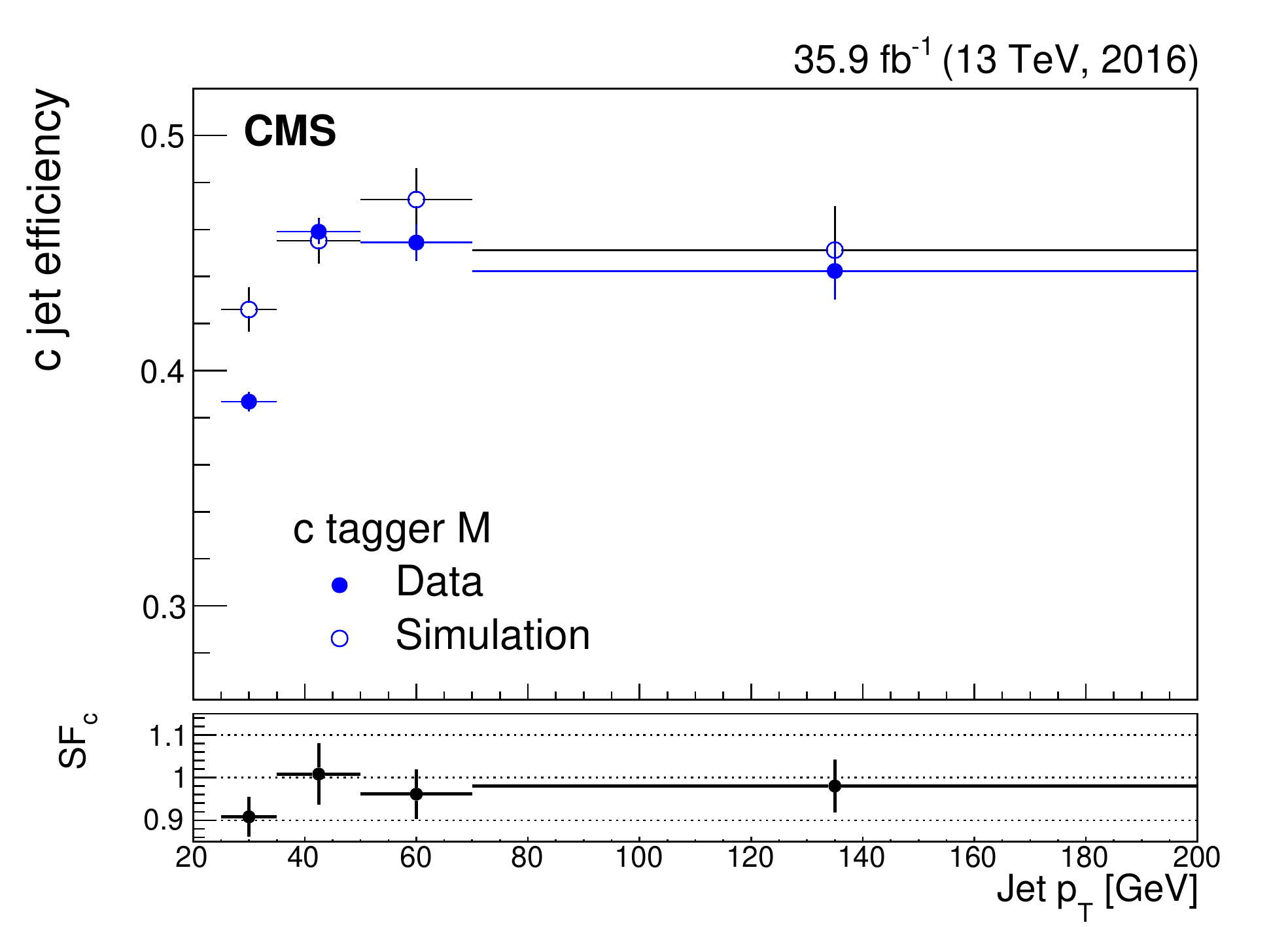}
    \caption{Efficiency for tagging {\cPqc} jets in data and simulation as a function of the jet \pt, and corresponding data-to-simulation scale factors (bottom panels) for the loose (left) and medium (right) working points of the {\cPqc} tagger.}
  \label{fig:eff_mu_vertex}
\end{figure}

The {\cPqc} jet tagging efficiency and the data-to-simulation scale factor $SF_{\cPqc}$ are computed as a function of jet \pt and presented in Fig.~\ref{fig:eff_mu_vertex} for the loose and medium working points of the {\cPqc} tagger. Scale factors for misidentifying {\cPqc} jets are also derived for the {\cPqb} tagging algorithms.

\subsubsection{Measurement relying on the single-lepton \texorpdfstring{\ttbar}{ttbar} events}
\label{sec:mauro}
If a {\PW} boson decays hadronically, the decay contains a {\cPqc} quark in about 50\% of the cases. Therefore, in a pure sample of single-lepton \ttbar events, about one event out of two will contain a {\cPqc} jet. Because of the particular decay chain of the top quark, the energy of up-type quarks from the {\PW} boson decay is, on average, larger than for down-type quarks. This property, verified in simulated \ttbar events, is used to obtain samples of jets enriched and depleted in {\cPqc} quarks. The {\cPqc} tagging efficiency is obtained by fitting the distribution of a variable in both of these samples simultaneously to the data, as will be explained in the following.

\begin{figure}[tbp]
\includegraphics[width=0.32\textwidth]{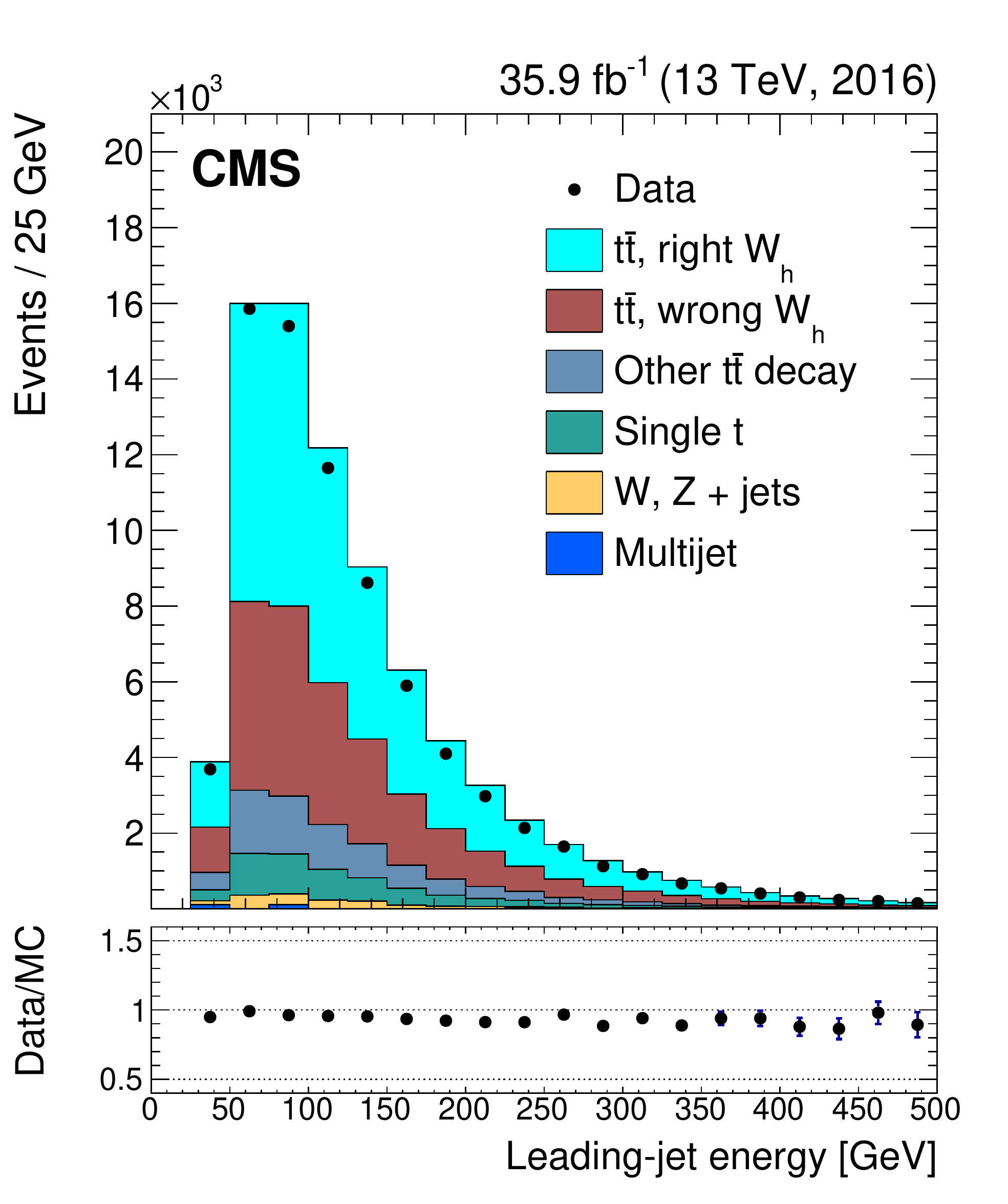}
\includegraphics[width=0.32\textwidth]{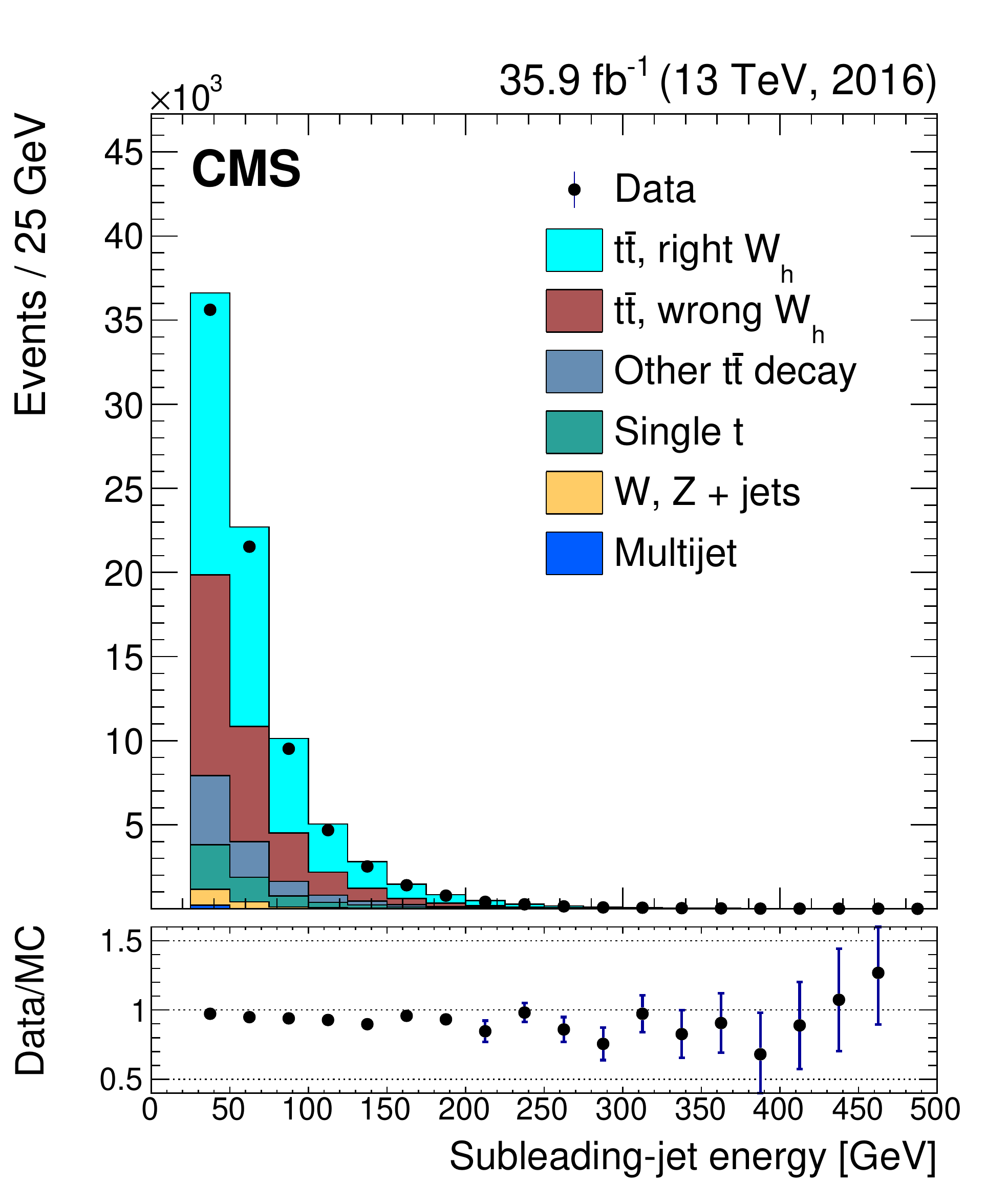}
\includegraphics[width=0.32\textwidth]{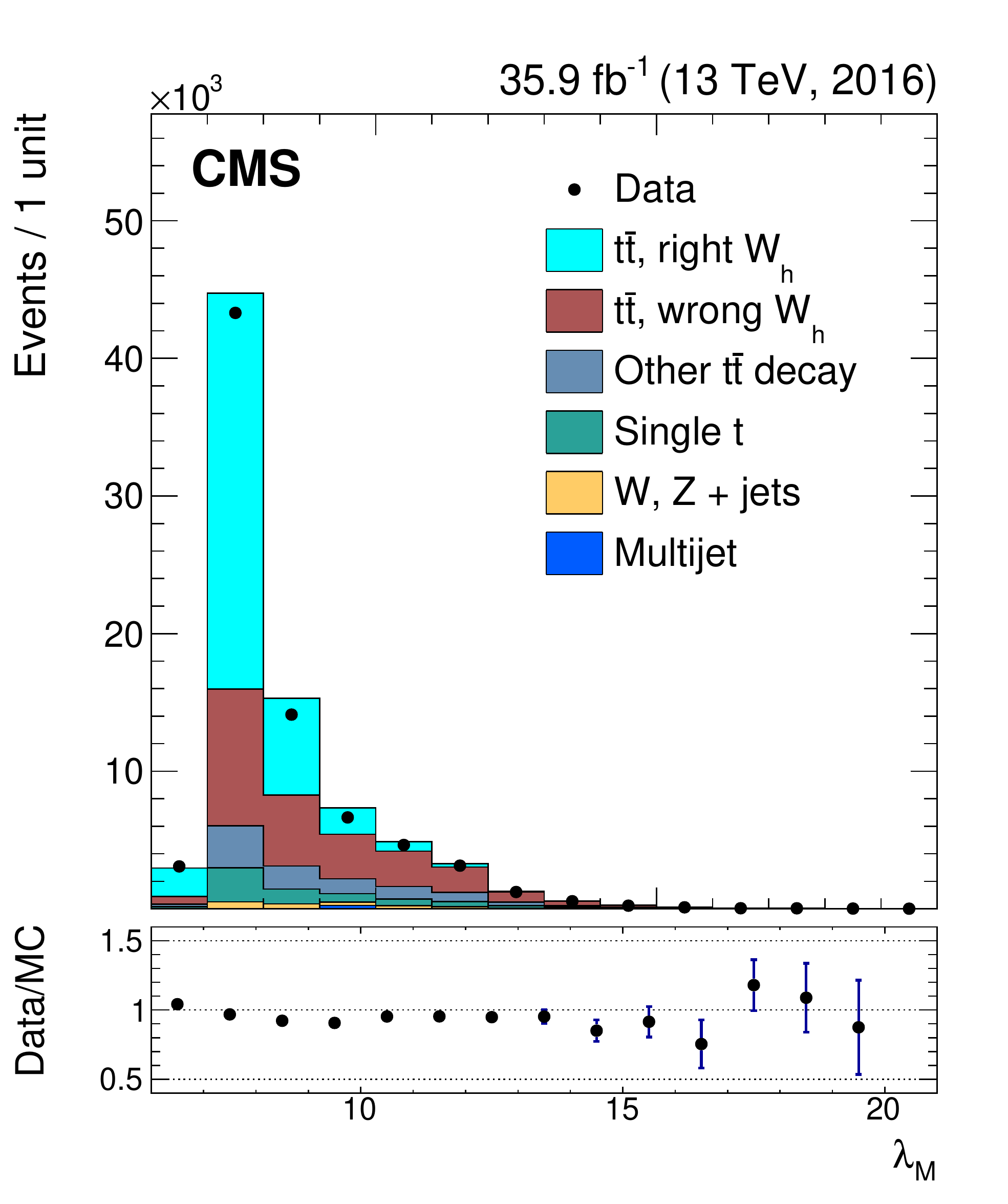}
\caption{Distributions of the leading- (left) and subleading- (middle) jet energy as well as of the mass discriminant \lm (right) after the full event selection, jet-quark assignment, and {\cPqb} tagging requirement on the two {\cPqb} jet candidates. }
\label{fig:sel_plots}
\end{figure}

Events are selected by requiring exactly one isolated muon satisfying the tight identification criteria and with a \pt exceeding 30\GeV~\cite{Chatrchyan:2012xi}. In addition, exactly four jets with $\pt > 30\GeV$ are required. All objects are required to be within the tracker acceptance. The background from multijet events is reduced to a negligible level by requiring the reconstructed transverse mass formed by the muon and \ptvecmiss, to be $\MT(\Pgm,{\ptmiss}) > 50\GeV$.
The \ttbar event is reconstructed by assigning the jets to the quarks from which they originate, using a mass discriminant \lm. This mass discriminant is defined as the 2D probability for the invariant mass of a correct combination of two jets to be consistent with the {\PW} boson mass, and the invariant mass of a correct three-jet combination to be consistent with the top quark mass. The jet-quark assignment for which the negative logarithm of \lm is minimal is chosen as the reconstructed \ttbar topology candidate for the event. The two jets assigned to the {\cPqb} quarks from the top quark decay are required to be {\cPqb}-tagged; one jet should pass the tight working point of the CSVv2 tagger and the other one its loose working point. By requiring those jets to be {\cPqb}-tagged only after the jet-quark assignment is done, a bias is avoided on the {\cPqc} jet tagging efficiency measurement.
Figure~\ref{fig:sel_plots} shows the distribution of \lm and of the highest (leading) and second-highest (subleading) energy for the two jets corresponding to the {\PW} decay after the full event selection. The \ttbar simulation is divided into three different subsamples:
\begin{itemize}
\item \textbf{\ttbar, right ${\PW}_{\text{h}}$:} The {\PW} boson is correctly reconstructed, hence the two jets are correctly assigned to the quarks from the {\PW} boson decay.
\item \textbf{\ttbar, wrong ${\PW}_{\text{h}}$:} The {\PW} boson is wrongly reconstructed, hence at least one of the two jets is not correctly assigned to the quarks from the {\PW} boson decay.
\item \textbf{Other \ttbar decay:} The generated event is not a single-lepton \ttbar event.
\end{itemize}
The non-\ttbar background is relatively small, with contributions from single top quark, {\Wj}, {\Zj}, and multijet production.

From Fig.~\ref{fig:sel_plots} it is clear that the \lm distribution has discrimination power to separate jets that are correctly associated with the {\PW} boson decay and jets for which this is not the case. Therefore, this distribution is used to measure the efficiency and data-to-simulation scale factor for {\cPqc} jets.

Four event categories are defined according to whether or not the jets that are assigned to the {\PW} boson decay (\ie the probe jets) pass the tagging working point for which the efficiency is to be measured:
\begin{itemize}
\item \textbf{Notag:} both probe jets fail the tagging requirement;
\item \textbf{Leadtag:} only the most energetic probe jet passes the tagging requirement;
\item \textbf{Subleadtag:} only the least energetic probe jet passes the tagging requirement; and
\item \textbf{Ditag:} both probe jets pass the tagging requirement.
\end{itemize}
For \ttbar events in the ``right ${\PW}_{\text{h}}$'' subsample, the number of events in the various categories can be written as:
\begin{linenomath}
\begin{equation}
\label{mathyields}
\begin{array}{l @{{}={}} c @{{}{}} l @{{}+{}} l @{{}+{}} l}
\ENN{{\text notag}} & \ENN{T} &((1-\EFF{1}{{\cPqc}})(1-\EFF{2}{LF})f_{\text{1}} & (1-\EFF{1}{LF})(1-\EFF{2}{{\cPqc}})f_{\text{2}} & (1-\EFF{1}{LF})(1-\EFF{2}{LF})(1-f_{\text{1}} - f_{\text{2}})), \\
\ENN{{\text leadtag}} & \ENN{T} &(\EFF{1}{{\cPqc}} (1-\EFF{2}{LF})f_{\text{1}} &    \EFF{1}{LF} (1-\EFF{2}{{\cPqc}})f_{\text{2}} & \EFF{1}{LF} (1-\EFF{2}{LF})(1-f_{\text{1}} - f_{\text{2}})), \\
\ENN{{\text subtag}} & \ENN{T} &((1-\EFF{1}{{\cPqc}})\EFF{2}{LF}    f_{\text{1}} & (1-\EFF{1}{LF})   \EFF{2}{{\cPqc}} f_{\text{2}} & (1-\EFF{1}{LF})   \EFF{2}{LF} (1-f_{\text{1}} - f_{\text{2}})), \\
\ENN{{\text ditag}} & \ENN{T} &( \EFF{1}{{\cPqc}} \EFF{2}{LF} f_{\text{1}} &    \EFF{1}{LF}    \EFF{2}{{\cPqc}} f_{\text{2}} &    \EFF{1}{LF}    \EFF{2}{LF} (1-f_{\text{1}} - f_{\text{2}})),  \\ \end{array}
\end{equation}
\end{linenomath}
with $\ENN{T}$ the total number of events, $f_{\text{1,2}}$ the fraction of leading (subscript 1) and subleading (subscript 2) {\cPqc} jets, and $\EFF{1,2}{{\cPqc},L}$ the tagging efficiencies for leading and subleading jets for {\cPqc} (superscript {\cPqc}) and light-flavour (superscript $\text{LF}$) quarks.
The indentation highlights the different components, namely the probability for a jet pair to be composed of ({\cPqc}, light), (light, {\cPqc}), and (light, light) jet flavours as (leading, subleading) jets from the {\PW} boson decay. The ({\cPqc}, {\cPqc}) pair is not present since it is unphysical.
Instead of measuring the efficiency $\EFF{1,2}{{\cPqc},LF}$, the data-to-simulation scale factor is measured. Therefore, $\EFF{1,2}{{\cPqc},LF}$ is replaced with $SF_{{\cPqc},l}\EFF{1,2}{{\cPqc},LF}(\text{MC})$ in Eq.~(\ref{mathyields}). In the latter expression $\EFF{1,2}{{\cPqc},LF}(\text{MC})$ is the efficiency obtained from simulation and $SF_{{\cPqc}}$ ($SF_{\text{l}}$) is the scale factor for {\cPqc} (light-flavour) jets. To reduce the number of unknown parameters, the value for $SF_{\text{l}}$ is taken to be the measured value using the negative-tag method presented in Section~\ref{sec:negtag}.

A maximum likelihood fit is performed on the binned \lm distributions using the signal and background distributions (templates) obtained from the simulated events. The measurement is performed inclusively, since the selected number of events is not sufficient for a precise measurement in bins of jet \pt. Systematic uncertainties are included in the fit as nuisance parameters that are profiled. Each nuisance parameter is floating with a Gaussian constraint around the central value with a standard deviation proportional to the systematic uncertainty. It is possible to group the systematic uncertainties in two sets based on their effect on the templates. The following systematic effects only affect the normalization of the templates:
\begin{itemize}
\item \textbf{Scale factor for light-flavour jets}: The data-to-simulation scale factor for the misidentification of light-flavour jets is varied within its uncertainty. This is the dominant uncertainty in the scale factor measurement for {\cPqc} jets.
\item \textbf{Cross sections of the simulated processes}: An uncertainty of 16, 50 and 20\% is assumed in the cross section of the \ttbar~\cite{Khachatryan:2016mnb}, single top quark~\cite{Sirunyan:2016cdg,Chatrchyan:2014tua}, and the combined {\Wj} and {\Zj}~\cite{Sirunyan:2017wgx,Khachatryan:2016ipq} processes, respectively.  The limited number of simulated {\Wj} and {\Zj} events requires an additional uncertainty in their yield, fully uncorrelated among the event categories.
\item \textbf{Integrated luminosity and pileup}: The uncertainty in the integrated luminosity measurement of 6.2\%~\cite{CMS-PAS-LUM-17-001} and on the number of additional pileup interactions are considered as yield uncertainties. These uncertainties as well as the uncertainty in the cross sections for the simulated processes are the same for each working point probed, and are applied to the related samples in a correlated way between categories.
\item \textbf{Scale factors for {\cPqb} tagging}: Since {\cPqb} tagging is applied for the event selection, the uncertainty in the {\cPqb} tagging data-to-simulation scale factor is considered as a systematic effect. The simulation has been processed with {\cPqb} jet scale factors shifted by their uncertainties. In case the {\cPqb}-tagged jets in the event selection are actually originating from {\cPqc} quarks, the scale factor is varied by a conservative 50\%. The size of the combined effect due to the uncertainty of correctly tagging {\cPqb} jets and wrongly tagging {\cPqc} or light-flavour jets for the event selection depends on the samples, the categories, and the working points considered. However, the effect of these uncertainties has limited impact on the final result, being fully correlated across samples and categories.
\end{itemize}
A potential source of systematic uncertainty for the normalization of the templates may arise from the uncertainty in the cross section of {\ttbar} events produced in association with heavy-flavour jets, which is constrained to within 35\%~\cite{Sirunyan:2017snr}. Such an uncertainty is covered by the systematic variation on the inclusive {\ttbar} production cross section and the uncertainty in the {\cPqb} tagging scale factor for the event selection, which is taken to be 50\% for jets arising from {\cPqc} quarks.

In addition to a possible impact on the normalization of the templates, the following systematic effects affect the shape of the templates:
\begin{itemize}
\item \textbf{Jet energy scale}: New templates are constructed by varying the jet energy scale by $\pm$1 standard deviation from its nominal value. The uncertainty is propagated to the fraction of {\cPqc} jets in leading and subleading jets.
\item \textbf{Jet energy resolution}: For the nominal efficiency measurement, the jet energies in the simulation are smeared according to a Gaussian function to accommodate the slightly worse resolution in data. The uncertainty in the jet energy resolution is propagated to the data-to-simulation scale factor measurement by varying the standard deviation of the Gaussian function by its uncertainty.
\item \textbf{Factorization and renormalization scales}: The factorization and renormalization scales used at matrix element and parton shower levels affect the number of additional jets from initial-state radiation (ISR) and final-state radiation (FSR), and may impact the fraction of leading and subleading {\cPqc} jets. The factorization and renormalization scales used at matrix element level are varied independently and simultaneously by factors of 2 and 0.5 with respect to their default values. Also the scale for ISR (FSR) in the parton shower is varied by a factor of 2 ($\sqrt{2}$) and 0.5 ($\sqrt{0.5}$)~\cite{Skands:2014pea}. A different way to assess the uncertainty in the modelling of ISR and FSR is to vary the ``hdamp'' parameter in \POWHEG. This parameter is used to limit the resummation of higher-order effects using a reference energy scale. The real emissions are reweighted by a step-function $h^2/(\pt^2+h^2)$, where $h$ is the hdamp parameter and $\pt$ is the transverse momentum of the top quark in the \ttbar rest frame. The hdamp parameter is varied between $0.5m_{\text{\cPqt}}$ and $2m_{\text{\cPqt}}$ to evaluate the uncertainty related to additional jets from ISR and FSR. The variations upwards and downwards having the largest impact on the templates, are used to repeat the data-to-simulation scale factor measurements independently for ISR and FSR. The deviation from the nominal scale factor value is taken as the uncertainty. Together with the uncertainties in the jet energy scale and resolution, the effect is 1\% for both the leading and subleading jets.
\item \textbf{Top quark mass}: The uncertainty in the top quark mass may affect the measurement of the data-to-simulation scale factor. The size of the uncertainty is estimated using alternative simulated samples with a mass that is shifted within the uncertainty in the measured value~\cite{Patrignani:2016xqp}.
\item \textbf{Parton distribution functions}: The uncertainties in the parton densities is evaluated in the same way as in Section~\ref{sec:Wc} and found to be negligible.
\item \textbf{Bin-by-bin statistical uncertainty}: Statistical uncertainties related to the single bin population in the templates have been addressed through bin-by-bin variations, \ie fully uncorrelated shape uncertainties in which only one bin of the template is shifted according to its uncertainty. In order to reduce the computational time required by the fit to converge, this uncertainty is only considered for template bins having an uncertainty larger than 5\% of the yield observed in the same bin, thus rejecting most of the low-yield backgrounds.
\end{itemize}

Table~\ref{tab:sfresults} summarizes the values of the measured data-to-simulation scale factors for all tagging requirements. The uncertainty in the scale factors in the table are a combination of the statistical and systematic uncertainties obtained from the fit.
\begin{table}
\centering
\topcaption{Measured data-to-simulation scale factors for {\cPqc} jets for various algorithms and working points in single-lepton \ttbar events. The uncertainty in the scale factor includes both the statistical and systematic uncertainties, while the last column shows the statistical uncertainty alone.}
 \begin{tabular}{ccc}
Working point & $SF_{\cPqc}$ & Statistical uncertainty \\
\hline
CSVv2 L & $0.89 \pm 0.05$ & $\pm$0.02 \\
CSVv2 M & $0.87 \pm 0.08$ & $^{+0.04}_{-0.03}$ \\
CSVv2 T & $1.15^{+0.35}_{-0.33}$ & $^{+0.15}_{-0.14}$ \\
{\cPqc} tagger L & $1.05 \pm 0.03$ & $\pm$0.01 \\
{\cPqc} tagger M & $0.93 \pm 0.05$ & $\pm$0.02 \\
{\cPqc} tagger T & $0.88^{+0.05}_{-0.04}$ & $\pm$0.02 \\
DeepCSV L & $0.98^{+0.05}_{-0.04}$ & $\pm$0.02 \\
DeepCSV M & $0.96 \pm 0.09$ & $\pm$0.04 \\
DeepCSV T & $0.87^{+0.37}_{-0.38}$ & $\pm$0.15 \\
cMVAv2 L & $0.87 \pm 0.04$ & $\pm$0.02 \\
cMVAv2 M & $0.76 \pm 0.09$ & $\pm$0.03 \\
cMVAv2 T & $0.86^{+0.31}_{-0.29}$ & $\pm$0.13 \\
\end{tabular}\label{tab:sfresults}
\end{table}

\subsubsection{Combination of the measured \texorpdfstring{{\cPqc}}{c} tagging efficiencies}
\label{sec:combSFc}
\begin{figure}[tbp]
\includegraphics[width=0.99\textwidth]{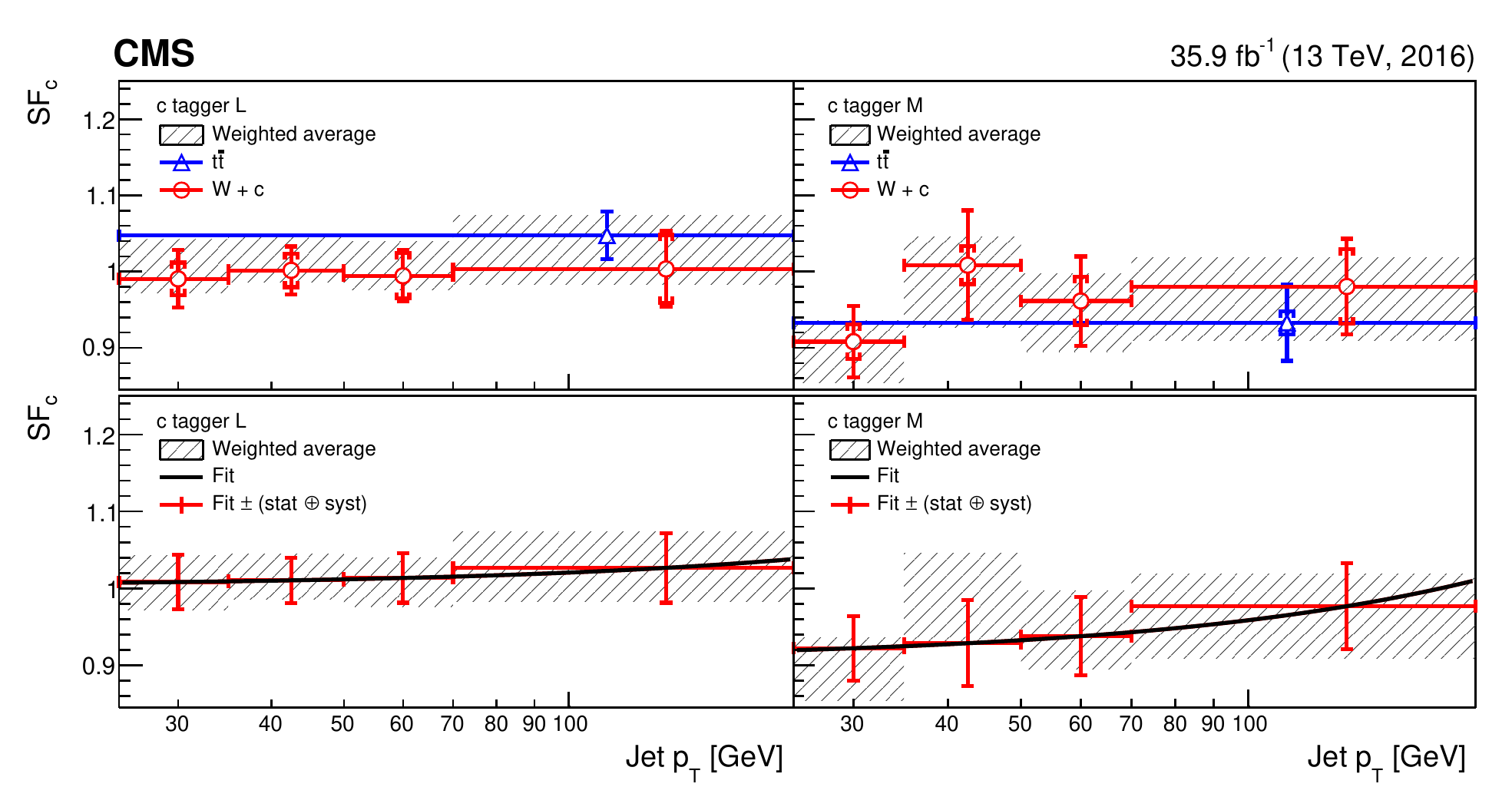}
\caption{Data-to-simulation scale factors for {\cPqc} jets for the loose (left) and medium (right) working points of the {\cPqc} tagger. The upper panels show the scale factors for {\cPqc} jets as a function of the jet \pt obtained with the two methods described in the text. The inner error bars represent the statistical uncertainty and the outer error bars the combined statistical and systematic uncertainty. The combined scale factor values with their overall uncertainty are displayed as a hatched area. The lower panels show the same combined scale factor values with superimposed the result of a fit function represented by the solid curve. The combined statistical and systematic uncertainty is centred around the fit result, represented by the points with error bars. The last bin includes the overflow entries.}
\label{fig:combSFc}
\end{figure}

In the previous sections two methods have been described to measure the {\cPqc} tagging efficiency. In this section, a combination of the measurements is performed. The combination is a weighted average taking into account the full covariance matrix for the uncertainties using the best linear unbiased estimator (BLUE) method~\cite{BLUE}. This technique was also used for combining the $SF_{\cPqb}$ measurements in Run 1~\cite{BTV12001}, but here it has been extended to fit all the jet \pt bins simultaneously~\cite{ExtBLUE}, treating more correctly the bin-to-bin correlations for the systematic uncertainties. Systematic uncertainties shared by the two measurements are treated as correlated in the combination. The averaging has been done using the finer jet \pt binning of the {\Wc} topology. The relative contribution of the single-lepton \ttbar measurement in each jet \pt bin is taken into account by assigning weights to this measurement corresponding to the fraction of jets from top quark decays expected in each of the jet \pt bins. These fractions are obtained from simulation. The result of the combination is shown in Fig.~\ref{fig:combSFc} for the loose and medium working points of the {\cPqc} tagger. In each panel the combined value is represented as the hatched area. The individual measurements are represented in the upper panel as markers with different colours. The inner thicker error bar represents the statistical uncertainty of the measurement. The lower panel includes a fit of the data-to-simulation scale factor dependence on the jet \pt, which is parameterized by a linear function.

For the {\cPqc} tagging algorithm, the relative precision on the data-to-simulation scale factors for {\cPqc} jets is 2\% (4\%) for the loose (tight) working point. For the {\cPqb} tagging algorithms, the relative precision is 3--5\% for the loose working points, and 10--38\% for the tight working points. Overall, the statistical uncertainty is 40--90\% of the total uncertainty.
	
\subsection{The \texorpdfstring{\cPqb}{b} jet identification efficiency}
\label{sec:SFb}
The data-to-simulation scale factor for {\cPqb} jets, $SF_{\cPqb}$, is obtained using a sample of jets enriched in {\cPqb} quark content, \eg by selecting multijet events with at least one jet containing a muon, or \ttbar events that contain two {\cPqb} jets from the decay of the two top quarks. To enhance the purity when selecting $\ttbar$ events, the decay of one or both of the {\PW} bosons into leptons is required. This section describes the various $SF_{\cPqb}$ measurements and their combination.

\subsubsection{Measurements relying on a muon-enriched topology}
Events are selected using various online criteria requiring the presence of two jets with at least one of those jets containing a muon. The different prescales of the various triggers are taken into account by reweighting the selected events according to the value of the prescale. Offline, the sample is enriched with events containing {\cPqb} jets by requiring that at least one jet has a muon with $\pt > 5\GeV$ and with $\Delta R<0.4$ from the jet axis, referred to as the ``muon jet''. The selected simulated events are reweighted to match the pileup profile observed in the data. The muon jet sample is used for three measurements, using the PtRel, LifeTime (LT), and System-8 methods~\cite{BTV12001}. As discussed in Section~\ref{sec:Wc}, the muon enrichment may introduce a bias for the efficiency measurement of taggers that rely on soft muon information, such as the cMVAv2, CvsB, and CvsL discriminators. Therefore, the methods described in this section are only used to derive data-to-simulation scale factors for the other taggers.

\paragraph{PtRel method}
\label{sec:ptrel}

The \pt of the muon relative to the jet axis, $\pt^{\text{rel}}$, is a variable that is able to discriminate between {\cPqb} jets and non-{\cPqb} jets. On average, this variable is expected to be larger for muons coming from the decay of {\cPqb} hadrons because of the large mass of these hadrons. Therefore, this variable can be used to measure the efficiency for tagging {\cPqb} jets with algorithms relying on track and secondary vertex variables. The fraction of {\cPqb} jets in data can be estimated by fitting the observed $\pt^{\text{rel}}$ distribution to the sum of the templates for the different jet flavours. The $\pt^{\text{rel}}$ templates for the different flavours are obtained from the simulated muon-enriched multijet samples. To reduce the fraction of non-{\cPqb} jets, the presence of a second jet is required away from the first one (``away jet'') with $\Delta R>1.5$ and exceeding a JBP discriminator value corresponding to the medium working point.

For light-flavour jets, a difference is observed between data and simulation in the distribution of the number of charged particles per jet. Therefore, the jets are reweighted with the ratio of the distribution of the observed number of charged particles in inclusive multijet data to that expected in simulation (without the muon enrichment). The template for {\cPqb} jets is corrected by applying a factor corresponding to the ratio of the $\pt^{\text{rel}}$ distribution in data to that in simulation for {\cPqb} jets passing the tight JP tagging requirement. The fraction of non-{\cPqb} jets in the JP-tagged samples is found to be of a few per cent and is subtracted. After this correction, we apply the algorithm working point for which the efficiency is to be measured. The observed $\pt^{\text{rel}}$ distribution is then fitted with the templates for the jet flavours to obtain the number of {\cPqb} jets passing ($N_{\cPqb}^{\text{tagged}}$) or failing ($N_{\cPqb}^{\text{vetoed}}$) the requirement. The {\cPqb} tagging efficiency in data is obtained as
\begin{linenomath}
\begin{equation}
\label{eq:effdata}
   \varepsilon_{\cPqb} = \frac{N_{\cPqb}^{\text{tagged}}}{N_{\cPqb}^{\text{vetoed}}+N_{\cPqb}^{\text{tagged}}}
\end{equation}
\end{linenomath}

\begin{figure}[hbtp]
  \centering
    \includegraphics[width=.49\textwidth]{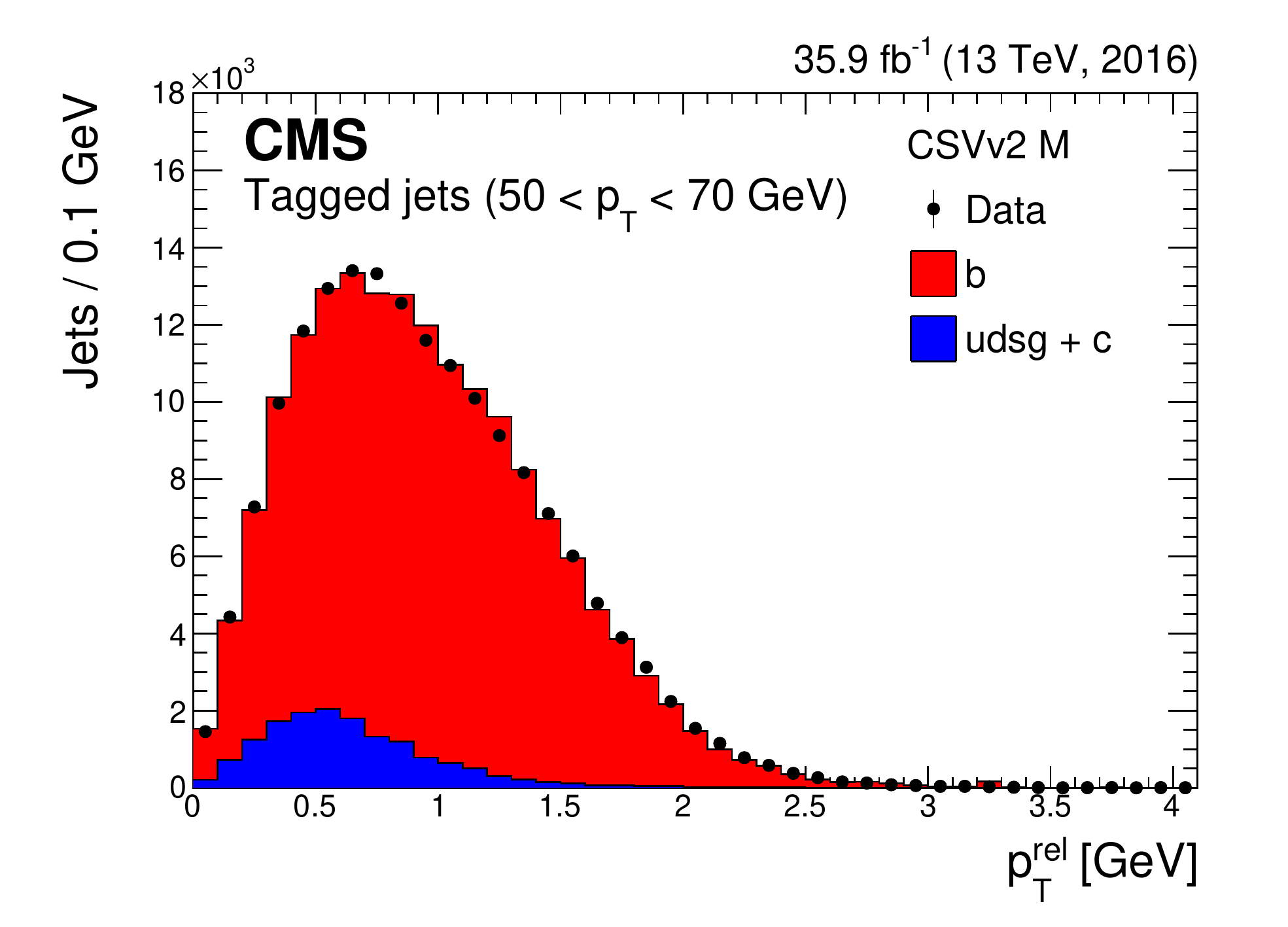}
    \includegraphics[width=.49\textwidth]{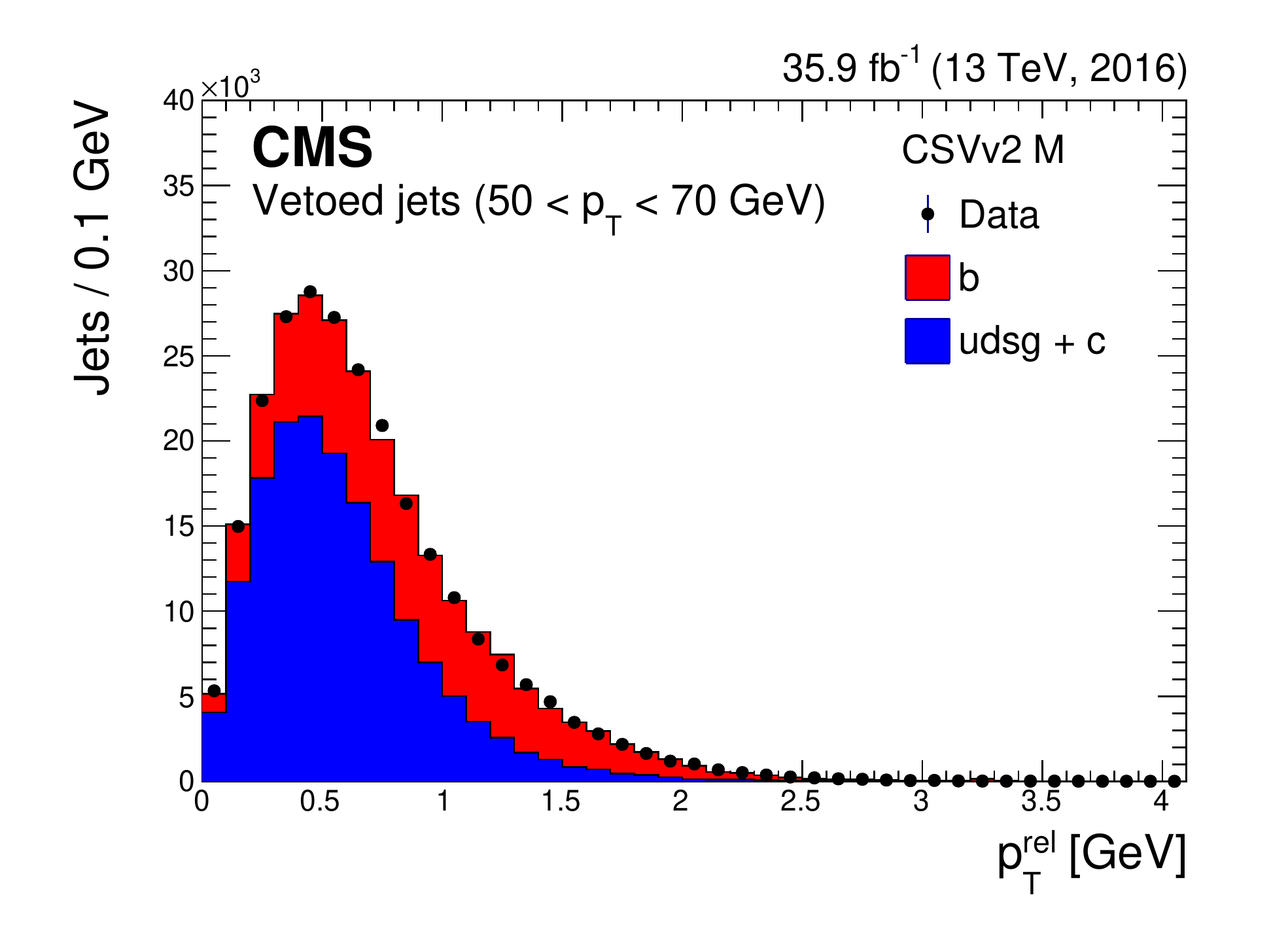}
    \caption{Fitted $\pt^{\text{rel}}$ distribution for muon jets passing (left) and failing (right) the medium working point of the CSVv2 algorithm. The distribution is shown for jets with $50 < \pt < 70\GeV$. The simulation is normalized to the observed number of events.}
    \label{Fig:PtRelFits}
\end{figure}

Examples of the fitted $\pt^{\text{rel}}$ distributions using jets passing and failing the medium working point of the CSVv2 algorithm and with $50 < \pt < 70\GeV$, are shown in Fig.~\ref{Fig:PtRelFits}.

\paragraph{LifeTime method}
\label{sec:LTmethod}

The muon jet sample used in the LT method is the same as for the PtRel method, except that the away jet is not required to be tagged. Also the strategy is similar to the PtRel method, but the fit is performed on the JP discriminator distribution. The track probabilities are calibrated using templates with negative impact parameter tracks in multijet events. The calibration is done separately on data and simulation to take into account a potential difference in the impact parameter resolution between both samples. The fraction of {\cPqb} jets is fitted including all shape systematic uncertainties via a correlation matrix. The tagging efficiency is then obtained as the ratio of the number of {\cPqb} jets obtained from the fit after and before applying the algorithm working point
\begin{linenomath}
\begin{equation}
\varepsilon_{\cPqb} = C_{\cPqb}\frac{N_{\cPqb}^{\text{tagged}}}{N_{\cPqb}}.
\end{equation}
\end{linenomath}
The factor $C_{\cPqb}$ is a correction factor, which takes into account the fraction of jets for which the JP discriminant can be computed. It is defined as
\begin{linenomath}
\begin{equation}
C_{\cPqb} = \frac{n_{{\cPqb},\text{MC}}^{\text{tag}}}{N_{{\cPqb},\text{MC}}^{\text{tag}}} \frac{N_{{\cPqb},\text{MC}}}{n_{{\cPqb},\text{MC}}},
\end{equation}
\end{linenomath}
with $N_{{\cPqb},\text{MC}}$ the number of {\cPqb} jets with JP information, $n_{{\cPqb},\text{MC}}$ the number of all selected {\cPqb} jets, $N_{{\cPqb},\text{MC}}^{\text{tag}}$ the number of {\cPqb} jets with JP information passing the algorithm working point for which the efficiency is being measured and $n_{{\cPqb},\text{MC}}^{\text{tag}}$ the number of {\cPqb} jets passing the tagging requirement for which the data-to-simulation scale factor is being measured. The fraction of jets without a JP discriminant value is maximum at very low jet \pt (8\%) and drops below 1\% using jets with $\pt> 120\GeV$.

\begin{figure}[hbtp]
  \centering
    \includegraphics[width=.49\textwidth]{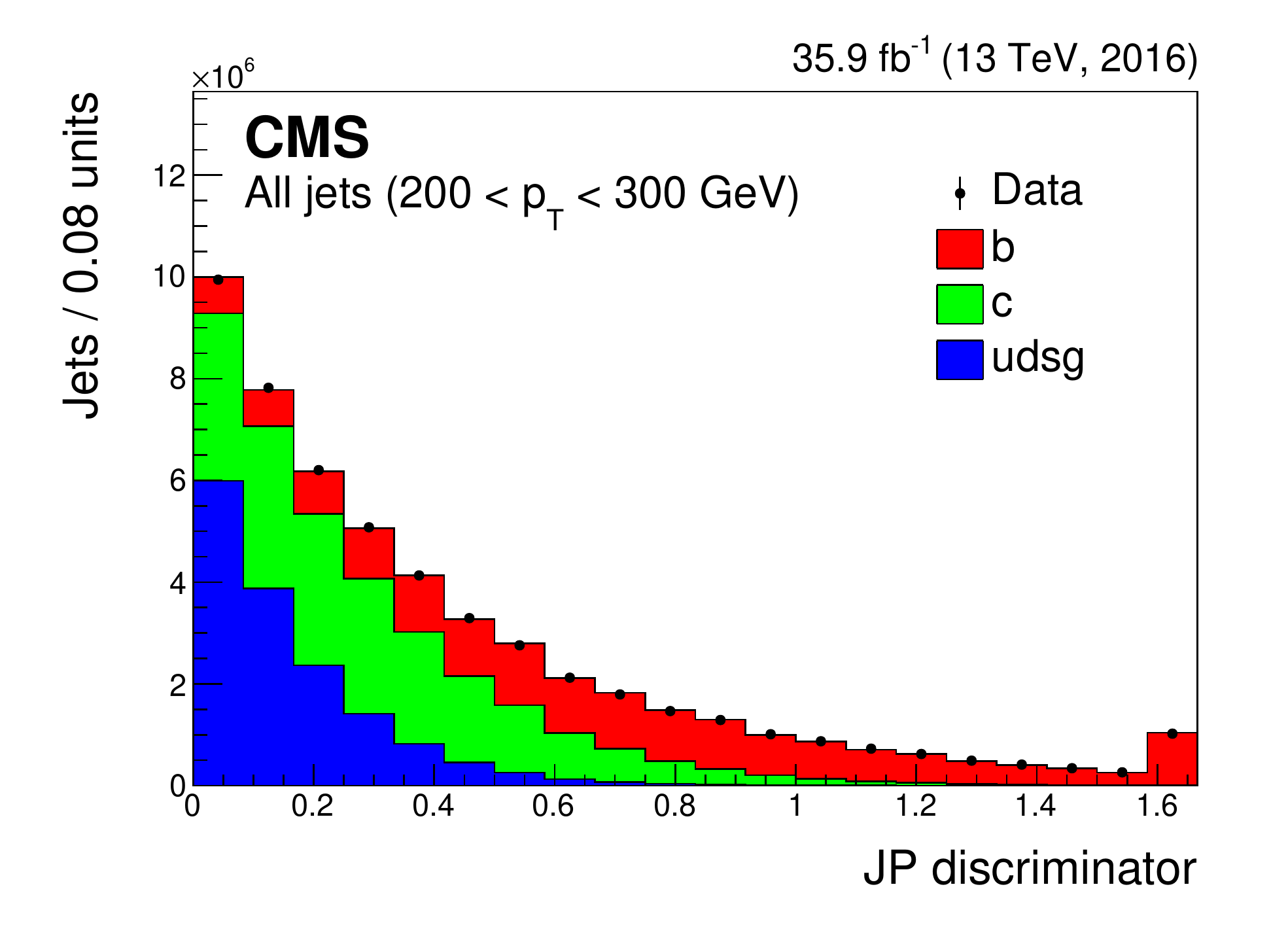}
    \includegraphics[width=.49\textwidth]{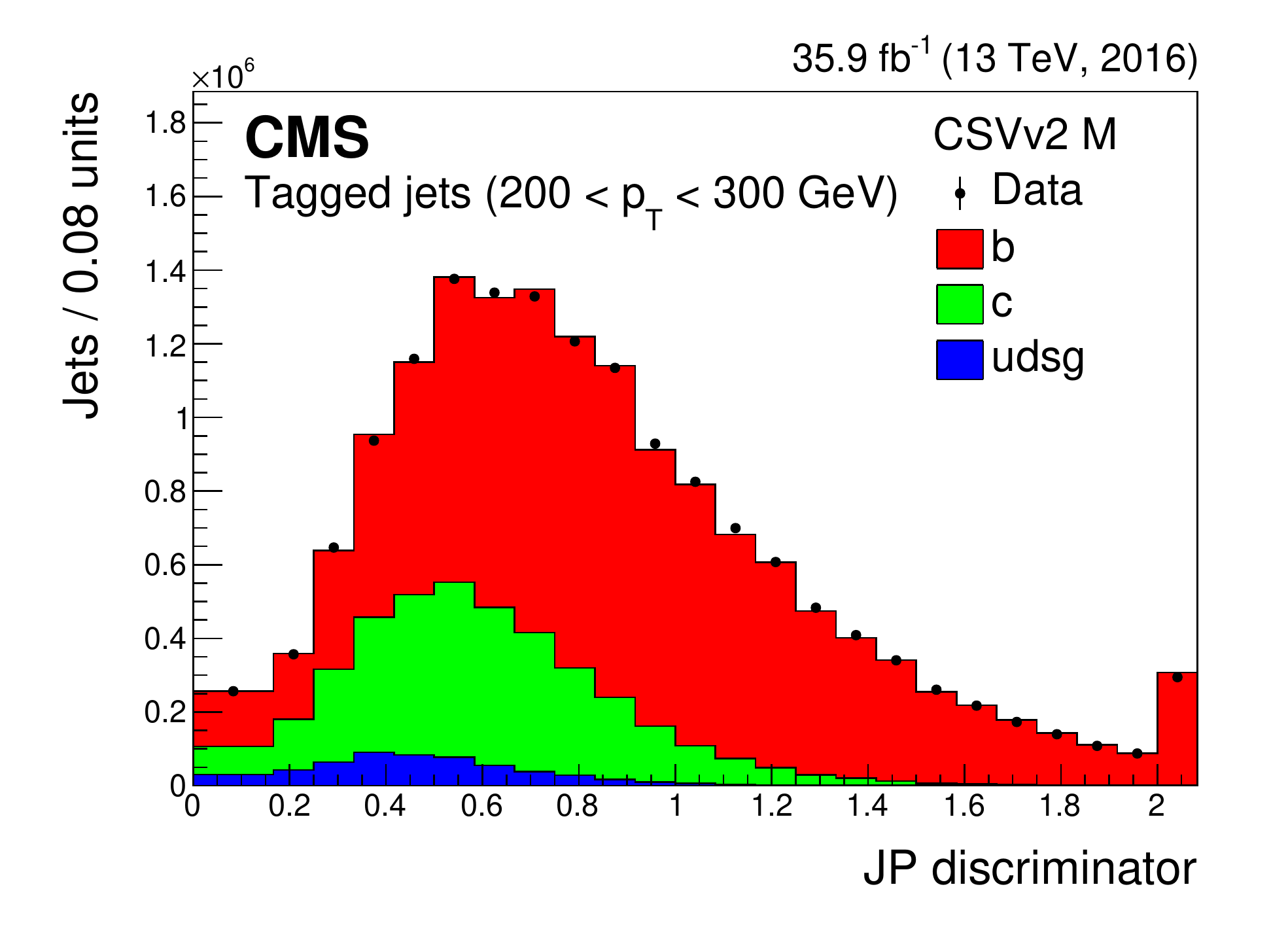}
    \caption{Fitted JP distribution for muon jets (left) and for the subsample of those jets passing the medium working point of the CSVv2 algorithm (right). The distribution is shown for jets with $200 < \pt < 300\GeV$. The simulation is normalized to the integrated luminosity for the data set. The last bin includes the overflow entries.}
    \label{Fig:LTTag}
\end{figure}

As an illustration, Fig.~\ref{Fig:LTTag} shows the fitted JP distributions using jets with $200 < \pt < 300\GeV$ before and after applying the medium working point of the CSVv2 algorithm.

\paragraph{System-8 method}

In contrast with the two methods described before, the System-8 method~\cite{Abazov:2010ab} does not rely on simulated templates of a discriminating variable. Instead, it is based on the usage of two weakly correlated {\cPqb} taggers and two samples containing muons within jets. The first {\cPqb} tagging requirement corresponds to the working point of the algorithm for which the efficiency is to be measured (tag); the second {\cPqb} tagging requirement is $\pt^{\text{rel}} > 0.8\GeV$. This requirement is weakly correlated with the working points for algorithms that do not rely on soft-muon information. The first sample consists of all events with a muon jet (sample $n$); the second sample is a subset where an away jet satisfies the medium working point of the JBP algorithm (sample $p$). For each combination of a sample with either zero, one of the two, or both tagging requirements applied, the observed number of jets can be written as the sum of the two ({\cPqb} and non-{\cPqb}) flavour contributions. The efficiency of the algorithm working point under study and the efficiency of the $\pt^{\text{rel}} > 0.8\GeV$ requirement are assumed to be factorizable modulo a correlation factor that is determined from simulated events. In total eight equations can be written, with eight unknown parameters, namely the {\cPqb} tagging efficiencies of the two requirements and the number of {\cPqb} and non-{\cPqb} jets in the two samples:
\begin{linenomath}
\begin{equation}
\label{eq:S8}
\begin{array}{lcrcr}
n & = & n_{\cPqb} & + & n_{\cPqc,\text{udsg}} \\
p & = & p_{\cPqb} & + & p_{\cPqc,\text{udsg}} \\
n^{\text{tag}} & = & \varepsilon^{\text{tag}}_{\cPqb}n_{\cPqb} & + & \varepsilon^{\text{tag}}_{\cPqc,\text{udsg}}n_{\cPqc,\text{udsg}} \\
p^{\text{tag}} & = & \beta\varepsilon^{\text{tag}}_{\cPqb}p_{\cPqb} & + & \alpha\varepsilon^{\text{tag}}_{\cPqc,\text{udsg}}p_{\cPqc,\text{udsg}} \\
n^{\pt^{\text{rel}}} & = & \varepsilon^{\pt^{\text{rel}}}_{\cPqb}n_{\cPqb} & + & \varepsilon^{\pt^{\text{rel}}}_{\cPqc,\text{udsg}}n_{\cPqc,\text{udsg}}\\
p^{\pt^{\text{rel}}} & = & \delta\varepsilon^{\pt^{\text{rel}}}_{\cPqb}p_{\cPqb} & + & \gamma\varepsilon^{\pt^{\text{rel}}}_{\cPqc,\text{udsg}}p_{\cPqc,\text{udsg}} \\
n^{\text{tag}, \pt^{\text{rel}}} & = & \kappa_{\cPqb}\varepsilon^{\text{tag}}_{\cPqb}\varepsilon^{\pt^{\text{rel}}}_{\cPqb}n_{\cPqb} & + & \kappa_{\cPqc,\text{udsg}}\varepsilon^{\text{tag}}_{\cPqc,\text{udsg}}\varepsilon^{\pt^{\text{rel}}}_{\cPqc,\text{udsg}}n_{\cPqc,\text{udsg}} \\
p^{\text{tag}, \pt^{\text{rel}}} & = & \kappa_{\cPqb}\beta\delta\varepsilon^{\text{tag}}_{\cPqb}\varepsilon^{\pt^{\text{rel}}}_{\cPqb}p_{\cPqb} & + & \kappa_{\cPqc,\text{udsg}}\alpha\gamma\varepsilon^{\text{tag}}_{\cPqc,\text{udsg}}\varepsilon^{\pt^{\text{rel}}}_{\cPqc,\text{udsg}}p_{\cPqc,\text{udsg}} \\
\end{array}
\end{equation}
\end{linenomath}
where $\alpha,\beta,\gamma,\delta,\kappa_{\cPqb}$ and $\kappa_{\cPqc,\text{udsg}}$ are the correlation factors. This system of eight equations is solved numerically. The solution has to pass some physical constraints, \eg the {\cPqb} tagging efficiency is required to be larger than the non-{\cPqb} tagging efficiency, and the fraction of {\cPqb} jets in the initial sample needs to be smaller than the fraction of non-{\cPqb} jets in the sample.

\paragraph{Systematic uncertainties}
\label{sec:muonAK4syst}

Various systematic uncertainties are taken into account that may affect the $SF_{\cPqb}$ measurement. For the three measurements based on the muon-enriched jet samples, the following systematic effects are considered:
\begin{itemize}
\item \textbf{Gluon splitting}: A variation in the fraction of {\cPqb} and {\cPqc} jets from gluon splitting may have an important impact on the {\cPqb} tagging efficiency since heavy-flavour jets from gluon splitting have a higher track multiplicity. The fraction of events with {\cPqb} jets from gluon splitting is varied by $\pm$25\%~\cite{bsplitting} to estimate the potential effect. For the tight working point of the taggers, this is one of the dominating uncertainties for the System-8 method. In the case of the LT method, the fraction of events with {\cPqc} jets from gluon splitting is also varied by this amount. For the System-8 and PtRel methods, the fraction of {\cPqc} jets in the non-{\cPqb} template is varied when evaluating other systematic effects.
\item \textbf{{\cPqb} quark fragmentation}: The modelling of the {\cPqb} quark fragmentation may affect the \pt distribution of the {\cPqb} jets in the sample. The size of this effect is estimated by varying the \pt of the primary {\cPqb} hadron in the muon jet by $\pm$5\%, which is the observed variation between the distribution of the energy fraction of the {\cPqb} jet carried out by the {\cPqb} hadron in \PYTHIA and \HERWIG. This variation between \PYTHIA and \HERWIG is typically larger than the variation observed between \PYTHIA and data.
\item \textbf{Branching fraction of ${\PD} \to {\Pgm} X$ and fragmentation of ${\cPqc} \to {\PD}$}: These systematic effects are evaluated in the same way as described in Section~\ref{sec:Wc}, with the exception that the PDG 2008 values~\cite{RPP2008} are used for the fragmentation rates. While the nominal values and uncertainties vary slightly in the PDG 2008 and 2016 references, they are fully consistent.
\item \textbf{{\PKzS} and {\PgL} decays (V$^0$)}: This systematic effect is evaluated in the same way as described in Section~\ref{sec:negtag}.
\item \textbf{Muon \pt and $\Delta R$}: The fraction of muons that reach the muon chambers depends on the muon \pt. The threshold on the muon \pt is varied between 5 and 8\GeV to assess the size of the systematic uncertainty. In addition, the dependence of the measured data-to-simulation scale factor on the $\Delta R$ requirement is tested by tightening the requirement to $\Delta R<0.3$. These systematic effects are among the dominant uncertainties for the System-8 method.
\item \textbf{Away jet tag}: The dependence of the {\cPqb} tagging efficiency on the away jet tagging requirement is studied by repeating the data-to-simulation scale factor measurement after changing the tagging requirement from the medium to the loose or tight working points. The largest deviation from the scale factor value obtained using the default away-jet tagging requirement is taken as the size of the systematic effect. This systematic effect is typically the dominant uncertainty for the PtRel method, and it is one of the dominating uncertainties for the System-8 method.
\item \textbf{JP correction factor $C_{\cPqb}$}: For the LT method, the fit is performed using only jets that have a JP discriminant value. The applicability of the measured data-to-simulation scale factor to all jets is ensured through the correction factor $C_{\cPqb}$. The systematic uncertainty associated with $C_{b}$ is defined as $(\delta C_{\cPqb})^{\text{SF}} = \pm \frac{1-C_{\cPqb}}{2}$. This systematic effect induces an uncertainty in the measured scale factor of a few per cent using jets in the lowest \pt bin and is negligible at high jet \pt.
\item \textbf{JP calibration}: The LT method relies on the calibrated JP discriminator distribution. For the nominal data-to-simulation scale factor value, the calibration of the impact parameter resolution derived from data is applied to the data, and the calibration derived from simulated events is applied to the simulation. However, a bias could be induced in the measurement if there are significant differences between data and simulation in the distribution of track impact parameter resolutions used. Therefore, an additional uncertainty is taken into account by applying the calibration derived on simulation, also on the data. The difference in the measured scale factor is included as additional systematic uncertainty. The inverse approach was also tested, \ie applying the JP calibration derived on data to both data and simulation. In that case, the shape changed in a similar way, yielding consistent results for the size of the systematic effect. The systematic effect due to the JP calibration is the dominating uncertainty for the LT method.
\item \textbf{JP bin-to-bin correlation}: For the LT method the systematic uncertainties are taken into account via a correlation matrix. This requires an assumption on the bin-to-bin correlation factors. To assess the impact of an uncertainty in these correlation factors, the data-to-simulation scale factors were remeasured when varying the bin-to-bin factors within $\pm$25\%. The size of the systematic effect is given by the maximal difference with the nominal SF value.
\item \textbf{Muon $\pt^{\text{rel}}$ requirement}: For the System-8 method, the default requirement of $\pt^{\text{rel}}>0.8\GeV$ on the muon is set to a value of 0.5 or 1.2\GeV. The largest deviation from the measured nominal data-to-simulation scale factor is taken as a systematic uncertainty.
\item \textbf{udsg-to-{\cPqc} jet ratio}: In the PtRel method the {\cPqc} and light-flavour jets are combined in a single template. The uncertainty in the ratio of light-flavour to {\cPqc} jets is changed by varying it by $\pm$30\% to cover the observed discrepancy in the fraction of light-flavour jets in inclusive and muon-enriched multijet events.
\item \textbf{Non-{\cPqb} jet template correction}: For the PtRel method, the non-{\cPqb} jet templates are corrected to accommodate the difference in the number of selected tracks for data and simulation. The difference between the measured data-to-simulation scale factors when applying these corrections or not is considered as the size of the systematic effect.
\item \textbf{{\cPqb} jet template correction}: Similarly as in the case for the non-{\cPqb} jet template, also for the {\cPqb} jet template the difference between the nominal data-to-simulation scale factor value and that measured without template correction is taken as an additional uncertainty for the PtRel method.
\item \textbf{Jet energy scale}: The impact of the uncertainty in the jet energy corrections is evaluated as described in Section~\ref{sec:Wc}.
\item \textbf{Pileup}: The effect of the uncertainty in the number of additional pileup interactions is evaluated as described in Section~\ref{sec:negtag}.
\end{itemize}

For the System-8 and PtRel methods the largest deviation from the nominal data-to-simulation scale factor value is taken as the size of the systematic effects. For the LT method, the shape variations are taken into account in the template fit. Table~\ref{tab:systSFbQCD} summarizes the list of systematic effects taken into account for each of the three methods to measure $SF_{\cPqb}$.
\begin{table}[ht]
\centering
\topcaption{Summary of the potential sources of systematic effects taken into account for the muon-enriched $SF_{\cPqb}$ measurements. The symbol ``x'' means that the uncertainty is considered, ``\NA{}'' means that it is negligible, and ``n/a'' that it is not applicable. The systematic effects are separated by horizontal lines according to the type of uncertainty. The first set indicates the modelling uncertainty of heavy-flavour jets in the simulation, the second set are uncertainties related to the selection requirements or to the method that is applied, and the third set covers any other type of uncertainty.}
\begin{tabular}{lccc}
Systematic effect & PtRel & LT & System-8\\
\hline
Gluon splitting to \bbbar & x & x & x\\
{\cPqb} quark fragmentation & x & x & x\\
Branching fraction of ${\PD} \to {\Pgm} X$ & n/a & x & n/a\\
${\cPqc} \to {\PD}$ fragmentation rate & n/a & x & n/a\\
{\PKzS} ({\PgL}) production fraction & n/a & x & n/a\\
\hline
Muon \pt and $\Delta R$ & x & \NA & x\\
Away jet tag & x & n/a & x\\
Fraction of jets with JP & n/a & x & n/a\\
JP calibration & n/a & x & n/a\\
JP bin-by-bin correlation & n/a & x & n/a\\
$\pt^{\text{rel}}$ requirement & n/a & n/a & x\\
udsg-to-{\cPqc} jet ratio & x & n/a & n/a\\
Non-{\cPqb} template correction & x & n/a & n/a\\
{\cPqb} template correction & x & n/a & n/a\\
\hline
JES & x & x & x\\
Pileup & x & \NA & x\
\end{tabular}
\label{tab:systSFbQCD}
\end{table}

\paragraph{Results}
\label{sec:muonjetresults}

The measurements of $SF_{\cPqb}$ obtained on muon-enriched multijet events are combined using the BLUE method as described in Section~\ref{sec:combSFc}. The weighted average is calculated taking into account the correlations between the three methods. The combination is performed as a function of the jet \pt, ranging from 20 to 1000\GeV. Jets with a higher \pt are included in the last bin. The PtRel and LT methods provide measurements on the full jet-\pt range from 20 to 1000\GeV, while the sensitivity of the System-8 method is limited to the lower part of the spectrum, $20 < \pt < 140\GeV$.

The PtRel, System-8, and LT methods are applied on the same events. However, the requirement on the second jet is different for each method. The fraction of events with {\cPqb} quarks that is in common between each pair of methods is obtained from simulated events, and is used to estimate the statistical correlation in the combination of the results. Systematic uncertainties that are in common for two or three methods are treated as correlated. Some of the systematic effects that induce a large uncertainty are however related to a specific method and are treated as uncorrelated.

\begin{figure}[hbtp]
  \centering
  \includegraphics[width=.99\textwidth]{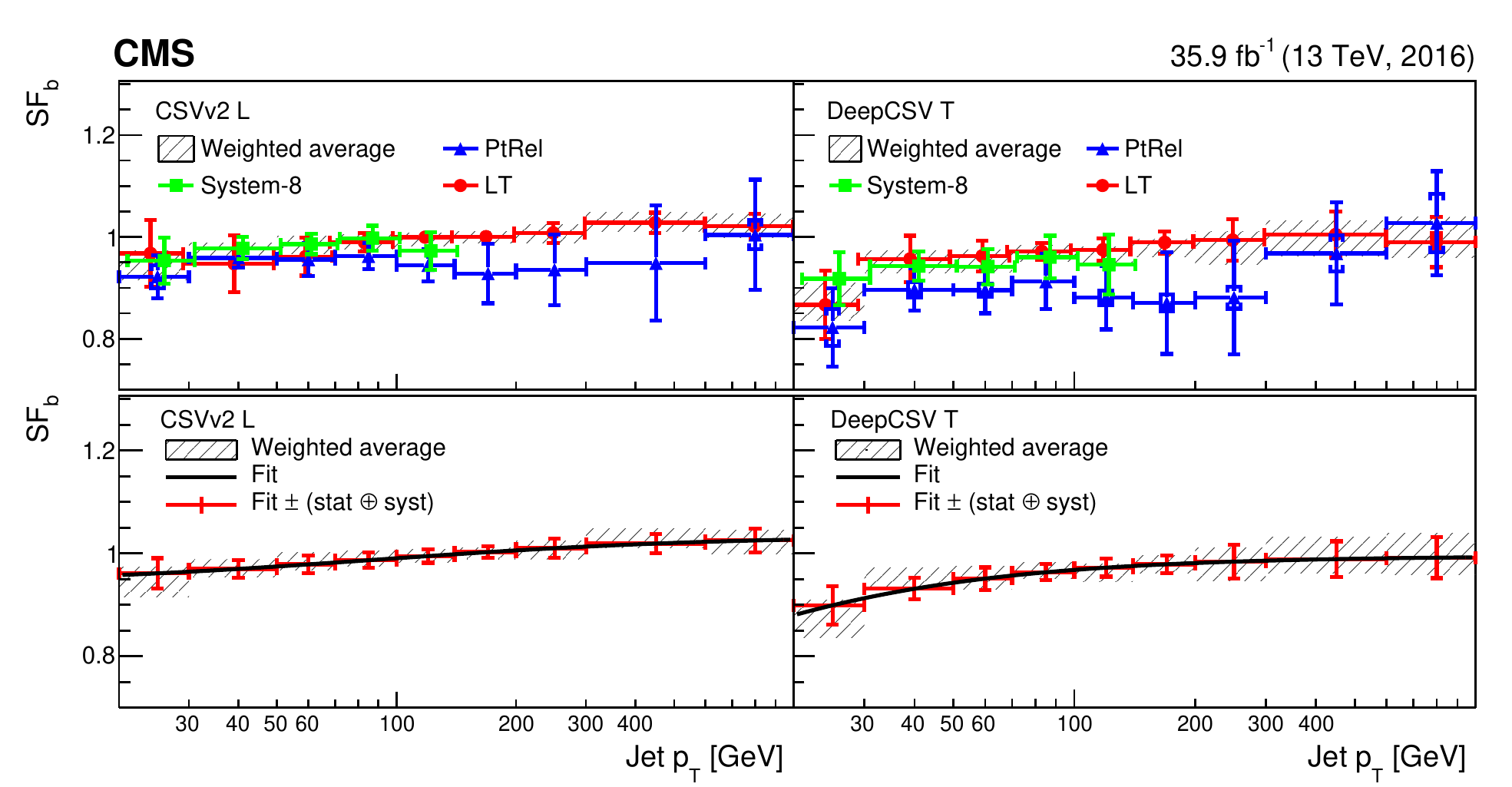}
    \caption{Data-to-simulation scale factors for {\cPqb} jets as a function of the jet \pt for the loose CSVv2 (left) and the tight DeepCSV (right) algorithms working points. The upper panels show the scale factors for tagging {\cPqb} as a function of the jet \pt measured with three methods in muon jet events. The inner error bars represent the statistical uncertainty and the outer error bars the combined statistical and systematic uncertainty. The combined scale factors with their overall uncertainty are displayed as a hatched area. The lower panels show the same combined scale factors with the result of a fit function (solid curve) superimposed. The combined scale factors with the overall uncertainty are centred around the fit result. To increase the visibility of the individual measurements, the scale factors obtained with various methods are slightly displaced with respect to the bin centre for which the measurement was performed. The last bin includes the overflow entries.}
    \label{fig:sfbqcd}
\end{figure}

The data-to-simulation scale factor measurements obtained with the PtRel, System-8, and LT methods for the loose (tight) working point of the CSVv2 (DeepCSV) algorithm as a function of the jet \pt are compared in the upper panel of Fig.~\ref{fig:sfbqcd}. For each point, the thick error bar corresponds to the statistical error and the thin one to the overall statistical and systematic uncertainty. The combined $SF_{\cPqb}$ value is displayed as a hatched area in both panels with its overall uncertainty. In the lower panel the result of a fit function is superimposed. The function used in the fit is $SF_{\cPqb}(\pt) = \alpha \frac{1 + \beta \pt}{1 + \gamma \pt}$, where $\alpha$, $\beta$, $\gamma$ are free parameters. The combined statistical and systematic uncertainty is centred around the fit result.
The measured data-to-simulation scale factors for the loose working point of the CSVv2 algorithm range from 0.96 to 1.03, and from 0.9 to 1.0 for the tight working point of the DeepCSV algorithm. The relative precision on the scale factors is 1--1.5\% using jets with $70 < \pt < 100\GeV$ and rises to 3--5\% at the highest considered jet \pt.

\subsubsection{Measurements relying on the dilepton \texorpdfstring{\ttbar}{ttbar} topology}
The {\cPqb} jet identification efficiency is also measured using dilepton \ttbar events, where two {\cPqb} jets are expected from the decay of the top quark pair.
Events are selected with exactly two isolated leptons (muons or electrons) fulfilling tight identification criteria~\cite{Khachatryan:2015hwa,Chatrchyan:2012xi} with opposite charge and \pt above 25\GeV. Events are selected if there are at least two jets with $\pt > 30\GeV$. All aforementioned objects are required to be in the tracker acceptance.

\paragraph{Kinematic selection method}
\label{sec:kin}

For the kinematic selection (Kin) method, events are further selected by requiring the presence of exactly one isolated electron and one isolated muon with opposite sign and with a dilepton invariant mass $M_{{\Pgm}{\Pe}} > 90\GeV$. These requirements significantly reduce the background from {\Zj} events. In addition, {\ptmiss} is required to be larger than 40\GeV.
While two jets are expected in dilepton \ttbar events, it is possible that more than two jets (or the wrong two jets) are selected because of, \eg ISR and FSR. A discriminator is constructed that is able to separate {\cPqb} jets and non-{\cPqb} jets. To avoid biasing the measurement of the {\cPqb} jet efficiency, only variables related to the kinematics of the event are used. For each jet $j$ in the event, the angular distance $\Delta R(\ell,j)$ is calculated between the jet and the two leptons. The following variables are calculated for each jet:
\begin{itemize}
\item $M(\ell, j)$: Invariant mass of the $(\ell,j)$ pair with the smallest and the largest $\Delta R(\ell,j)$.
\item $\Delta\eta(\ell,j)$ and $\Delta\phi(\ell,j)$: Difference in pseudorapidity and in azimuthal angle between the lepton and jet for the $(\ell,j)$ pair with the smallest and the largest $\Delta R(\ell,j)$.
\item $\Delta\eta(\ell\ell,\ell j)$ and $\Delta\phi(\ell\ell,\ell j)$: Difference in pseudorapidity and azimuthal angle between the dilepton system and an $(\ell,j)$ pair for the $(\ell,j)$ pair with the smallest and the largest $\Delta R(\ell,j)$.
\item $\Delta\eta(\ell\ell,j)$ and $\Delta\phi(\ell\ell,j)$: Difference in pseudorapidity and azimuthal angle between the dilepton system and the jet.
\end{itemize}
The variables in the first two items are sensitive to $(\ell,j)$ pairs originating from the same top quark decay, while the variables in the two latter items use the correlation between the spin of the top quark and the top antiquark that is present in \ttbar events~\cite{Mahlon:2010gw}.
The 12 variables listed above are combined with a BDT using the TMVA package~\cite{TMVA}. Prior to the training on simulated \ttbar events, jets in the event are classified according to their rank when ordered according to decreasing \pt. In particular, the training is performed in three different categories for the leading, subleading, and other jets. This classification helps to better use the correlations between the variables for the signal and background, in particular for events with a high jet multiplicity. The parameters of the BDT, such as the number of trees, the depth and the shrinkage factor of the gradient learning algorithm, were roughly optimized to obtain a smooth background shape at large discriminator values without reducing the discriminating power.

A binned likelihood fit is performed on the kinematic discriminator of jets passing and failing the {\cPqb} tagging requirement, inclusively for all jets together. For each flavour $f$, the total number of jets $N_f$ can be expressed as a function of the tagging efficiency in simulation, $\varepsilon_f^{\text{MC}}$, and the data-to-simulation scale factor $SF_f$:
\begin{linenomath}
\begin{equation}
\begin{cases}
N_f^{\text{tagged}}= SF_f \varepsilon_f^{\text{MC}} N_f = SF_f N_f^{\text{MC, tagged}}\\
N_f^{\text{vetoed}}= (1-SF_f \varepsilon_f^{\text{MC}}) N_f =\frac{(1-SF_f \varepsilon_f^{\text{MC}})}{ SF_f \varepsilon_f^{\text{MC}}} N_f^{\text{MC, tagged}},
\end{cases}
\label{eq:kinpopulations}
\end{equation}
\end{linenomath}
where $N_f^{\text{MC, tagged}}$ is the expected number of jets of flavour $f$ passing the requirement determined from simulation. The templates for light-flavour and {\cPqc} jets are similar. The measured value of the mistag data-to-simulation scale factor, as presented in Section~\ref{sec:negtag}, is used to correct for the different misidentification probability in the data and simulation. It is not necessary to use a dedicated scale factor for {\cPqc} jets since the fraction of {\cPqc} jets is expected to be less than 1\% and fully covered by the systematic uncertainties.
The scale factor for {\cPqb} jets is the only free parameter to be determined from the fit. The fit is performed simultaneously in bins of jet multiplicity, with up to four jets. For convenience, the discriminator values are transformed from $[-1,1]$ to $[-1,1]+ 2 (N_{\text{jets}}-2)$.

Several sources of systematic uncertainties are considered:
\begin{itemize}
\item \textbf{Factorization and renormalization scales}: The uncertainty in the factorization and renormalization scales is evaluated in the same way as in Section~\ref{sec:mauro}, except that the scale for FSR in the parton shower is varied by a factor of two up and down, and not by a factor of $\sqrt{2}$. In addition, both the variation of the scale in the parton shower as well as the variation of the hdamp parameter in \POWHEG are taken into account to assess the impact of ISR and FSR instead of using the largest variation. Although these systematic uncertainties are correlated, they are conservatively treated as uncorrelated.
\item \textbf{Cross section of background processes}: The cross section of each non-\ttbar background process is varied by 30\% to assess the systematic effect due to the uncertainty in the background contributions.
\item \textbf{Top quark mass}: The uncertainty in the top quark mass is evaluated in the same way as in Section~\ref{sec:mauro}.
\item \textbf{Scale factor for non-{\cPqb} jets}: The data-to-simulation scale factor for light-flavour jets, $SF_{\text{l}}$, is applied to correct the expected fraction of light-flavour and {\cPqc} jets. To evaluate the uncertainty related to $SF_{\text{l}}$, the value is changed to $SF_{\text{l}} \pm 1\sigma$, where $\sigma$ represents the uncertainty in $SF_{\text{l}}$. The effect of this variation on the measured value of $SF_{\cPqb}$ is taken as the size of the uncertainty due to $SF_{\text{l}}$.
\item \textbf{Jet energy scale}: The uncertainty in the jet energy scale is assessed in the same way as in Section~\ref{sec:mauro}. The variation in jet momentum is simultaneously propagated to the {\ptmiss} value for a consistent approach.
\item \textbf{Jet energy resolution}: The uncertainty in the jet energy resolution is assessed in the same way as in Section~\ref{sec:mauro}.
\item \textbf{Selection efficiency}: The uncertainty in the lepton identification and isolation efficiency is propagated to the measurement by reweighting the simulation using a lepton efficiency scale factor that is shifted up or down by one standard deviation with respect to the nominal value.
\item \textbf{Pileup}: The uncertainty in the pileup modelling is assessed as described in Section~\ref{sec:negtag}.
\end{itemize}
The systematic effect induced by the uncertainty in the parton distribution functions is negligible.

\begin{figure}[hbtp]
  \centering
     \includegraphics[width=0.49\textwidth]{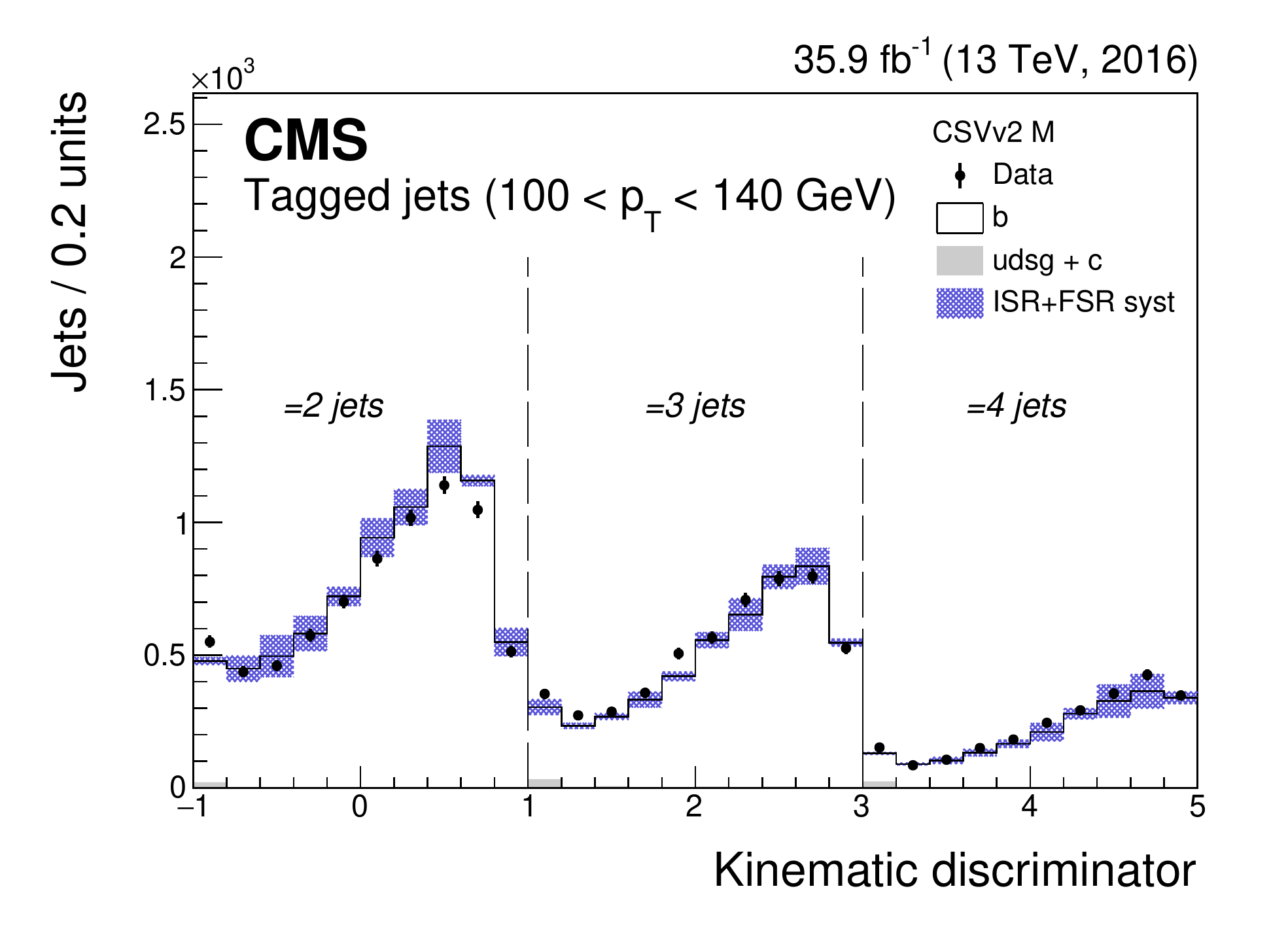}
     \includegraphics[width=0.49\textwidth]{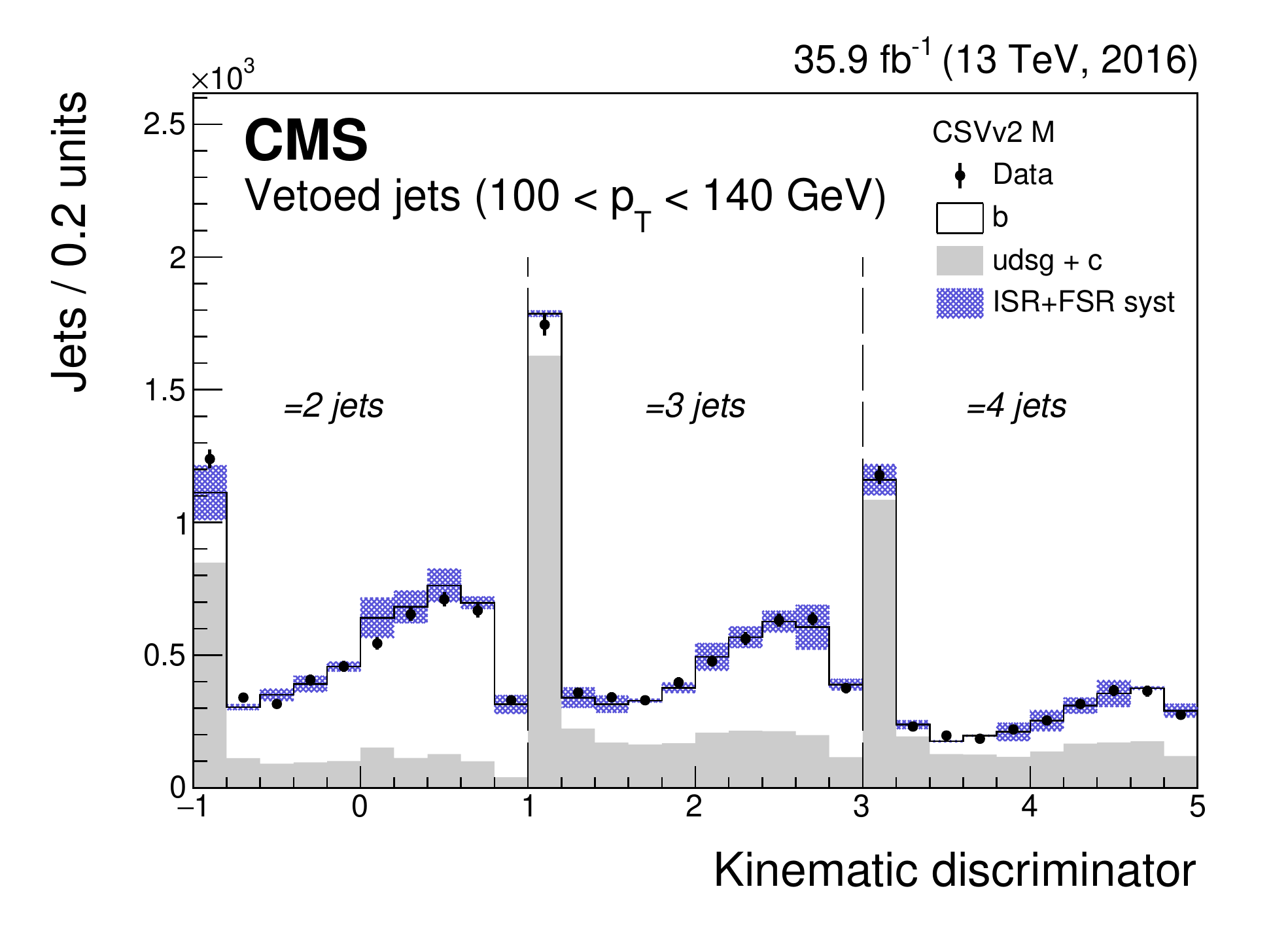}
    \caption{Fitted distribution of the kinematic discriminator for jets with $100<\pt<140\GeV$ passing (left) and failing (right) the medium working point of the CSVv2 algorithm. The discriminator distribution is shown in bins of jet multiplicity with the discriminator output transformed from $[-1,1]$ to $[-1,1]+ 2 (N_{\text{jets}}-2)$. The dominant systematic uncertainty due to initial- and final-state radiation is represented by the band.}
    \label{fig:kindiscr2}
\end{figure}

\begin{figure}[hbtp]
  \centering
     \includegraphics[width=0.49\textwidth]{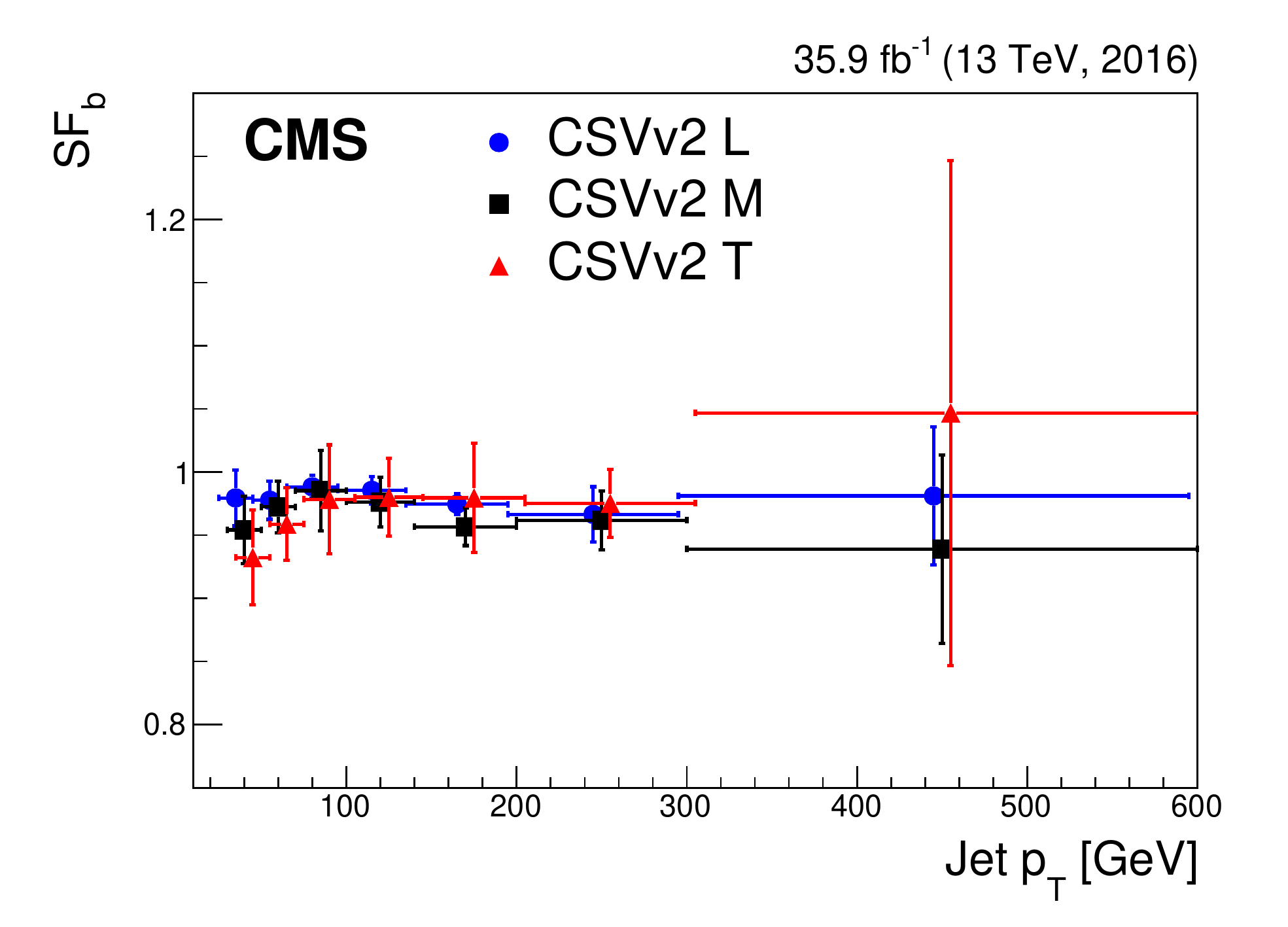}
     \includegraphics[width=0.49\textwidth]{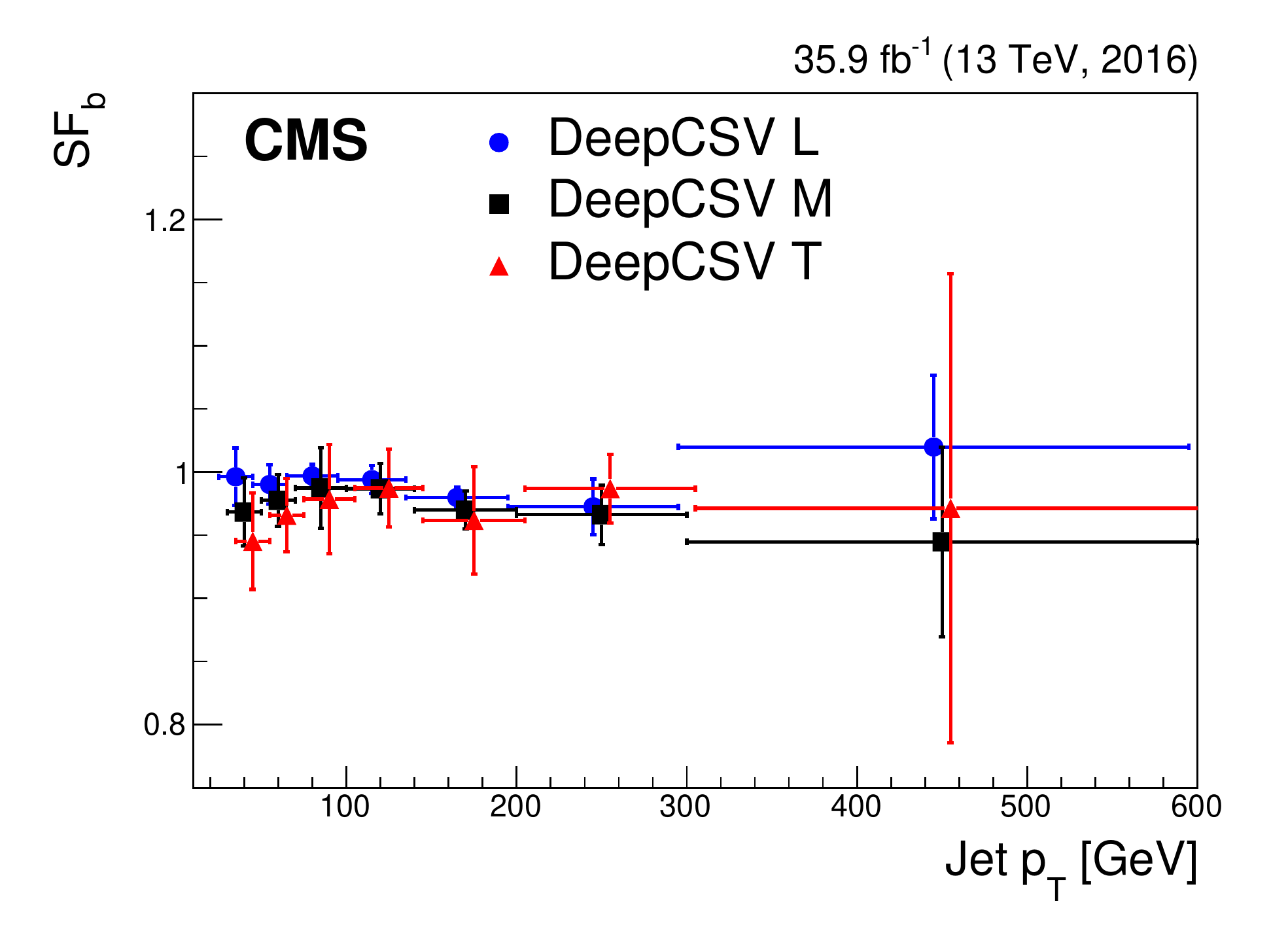}
    \caption{Data-to-simulation scale factors for {\cPqb} jets obtained with the Kin method as a function of the jet \pt for the three CSVv2 (left) and the three DeepCSV (right) working points. The uncertainty corresponds to the combined statistical and systematic uncertainty. For clarity, the points for the loose and tight tagging requirement are shifted by $-5$ and $+5$\GeV with respect to the bin centre.}
    \label{fig:kinfitcsv2}
\end{figure}

To determine the dependence of the data-to-simulation scale factor on the jet \pt, independent fits are performed in mutually exclusive bins of jet \pt. An example of the fitted distribution in the jet \pt range between 100 and 140\GeV using jets passing and failing the medium working point of the CSVv2 algorithm is shown in left and right panels of Fig.~\ref{fig:kindiscr2}, respectively. Discrepancies between the data and simulation are covered by the combined statistical and systematic uncertainty. The scale factor as a function of the jet \pt is shown in Fig.~\ref{fig:kinfitcsv2} for the three working points of the CSVv2 and DeepCSV algorithms.

\paragraph{Two-tag counting method}

The two-tag counting (TagCount) method is mainly used as a cross check of the Kin method. While the Kin method is able to determine data-to-simulation scale factors at higher jet \pt, the TagCount method is a simple and robust approach to assess the size of the scale factors. The dilepton \ttbar events are selected by requiring the dilepton invariant mass $M_{\ell \ell}>12\GeV$. If the two leptons have the same flavour, the contribution from {\Zj} events is reduced by applying a veto around the Z boson mass, $|M_{\ell \ell}-M_{\PZ}|>10$\GeV,  and requiring ${\ptmiss} > 50\GeV$. In addition, each event is required to have exactly two jets.

The {\cPqb} jet identification efficiency, $\varepsilon_{\cPqb}$, can be obtained by counting the number of events with two {\cPqb}-tagged jets in the selected sample of events:
\begin{linenomath}
\begin{equation}
	N_{\text{2 {\cPqb}-tagged}} - N_{\text{2 {\cPqb}-tagged}}^{\text{non-{\cPqb} jet}} = \varepsilon_{\cPqb}^2 n_{\text{2 {\cPqb} jets}}, \label{eq:1}
\end{equation}
\end{linenomath}
where $N_{\text{2 {\cPqb}-tagged}}$ is the number of events with two {\cPqb}-tagged jets from data, $N_{\text{2 {\cPqb}-tagged}}^{\text{non-{\cPqb} jet}}$ is the number of events with two {\cPqb}-tagged jets with at least one of them being a light-flavour or {\cPqc} jet, and $n_{\text{2 {\cPqb} jets}}$ is the number of events with two true {\cPqb} jets. This equation can be solved for $\varepsilon_{\cPqb}$ if $N_{\text{2 {\cPqb}-tagged}}^{\text{non-{\cPqb} jet}}$ and $n_{\text{2 {\cPqb} jets}}$ are known. To reduce the dependence on the \ttbar production cross section, the equation is divided by the number of selected events,
\begin{linenomath}
\begin{equation}
 \varepsilon_{\cPqb}=\sqrt{\frac{F_{\text{2 {\cPqb}-tagged}} - F_{\text{2 {\cPqb}-tagged}}^{\text{non-{\cPqb} jet}}}{f_{\text{2 {\cPqb} jets}}}}
\end{equation}
\end{linenomath}
where $F_{\text{2 {\cPqb}-tagged}}$ is the fraction of events with two {\cPqb}-tagged jets, $F_{\text{2 {\cPqb}-tagged}}^{\text{non-{\cPqb} jet}}$ is the fraction of events with two {\cPqb}-tagged jets of which at least one is a non-{\cPqb} jet, and $f_{\text{2 {\cPqb} jets}}$ is the fraction of events with two true {\cPqb} jets. The two latter fractions are obtained from simulation. When the method is used to measure the efficiency as a function of the jet \pt, the two tagged jets are required to be in the same jet \pt bin.

While the method is sensitive to the uncertainties in the predicted fraction of events with non-{\cPqb} jets $F_{\text{2 {\cPqb}-tagged}}^{\text{non-{\cPqb} jet}}$, using the fraction of events ensures that systematic uncertainties related to the number of \ttbar events cancel out. The dominant uncertainties originate from the normalization of background events and the fraction of non-{\cPqb} jet events in the bin with two {\cPqb}-tagged jets. The following systematic effects were studied:
\begin{itemize}
\item \textbf{The fraction of non-{\cPqb} jets ($F_{\text{2 {\cPqb}-tagged}}^{\text{non-{\cPqb} jet}}$)}: A conservative variation of 50\% is used to estimate the uncertainty in the fraction of non-{\cPqb} jets. This represents the leading uncertainty in the final data-to-simulation scale factor for the loose working point of the {\cPqb} jet identification algorithms.
\item \textbf{Background yield}: The effect of the uncertainty in the background estimation for the {\Zj} background obtained from data is evaluated by varying its normalization by 50\%. For the background yields that are estimated from the simulation, an uncertainty of 30\% is assumed. This uncertainty is the subleading source of uncertainty.
\item \textbf{Factorization and renormalization scales}: The uncertainty in the factorization and renormalization scales is assessed as described in Section~\ref{sec:mauro}, except for the scale for FSR in the parton shower that is varied by a factor of two up and down, and not by factor of $\sqrt{2}$.
\item \textbf{Jet energy scale}: The uncertainty in the jet energy scale is propagated to an uncertainty in the data-to-simulation scale factor as described in Section~\ref{sec:Wc}.
\item \textbf{Jet energy resolution}: The uncertainty in the jet energy resolution is addressed as described in Section~\ref{sec:mauro}.
\item \textbf{Pileup}: The systematic effect related to the uncertainty in the number of pileup interactions is evaluated as described in Section~\ref{sec:negtag}.
\end{itemize}
The systematic effects related to the uncertainty in the top quark mass and the parton distribution functions are negligible compared to the impact of the uncertainty in the background yield and the number of non-{\cPqb} jets.

\begin{figure}[hbtp]
  \centering
     \includegraphics[width=0.49\textwidth]{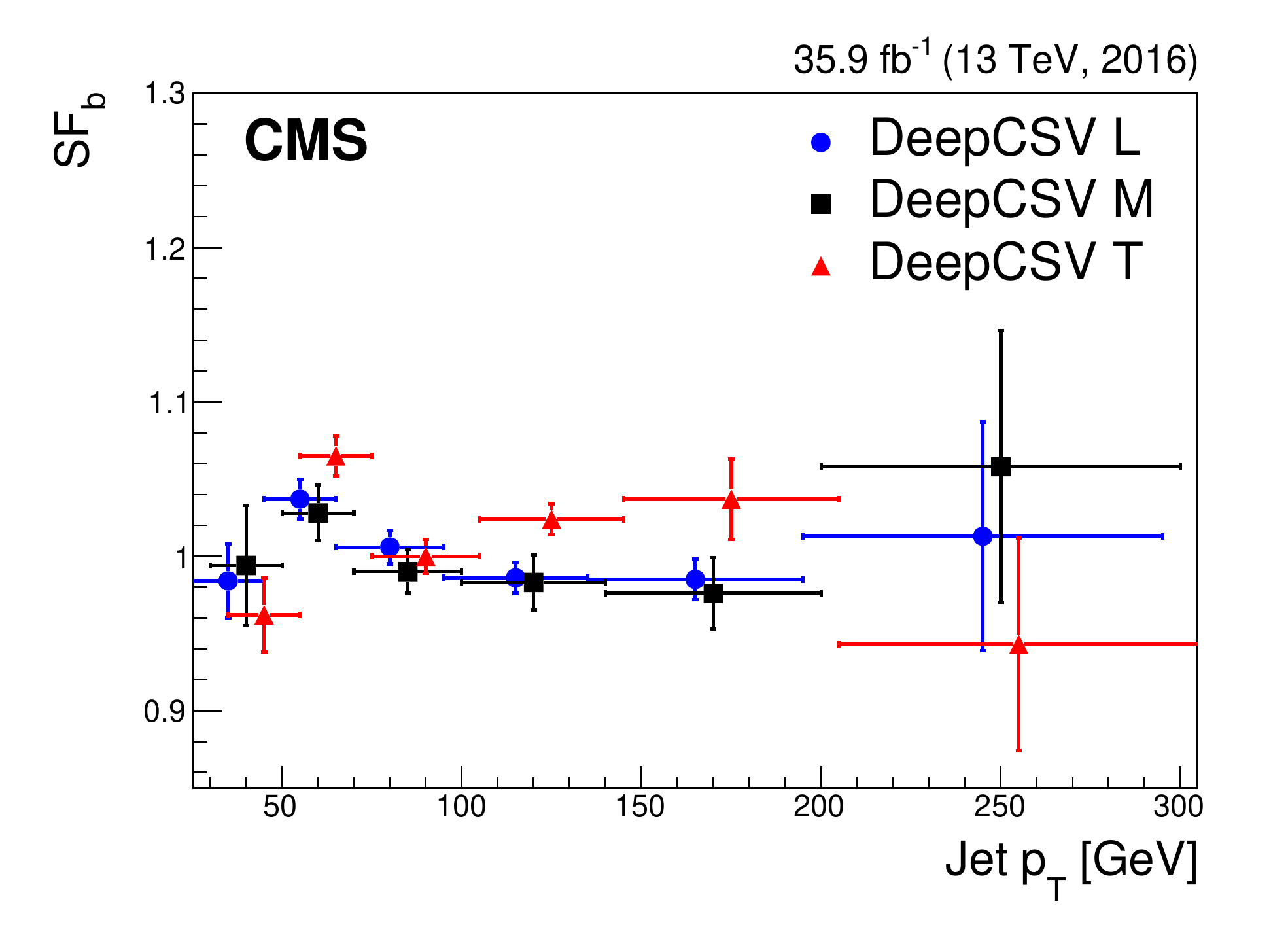}
     \includegraphics[width=0.49\textwidth]{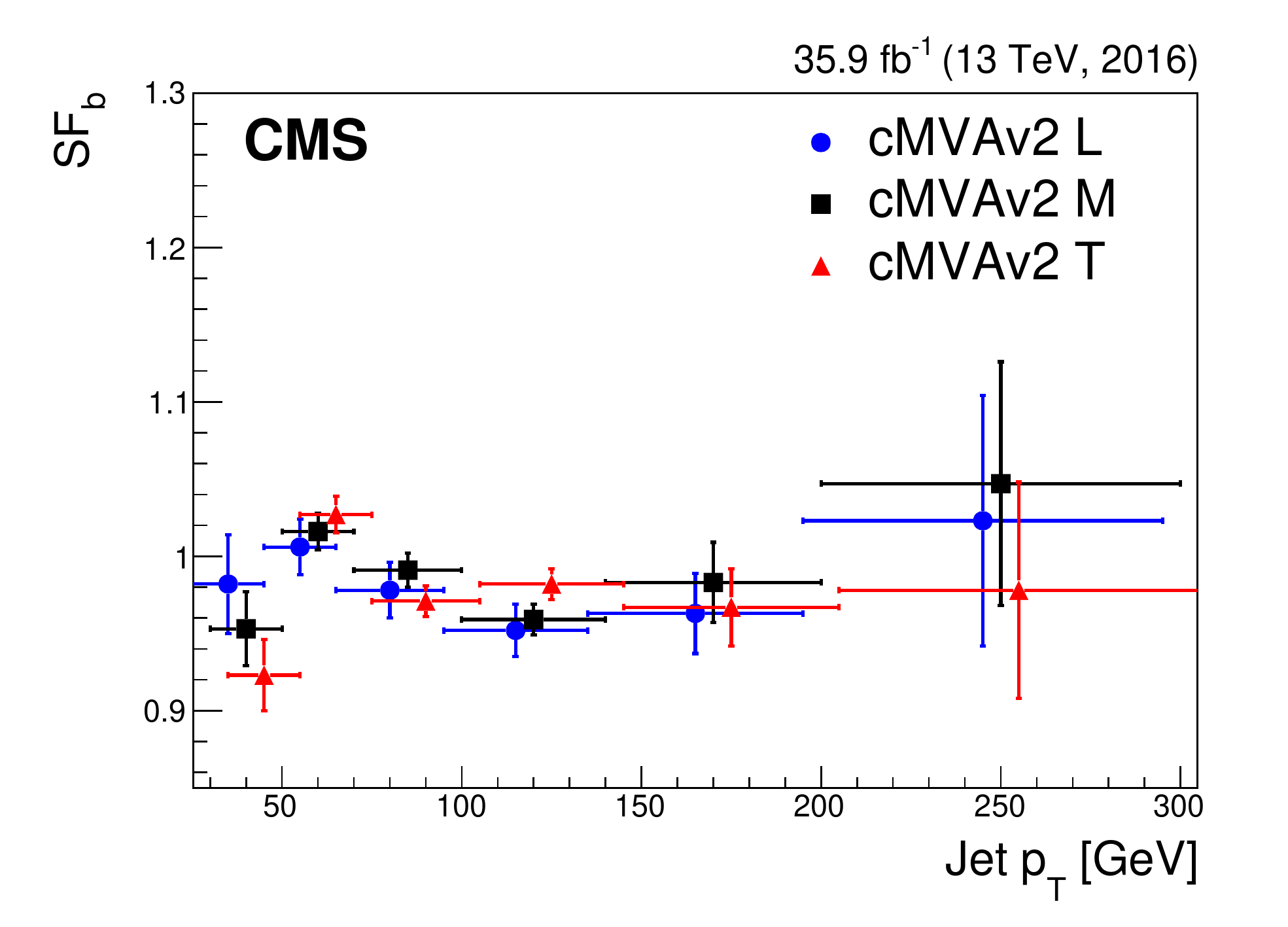}
    \caption{Data-to-simulation scale factors for {\cPqb} jets measured with the TagCount method as a function of the jet \pt for the three DeepCSV (left) and cMVAv2 (right) working points. The uncertainty corresponds to the combined statistical and systematic uncertainty. For clarity, the points for the loose and tight tagging requirement are shifted by $-5$ and $+5$\GeV with respect to the bin centre.}
    \label{fig:2tagresult}
\end{figure}

The {\cPqb} jet identification efficiency is determined in bins of jet \pt and the corresponding data-to-simulation scale factors are shown in Fig.~\ref{fig:2tagresult}.
Large bin-to-bin variations are observed for low-\pt jets, in particular for the tight working points of the taggers, for which the statistical uncertainty dominates.

\subsubsection{Tag-and-probe technique using single-lepton \texorpdfstring{\ttbar}{ttbar} events}
\label{sec:tagandprobe}
In addition to dilepton \ttbar events, one can also use the \ttbar topology where only one of the {\PW} bosons decays to leptons. In this case, two {\cPqb} jets are expected from the top quark decays as well as two non-{\cPqb} jets from the decay of one of the {\PW} bosons. The decay chain ${\cPqt}\to{\cPqb}{\PW}\to{\cPqb}{\cPq}{\cPaq}$ is referred to as the hadronic side, while ${\cPqt}\to{\cPqb}{\PW}\to{\cPqb}{\ell}{\Pgn}$ is the leptonic side.
The event selection criteria are similar to those described in Section~\ref{sec:mauro}, requiring exactly one isolated muon or electron with $\pt > 30\GeV$ and satisfying tight identification criteria~\cite{Khachatryan:2015hwa,Chatrchyan:2012xi} and exactly four jets with $\pt>30$\GeV to reduce the possible number of jet-quark assignments.

To enhance the {\cPqb} quark content in the jet sample on which the {\cPqb} tagging efficiency will be determined, the jets need to be correctly assigned to the quarks from which they originate. To achieve this, a likelihood method is used that is described in detail in Ref.~\cite{Khachatryan:2016mnb}. The reconstruction of the \ttbar topology is enhanced by determining first the four-momentum of the neutrino $p_{\Pgn}$ using the {\PW} boson and top quark mass constraints, $(p_{\Pgn} + p_{\ell})^2 = m_{\PW}^2$ and $(p_{\Pgn} + p_{\ell} + p_{\cPqb,\ell})^2 = m_{\cPqt}^2$, with $p_{\ell}$ and $p_{\cPqb,\ell}$ being the four-momenta of the charged lepton and of the {\cPqb} jet candidate on the leptonic side, respectively. If both equations need to be satisfied, the possible solutions are found on an ellipsoid in the 3D momentum space of the neutrino. For each solution, the distance $D_{\Pgn}$ is computed between the ellipse projection on the transverse plane and the \ptvecmiss vector. The solution of $p_{\Pgn}$ for which this distance is minimal, $D_{\Pgn\text{,min}}$, is used. More details on this procedure and its performance can be found in Ref.~\cite{Betchart:2013nba}.
Once the neutrino momentum is defined, the jets are assigned to the quarks by choosing the jet-quark assignment that minimizes the negative logarithm of the likelihood $\lambda$. For each permutation $-\log(\lambda)$ is obtained as:
\begin{linenomath}
\begin{equation}
-\log(\lambda)= -\log(\lm)-\log(\lambda_{\Pgn}),
\end{equation}
\end{linenomath}
where $\lm=\mathcal{P}_m(m_2,m_3)$ is the 2D probability distribution of the invariant mass of the correctly reconstructed {\PW} boson on the hadronic side ($m_2$) and the invariant mass of the correctly reconstructed top quark on the hadronic side ($m_3$), that was already introduced in Section~\ref{sec:mauro}. Similarly, $\lambda_{\Pgn}=\mathcal{P}_{\Pgn}(D_{\Pgn\text{,min}})$ is the probability distribution of $D_{\Pgn\text{,min}}$ for the correct assignment of the {\cPqb} jet on the leptonic side. While \lm is sensitive to the correct reconstruction of the top quark on the hadronic side, $\lambda_{\Pgn}$ is sensitive to the correct reconstruction of the top quark on the leptonic side.

Once the jets are assigned to the quarks, a tag-and-probe (TnP) technique is applied to determine the {\cPqb} tagging efficiency from data. As a tagging requirement, the medium working point of the CSVv2 algorithm is applied to either the {\cPqb} jet on the hadronic or leptonic side while the {\cPqb} jet from the other side is used as probe. The event is rejected if the tagging requirement is not satisfied. The probe jets are used to determine the {\cPqb} tagging efficiency of a given working point for each tagger under consideration. To achieve that, the distributions of $-\log(\lambda)$ and {\ptmiss} for probe jets passing and failing the tagging requirement are fitted with their expected templates to determine their number in data for the correctly-reconstructed \ttbar events. During the fit, the normalization of the template for the non-\ttbar background is naturally constrained by the {\ptmiss} distribution during the simultaneous fit. The {\cPqb} tagging efficiency in data is then obtained from the fitted fraction of probe jets passing the tagging requirement with respect to all probe jets, as in Eq.~(\ref{eq:effdata}).
To increase the number of probe jets and to avoid a possible bias in the measurement, each {\cPqb} jet is used once as tag and once as probe. While the measurements are performed separately with either the {\cPqb} jet from the hadronic or the leptonic side as probe jet, they are afterwards combined by treating all systematic uncertainties as correlated. The measurement is performed in bins of jet \pt.

\begin{figure}[hbtp]
  \centering
     \includegraphics[width=0.49\textwidth]{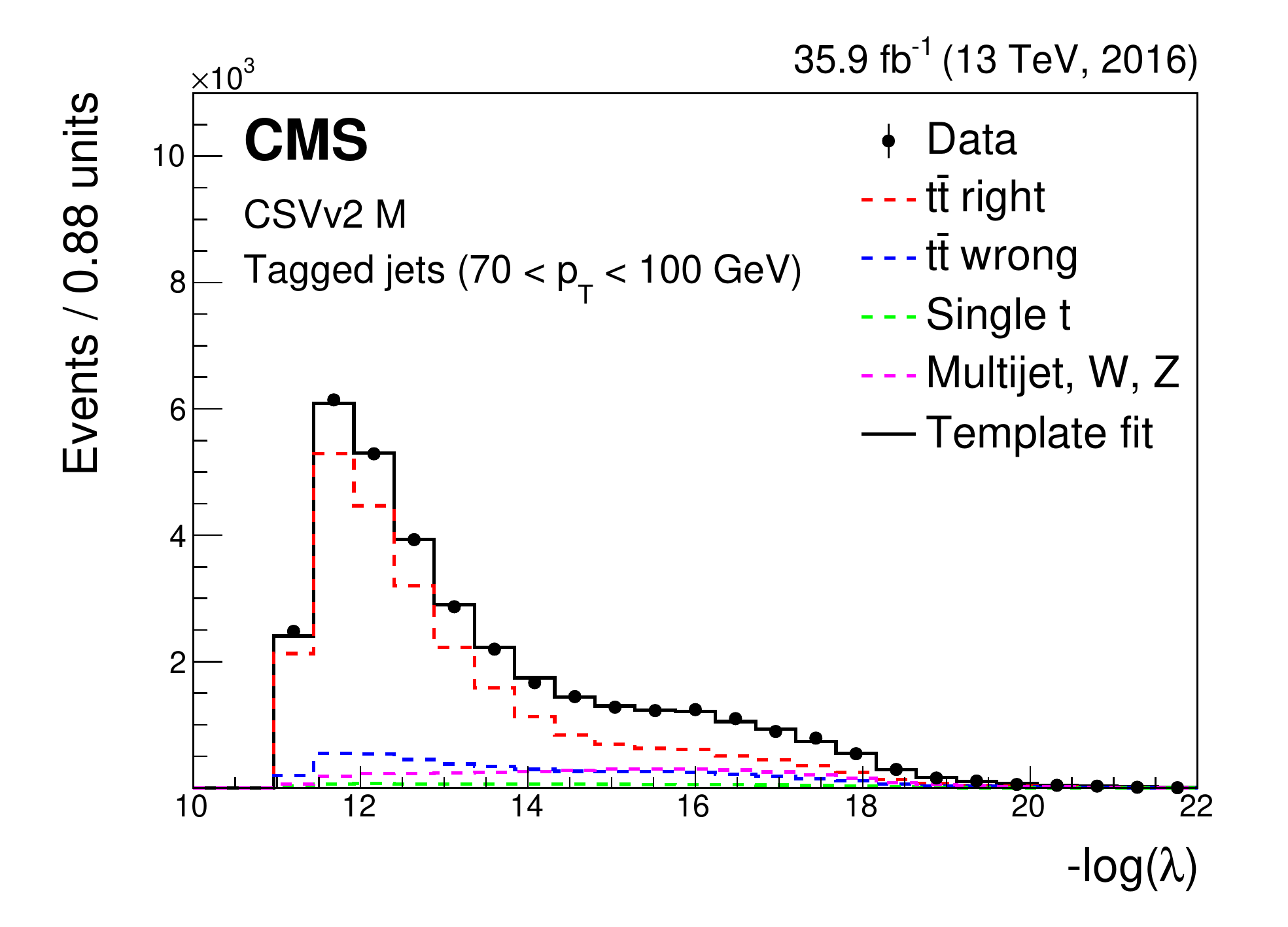}
     \includegraphics[width=0.49\textwidth]{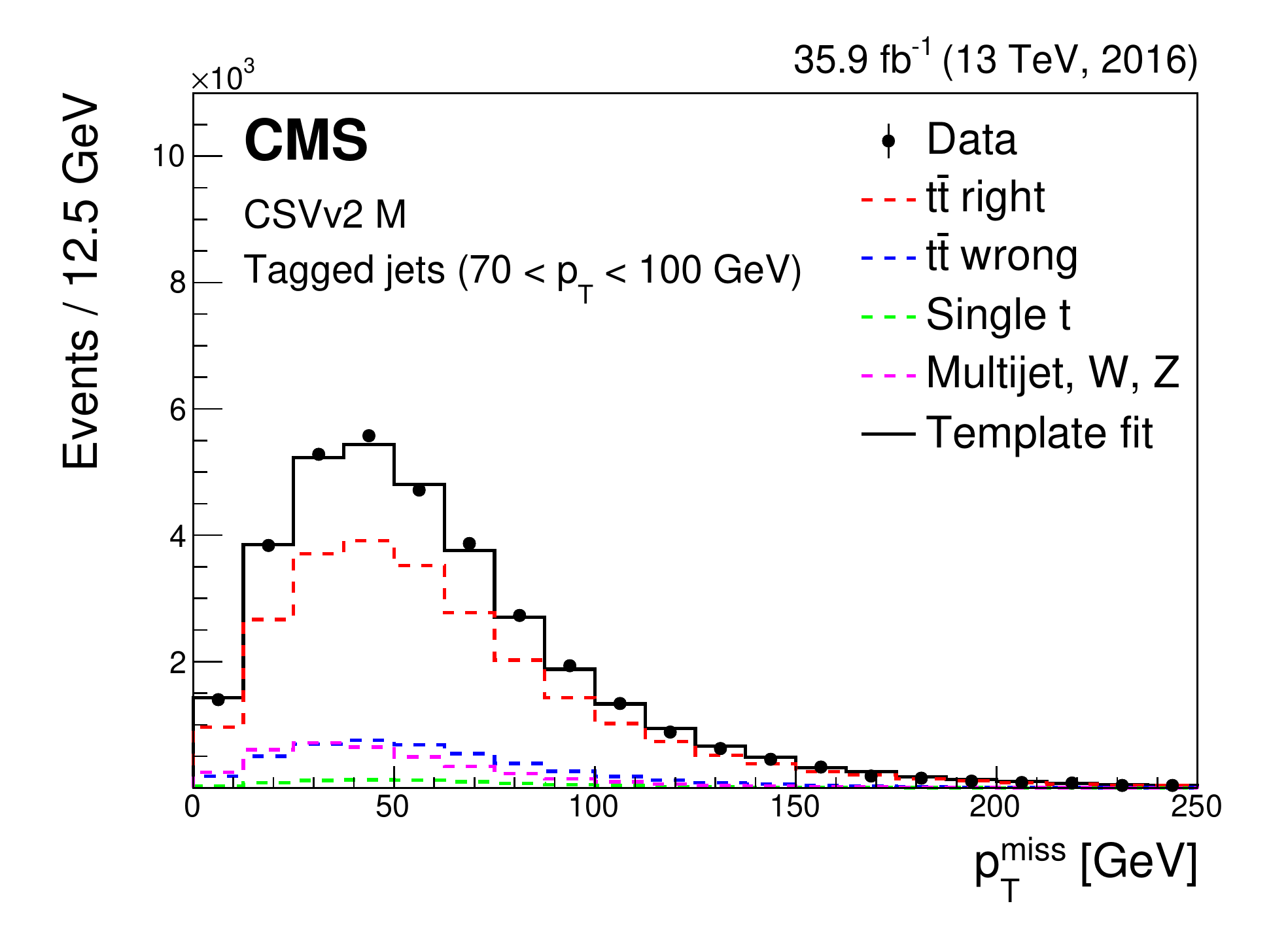}\\
     \includegraphics[width=0.49\textwidth]{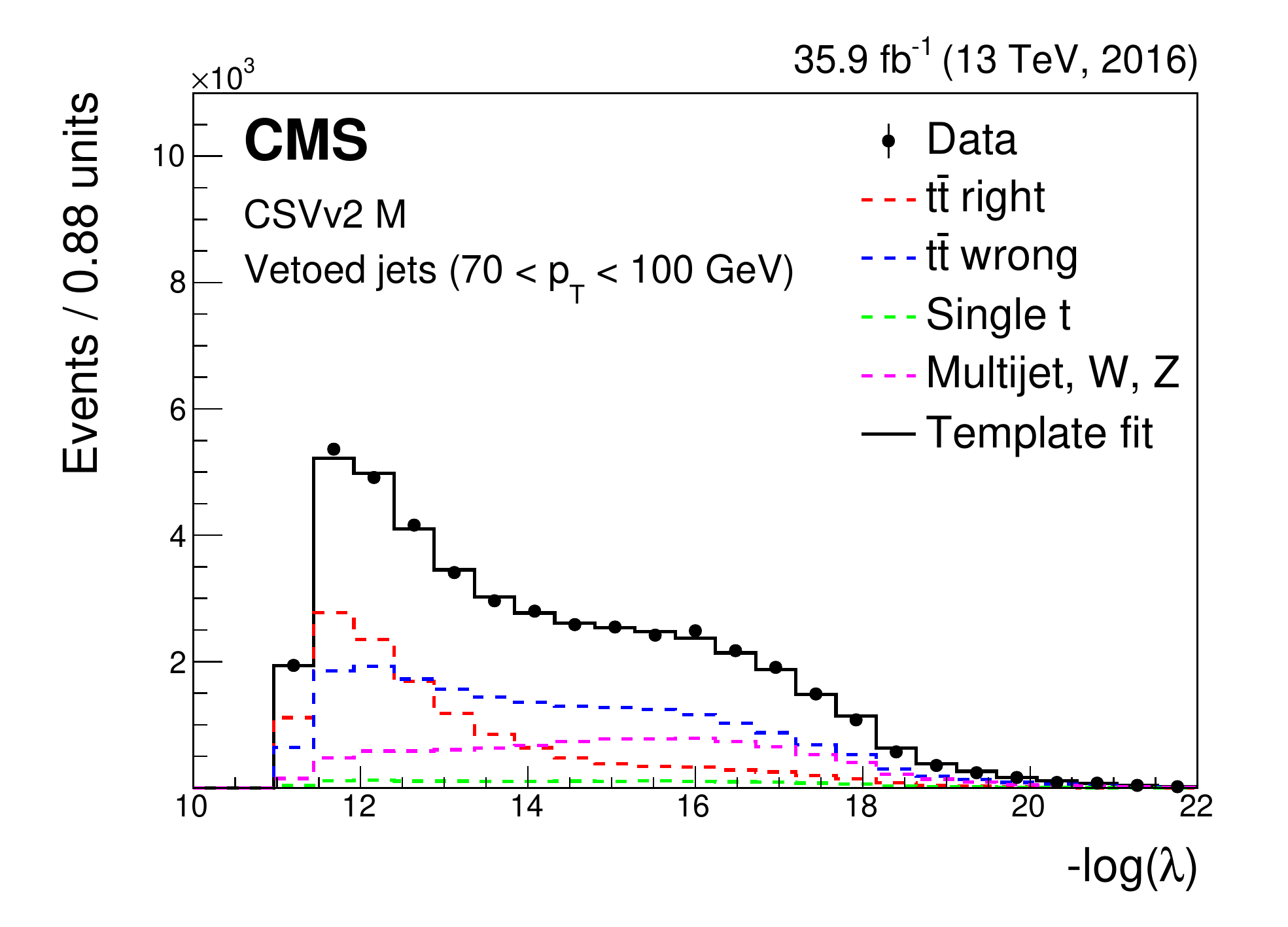}
     \includegraphics[width=0.49\textwidth]{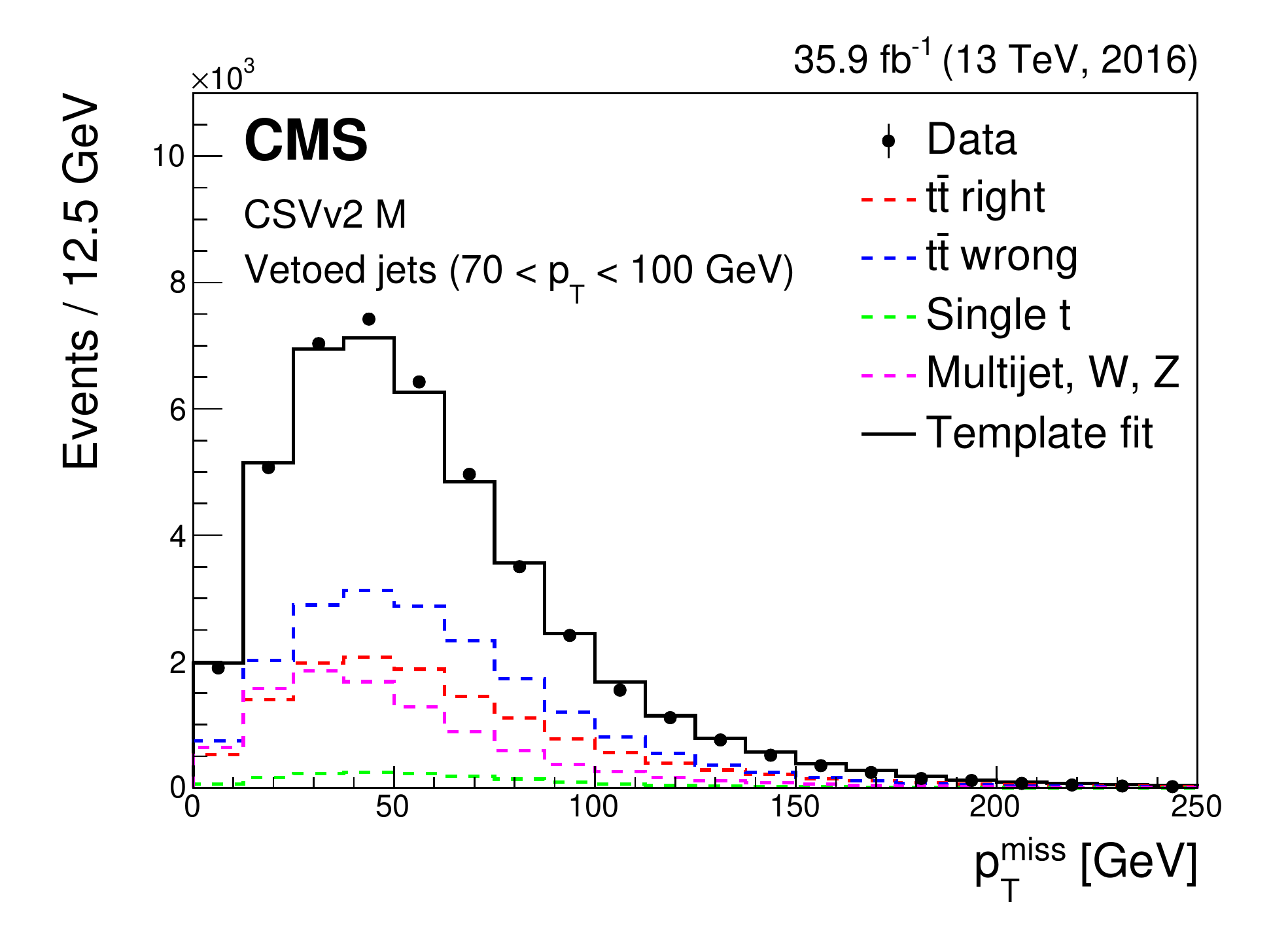}
    \caption{Distributions of fitted $-\log(\lambda)$ (left) and {\ptmiss} (right) for jets from the \ttbar leptonic side with $70 <\pt< 100\GeV$ passing (upper) and failing (lower) the medium working point of the CSVv2 algorithm.}
    \label{fig:tagandprobetemplatefit}
\end{figure}

Figure~\ref{fig:tagandprobetemplatefit} shows an example of the fitted $-\log(\lambda)$ and {\ptmiss} distributions for probe jets from the leptonic side, with $70 < \pt < 100\GeV$, passing and failing the medium working point of the CSVv2 algorithm.
The template distributions for correctly and wrongly reconstructed \ttbar events are obtained from simulation. Also the template distribution for single top quark events is taken from simulation and in addition, its normalization is constrained within 20\% of the expected standard model yield. The non-\ttbar background is composed of multijet, {\Zj}, and {\Wj} events and the combined template for these background processes is derived from data in a control region. The control region contains the events for which the jet with the highest CSVv2 discriminator value is below 0.6.

\begin{figure}[hbtp]
  \centering
     \includegraphics[width=0.49\textwidth]{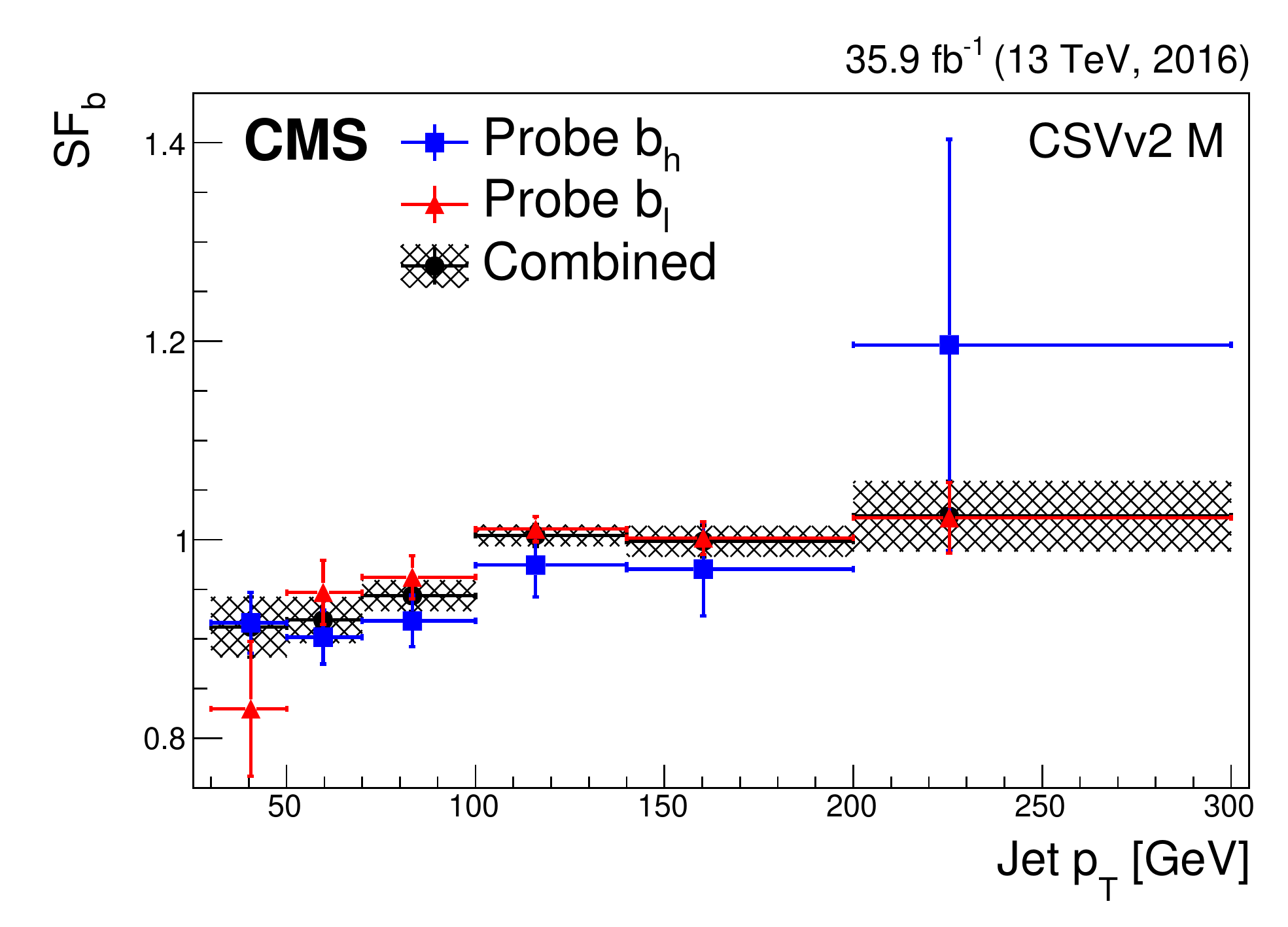}
     \includegraphics[width=0.49\textwidth]{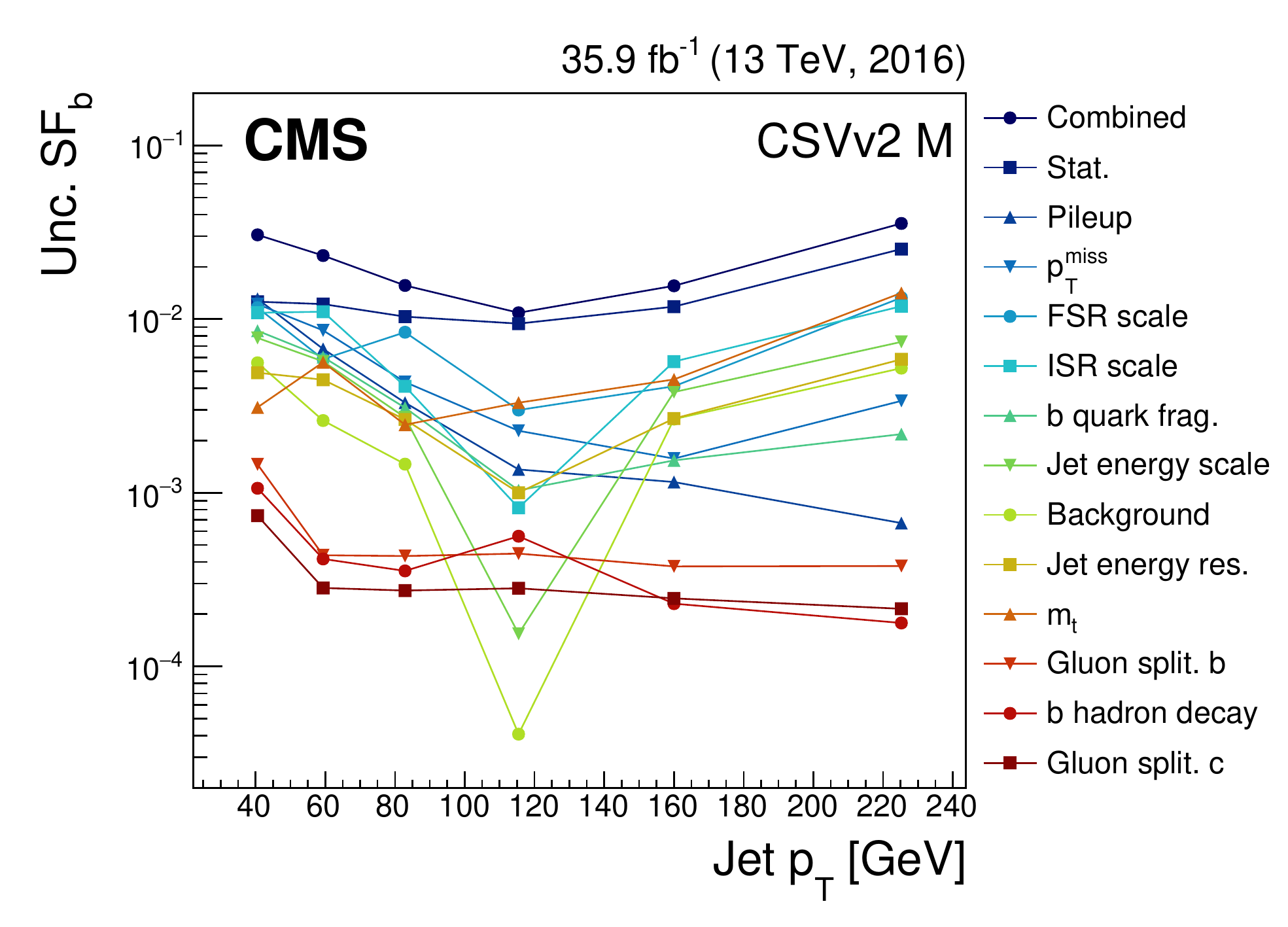}
    \caption{Data-to-simulation scale factors for {\cPqb} jets from the hadronic or leptonic side of the single-lepton \ttbar decay as well as for their combination, as a function of the jet \pt for the medium working point of the CSVv2 tagger (left). The error bars represent the combined statistical and systematic uncertainty. Size of the individual uncertainties in the combined scale factors for the CSVv2 medium working point (right).}
    \label{fig:tagandproberesult}
\end{figure}

\begin{figure}[hbtp]
  \centering
     \includegraphics[width=0.49\textwidth]{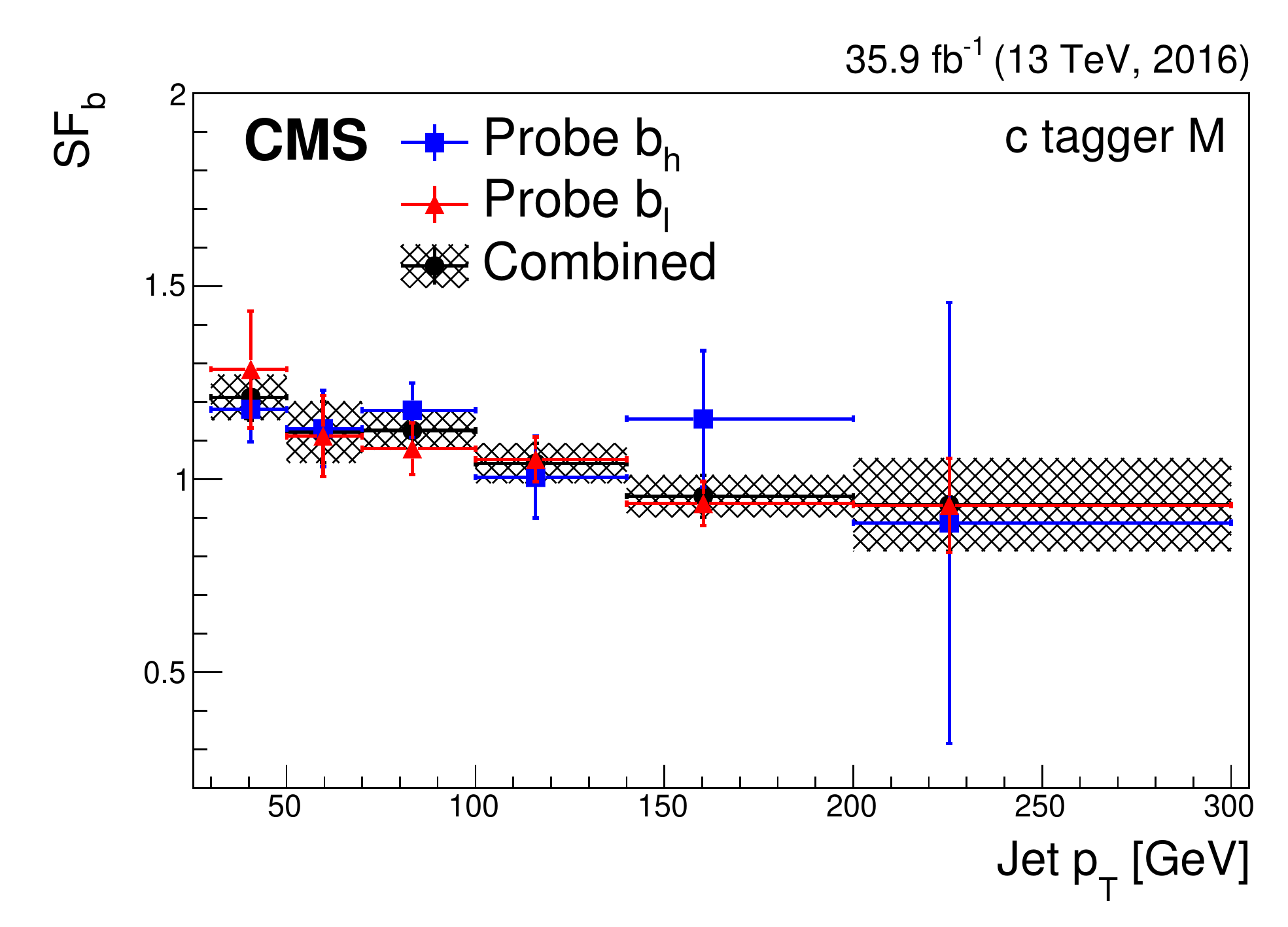}
     \includegraphics[width=0.49\textwidth]{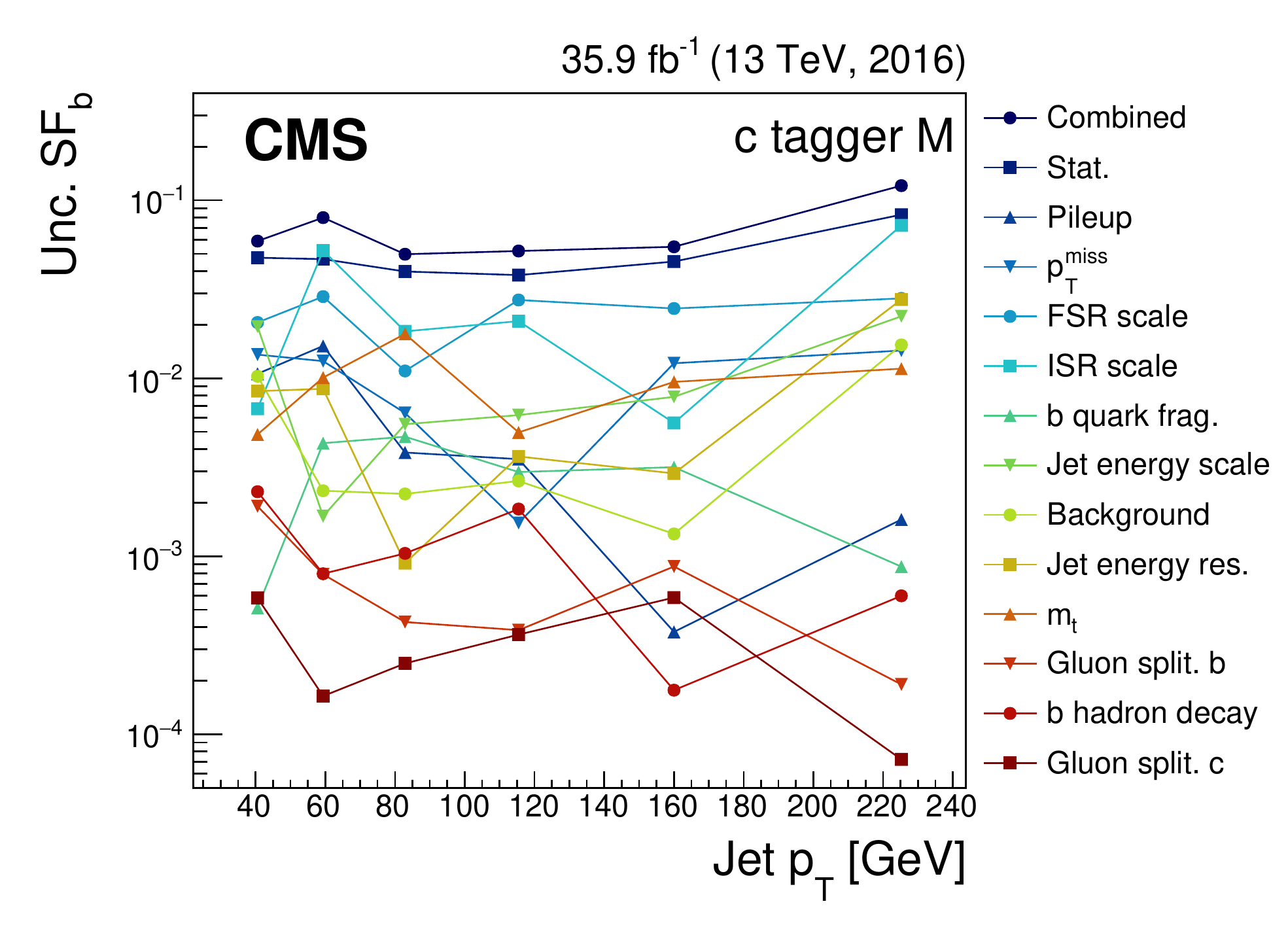}
    \caption{Data-to-simulation scale factors for {\cPqb} jets from the hadronic or leptonic side of the single-lepton \ttbar decay as well as for their combination, as a function of the jet \pt for the medium working point of the {\cPqc} tagger (left). The error bars represent the combined statistical and systematic uncertainty. Size of the individual uncertainties in the combined scale factors for the medium working point of the {\cPqc} tagger (right).}
    \label{fig:tagandproberesult2}
\end{figure}

Several sources of systematic effects may impact the measurement of the {\cPqb} tagging efficiency. These effects are related to the data-taking conditions or to the uncertainty in the object reconstruction, affecting the selection of events and reconstruction of the \ttbar topology. On the other hand, systematic effects are related to the modelling of the \ttbar production and decay. In particular the following sources of systematic effects have been taken into account:
\begin{itemize}
\item \textbf{Factorization and renormalization scales}: The uncertainty due to the factorization and renormalization scales is assessed as described in Section~\ref{sec:mauro}.
\item \textbf{Top quark mass}: The uncertainty in the top quark mass is propagated to the data-to-simulation scale factor measurement as described in Section~\ref{sec:mauro}.
\item \textbf{Background}: The non-\ttbar background template is derived using events for which the jet with the highest CSVv2 discriminator value is below 0.6. The systematic effect due to this requirement is evaluated by varying its value to less than 0.3, or to values between 0.4 and 0.7. Although these alternative selections result in a different relative fraction of {\cPqb} and non-{\cPqb} jets, as well as in a different background composition, the overall template shape and the fitted value for the number of correctly reconstructed \ttbar events is stable.
\item \textbf{Gluon splitting}: The uncertainty in the gluon splitting into a heavy quark pair is estimated by reweighting events with at least one additional heavy quark that is not originating from the \ttbar decay. Events with an additional {\cPqc} and {\cPqb} quark are reweighted by $\pm$15\%~\cite{csplitting} and $\pm$25\%~\cite{bsplitting}, respectively. As can be seen in Fig.~\ref{fig:tagandproberesult} (right) the effect is relatively small.
\item \textbf{{\cPqb} quark fragmentation}: The uncertainty in the {\cPqb} quark fragmentation function is estimated by varying the Bowler--Lund parameterization within the tune uncertainties. In particular, the parameter \texttt{StringZ:rFactB} in \PYTHIA is varied by $+0.184$ and $-0.197$ to obtain alternative distributions for the ratio of the {\cPqb} hadron \pt to the jet \pt. The \ttbar simulation is then reweighted using these functions and the impact on the measured data-to-simulation scale factor is taken as the size of the systematic effect.
\item \textbf{Branching fraction of ${\PB} \to \ell X$}: The systematic uncertainty induced by the values of the branching fractions of the semileptonic decay of {\cPqb} hadrons may affect the {\cPqb} jet energy response. It is evaluated by reweighting the fractions to the values in Ref.~\cite{Patrignani:2016xqp}. In particular, the branching fraction to leptons is varied by 2.7\% for {\PBz}, by 8\% {\PBs}, by 2.5\% for {\PBp}, and by 21\% for $\Lambda_B$. As can be seen in Figs.~\ref{fig:tagandproberesult} (right) and~\ref{fig:tagandproberesult2} (right), the impact of this variation, labelled ``b hadron decay'', is negligible compared to the other systematic effects.
\item \textbf{Jet energy scale}: The impact of the uncertainty in the jet energy scale and its propagation to {\ptmiss} is assessed as described in Section~\ref{sec:mauro}.
\item \textbf{Jet energy resolution}: The uncertainty in the jet energy resolution is propagated to the data-to-simulation scale factor measurement as described in Section~\ref{sec:mauro}.
\item \textbf{\ptmiss}: The uncertainty in the lepton, photon, and unclustered energy is estimated by changing {\ptmiss} within its uncertainty and repeating the measurement.
\item \textbf{Pileup}: The uncertainty in the pileup modelling is assessed as described in Section~\ref{sec:negtag}.
\end{itemize}

The TnP method is applied to derive data-to-simulation scale factors for the three working points of the CSVv2, DeepCSV, cMVAv2, and {\cPqc} taggers. An example of the size of the systematic uncertainties as a function of the jet \pt is shown in Fig.~\ref{fig:tagandproberesult} (right). In the same figure the scale factor $SF_{\cPqb}$ as a function of the jet \pt is also shown for the medium working point of the CSVv2 algorithm. As discussed previously, $SF_{\cPqb}$ is derived separately for {\cPqb} jets from the hadronic or leptonic side of the single-lepton \ttbar decay. As expected, both results are consistent over the full jet \pt range. To reduce the overall uncertainty, the results are combined using the BLUE method, assuming fully correlated systematic uncertainties and uncorrelated statistical uncertainties.

Figure~\ref{fig:tagandproberesult2} (left) shows the data-to-simulation scale factors for {\cPqb} jets for the medium working point of the {\cPqc} tagger as a function of the jet \pt. Since the probability to tag non-{\cPqb} jets is higher for the {\cPqc} tagger than for the {\cPqb} taggers, the systematic uncertainties will be larger. On the other hand, since the probability to tag {\cPqb} jets with the {\cPqc} tagger is also smaller compared to the {\cPqb} tagger, also the statistical uncertainty increases. This can be seen in Fig.~\ref{fig:tagandproberesult2} (right); while the statistical uncertainty on the measured scale factors still dominates, the systematic uncertainties are significantly larger compared to Fig.~\ref{fig:tagandproberesult} (right). As a result, the total uncertainty for the scale factors for {\cPqb} jets is larger for the {\cPqc} tagger than for the {\cPqb} taggers.

\subsubsection{Combination of the data-to-simulation scale factors from multijet and \texorpdfstring{\ttbar}{ttbar} events}
For the CSVv2 and DeepCSV taggers, the data-to-simulation scale factors measured with the muon-enriched multijet events are combined with the ones measured in \ttbar events using the Kin and TnP methods. Since the {\cPqc} tagger and the cMVAv2 tagger rely on the information from muons from the {\cPqb} hadron decay, the scale factors are only measured with \ttbar events since the muon enrichment of the multijet sample may bias the scale factor measurement. Since the Kin method relies on dilepton \ttbar events and the TnP method on single-lepton \ttbar events, the two scale factor measurements are statistically independent. Similarly as for the combination of the scale factors on the muon-enriched sample, the correlations between the systematic uncertainties are taken into account when combining all measurements with the BLUE method. In particular, when combining the scale factors measured with the TnP and Kin methods, the systematic uncertainty associated to final-state radiation for the TnP method is assessed in the same way as done for the Kin method.

\begin{figure}[hbtp]
  \centering
  \includegraphics[width=.99\textwidth]{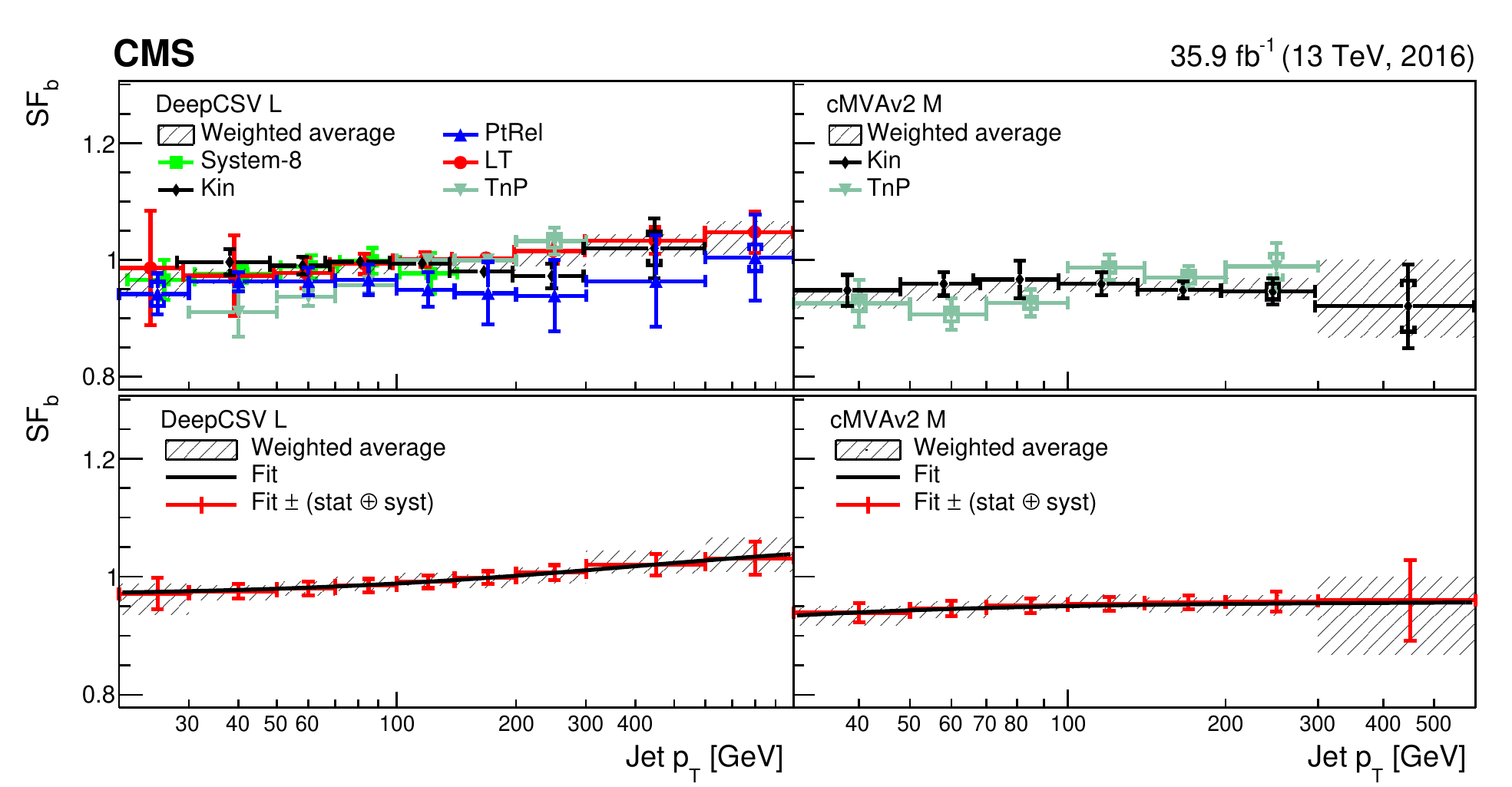}
    \caption{Data-to-simulation scale factors for {\cPqb} jets as a function of jet \pt for the loose DeepCSV (left) and the medium cMVAv2 (right) algorithms working points. The upper panels show the scale factors for tagging {\cPqb} as function of jet \pt measured with the various methods. The inner error bars represent the statistical uncertainty, and the outer error bars the combined statistical and systematic uncertainty. The combined scale factors with their overall uncertainty are displayed as a hatched area. The lower panels show the same combined scale factors with the result of a fit function (solid curve) superimposed. The combined scale factors with the overall uncertainty are centred around the fit result. To increase the visibility of the individual measurements, the scale factors obtained with various methods are slightly displaced with respect to the bin centre for which the measurement was performed. The last bin includes the overflow entries.}
    \label{fig:SFbCombttbar}
\end{figure}

\begin{figure}[hbtp]
  \centering \includegraphics[width=.55\textwidth]{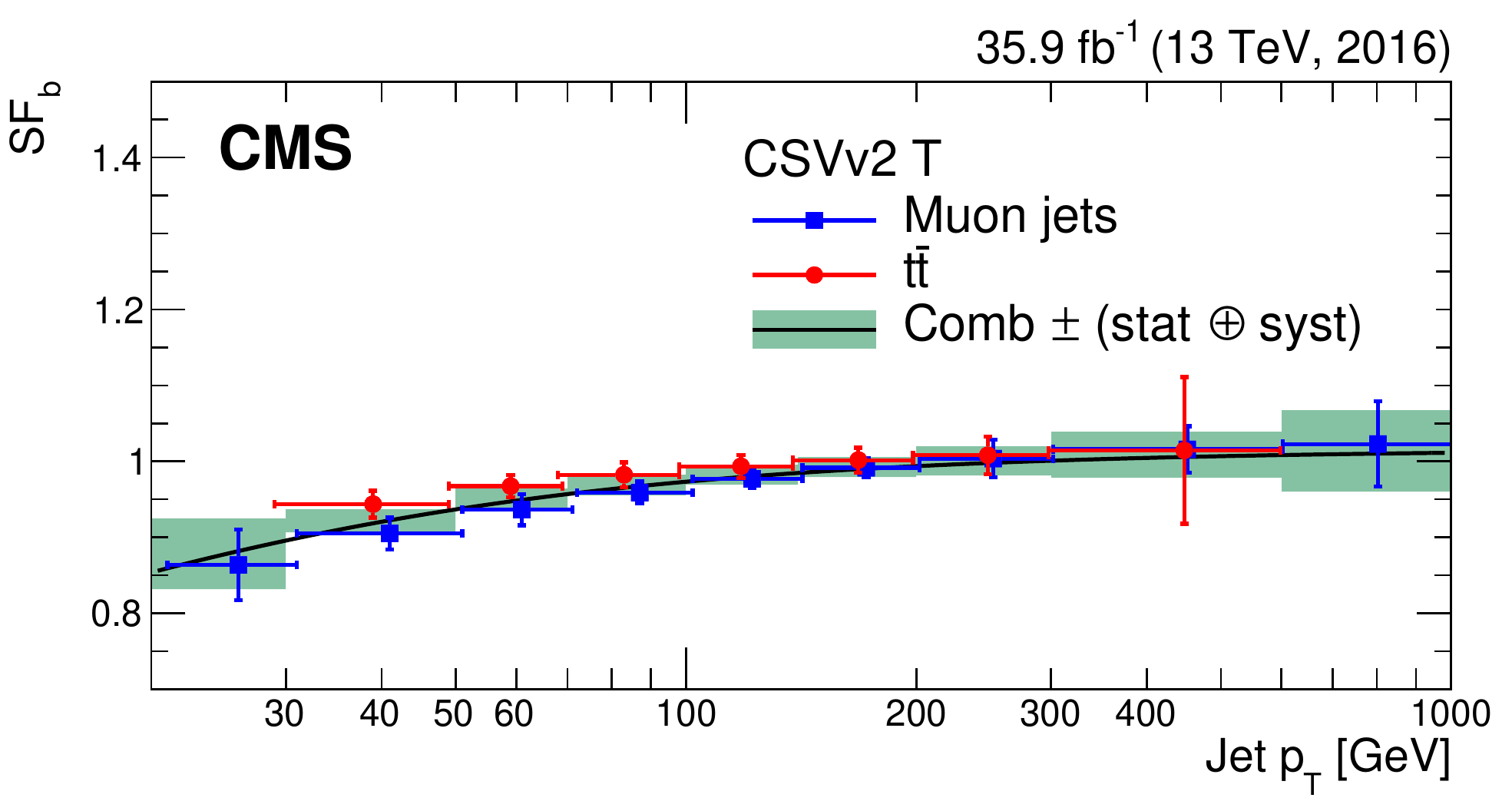}
    \caption{Data-to-simulation scale factors for {\cPqb} jets as a function of jet \pt measured in muon-enriched multijet and \ttbar events for the tight working point of the CSVv2 tagger. The green area shows the combined scale factors with their overall uncertainty, including an additional 1\% uncertainty to cover any residual sample dependence, fitted to the superimposed solid curve. For visibility purposes, the scale factors are slightly displaced with respect to the bin centre for which the measurement was performed. The last bin includes the overflow entries.}
    \label{fig:SFbCombttbar2}
\end{figure}

Figure~\ref{fig:SFbCombttbar} shows the combination of \ttbar measurements for the medium working point of the cMVAv2 tagger (right), and for the loose working point of the DeepCSV algorithm (left).
As an illustration of the consistency between the measurements performed on \ttbar and muon-enriched multijet events, the data-to-simulation scale factors are shown for the tight working point of the CSVv2 tagger in Fig.~\ref{fig:SFbCombttbar2}. Within the uncertainty, no sample dependence is observed. As a conservative estimate to cover any residual sample dependence, a 1\% systematic uncertainty is included when combining the measurements. Both in Figs.~\ref{fig:SFbCombttbar} and~\ref{fig:SFbCombttbar2} the fit function is parameterized as described in Section~\ref{sec:muonjetresults} for jets with $20<\pt<1000\GeV$. For jets with $\pt>1000\GeV$ the uncertainty of the scale factor is doubled.
For all taggers and for an average jet \pt found in \ttbar events the data-to-simulation scale factors vary from about 0.99 for the loose working point to 0.95 for the tight working point. The achieved relative precision on the scale factor for {\cPqb} jets is 1 to 1.5\% using jets with $70 <\pt< 100\GeV$ and rises to 3--5\% at the highest considered jet \pt. Overall, the statistical uncertainty is 15--30\% of the total uncertainty.

In some physics analyses of precision measurements, a correlation is present between the quantity to measure and the method to derive the {\cPqb} tagging scale factors. An example is the measurement of the \ttbar production cross section in an analysis requiring one or more {\cPqb}-tagged jets. In that case, the scale factors derived from \ttbar events are correlated with the production cross section to be measured and the scale factor measured with muon-enriched multijet events should be used.

\subsection{Measurement of the data-to-simulation scale factors as a function of the discriminator value}
\label{sec:SFiter}
The last method to measure data-to-simulation scale factors is a technique of iterative fitting (IterativeFit) first described in Ref.~\cite{Khachatryan:1748396}, which aims at correcting the full discriminator shape. This method is designed to meet the needs of analyses in which the full distribution of the {\cPqb} tagging discriminator values is used instead of applying a working point of the algorithm to select jets or events. If the full discriminator distribution is used, the distribution using jets in simulated events has to be corrected to match the one observed in data. Scale factors for both {\cPqb} and light-flavour jets are derived as a function of the discriminator value in bins of jet \pt and $\eta$. An iterative procedure is used based on a tag-and-probe technique to measure the scale factors for both {\cPqb} and light-flavour jets simultaneously. The scale factors are derived from events with two oppositely charged leptons (electron or muon) within the tracker acceptance and satisfying the tight identification and isolation requirements~\cite{Khachatryan:2015hwa,Chatrchyan:2012xi}. The leading (subleading) lepton is required to have $\pt > 25~(15)\GeV$. Exactly two jets are required with $\pt > 20\GeV$ and to lie within the tracker acceptance.

The data-to-simulation scale factors for {\cPqb} jets are derived from events passing the above requirements. In addition, for the events with same-flavour leptons, the dilepton invariant mass is required to be away from the {\PZ} boson mass, $|M_{\ell \ell} - M_{\PZ}| > 10$\GeV, and $\ptmiss>30\GeV$. These two requirements reduce the contribution from {\Zj} events. The tag jet should pass the medium working point of the algorithm for which the scale factor is to be measured. The other jet is used as probe. After these criteria have been applied, the simulated event sample is composed of 87\% \ttbar, 6\% single top and 7\% {\Zj} events. Other backgrounds are reduced to a negligible level.

The data-to-simulation scale factors for light-flavour jets are measured with {\Zj} events selected among the same-flavour dilepton events with a dilepton invariant mass close to that of the \PZ boson, $|M_{\ell \ell} - M_Z| < 10$\GeV, and inverting the requirement on \ptmiss. A {\cPqb} jet veto is applied on the tag jet using the loose working point of the tagger for which the scale factor is to be measured. After the event selection, the sample is very pure in {\Zj} events (99.9\%).

After the event selection and tagging or vetoing one of the two jets, the data-to-simulation scale factors are measured using the other jet in the event as the probe. The scale factors are extracted by first normalizing the {\cPqb} tagging discriminator distribution of the probe jets in simulation to that observed in data. Then, when measuring the scale factor for {\cPqb} jets, the contribution from non-{\cPqb} jets is subtracted using the simulated events. Similarly, when measuring the scale factor for light-flavour jets, the expected contributions from {\cPqb} and {\cPqc} jets are subtracted. The scale factor is determined separately in exclusive bins of the {\cPqb} tagging discriminator distribution, $\pt$, and $\eta$ (for light-flavour jets). Since the scale factors for light-flavour jets have an impact on the measured scale factors for {\cPqb} jets, an iterative procedure is performed. In the first iteration no scale factor is applied, while for the next iteration the background is subtracted using the scale factors obtained in the previous iteration. The iterative procedure stops once the scale factors obtained in the current iteration are stable with respect to those obtained in the previous iteration. Convergence is typically achieved after three iterations. When estimating the scale factor for {\cPqb} jets and light-flavour jets, the scale factor for {\cPqc} jets is set to unity with an uncertainty that is twice the uncertainty in the scale factor for {\cPqb} jets.

For the IterativeFit method, the following list of systematic uncertainties is considered. This list covers possible shape discrepancies between data and simulation for the tagger discriminator distribution.
\begin{itemize}
\item \textbf{Sample purity}: Several systematic uncertainties impact the sample purity. These need to be taken into account when measuring the data-to-simulation scale factor for light-flavour or {\cPqb} jets. The sample purity may be affected by background processes or the modelling of the signal in the simulation, e.g. related to the production of additional jets in association with the top quark pair when measuring the scale factor for {\cPqb} jets. All sources of systematic uncertainties influencing the sample purity are combined in a single systematic uncertainty. For the scale factor for light-flavour jets, the expected contribution from processes other than {\Zj} is negligible. However, the sample purity can be contaminated by heavy-flavour jets produced in association with the {\PZ} boson. The fraction of heavy-flavour jets in the sample is conservatively varied upwards and downwards by 20\% when calculating $SF_{\text{l}}$. For $SF_{\cPqb}$, the dominant contribution originates from \ttbar events. The dilepton \ttbar events are selected requiring exactly two jets, consistent with the two {\cPqb} jets expected from the \ttbar decay. However, because of ISR and FSR and the acceptance of the event selection, also non-{\cPqb} jets are selected. The rate of \ttbar events produced with $\geq 2$ additional partons varies within up to 20\%. Therefore, the fraction of non-{\cPqb} jets is varied by this amount to evaluate the uncertainty in the purity of the sample.
\item \textbf{Jet energy scale}: The uncertainty in the jet energy scale is assessed in the same way as described in Section~\ref{sec:Wc}.
\item \textbf{Statistical uncertainty}: An uncertainty arises due to the limited number of entries in each bin of the discriminator distribution, resulting in statistical fluctuations in certain regions, \eg at high discriminator values for light-flavour jets and at low discriminator values for {\cPqb} jets.  Linear and quadratic functions, $f_{1}(x) = 1 - 2x$ and $f_{2}(x) = 1 - 6x(1-x)$, are introduced, where $x$ corresponds to the central value of a discriminator bin.
The linear function parameterizes the effect of statistical fluctuations that would tilt the discriminator distribution. In contrast, the quadratic function represents fluctuations that would increase or decrease the data-to-simulation scale factor in the centre of the discriminator distribution compared to the low and high discriminator values. To assess the size of the systematic uncertainty related to statistical fluctuations, the scale factor value is varied according to $\pm\sigma(x)f_{i}(x)$, where $\sigma(x)$ is the statistical uncertainty in the scale factor in that bin. The scale factors are refitted after applying these variations, resulting in two independent functions that span an envelope around the nominal scale factor function for each of the two types of statistical fluctuations.
\item \textbf{Treatment of $SF_{\cPqc}$}: For {\cPqc} jets the data-to-simulation scale factor, $SF_{\cPqc}$, is set to unity. The uncertainty in this value is obtained by doubling the aforementioned relative uncertainties in the scale factor for {\cPqb} jets and adding them in quadrature to obtain a relative uncertainty in $SF_{\cPqc}$. Similarly as for the statistical uncertainty, two separate uncertainties are constructed using linear and quadratic functions $f_{i}(x)$. The scale factor value is then varied according to $\pm\sigma(x)f_{i}(x)$, where $\sigma(x)$ is the relative uncertainty in $SF_{\cPqc}$. These linear and quadratic variations of $SF_{\cPqc}$ are applied independently from the other uncertainties after which the scale factors are refitted to obtain the functions corresponding to the uncertainty in $SF_{\cPqc}$.
\end{itemize}

\begin{figure}[htbp]
  \centering
   \includegraphics[width=0.49\textwidth]{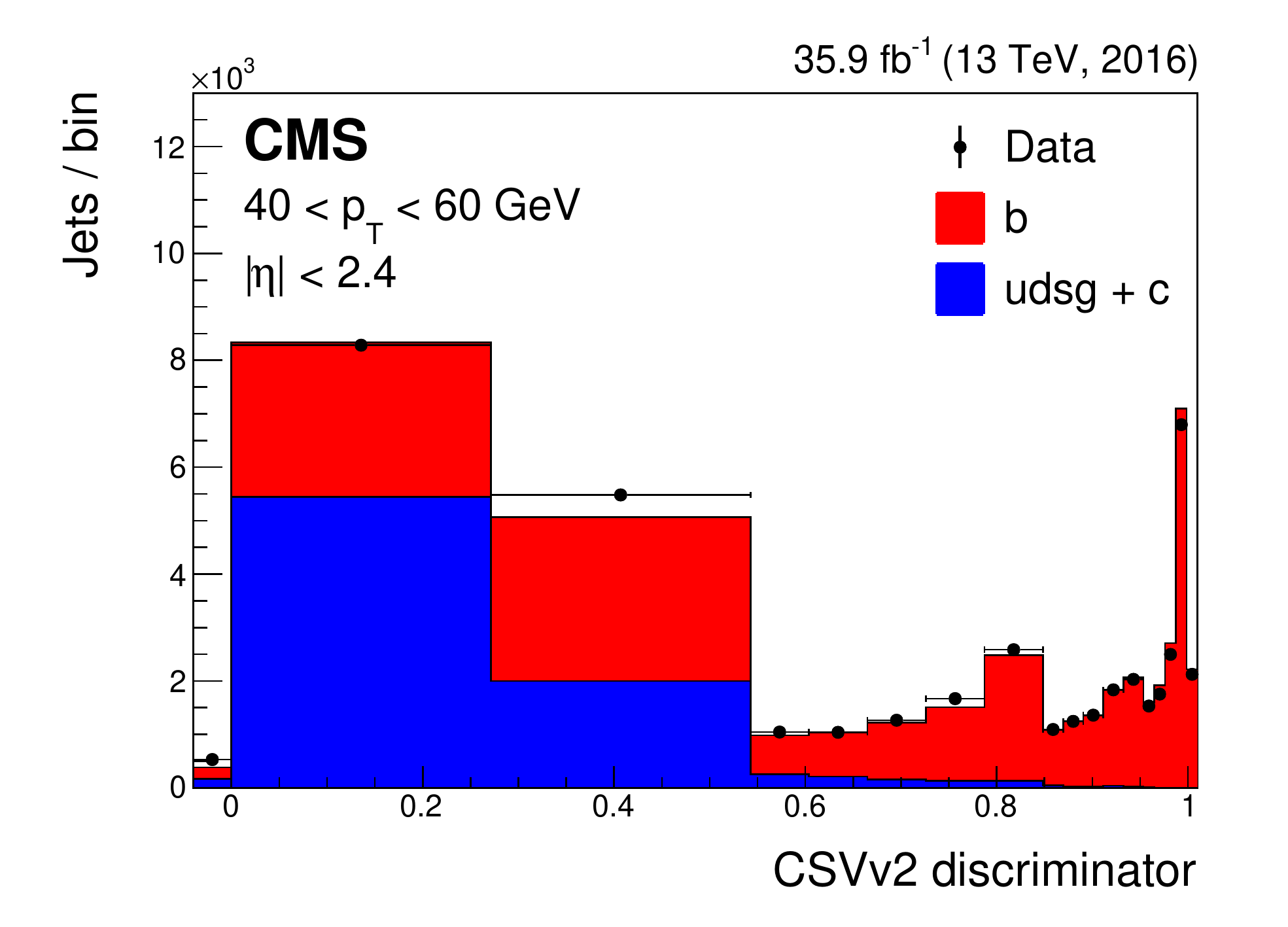}
   \includegraphics[width=0.49\textwidth]{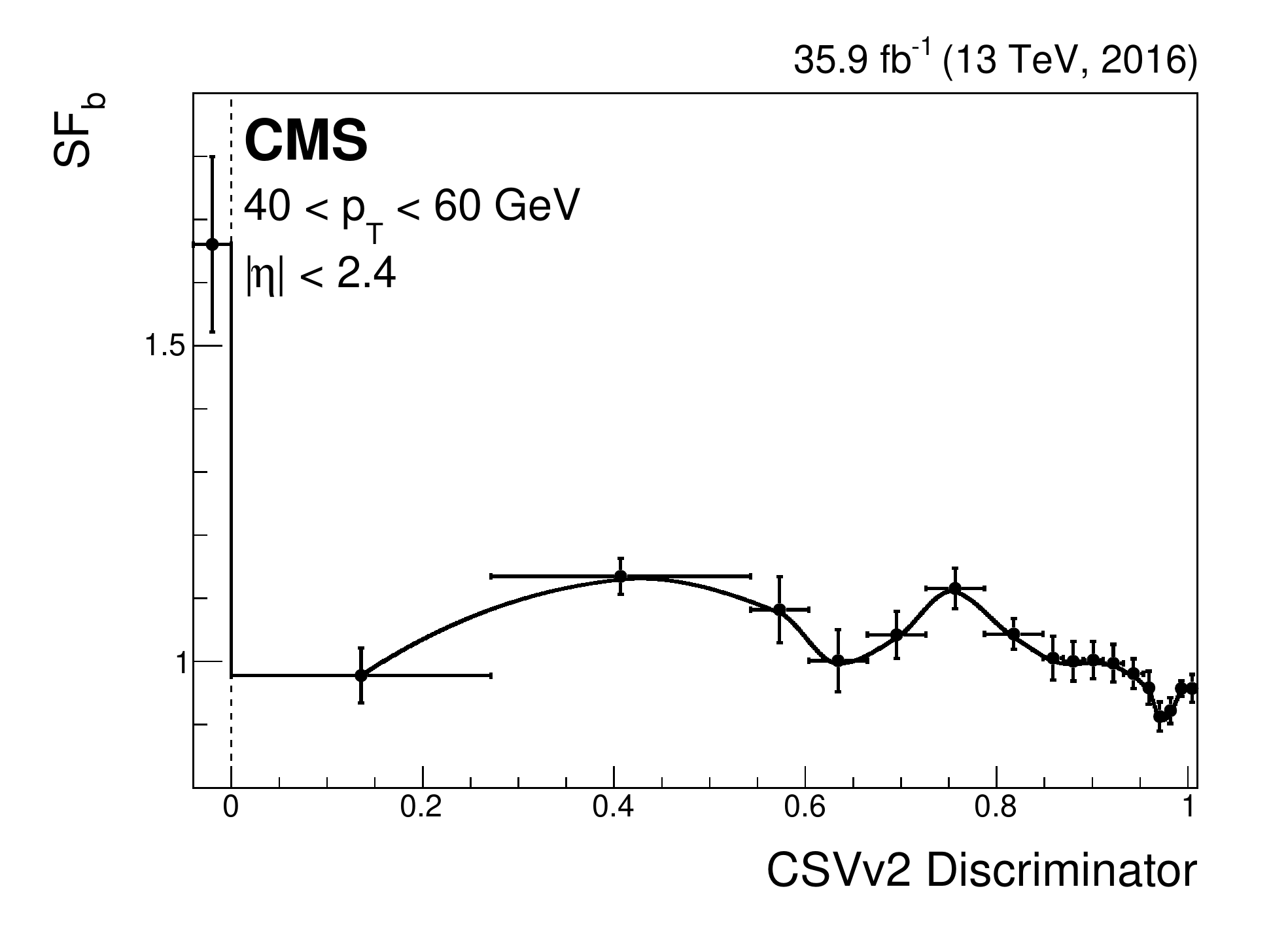}
    \caption{Distribution of the CSVv2 discriminator values for jets with $40 <\pt < 60\GeV$ before the data-to-simulation scale factors are applied in the \ttbar dilepton sample (left). The simulation is normalized to the number of entries in data. Measured scale factors for {\cPqb} jets as a function of the CSVv2 discriminator value (right). The line is an interpolation between the scale factors measured in each bin of the CSVv2 discriminator distribution. The bin below 0 contains the jets with a default discriminator value.}
    \label{fig:reweighHF}
\end{figure}

\begin{figure}[htbp]
  \centering
   \includegraphics[width=0.49\textwidth]{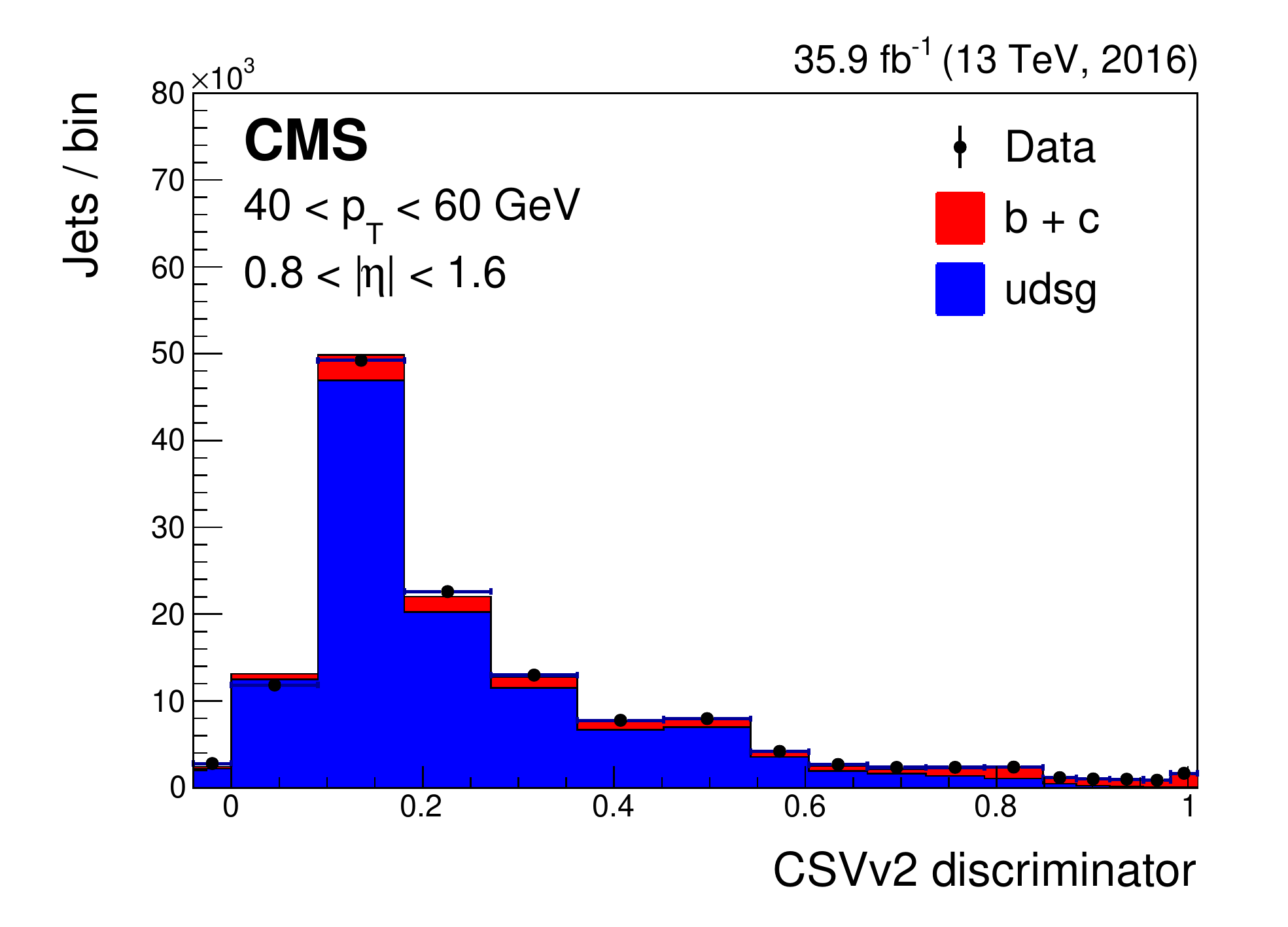}
   \includegraphics[width=0.49\textwidth]{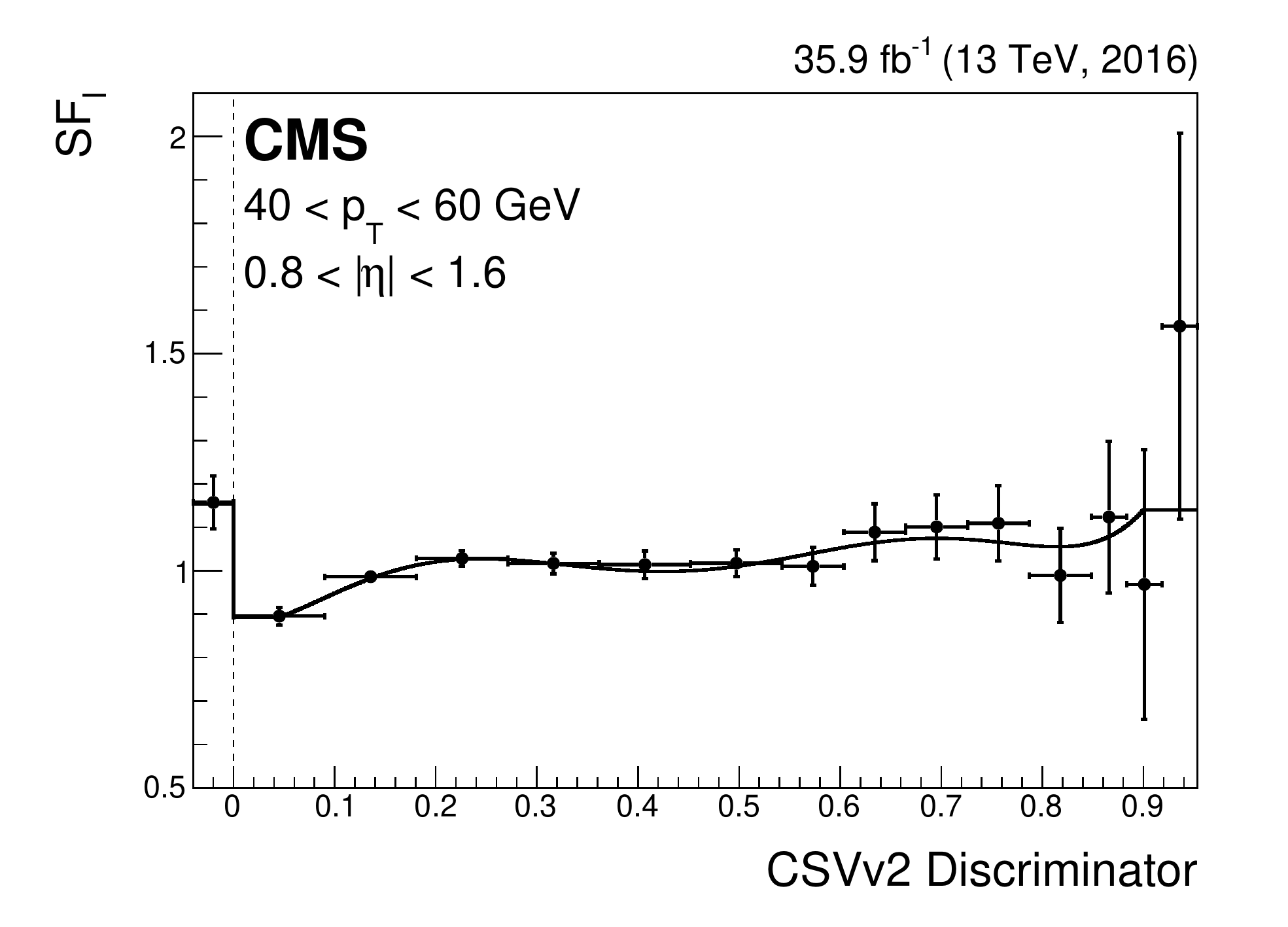}
   \caption{Distribution of the CSVv2 discriminator values for jets with $40 <\pt < 60\GeV$ and $0.8<\abs{\eta}<1.6$ before the data-to-simulation scale factors are applied in the {\Zj} sample (left). The simulation is normalized to the number of entries for data. Measured scale factors for light-flavour jets as a function of the CSVv2 discriminator value (right). The line represents a polynomial fit to the scale factors measured in each bin of the CSVv2 discriminator distribution. The bin below 0 contains the jets with a default discriminator value.}
    \label{fig:reweighLF}
\end{figure}

Figures~\ref{fig:reweighHF} and ~\ref{fig:reweighLF} show an example of the distribution for the CSVv2 tagger and the derived data-to-simulation scale factors using jets with $40 <\pt < 60\GeV$ in a topology enriched in {\cPqb} jets and a topology enriched in light-flavour jets, respectively. The scale factors are parameterized as a function of the CSVv2 discriminator value. The scale factor for light-flavour jets as a function of the discriminator value is fitted with a sixth-order polynomial function. For the scale factor for {\cPqb} jets, no satisfactory parameterization was found. Therefore, a smooth function is obtained by interpolating between the scale factors measured in bins of the CSVv2 discriminator distribution. No interpolation is done between the bin below 0, which includes jets with a negative CSVv2 discriminator value, and the first bin above 0.

\begin{figure}[htbp]
  \centering
   \includegraphics[width=0.49\textwidth]{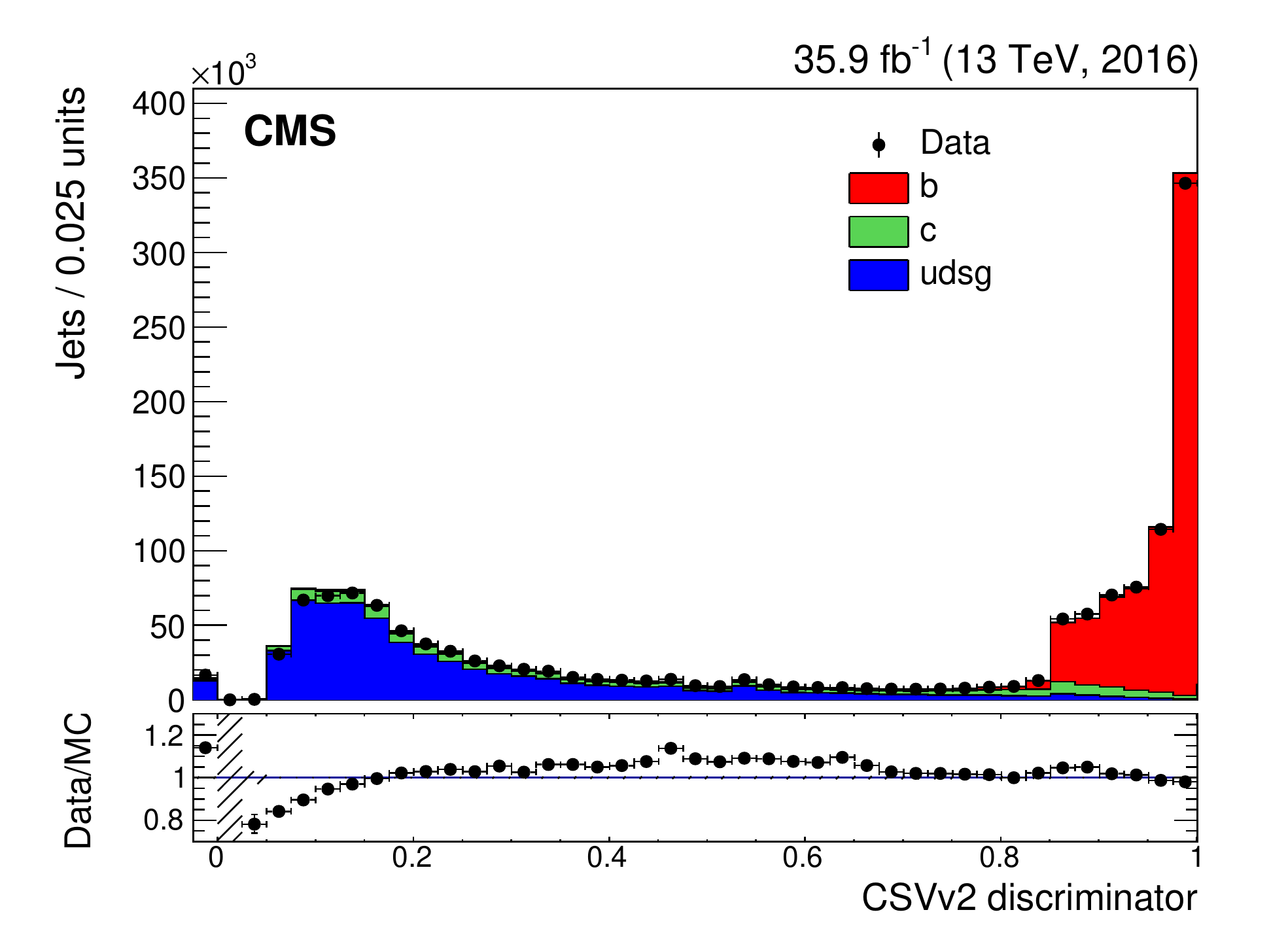}
   \includegraphics[width=0.49\textwidth]{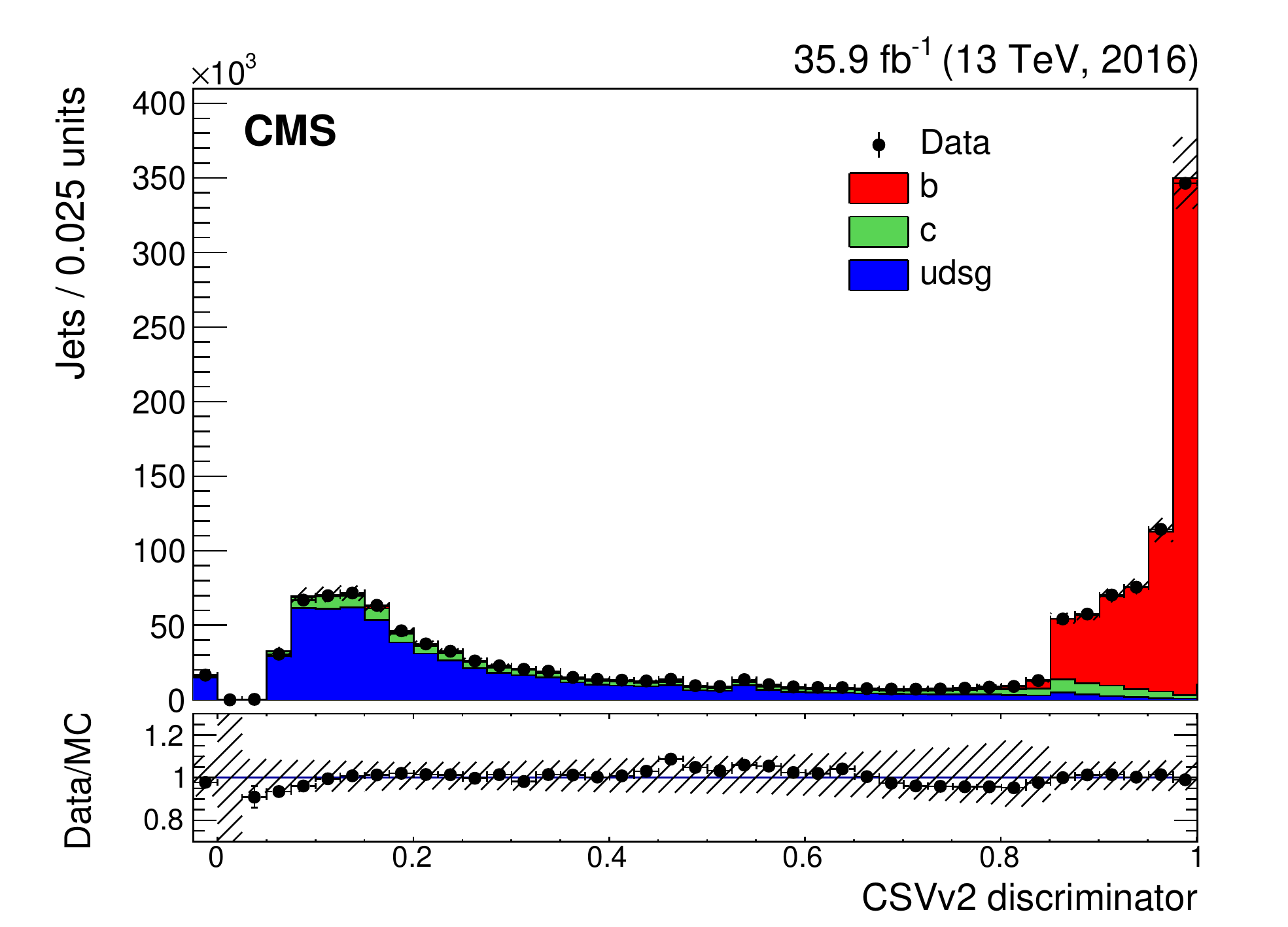}
    \caption{Distribution of the CSVv2 discriminator values for the single-lepton \ttbar sample. Exactly four jets are required, two of which passing the medium working point of the CSVv2 algorithm. The values of the discriminator are shown before (left) and after (right) applying the data-to-simulation scale factors derived with the IterativeFit method. The hatched band around the ratios shows the statistical uncertainty (left), and the total uncertainty (right) in the measured scale factors. The simulation is normalized to the total number of data events. The bin below 0 contains the jets with a default discriminator value.}
    \label{fig:reweighControl}
\end{figure}

The data-to-simulation scale factors obtained with the IterativeFit method have been validated in various control regions. One example is the validation in a control region dominated by single-lepton \ttbar events. The flavour composition in this control region is very different from both the dilepton \ttbar and {\Zj} topologies used to derive the scale factors, thereby providing a powerful cross check. Events are selected requiring an isolated electron or muon with $\pt > 30\GeV$ and $\abs{\eta}<2.1$ and exactly four jets with $\pt > 30\GeV$, of which exactly two are {\cPqb} tagged according to the medium working point of the CSVv2 algorithm. The distribution of the CSVv2 discriminator values is shown in Fig.~\ref{fig:reweighControl} for all the jets in the control region. The agreement between the data and simulation improves significantly after applying the measured scale factors, and the remaining fluctuations are covered by the systematic uncertainties.

\subsection{Comparison of the measured data-to-simulation scale factors}
\label{sec:SFconcl}
\begin{figure}[htbp]
  \centering
   \includegraphics[width=0.49\textwidth]{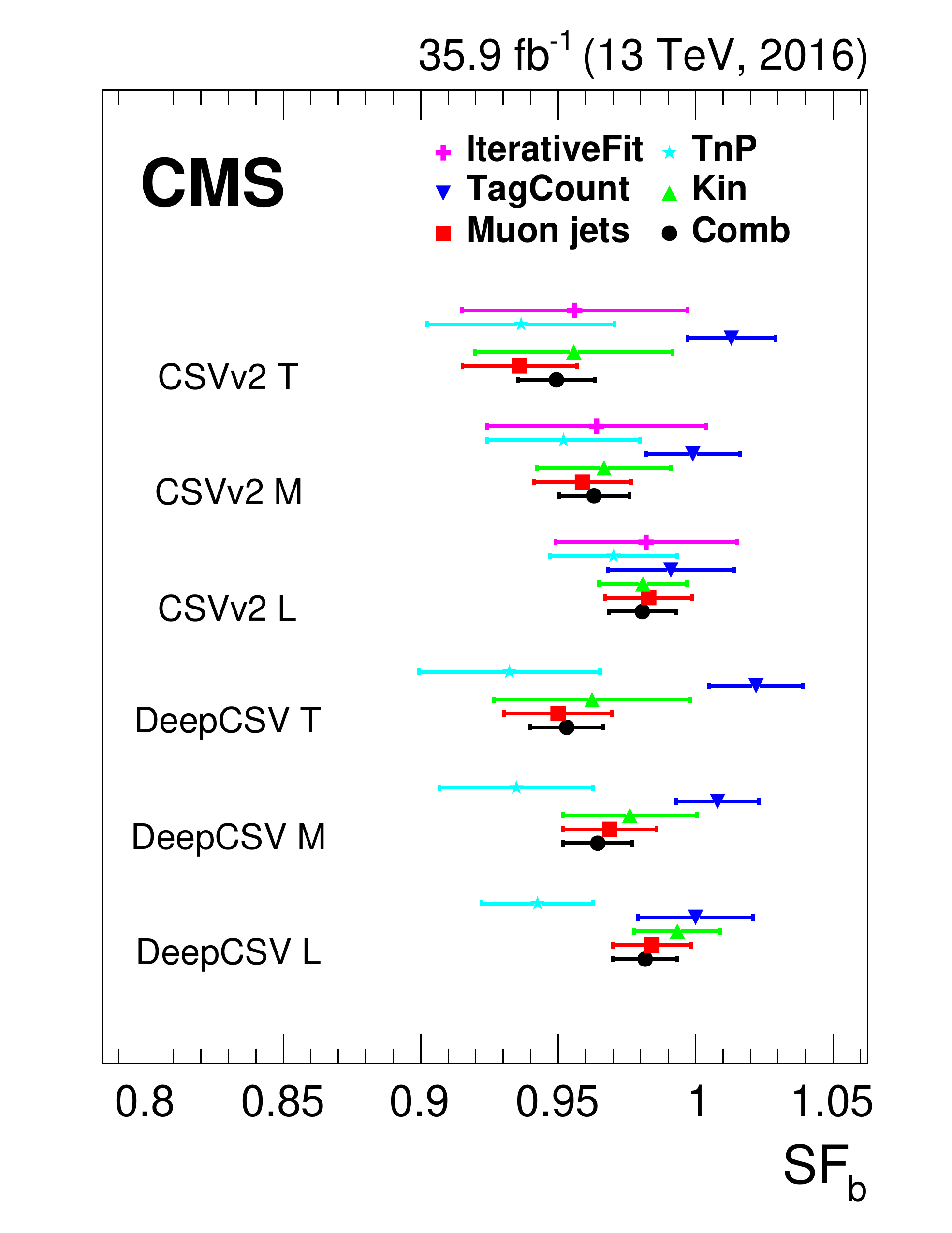}
   \includegraphics[width=0.49\textwidth]{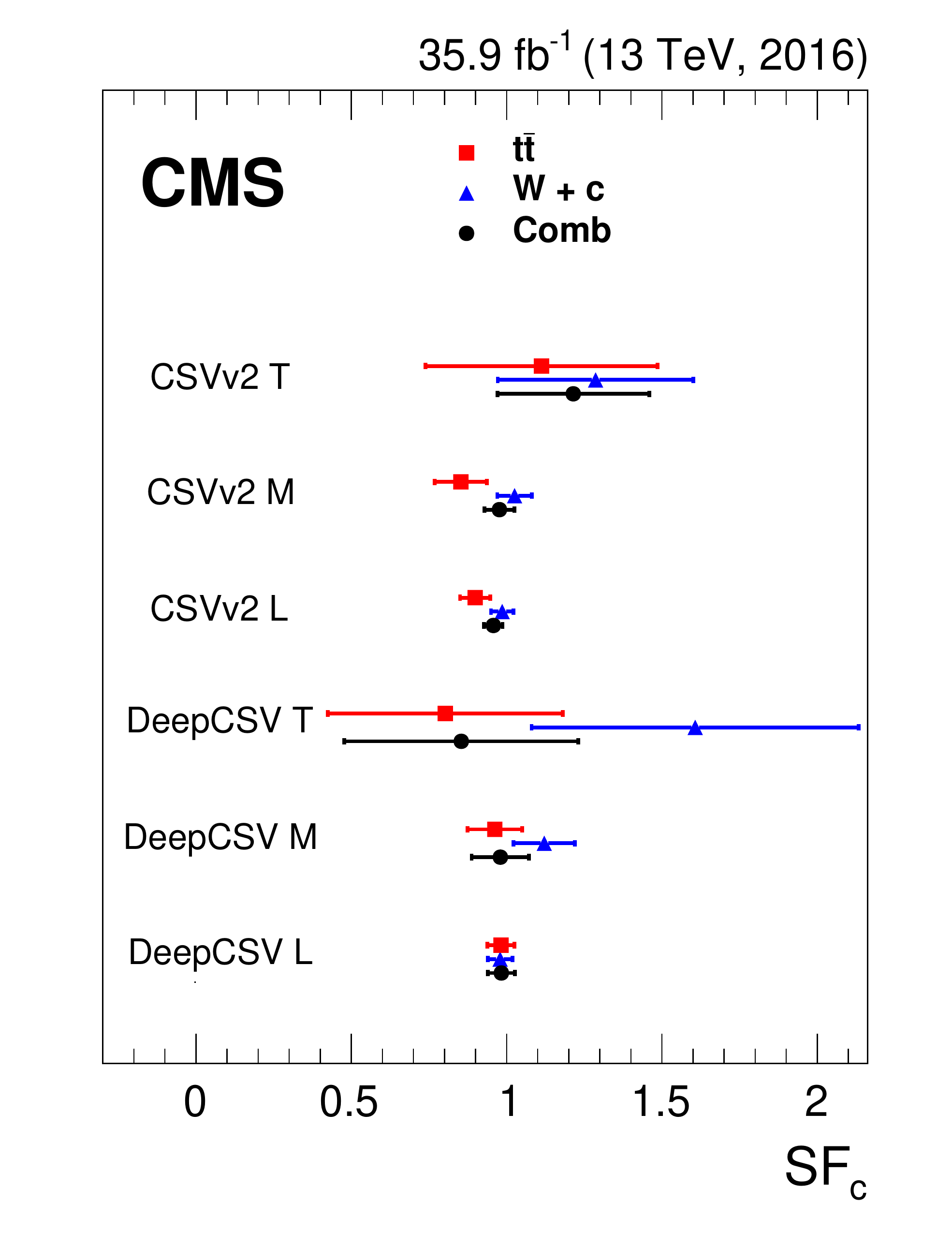}\\
   \includegraphics[width=0.49\textwidth]{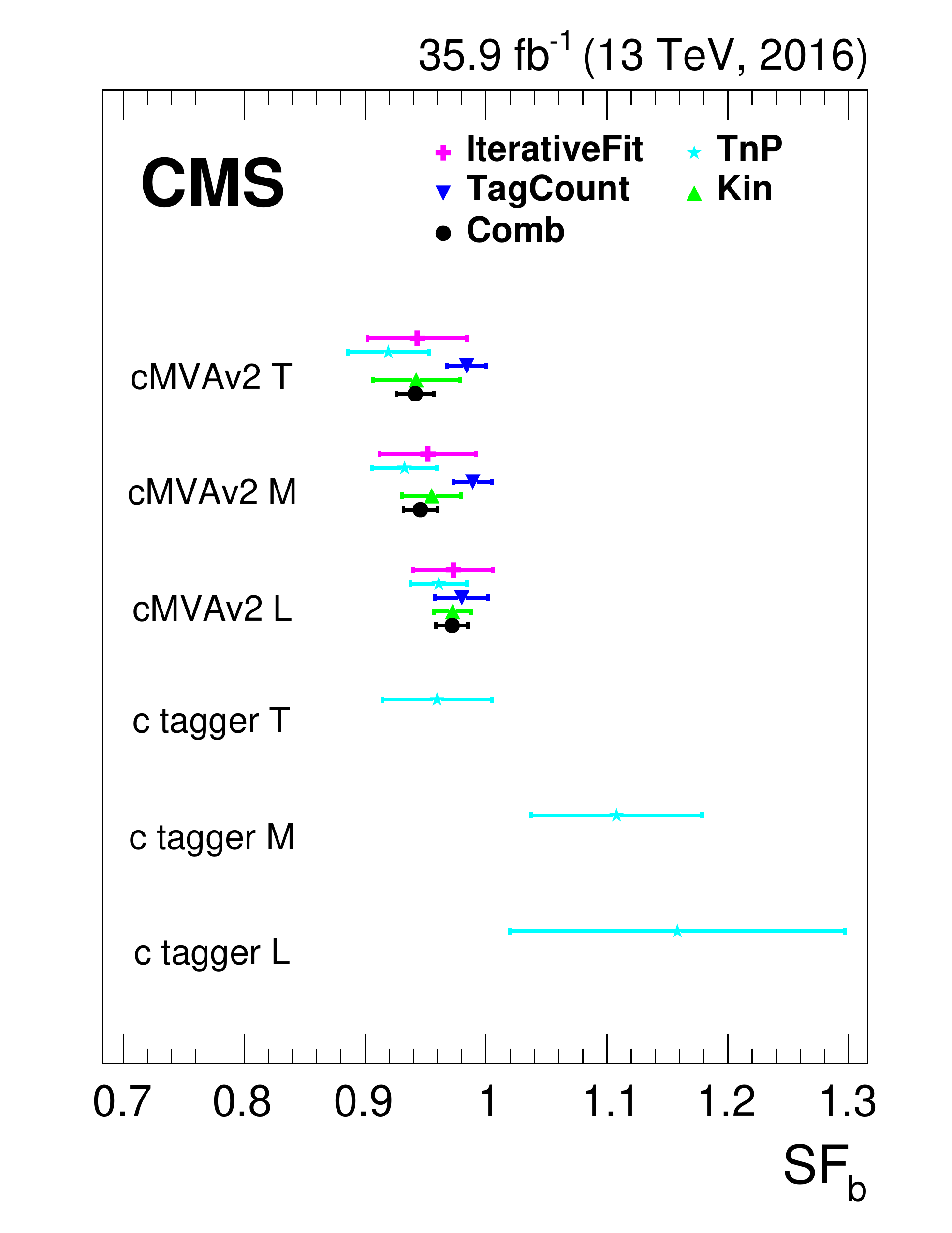}	
   \includegraphics[width=0.49\textwidth]{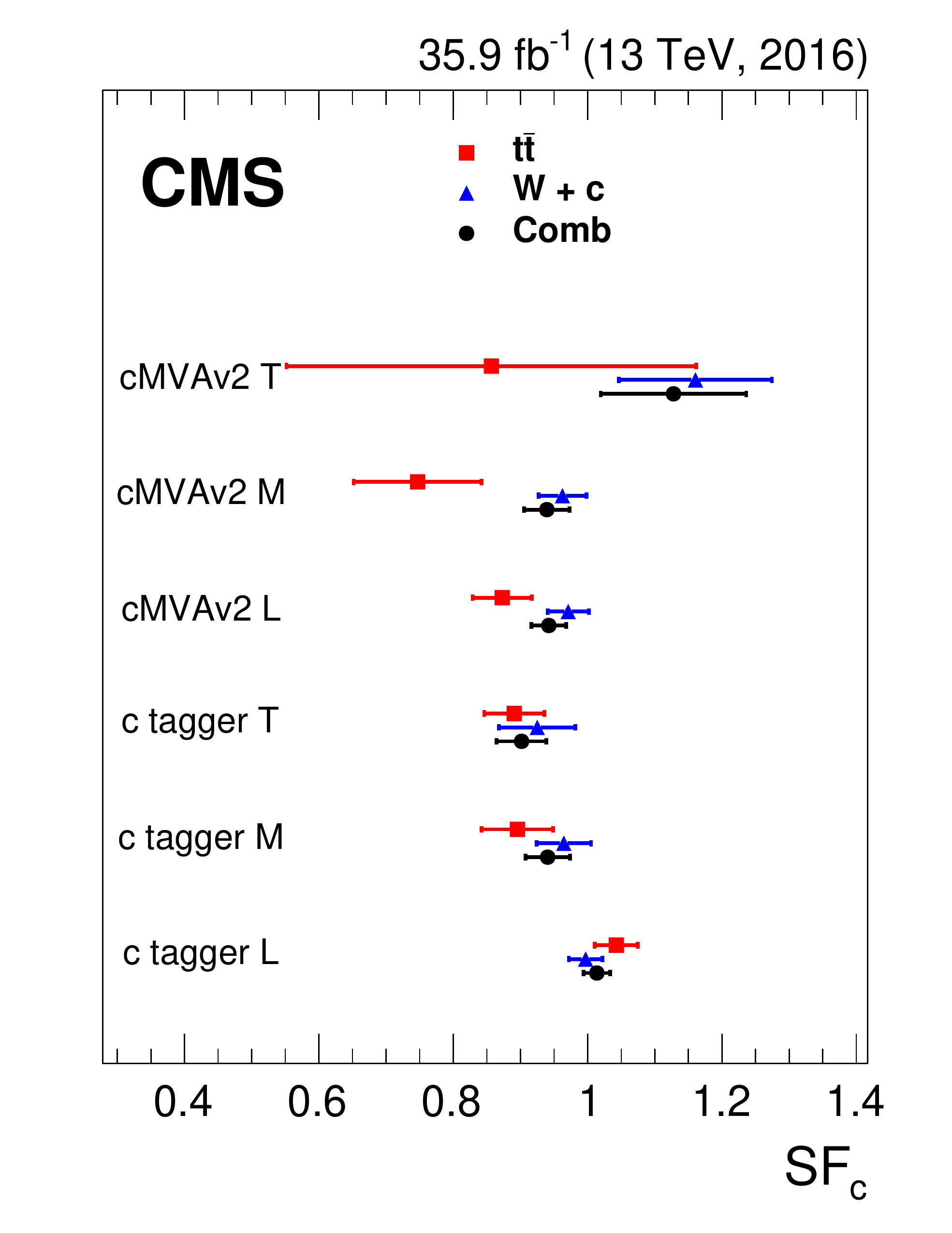}	
    \caption{Comparison of the data-to-simulation scale factors derived with various methods and their combination, for {\cPqb} (left) and {\cPqc} (right) jets. The scale factors measured with the different methods agree within their uncertainties. For the left panels, the combination includes all measurements with the exception of the IterativeFit and the TagCount methods.  }
    \label{fig:comparemethods}
\end{figure}

In most cases, the measured data-to-simulation scale factors for heavy- (light-) flavour jets are smaller (larger) than unity. This is expected because the quantities of relevance for heavy-flavour jet identification are not perfectly modelled by the simulation. The scale factors derived with the various methods are compared to each other after averaging the measured scale factor following the \pt spectrum for \ttbar events. Figure~\ref{fig:comparemethods} compares the measured scale factors. However, this figure should not be used to decide which method performs best, since, e.g. for the TagCount method the scale factors were remeasured inclusively over the jet \pt range, resulting in a smaller uncertainty than when the weighted average is used over the measurements in bins of jet \pt. This is because for the measurement as a function of the jet \pt the two tagged jets are required to be in the same jet \pt bin, resulting in a loss of events compared to the inclusive measurement. Moreover, to allow a comparison, the scale factors for the IterativeFit method are remeasured using only one bin above the discriminator value corresponding to the working point for which the scale factor is derived. As can be seen from Fig.~\ref{fig:comparemethods}, the measured scale factors are consistent within their uncertainties. Only for the tight working point of the CSVv2 and DeepCSV taggers there is a hint of tension between the TagCount method and the other methods. This is explained by the fact that the central value of the TagCount method is quite sensitive to the background subtraction and the sample purity.
The scale factor for {\cPqb} jets for the cMVAv2 and {\cPqc} tagger working points is not measured with muon-enriched multijet events to avoid a bias due to the muon information used in these taggers.
The right panels in Fig.~\ref{fig:comparemethods} show that the precision on the scale factors for {\cPqc} jets for the loose and medium working points of the {\cPqb} taggers, is on the same level as the precision reached on the scale factors for {\cPqb} jets, for jets with a \pt distribution as expected in \ttbar events. For the tight working point of the {\cPqb} taggers, the uncertainty in the average scale factor is relatively large because of the low number of {\cPqc} jets passing the tagging requirement. Similarly, as can be seen from the lower left panel in Fig.~\ref{fig:comparemethods} the uncertainty in the average scale factor for {\cPqb} jets for the {\cPqc} tagger working points is larger compared to the corresponding uncertainty for the working points of the {\cPqb} taggers, because of two reasons. First, the uncertainty for the {\cPqc} tagger tight working point is large because of the low efficiency for {\cPqb} jets to pass this tagging requirement (Section~\ref{sec:perfak4c}), resulting in a relatively large statistical uncertainty. Second, the uncertainty for the {\cPqc} tagger loose and medium working points is large due to the larger contribution from light-flavour jets resulting in a larger systematic uncertainty. It was also checked that the scale factor for light-flavour jets obtained with the IterativeFit method is consistent with the one obtained using the negative tag method.

\section{Measurement of the tagging efficiency for boosted topologies}
\label{sec:boostedeff}
In Section~\ref{sec:boostedalgos}, the performance of {\cPqb} tagging algorithms in boosted topologies was discussed and the double-{\cPqb} tagger was presented to identify boosted particles decaying to two {\cPqb} quarks. This section summarizes the efficiency measurements for {\cPqb} tagging in boosted topologies. In Section~\ref{sec:boostedcomm} the data are compared to the simulation for two topologies: a sample of muon-enriched subjets of AK8 jets and a sample of double-muon-tagged AK8 jets. Section~\ref{sec:subjetSF} discusses the methods to measure the efficiency for {\cPqb} tagging subjets with the CSVv2 tagger. The efficiency measurement of the double-{\cPqb} tagger is presented in Section~\ref{sec:doublebSF}. In both cases, the data-to-simulation scale factors are measured as a function of the jet \pt. At this stage, the size of the jet sample is not yet large enough to provide also scale factors as a function of the jet $\abs{\eta}$.

\subsection{Comparison of data with simulation}
\label{sec:boostedcomm}
The data are compared to the simulation using jets in boosted topologies. Jets are selected from events satisfying the following description:
\begin{itemize}
\item \textbf{Muon-enriched boosted subjets sample}: A sample of muon-enriched multijet events is obtained using a combination of single-jet (AK4 and AK8) triggers requiring a muon in the jet. The data are compared to the simulation for soft-drop subjets (Section~\ref{sec:boostedalgos}) of AK8 jets with \pt$>350$\GeV and within the tracker acceptance. The subjets are required to contain at least one muon with $\pt>7$\GeV and $\Delta R<0.4$. In addition, to reduce the contribution from prompt muons, the ratio of the \pt of the muon to that of the jet is required to be smaller than 0.5. The subjet \pt distribution in simulation is reweighted to match the observed distribution.
\item \textbf{Double-muon-tagged boosted jet sample}: A second sample of muon-enriched multijet events is obtained by combining the triggers used to select the previous sample with dijet triggers with a lower jet \pt threshold, and by requiring a muon in each of the two jets. In this way, the sample contains also AK8 jets with $250 <\pt<350\GeV$. Each subjet is required to contain a muon with $\pt>7$\GeV and $\Delta R<0.4$. The sum of the \pt of the two muons with respect to the \pt of the AK8 jet is required to be less than 0.6. Some of the triggers are prescaled. The \pt distribution of the AK8 jet in the simulation is reweighted to match the observed distribution in data.
\end{itemize}

\begin{figure}[hbtp]
  \centering
    \includegraphics[width=0.49\textwidth]{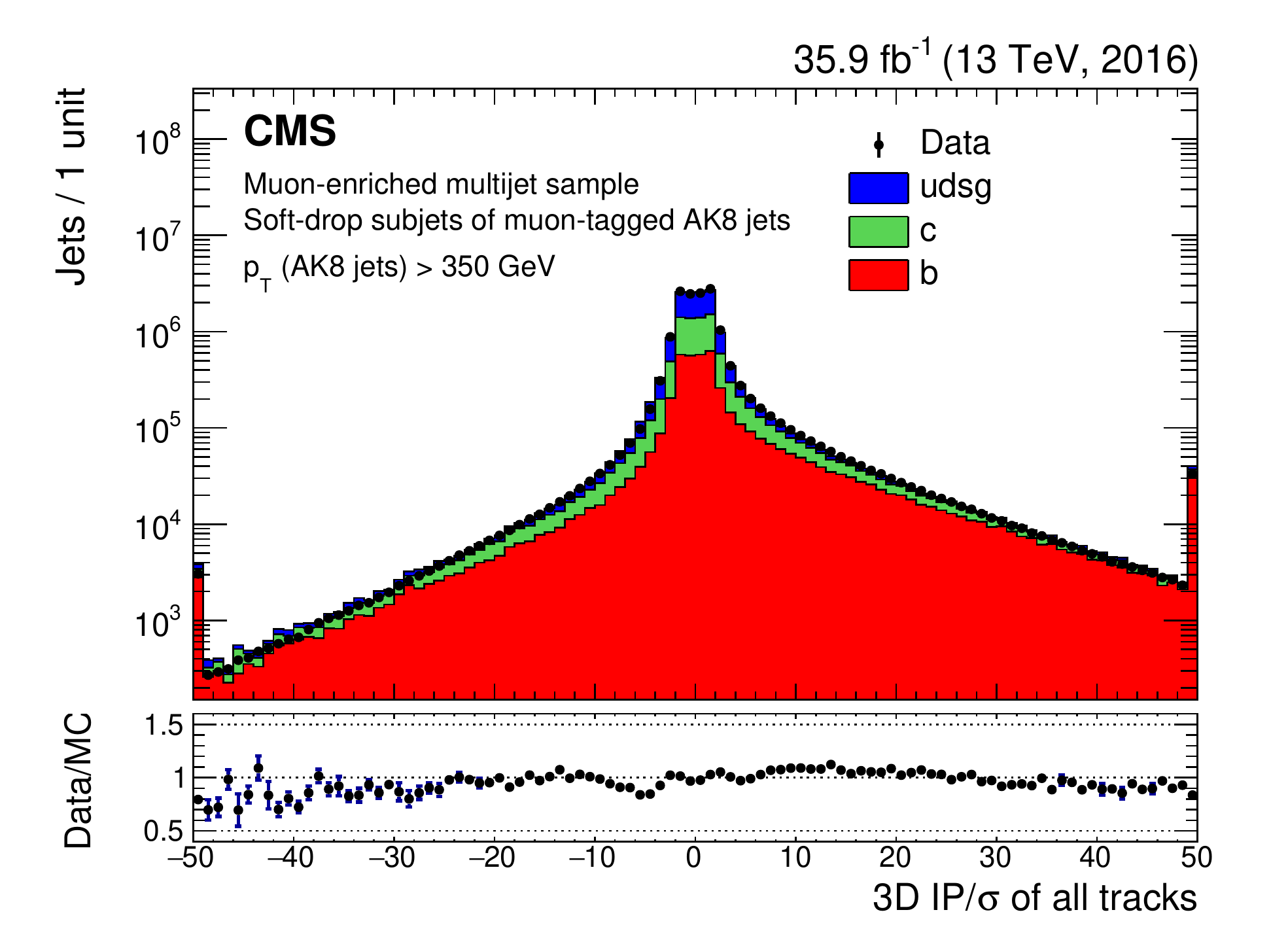}
    \includegraphics[width=0.49\textwidth]{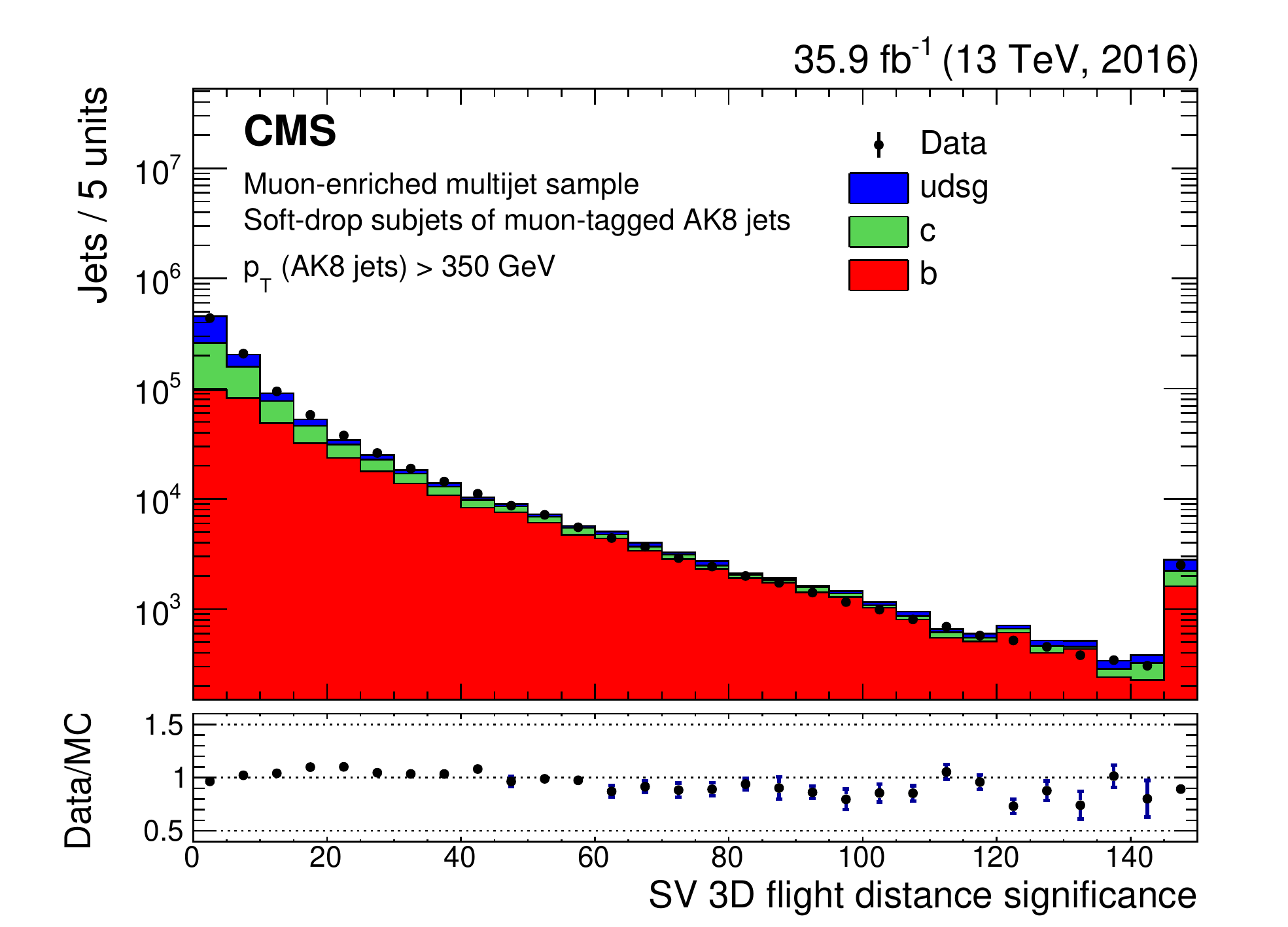}
    \includegraphics[width=0.49\textwidth]{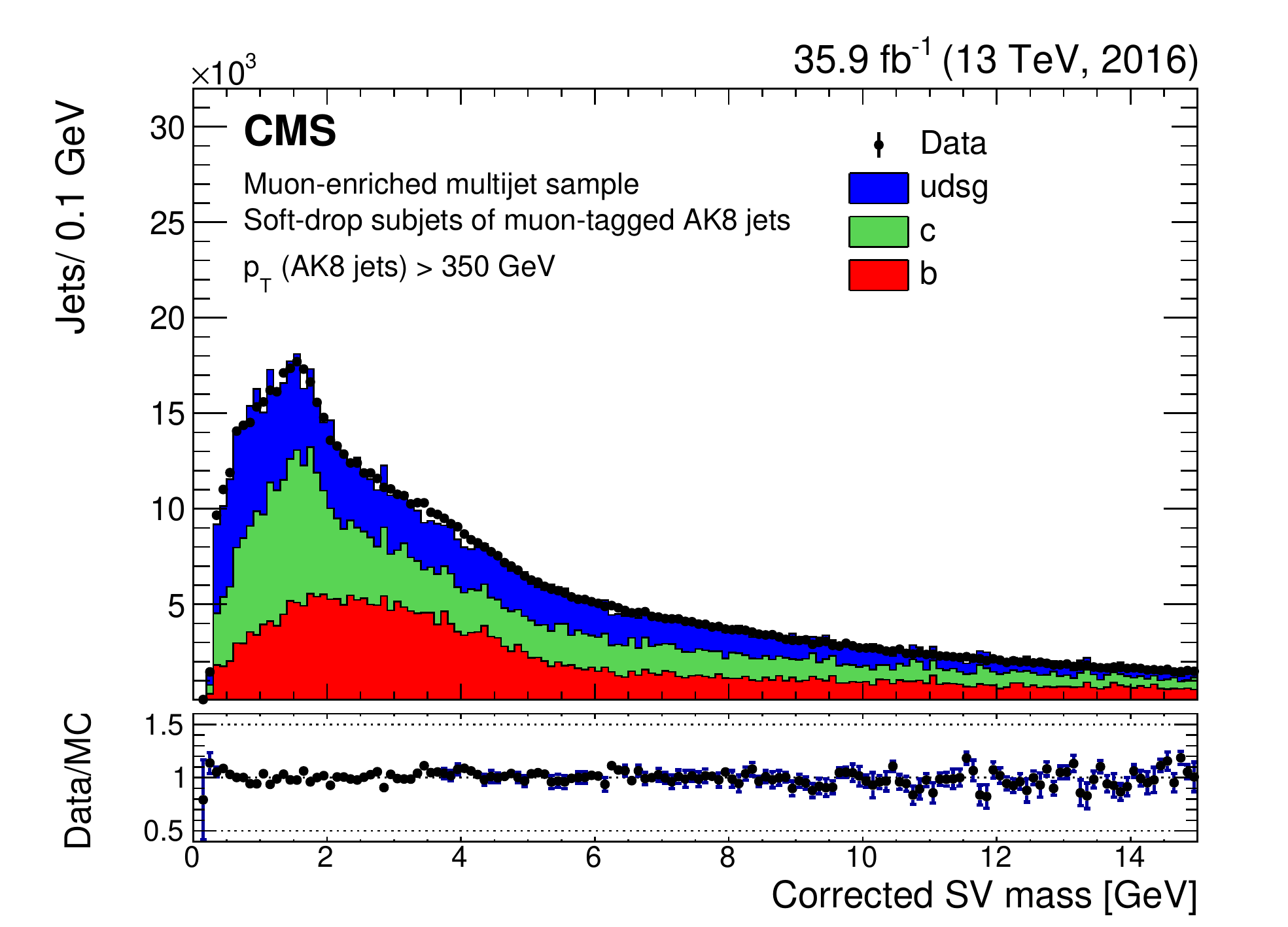}
    \includegraphics[width=0.49\textwidth]{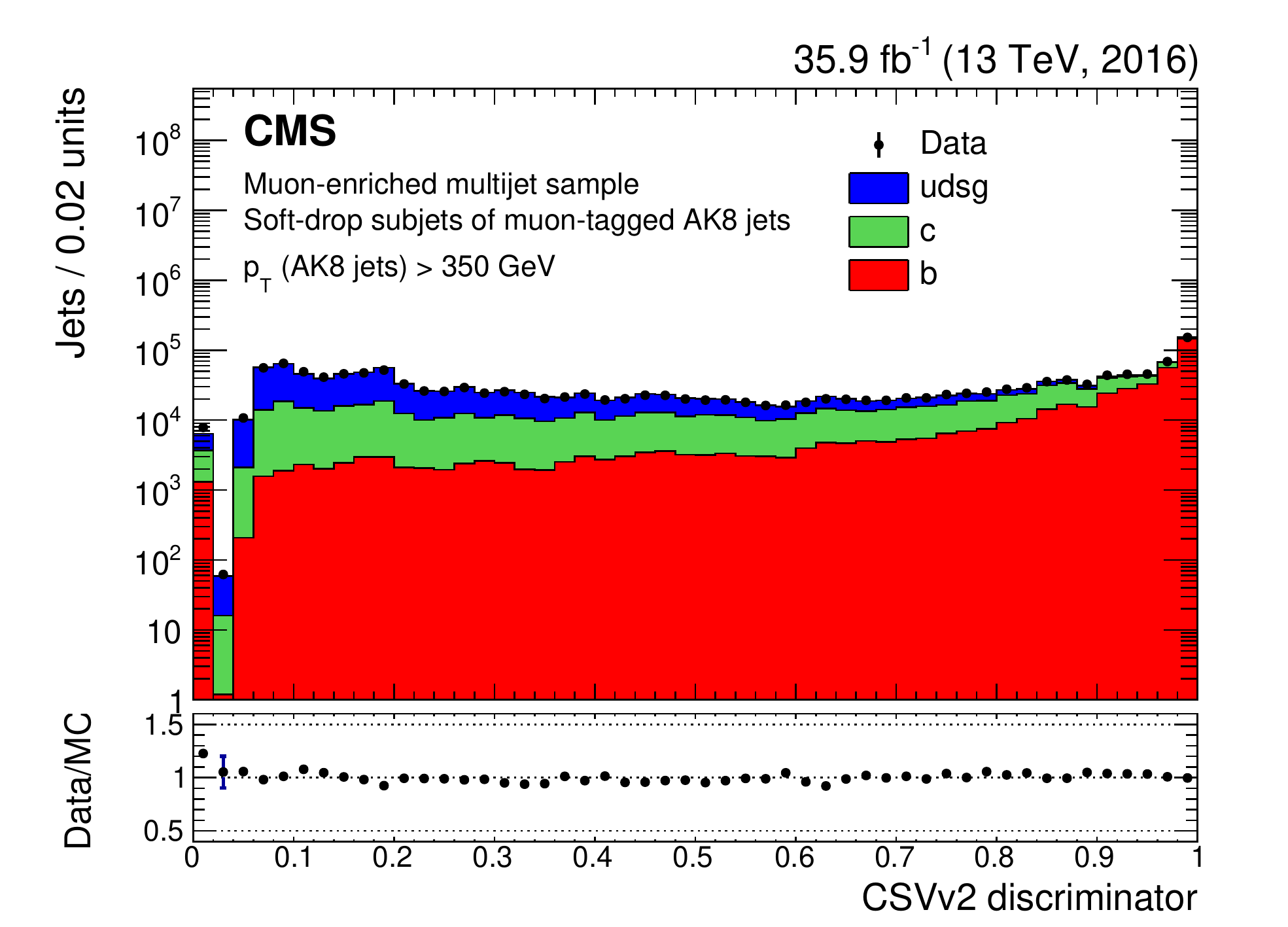}
    \caption{Distribution of the 3D impact parameter significance of the tracks (upper left), the secondary vertex 3D flight distance significance (upper right), the corrected secondary vertex mass (lower left), and the CSVv2 discriminator (lower right) for muon-tagged subjets of AK8 jets with $\pt > 350\GeV$. The simulated contributions of each jet flavour are shown with a different colour. The total number of entries in the simulation is normalized to the number of observed entries in data. The first and last bin of each histogram contain the underflow and overflow entries, respectively.}
    \label{fig:boostedcommsubjets}
\end{figure}

\begin{figure}[hbtp]
  \centering
    \includegraphics[width=0.49\textwidth]{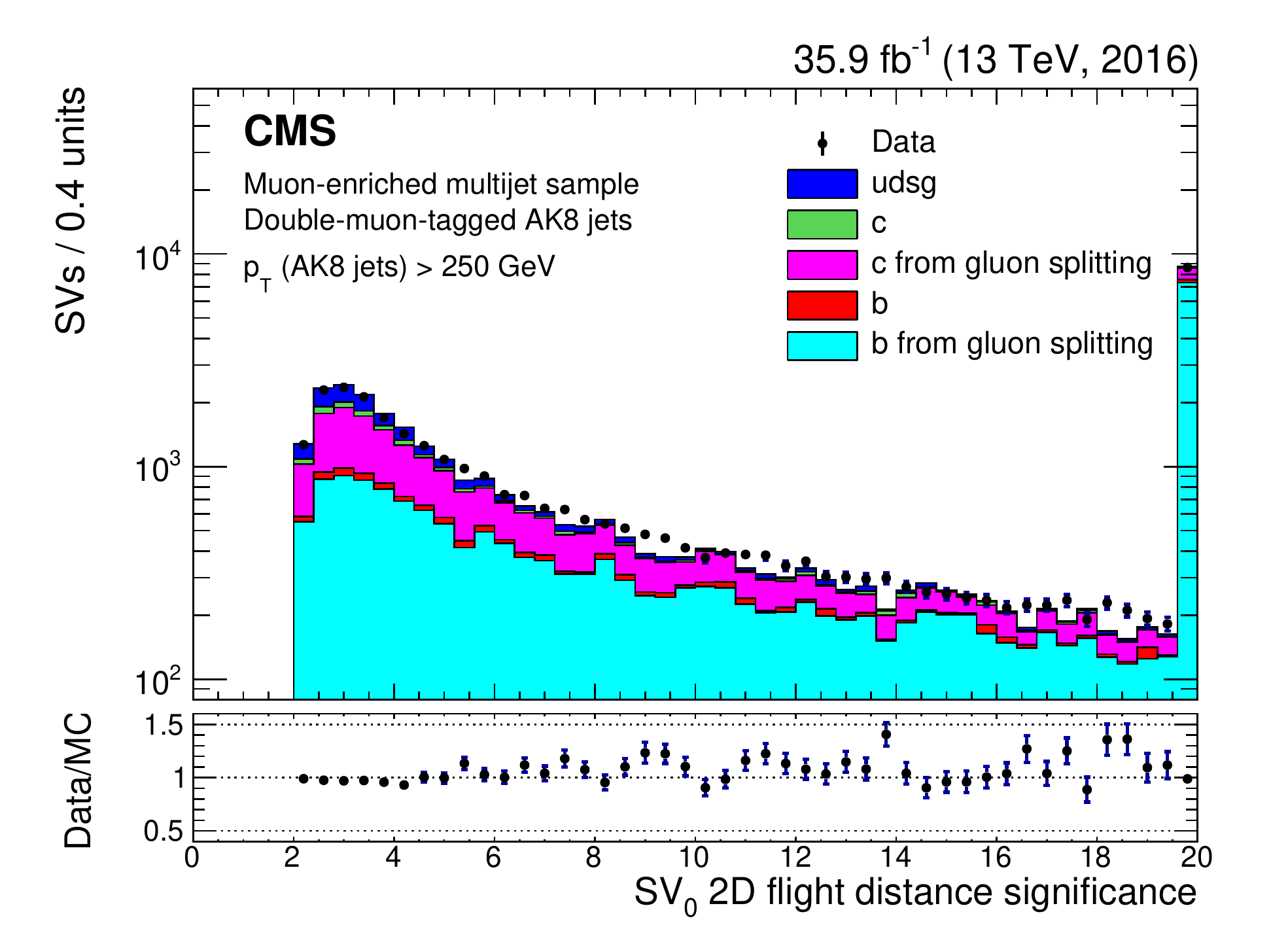}
    \includegraphics[width=0.49\textwidth]{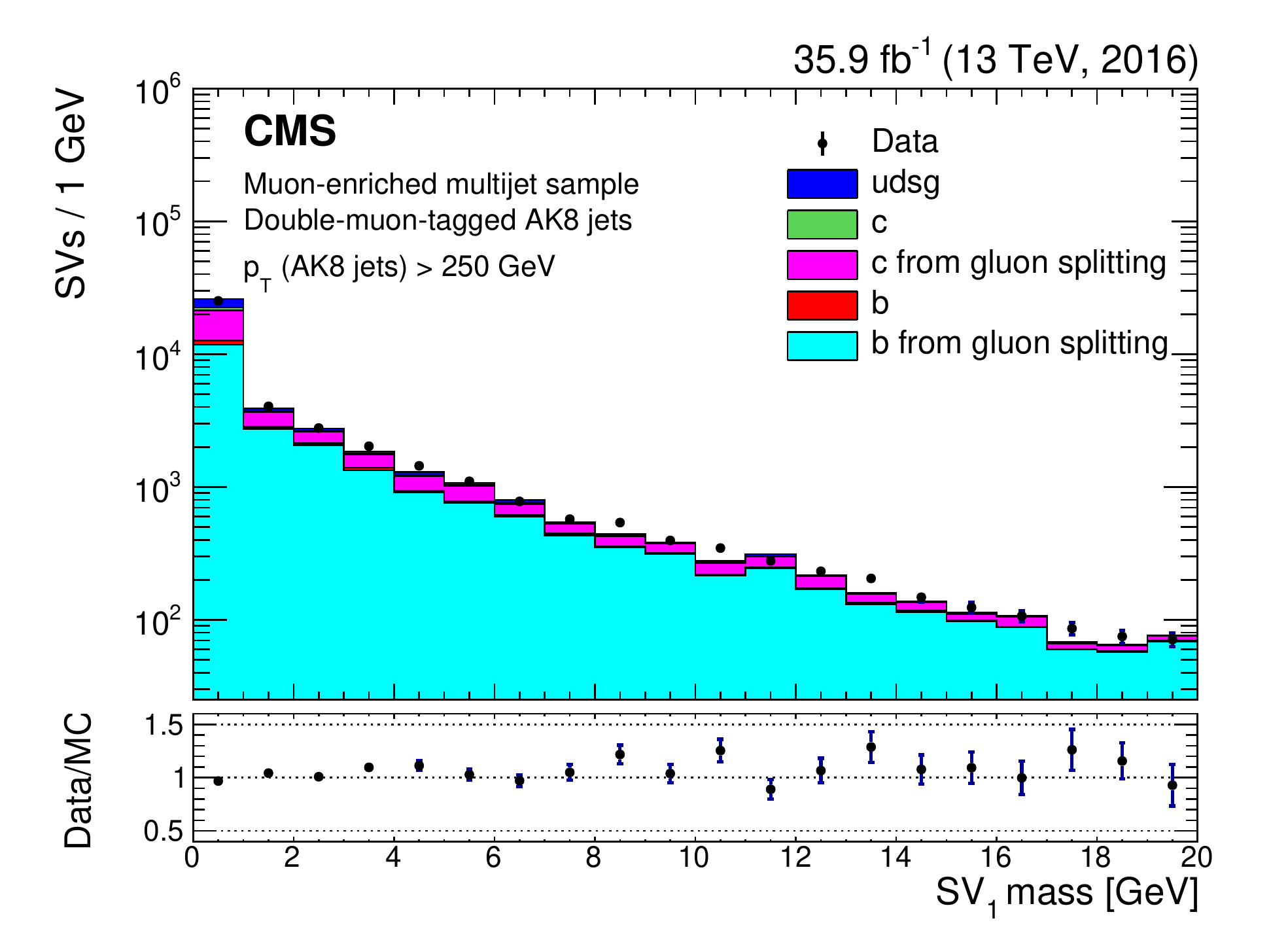}
    \includegraphics[width=0.49\textwidth]{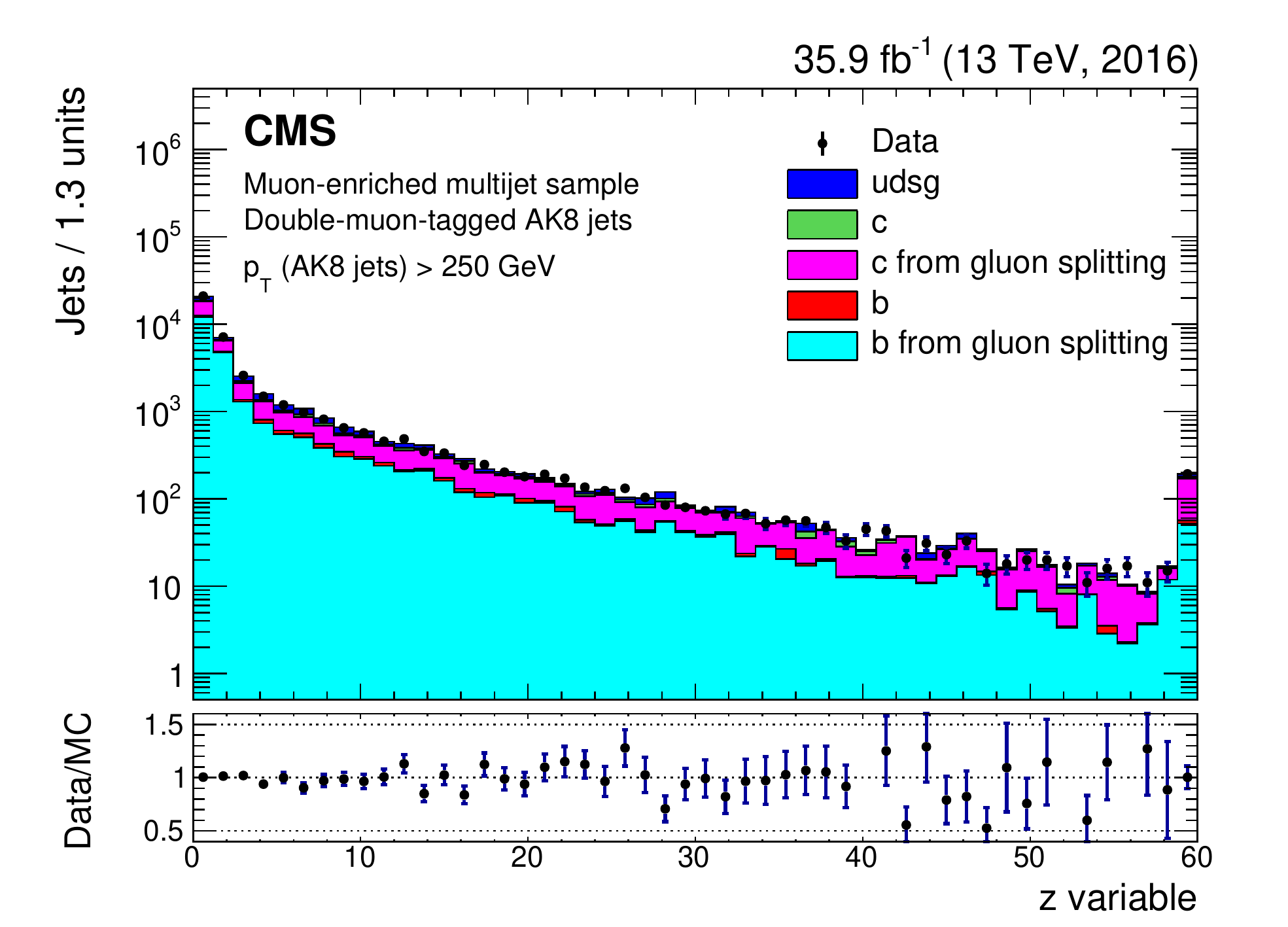}
    \includegraphics[width=0.49\textwidth]{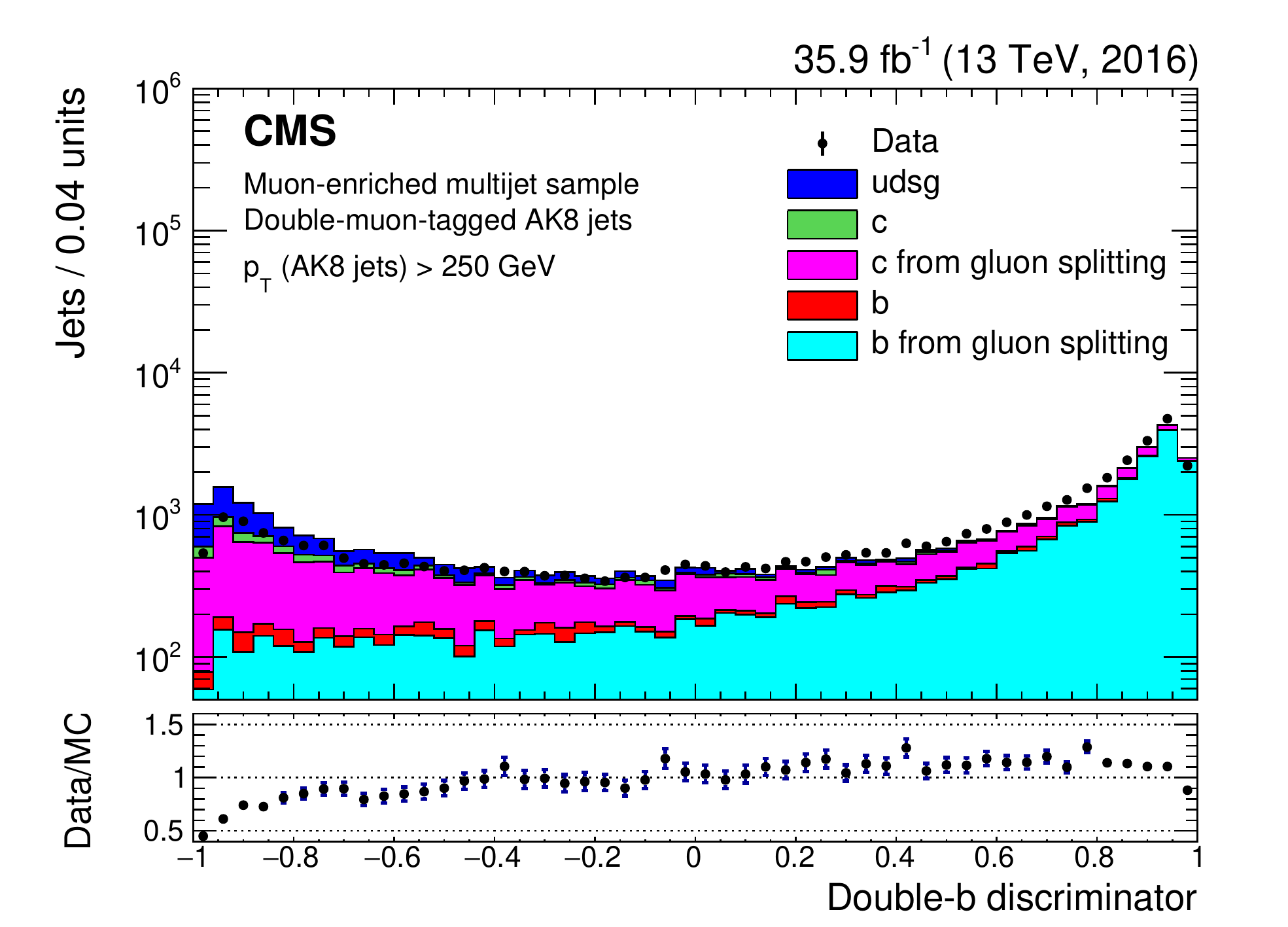}
    \caption{Distribution of the 2D flight distance significance of the secondary vertex associated with the first $\tau$ axis (upper left), the mass of the secondary vertex associated with the second $\tau$ axis (upper right), the $z$ variable (lower left), and the double-{\cPqb} discriminator (lower right) for double-muon-tagged AK8 jets with $\pt > 250\GeV$. The simulated contributions of each jet flavour are shown with a different colour. The total number of entries in the simulation is normalized to the number of observed entries in data. The first and last bin of the upper and lower right histograms contain the underflow and overflow entries, respectively.}
    \label{fig:boostedcommdoublemu}
\end{figure}

In Fig.~\ref{fig:boostedcommsubjets} the data are compared to the simulation for subjets in the muon-enriched sample. The distributions of a few selected input variables are shown as well as the CSVv2 discriminator output distribution. The agreement is reasonable, with variations of up to 20\%. Similarly, Fig.~\ref{fig:boostedcommdoublemu} shows the simulation and data for double-muon-tagged AK8 jets. Some of the input variables of the double-{\cPqb} tagger are shown as well as the discriminator output distribution itself.

\subsection{Efficiency for subjets}
\label{sec:subjetSF}
\subsubsection{Misidentification probability }
The CSVv2 algorithm is used when applying {\cPqb} jet identification on subjets of AK8 jets. Data-to-simulation scale factors for light-flavour subjets from AK8 jets are derived with the negative-tag method used to measure the scale factors for light-flavour jets in Section~\ref{sec:negtag}. A sample of inclusive multijet events is selected using single-jet triggers with different \pt thresholds ranging from 140 to 500\GeV. The AK8 jet is required to have an offline reconstructed soft-drop jet mass between 50 and 200\GeV, where the jet mass is obtained from the invariant mass of the two subjets. The scale factors are measured for the loose and medium working points of the CSVv2 taggers using subjets with $\pt > 20\GeV$ within the tracker acceptance. The same sources of systematic effects are taken into account as for the scale factor measurement for AK4 light-flavour jets.

\begin{figure}[hbtp]
  \centering
    \includegraphics[width=.49\textwidth]{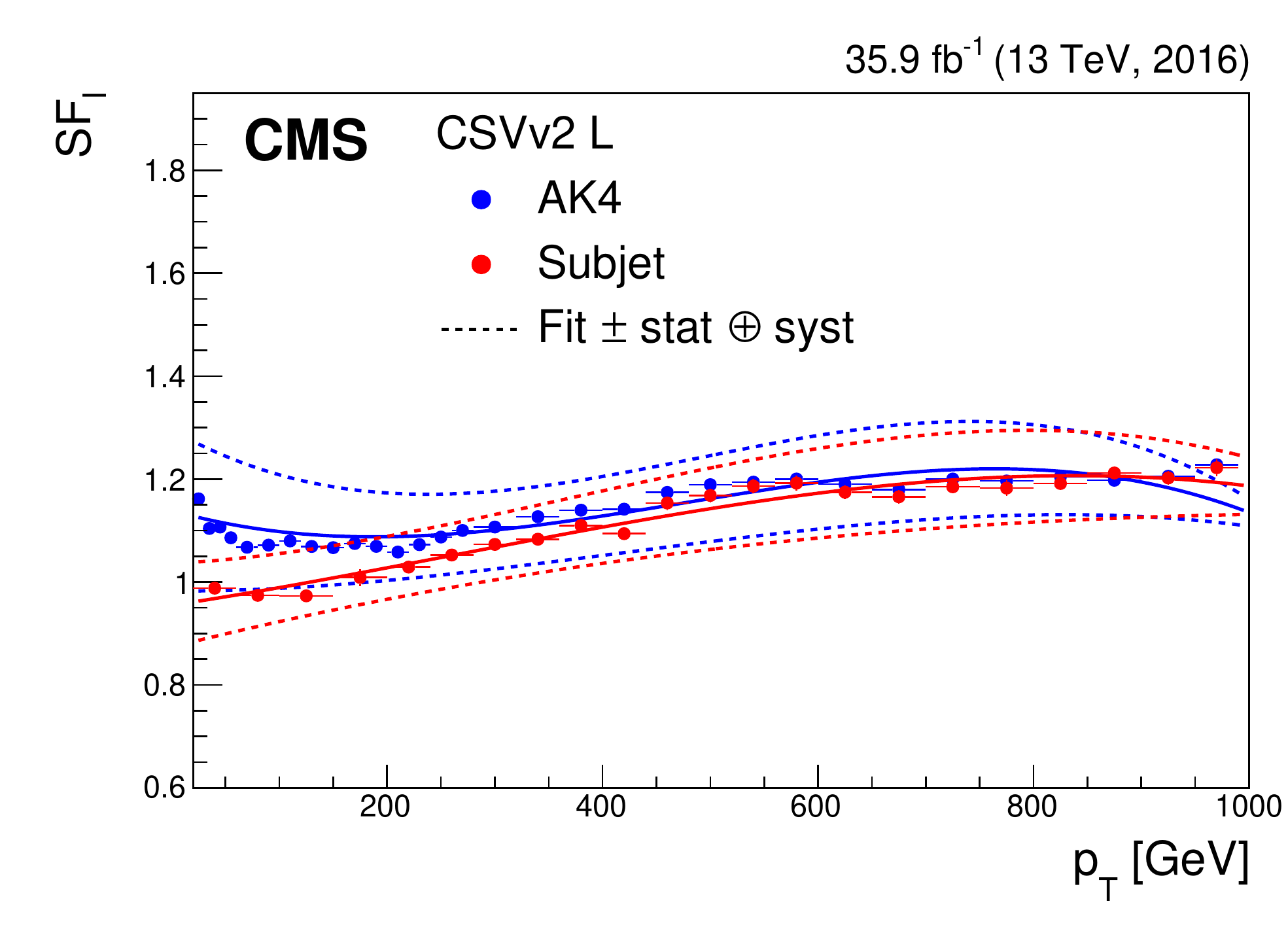}
    \includegraphics[width=.49\textwidth]{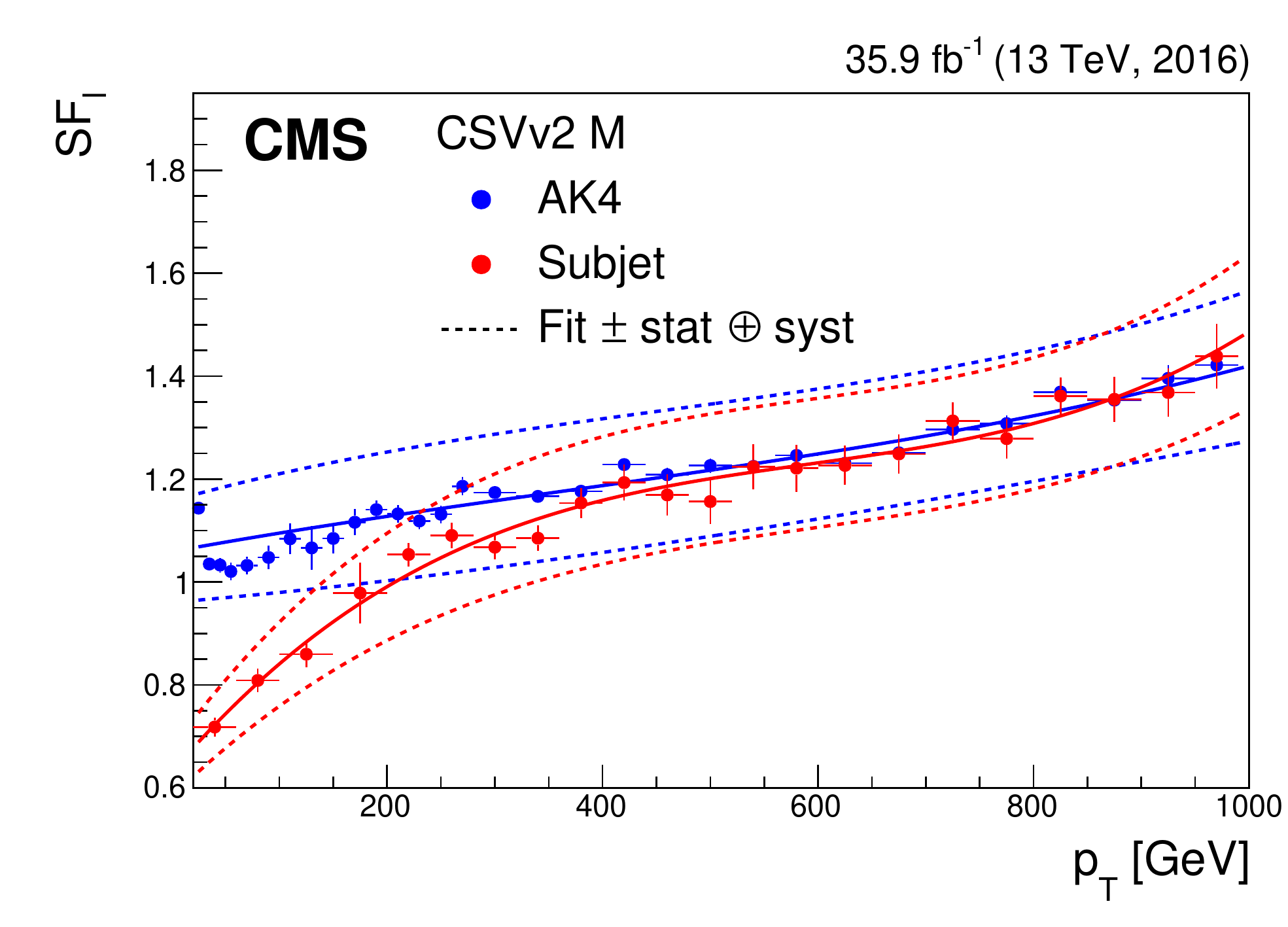}
    \caption{Data-to-simulation scale factors for light-flavour subjets of AK8 jets as a function of the subjet \pt, as well as for AK4 jets as a function of jet \pt, for the loose (left) and medium (right) working points of the CSVv2 algorithm. The solid curve is the result of a fit to the scale factors, and the dashed lines represent the overall statistical and systematic uncertainty of the measurements. }
    \label{fig:NegTagSFSub_CSVv2}
\end{figure}

The measured data-to-simulation scale factors are shown in Fig.~\ref{fig:NegTagSFSub_CSVv2} for the loose and medium working points of the CSVv2 algorithm as a function of the subjet \pt. The measurement is compared to the corresponding AK4 jet scale factors, and within the uncertainty both scale factors agree for jets with $\pt > 200\GeV$. The difference for low jet \pt is because of the very different environment for low-\pt subjets in a boosted AK8 jet compared to low-\pt AK4 jets.

\subsubsection{Measurement of the {\cPqb} tagging efficiency}
\label{sec:LTsubjet}
\begin{figure}[hbtp]
  \centering
    \includegraphics[width=0.49\textwidth]{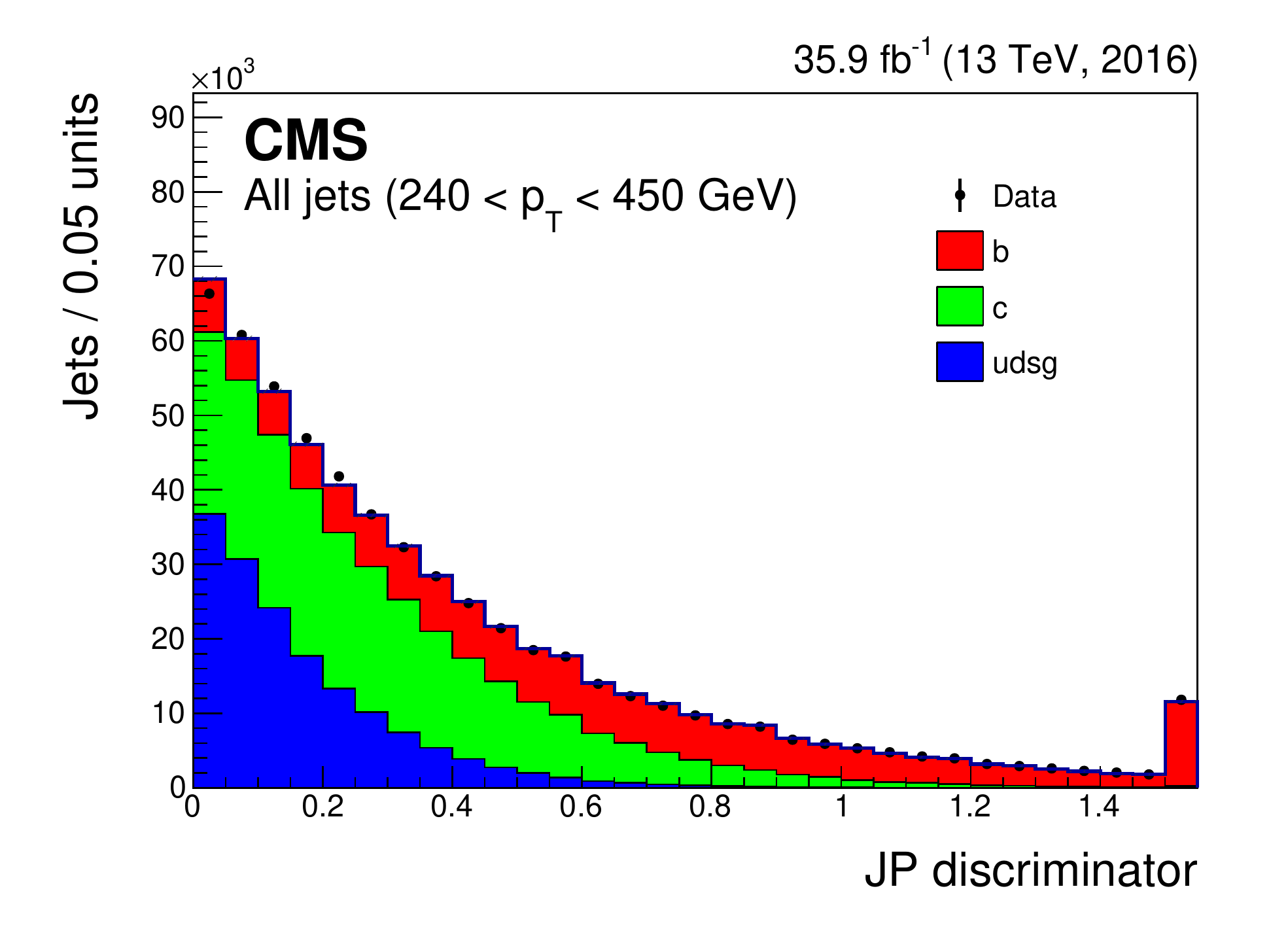}
    \includegraphics[width=0.49\textwidth]{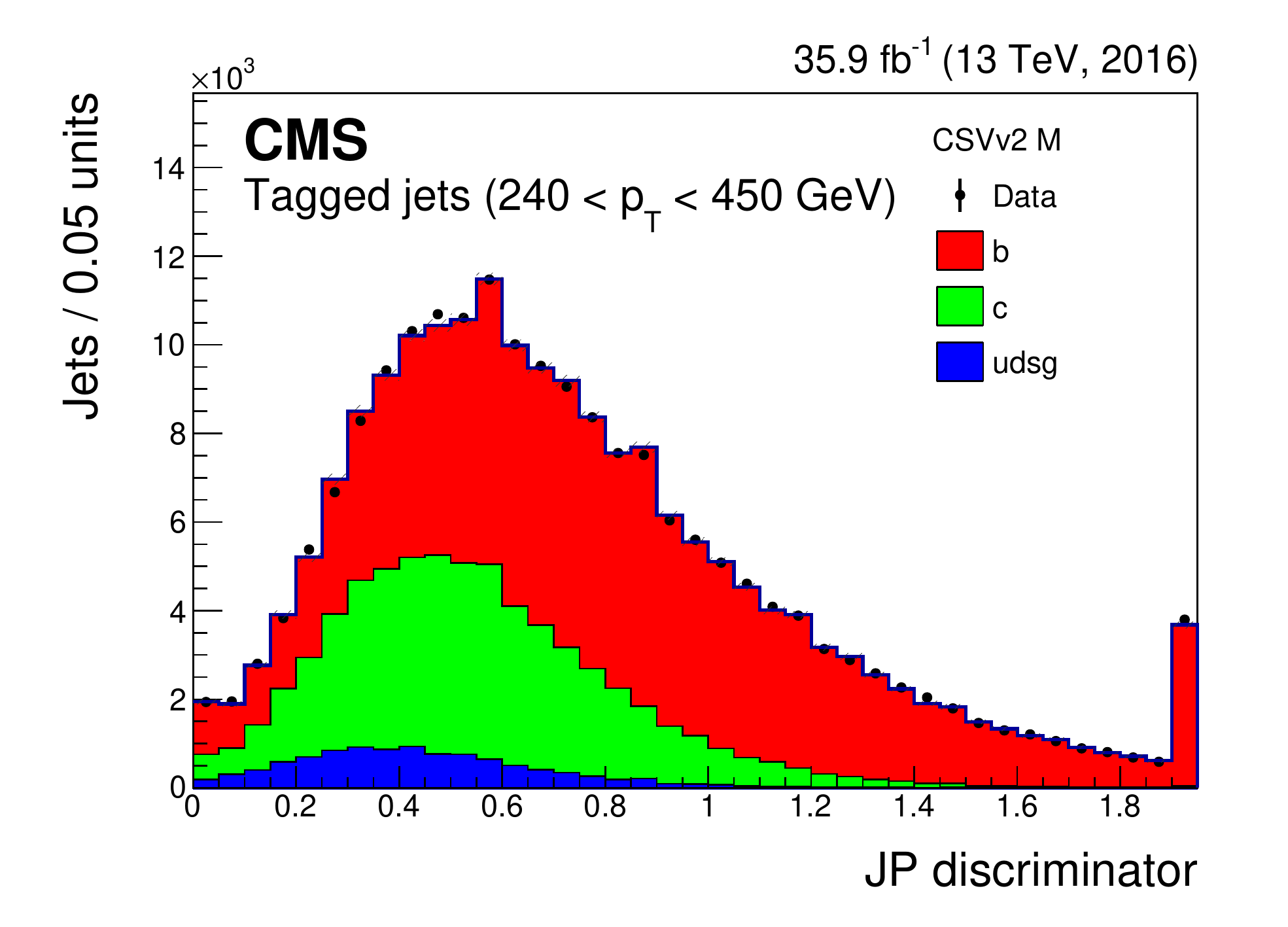}
    \caption{Fitted JP discriminator distribution for all soft-drop subjets with $240 < \pt<450\GeV$ (left) and for the subsample of those subjets passing the medium working point of the CSVv2 algorithm (right). The last bin contains the overflow entries.}
    \label{fig:LTFitBoost}
\end{figure}

The data-to-simulation scale factors for subjets originating from {\cPqb} quarks are measured on subjets of AK8 jets using the selection requirements described in Section~\ref{sec:boostedcomm}. The LifeTime LT method presented in Section~\ref{sec:LTmethod} is applied to measure the scale factors for the loose and medium working points of the CSVv2 algorithm. The templates of the JP distribution for the various flavours obtained from simulation are fitted to the distribution observed in the data before and after applying the tagging requirement. An example of the fitted JP distribution for subjets with $240 <\pt< 450\GeV$ is shown in Fig.~\ref{fig:LTFitBoost} for all subjets and for subjets passing the medium working point of the CSVv2 algorithm.
The systematic uncertainties associated with the scale factor measurements are the same as evaluated for AK4 jets discussed in Section~\ref{sec:muonAK4syst}. Compared to the measurements in Section~\ref{sec:LTmethod}, the calibration of the track probabilities used in the resolution function of the JP algorithm (Section~\ref{sec:JP}) is performed differently. In particular, for the nominal scale factor values considered here, the calibration of the track probabilities is derived from simulation and applied to both data and simulation. The systematic effect is evaluated from the difference between the nominal scale factor and that obtained by applying to the data the calibration of the track probabilities derived from the data. The uncertainty due to jets without a JP discriminator value is found to be negligible because of the higher jet \pt.

\begin{figure}[hbtp]
  \centering
    \includegraphics[width=0.49\textwidth]{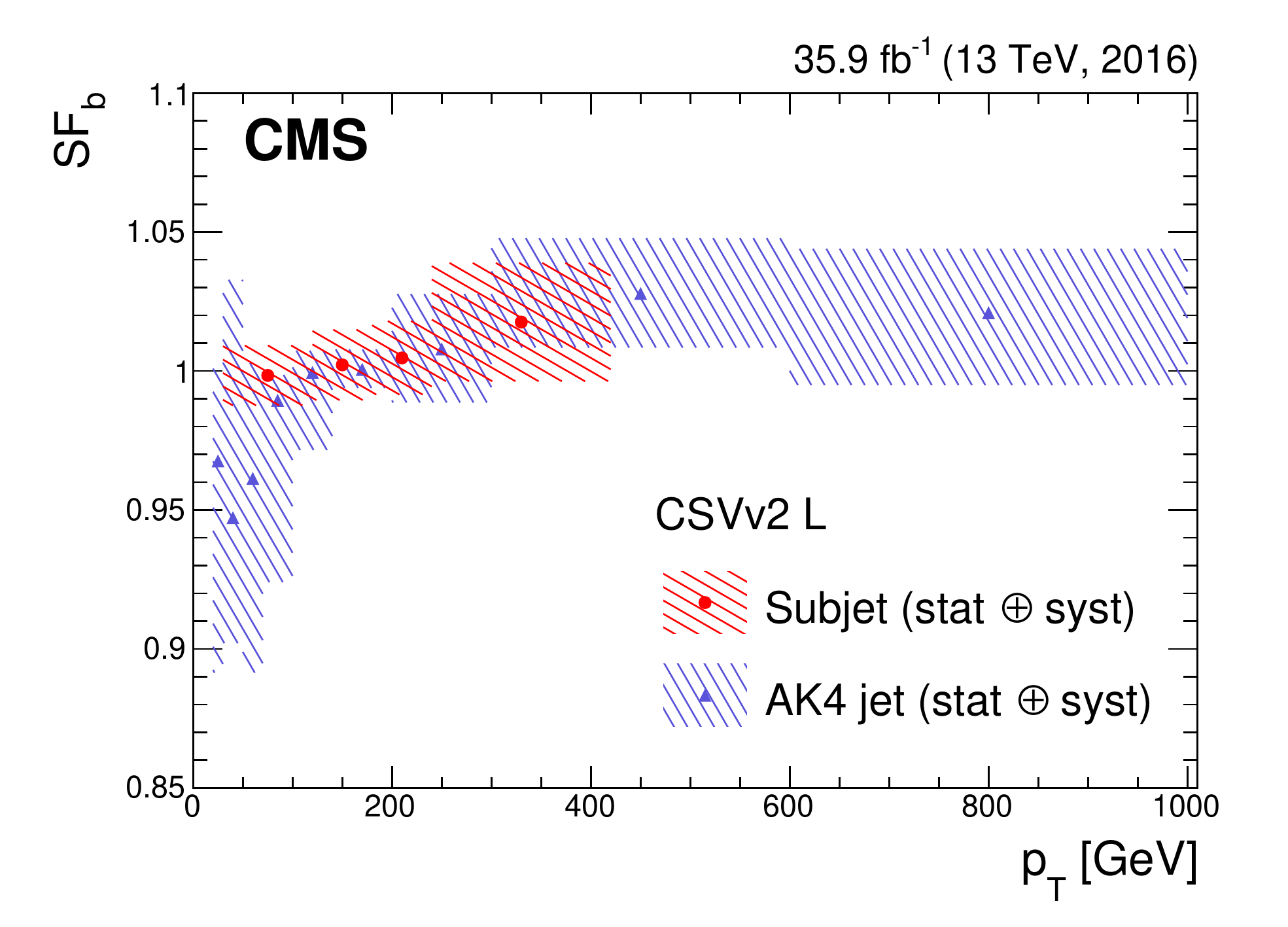}
    \includegraphics[width=0.49\textwidth]{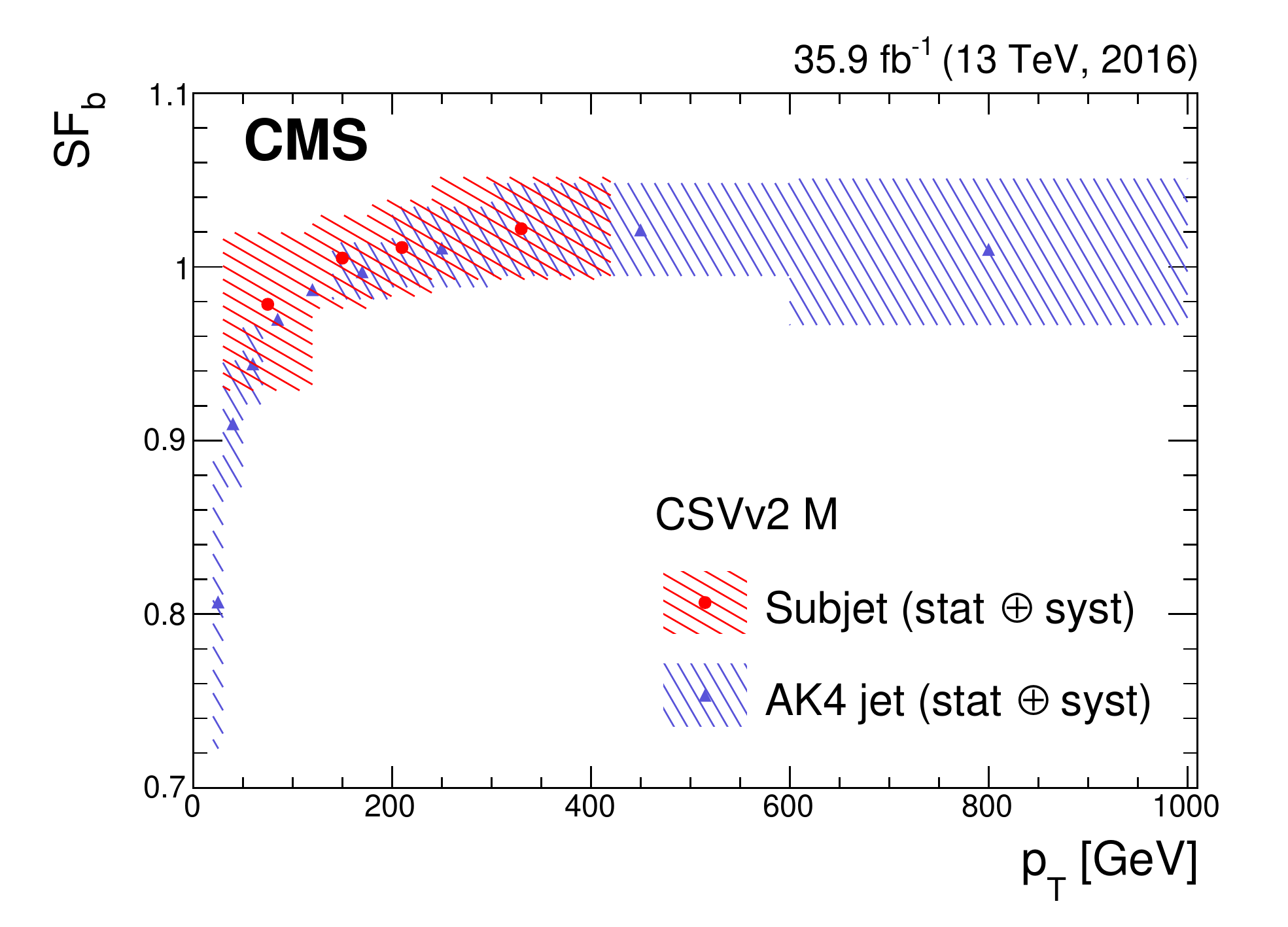}
    \caption{Data-to-simulation scale factors for {\cPqb} subjets of AK8 jets as a function of the subjet \pt, as well as for AK4 jets as a function of jet \pt, for the loose (left) and medium (right) working points of the CSVv2 algorithm. The hatched band around the scale factors represents the overall statistical and systematic uncertainty of the measurements.}
    \label{fig:LTSFSub_CSVv2_AK4}
\end{figure}

The measured data-to-simulation scale factors for the loose and medium working points of the CSVv2 tagger are presented as function of the subjet \pt in Fig.~\ref{fig:LTSFSub_CSVv2_AK4}. As a comparison, the scale factors for AK4 jets obtained with the LT method are also shown. The scale factors for AK4 jets and subjets are consistent within their uncertainties.

\subsection{Efficiency of the double-\texorpdfstring{{\cPqb}}{b} tagger}
\label{sec:doublebSF}
\subsubsection{Measurement of the double-{\cPqb} tagging efficiency}
To measure the efficiency of the four working points of the double-{\cPqb} tagger defined in Section~\ref{sec:doubleb}, a pure sample of boosted {{\cPqb}{\cPaqb}} jets needs to be selected from data. The measurement is performed using a sample of high-\pt jets enriched with $\cPg\to{\cPqb}{\cPaqb}$ jets. The enrichment is achieved by requiring each AK8 jet to be double-muon tagged, as described in Section~\ref{sec:boostedcomm}. While additional systematic uncertainties arise from using {{\cPqb}{\cPaqb}} jets from gluon splitting, the statistical uncertainty of a measurement performed on boosted $\PH\to{\cPqb}{\cPaqb}$ jets would be too large. Also $\PZ\to{\cPqb}{\cPaqb}$ events cannot be easily used because of the difficulty to obtain a pure sample of those events. Using the simulation, it has been verified that the $\cPg\to{\cPqb}{\cPaqb}$ jets can be used as a proxy for the $\PH\to{\cPqb}{\cPaqb}$ jets signal. Indeed, after the selection, the distributions of the double-{\cPqb} tagger discriminator values and its input variables were compared for simulated $\cPg\to{\cPqb}{\cPaqb}$ and $\PH\to{\cPqb}{\cPaqb}$ jets. Since a different shape was observed for the discriminator distribution, the $\cPg\to{\cPqb}{\cPaqb}$ events were reweighted using the distribution of the $z$ variable and the secondary vertex energy ratio, which are the variable distributions with the largest shape difference. The data-to-simulation scale factors were then computed using either the reweighted $\cPg\to{\cPqb}{\cPaqb}$ simulation or the original $\cPg\to{\cPqb}{\cPaqb}$ simulation. Both scale factors were found to be compatible, which confirms that the $\cPg\to{\cPqb}{\cPaqb}$ events allow for an unbiased measurement of the efficiency.

\begin{figure}[hbtp]
  \centering
    \includegraphics[width=0.49\textwidth]{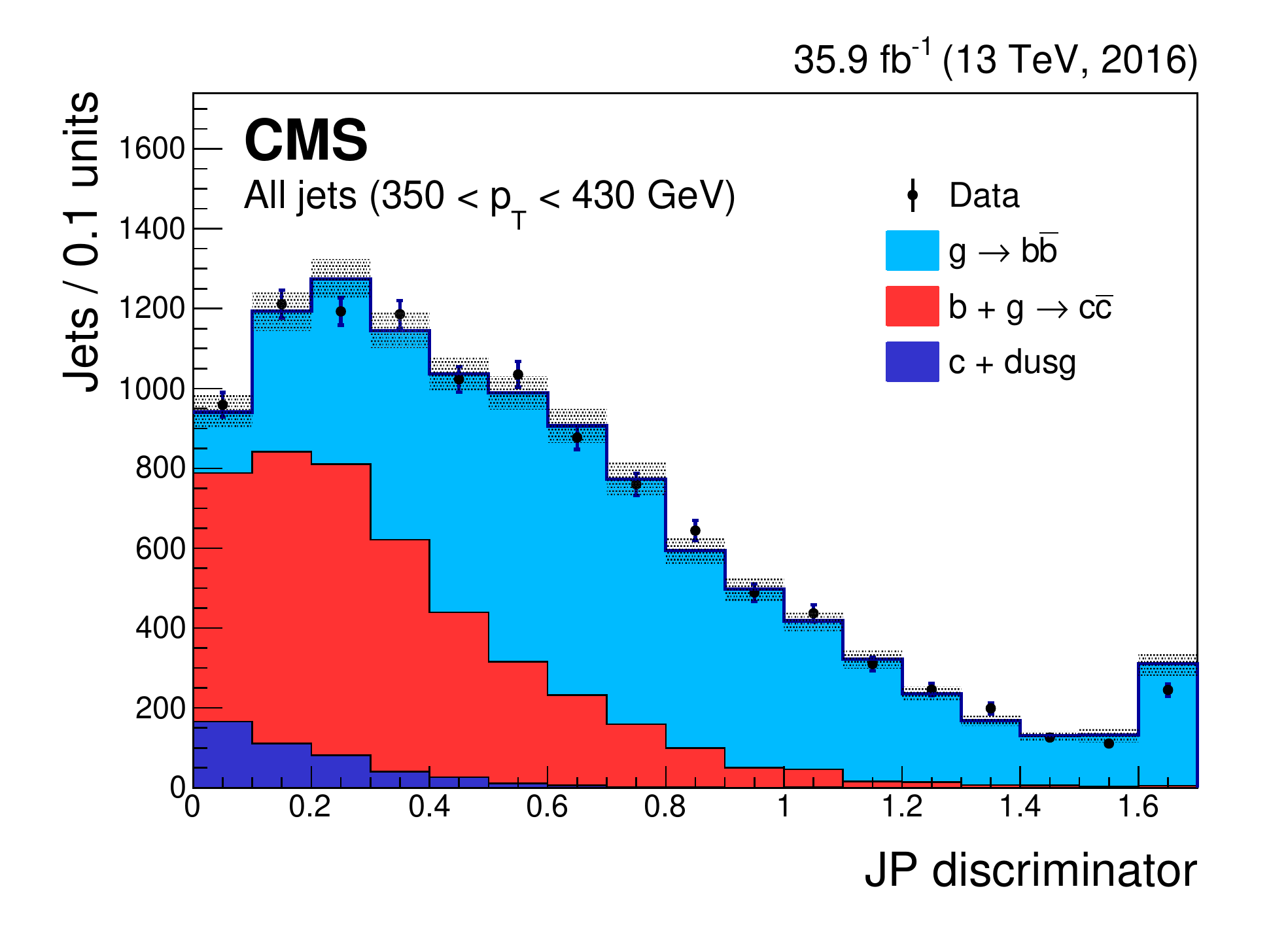}
    \includegraphics[width=0.49\textwidth]{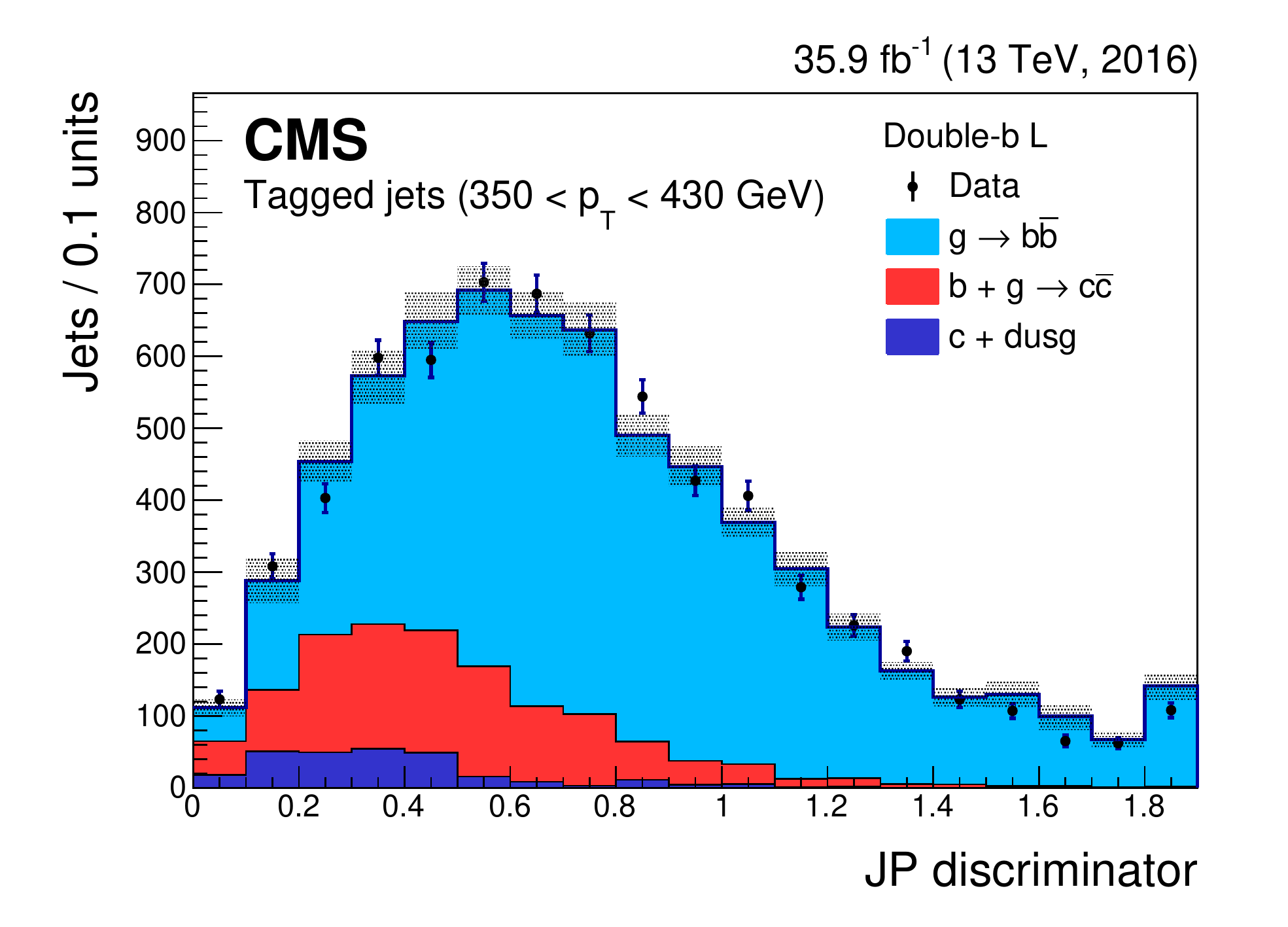}
    \caption{Fitted JP discriminator distribution for all soft-drop subjets with $350 < \pt<430\GeV$ (left) and for the subsample of those subjets passing the loose working point of the double-b algorithm (right). The shaded area represents the statistical and systematic uncertainties in the templates obtained from simulation. The last bin contains the overflow entries.}
    \label{Fig:LTFitDoubleMu}
\end{figure}

The efficiency and the corresponding data-to-simulation scale factor $SF_{\text{double-{\cPqb}}}$ is measured using data for the working points of the double-{\cPqb} tagger defined in Section~\ref{sec:boostedalgos}. The measurement is performed using the LT method, presented in Section~\ref{sec:LTmethod} and also used in Section~\ref{sec:LTsubjet}. The expected templates of the JP discriminator after the tagging requirement consist of two contributions, one arising from $\cPg\to{\cPqb}{\cPaqb}$ jets and one from jets not stemming from this process (background jets). These two templates are used to fit the fraction of each contribution to the JP discriminator in data. The fit is performed in three bins of jet \pt for the loose, medium-1, and medium-2 working points, and in two bins of jet \pt for the tight working point.
An example of the fitted distributions is shown in Fig.~\ref{Fig:LTFitDoubleMu} for AK8 jets with $350 <\pt< 430\GeV$ before and after applying the loose working point of the double-{\cPqb} algorithm. The background jets are shown separately for {\cPqb} and $\cPg\to{\cPqc}{\cPaqc}$ jets and for {\cPqc} and light-flavour jets. However, the templates of these two components are merged for the tagged jet sample when performing the fit.

The measurement is sensitive to the flavour composition of the background sample. The uncertainty due to the flavour composition is estimated by varying by $\pm$50\% the normalization of each flavour in the background templates. As a cross check, the potential systematic effect of merging all background jets in a single template is assessed by remeasuring the data-to-simulation scale factor using a separate template for each flavour in the fit. The systematic uncertainty due to the template variation results in a systematic uncertainty of up to 2.3\% in the measured scale factor. The uncertainty related to the track probability calibration for the resolution function used in the JP discriminator is evaluated as described in Section~\ref{sec:LTsubjet}, and results in an uncertainty of 2.9\% in the measured scale factors. The impact of the uncertainty in the number of pileup interactions results in an uncertainty of 1.3\% in the scale factors. The following systematic uncertainties were found to be negligible: bin-by-bin correlations, jet energy corrections, the number of tracks, the branching fractions for {\cPqc} hadrons to muons, the {\cPqb} fragmentation function, the fragmentation rate of a {\cPqc} quark to various {\PD} mesons, and the {\PKzS} and {\PgL} production fractions.

\begin{figure}[hbtp]
  \centering
    \includegraphics[width=0.49\textwidth]{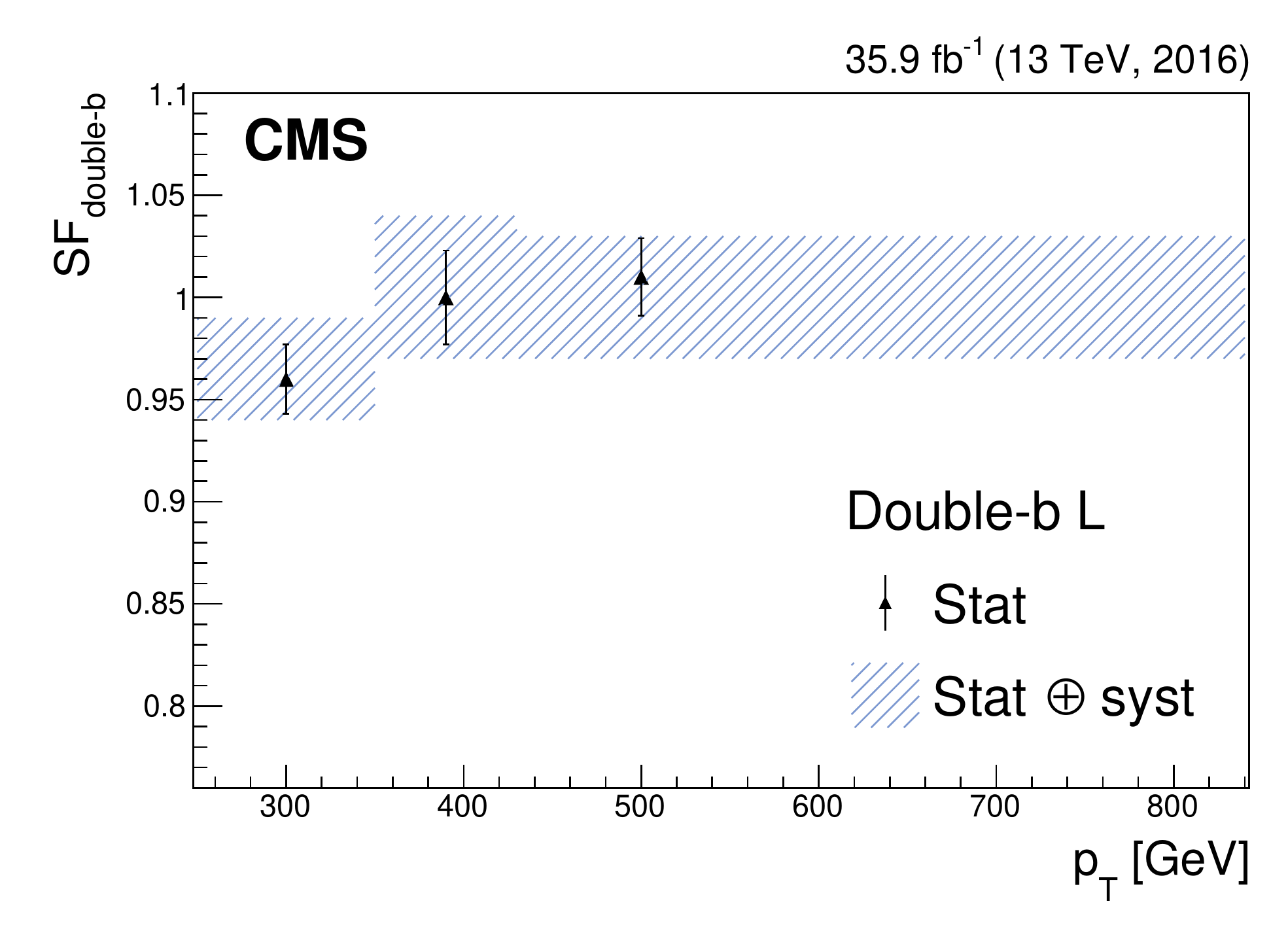}
    \includegraphics[width=0.49\textwidth]{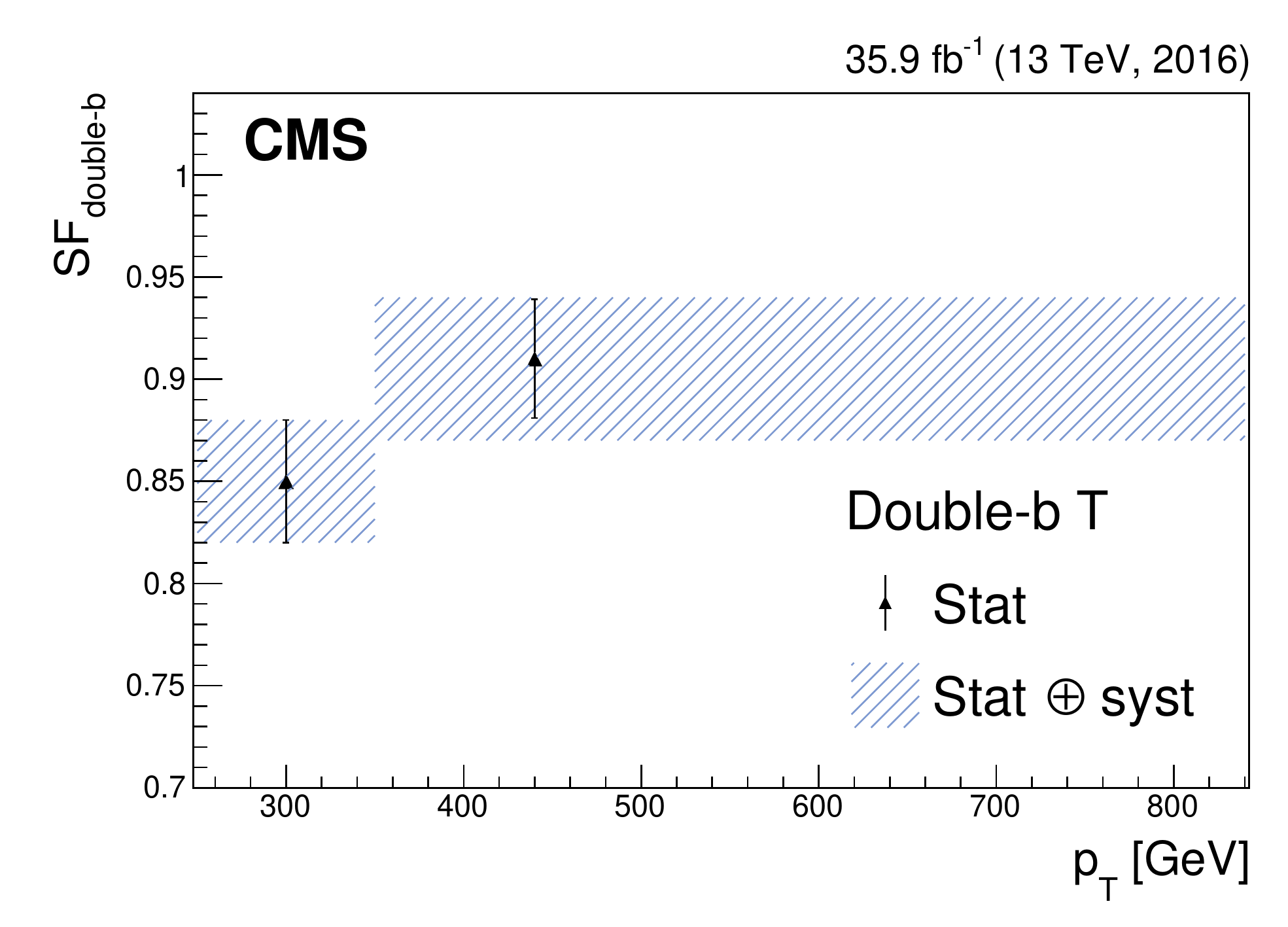}
\caption{Data-to-simulation scale factors for correctly identifying two {\cPqb} jets in an AK8 jet as a function of the jet \pt for the loose (left) and tight (right) working points of the double-{\cPqb} tagger. The hatched band around the scale factors represents the overall statistical and systematic uncertainty in the measurement. Jets with $\pt>840\GeV$ are included in the last \pt bin.}
    \label{Fig:LTSF}
\end{figure}

The data-to-simulation scale factor $SF_{\text{double-{\cPqb}}}$ is presented in Fig.~\ref{Fig:LTSF} for two working points of the double-{\cPqb} tagger.
The measurement is performed using jets with $\pt > 250\GeV$. Jets with $\pt>840\GeV$ are included in the last bin. The scale factor is positioned at the average jet \pt value of the jets populating that bin.

\subsubsection{Measurement of the misidentification probability for top quarks}
The probability to misidentify a boosted top quark jet corresponding to the decay ${\cPqt}\to{\cPqb}{\PW}\to{\cPqb}{\cPq}{\cPaq}$ for the four working points of the double-{\cPqb} tagger is estimated from the data. Semileptonic \ttbar events are selected by requiring exactly one isolated muon with $\pt>50\GeV$ and $\abs{\eta} < 2.1$. The muon is used to define two hemispheres in the event. The leptonic hemisphere is defined as $|\phi_{\text{jet}}-\phi_\Pgm|<\frac{2}{3}\pi$, and the hadronic hemisphere is its complement. At least one AK4 jet is required in each hemisphere, with $\pt > 30\GeV$ and within the tracker acceptance. In addition, the AK4 jet in the leptonic hemisphere should pass the loose working point of the CSVv2 algorithm. At least one AK8 jet is required in the hadronic hemisphere with $\pt>250\GeV$, $\abs{\eta}<2.4$, and a pruned jet mass between 50 and 200\GeV. The N-subjettiness parameters $\tau_{1}$ and $\tau_{2}$ (Section~\ref{sec:boostedalgos}) should satisfy the condition $\tau_{2}/\tau_{1} < 0.6$. If more than one such jet is present, the one with the highest \pt is considered. The aforementioned selection is referred to as the ``2-prong'' selection.

\begin{figure}[!htb]
 \centering
   \includegraphics[width=0.49\textwidth]{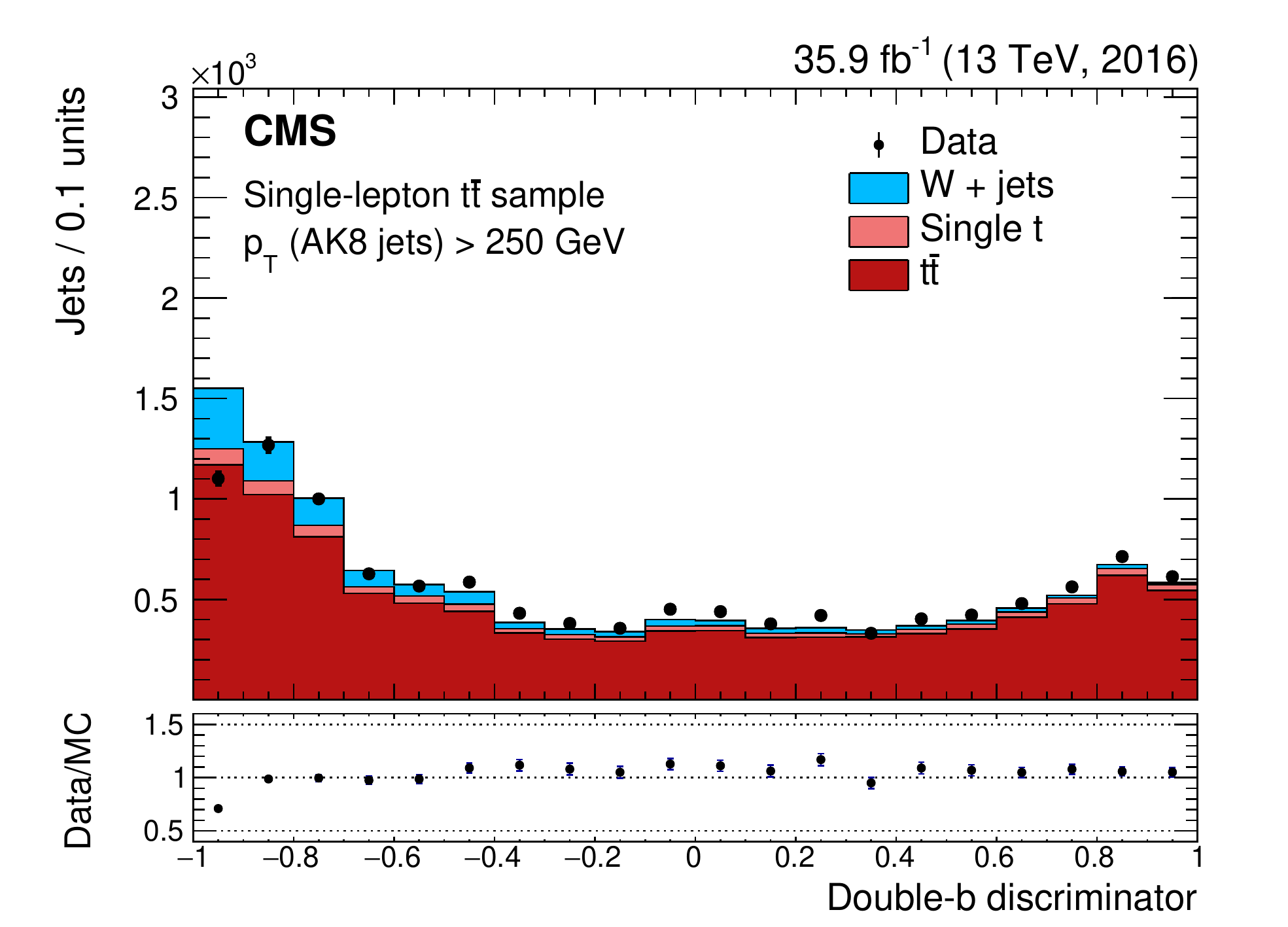}
   \includegraphics[width=0.49\textwidth]{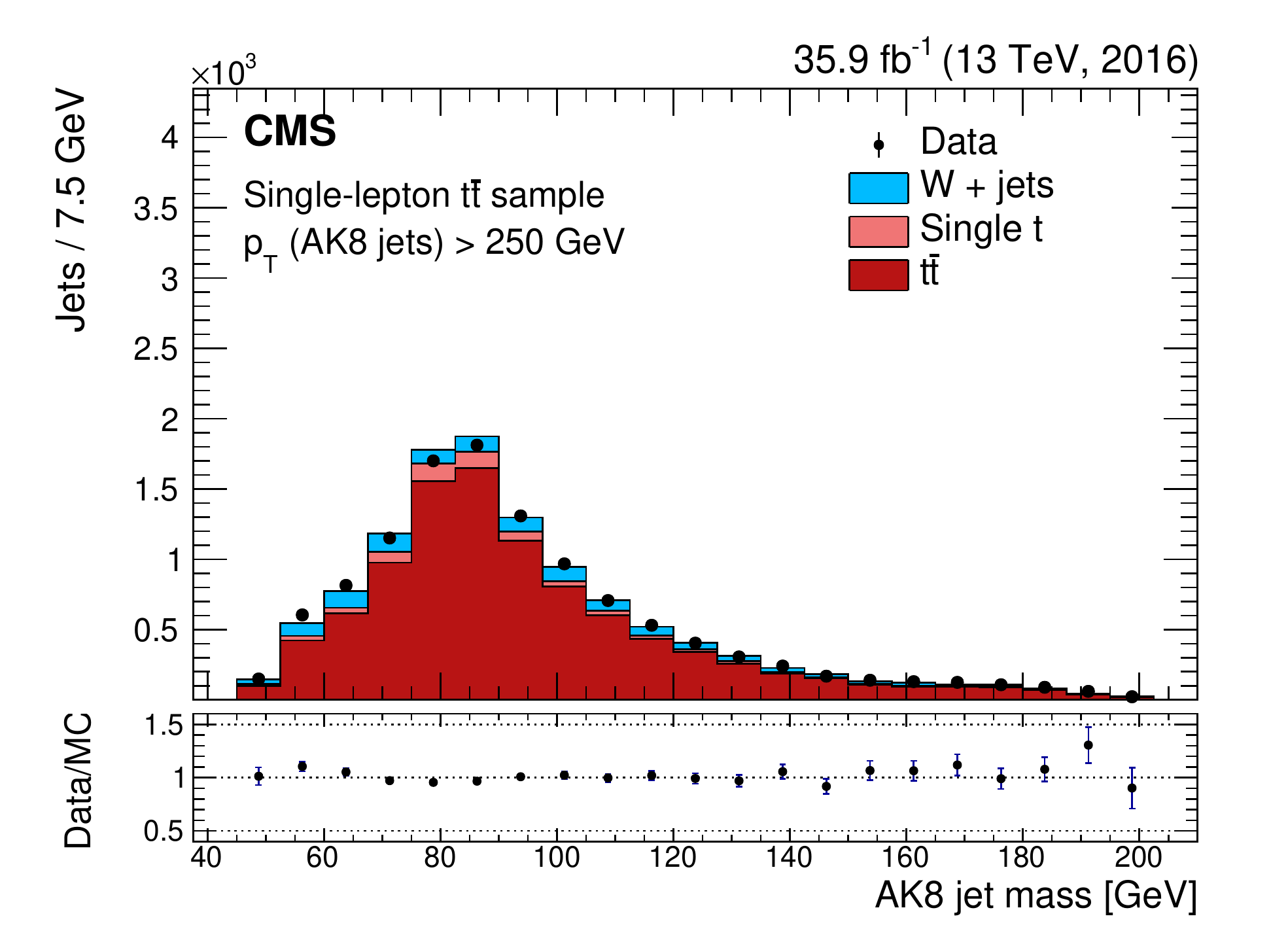}
 \caption{Distribution of the double-{\cPqb} tagger discriminator (left) and pruned jet mass (right) for AK8 jets passing the 2-prong event selection as described in the text. The simulation is normalized to the observed number of events. }
 \label{fig:ttbarsemiplots}
\end{figure}

After the event selection, the simulated events are normalized to the yield observed in the data. Figure~\ref{fig:ttbarsemiplots} shows the distribution of the double-{\cPqb} discriminator and the pruned jet mass for the selected 2-prong events. The purity of the sample is high and the AK8 jet mass distribution is consistent with the decay of the {\PW} boson to quarks.

The probability to misidentify a boosted top quark jet in data is obtained as follows:
\begin{linenomath}
\begin{equation}
\varepsilon_{\text{top}}= \frac{N_{\text{{\cPqb}{\cPaqb}-tagged}}^{\text{data}} - N_{ \text{{\cPqb}{\cPaqb}-tagged}}^{\text{bkg,MC}}}{ N^{\text{data}} - N^{\text{bkg,MC}} },
\end{equation}
\end{linenomath}
where $N_{\text{{\cPqb}{\cPaqb}-tagged}}^{\text{data}}$ and $N^{\text{data}}$ are the number of events with a tagged AK8 jet in data and the total number of events in data, respectively. Similarly, $N_{ \text{{\cPqb}{\cPaqb}-tagged}}^{\text{bkg,MC}}$ and $N^{\text{bkg,MC}}$ are the simulated number of background events with a tagged AK8 jet and the number of simulated background events before applying the working point of the double-{\cPqb} tagger, respectively. The data-to-simulation scale factors are measured both inclusively and in bins of the AK8 jet \pt. The main systematic effect arises from the normalization of the background processes. An uncertainty of 30\% is assigned to the cross section of each background contribution. An additional systematic uncertainty is related to the reweighting of the top quark \pt spectrum. The shape of the \pt distribution for top quarks in data is observed to be softer than in the simulation~\cite{Khachatryan:2053952,Khachatryan:2016641}. For the nominal scale factor measurements, a reweighting procedure is applied to correct for the observed difference. To assess the size of any systematic effect due to the reweighting, the uncertainty is obtained as the difference between the nominal scale factor values and the scale factors obtained when repeating the measurement without applying the reweighting procedure. The systematic uncertainty is found to be 1--2\%.

\begin{figure}[!htb]
 \centering
   \includegraphics[width=0.49\textwidth]{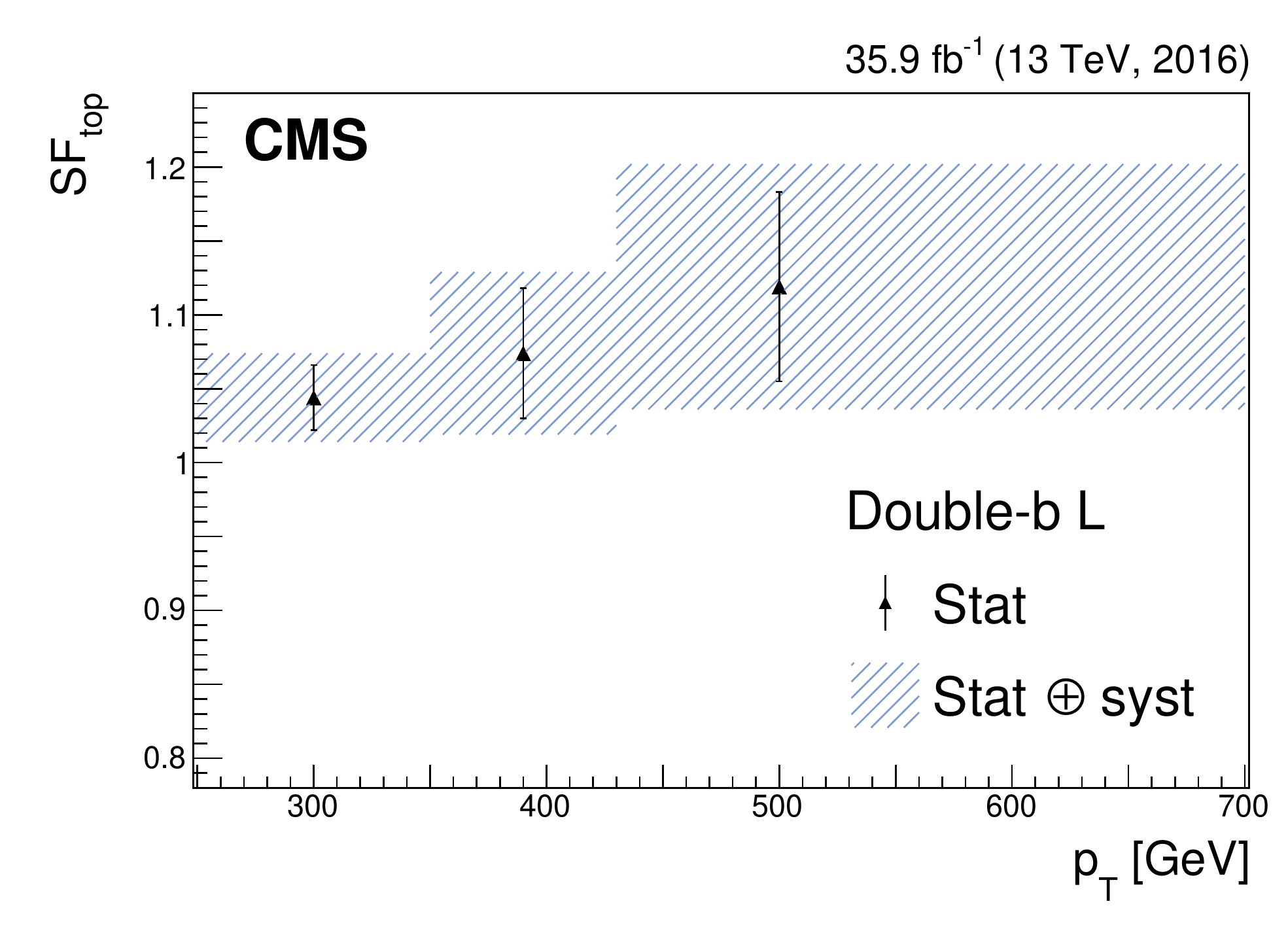}
   \includegraphics[width=0.49\textwidth]{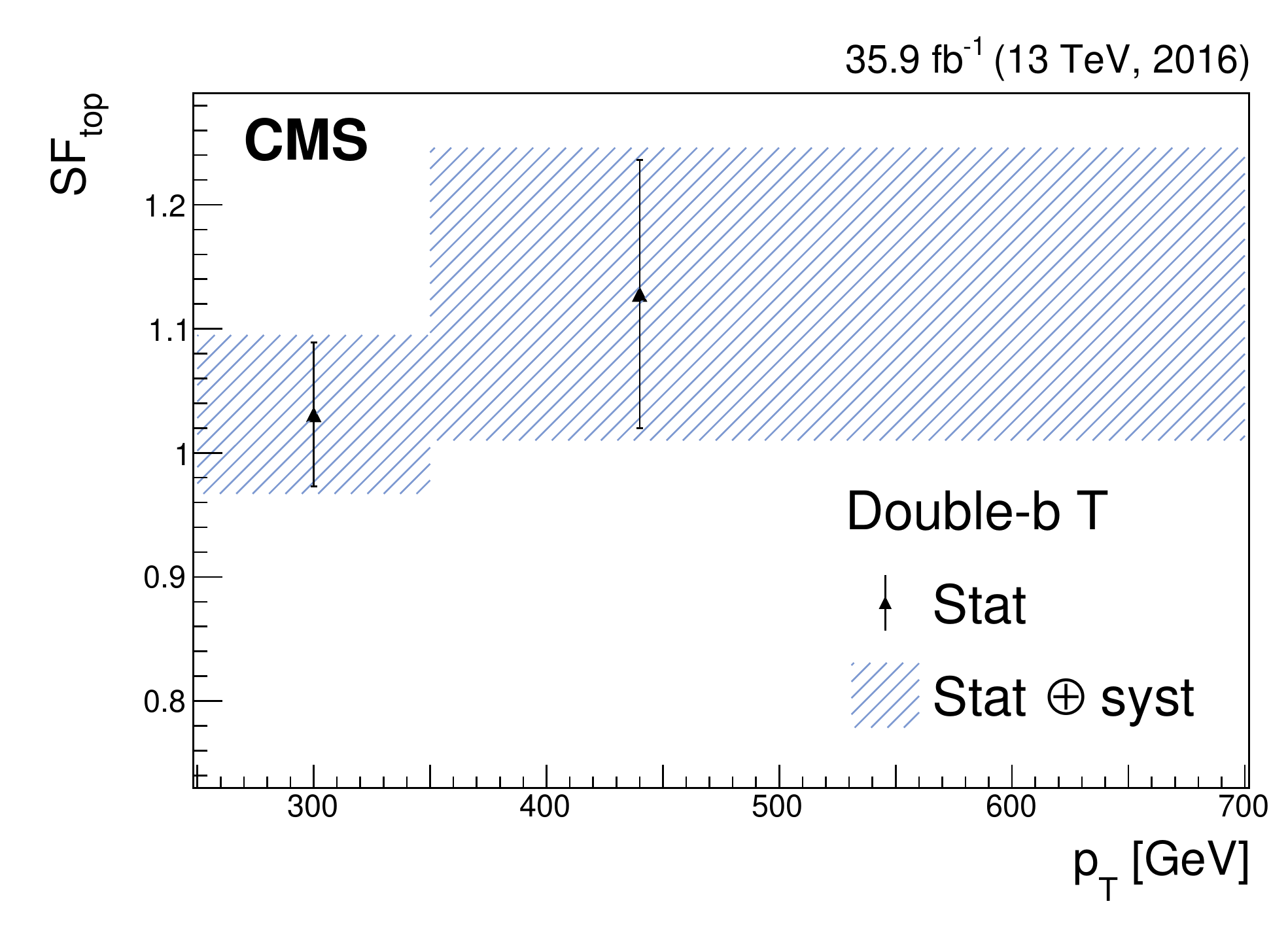}
 \caption{Data-to-simulation scale factors for misidentifying a top quark jet as a function of the jet \pt for the loose (left) and tight (right) working points of the double-{\cPqb} tagger. The hatched band around the scale factors represents the overall statistical and systematic uncertainty in the measurement. Jets with $\pt>840\GeV$ are included in the last \pt bin.}
 \label{fig:ttbarsemiSF}
\end{figure}

The data-to-simulation scale factors for the misidentification of boosted top quark jets for two of the working points of the double-{\cPqb} tagger are shown as a function of the jet \pt in Fig.~\ref{fig:ttbarsemiSF}. The scale factors are positioned at the average jet \pt value of the jets populating that bin.

\section{Summary}

A variety of discriminating variables and algorithms used by the CMS experiment for the identification of heavy-flavour (charm and bottom) jets in proton-proton collisions at 13 TeV have been reviewed. Detailed simulation studies have allowed the reoptimization of existing {\cPqb} tagging algorithms and, in addition, new algorithms have been developed for the first time to identify {\cPqc} jets, as well as {\cPqb\cPaqb} jets in events with boosted topologies. The performance of these heavy-flavour jet identification algorithms has been studied with simulations of different final states with heavy- and light-flavour quarks. The efficiency to correctly identify {\cPqb} jets in resolved {\ttbar} events is 68\% at a misidentification probability for light-flavour jets of 1\%, which is an improvement of 15\% in relative efficiency compared to the best performing algorithm used during LHC Run 1.

The variables and discriminators have been also compared to the data collected by the CMS experiment in 2016 for various event topologies enriched in heavy- or light-flavour jets. Various methods have been presented to determine the data-to-simulation scale factors for the heavy-flavour jet identification efficiency, as well as for the probability to misidentify light-flavour jets. A precision of a few per cent is obtained in the tagging efficiency for {\cPqb} jets with $30 < \pt < 300\GeV$. For {\cPqb} jets with $\pt > 500\GeV$, the precision is of the order of 5\%. For scale factors measured in boosted topologies and for {\cPqc} jets in resolved topologies, the total uncertainty is 5--10\%, and the statistical uncertainty in the tagging efficiency dominates over the full jet \pt range.

With the increasing integrated luminosity delivered by the LHC, the precision of the data-to-simulation scale factors for the specified topologies, jet flavours, and \pt ranges will increase further. Differential studies of the heavy-flavour identification performances as a function of jet pseudorapidity, and of the number of multiple proton-proton interactions in the same bunch crossing, will also become viable.

\begin{acknowledgments}
\hyphenation{Bundes-ministerium Forschungs-gemeinschaft Forschungs-zentren Rachada-pisek} We congratulate our colleagues in the CERN accelerator departments for the excellent performance of the LHC and thank the technical and administrative staffs at CERN and at other CMS institutes for their contributions to the success of the CMS effort. In addition, we gratefully acknowledge the computing centres and personnel of the Worldwide LHC Computing Grid for delivering so effectively the computing infrastructure essential to our analyses. Finally, we acknowledge the enduring support for the construction and operation of the LHC and the CMS detector provided by the following funding agencies: the Austrian Federal Ministry of Science, Research and Economy and the Austrian Science Fund; the Belgian Fonds de la Recherche Scientifique, and Fonds voor Wetenschappelijk Onderzoek; the Brazilian Funding Agencies (CNPq, CAPES, FAPERJ, and FAPESP); the Bulgarian Ministry of Education and Science; CERN; the Chinese Academy of Sciences, Ministry of Science and Technology, and National Natural Science Foundation of China; the Colombian Funding Agency (COLCIENCIAS); the Croatian Ministry of Science, Education and Sport, and the Croatian Science Foundation; the Research Promotion Foundation, Cyprus; the Secretariat for Higher Education, Science, Technology and Innovation, Ecuador; the Ministry of Education and Research, Estonian Research Council via IUT23-4 and IUT23-6 and European Regional Development Fund, Estonia; the Academy of Finland, Finnish Ministry of Education and Culture, and Helsinki Institute of Physics; the Institut National de Physique Nucl\'eaire et de Physique des Particules~/~CNRS, and Commissariat \`a l'\'Energie Atomique et aux \'Energies Alternatives~/~CEA, France; the Bundesministerium f\"ur Bildung und Forschung, Deutsche Forschungsgemeinschaft, and Helmholtz-Gemeinschaft Deutscher Forschungszentren, Germany; the General Secretariat for Research and Technology, Greece; the National Scientific Research Foundation, and National Innovation Office, Hungary; the Department of Atomic Energy and the Department of Science and Technology, India; the Institute for Studies in Theoretical Physics and Mathematics, Iran; the Science Foundation, Ireland; the Istituto Nazionale di Fisica Nucleare, Italy; the Ministry of Science, ICT and Future Planning, and National Research Foundation (NRF), Republic of Korea; the Lithuanian Academy of Sciences; the Ministry of Education, and University of Malaya (Malaysia); the Mexican Funding Agencies (BUAP, CINVESTAV, CONACYT, LNS, SEP, and UASLP-FAI); the Ministry of Business, Innovation and Employment, New Zealand; the Pakistan Atomic Energy Commission; the Ministry of Science and Higher Education and the National Science Centre, Poland; the Funda\c{c}\~ao para a Ci\^encia e a Tecnologia, Portugal; JINR, Dubna; the Ministry of Education and Science of the Russian Federation, the Federal Agency of Atomic Energy of the Russian Federation, Russian Academy of Sciences, the Russian Foundation for Basic Research and the Russian Competitiveness Program of NRNU ``MEPhI"; the Ministry of Education, Science and Technological Development of Serbia; the Secretar\'{\i}a de Estado de Investigaci\'on, Desarrollo e Innovaci\'on, Programa Consolider-Ingenio 2010, Plan de Ciencia, Tecnolog\'{i}a e Innovaci\'on 2013-2017 del Principado de Asturias and Fondo Europeo de Desarrollo Regional, Spain; the Swiss Funding Agencies (ETH Board, ETH Zurich, PSI, SNF, UniZH, Canton Zurich, and SER); the Ministry of Science and Technology, Taipei; the Thailand Center of Excellence in Physics, the Institute for the Promotion of Teaching Science and Technology of Thailand, Special Task Force for Activating Research and the National Science and Technology Development Agency of Thailand; the Scientific and Technical Research Council of Turkey, and Turkish Atomic Energy Authority; the National Academy of Sciences of Ukraine, and State Fund for Fundamental Researches, Ukraine; the Science and Technology Facilities Council, UK; the US Department of Energy, and the US National Science Foundation.

Individuals have received support from the Marie-Curie programme and the European Research Council and Horizon 2020 Grant, contract No. 675440 (European Union); the Leventis Foundation; the A. P. Sloan Foundation; the Alexander von Humboldt Foundation; the Belgian Federal Science Policy Office; the Fonds pour la Formation \`a la Recherche dans l'Industrie et dans l'Agriculture (FRIA-Belgium); the Agentschap voor Innovatie door Wetenschap en Technologie (IWT-Belgium); the Ministry of Education, Youth and Sports (MEYS) of the Czech Republic; the Council of Scientific and Industrial Research, India; the HOMING PLUS programme of the Foundation for Polish Science, cofinanced from European Union, Regional Development Fund, the Mobility Plus programme of the Ministry of Science and Higher Education, the National Science Center (Poland), contracts Harmonia 2014/14/M/ST2/00428, Opus 2014/13/B/ST2/02543, 2014/15/B/ST2/03998, and 2015/19/B/ST2/02861, Sonata-bis 2012/07/E/ST2/01406; the National Priorities Research Program by Qatar National Research Fund; the Programa Severo Ochoa del Principado de Asturias; the Thalis and Aristeia programmes cofinanced by EU-ESF and the Greek NSRF; the Rachadapisek Sompot Fund for Postdoctoral Fellowship, Chulalongkorn University and the Chulalongkorn Academic into Its 2nd Century Project Advancement Project (Thailand); the Welch Foundation, contract C-1845; and the Weston Havens Foundation (USA).
\end{acknowledgments}

\bibliography{auto_generated}

\appendix
\section{Parameterization of the efficiency}
\label{sec:app}
\begin{figure}[hbtp]
  \centering
    \includegraphics[width=0.6\textwidth]{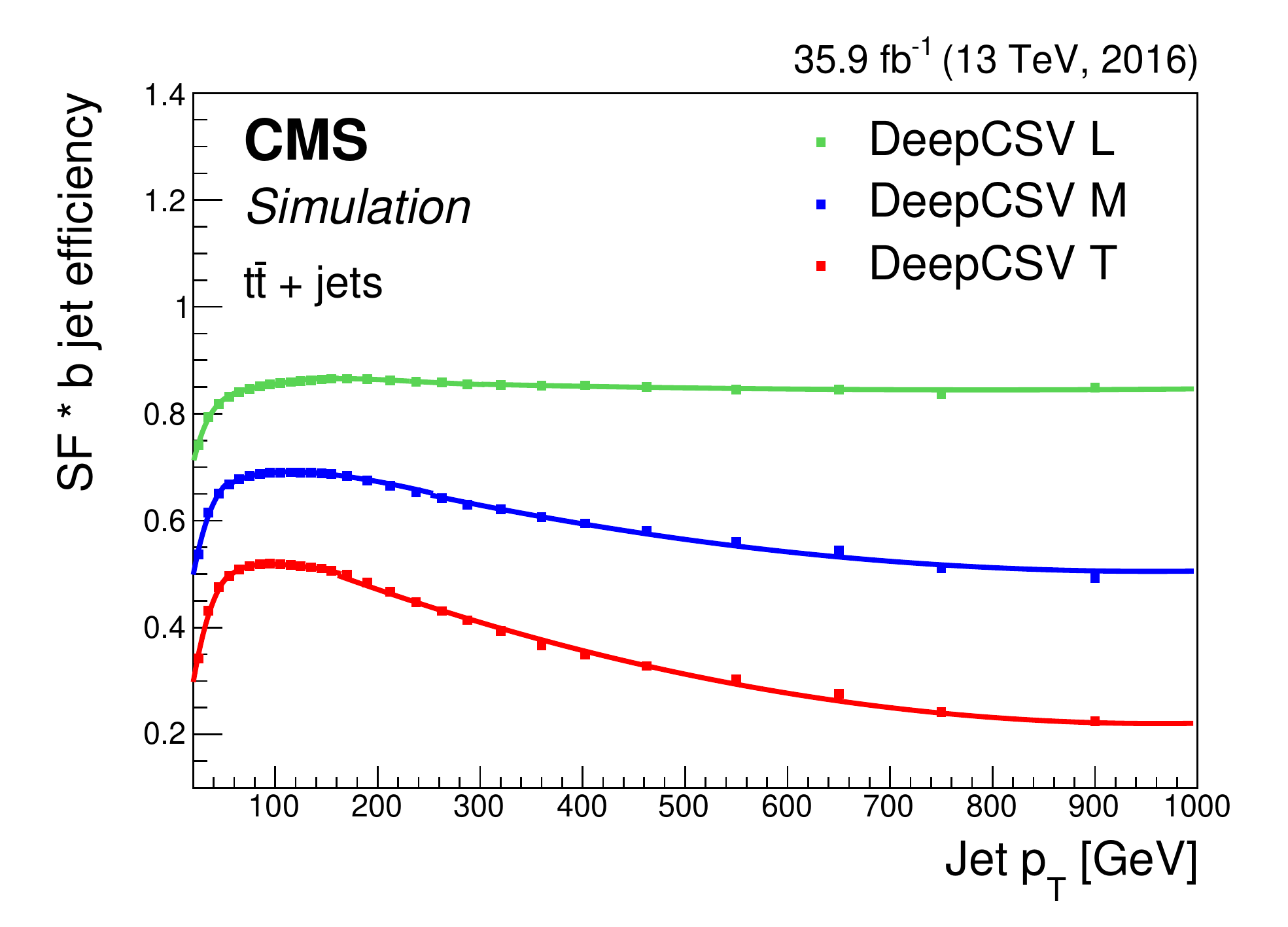}\\
    \includegraphics[width=0.6\textwidth]{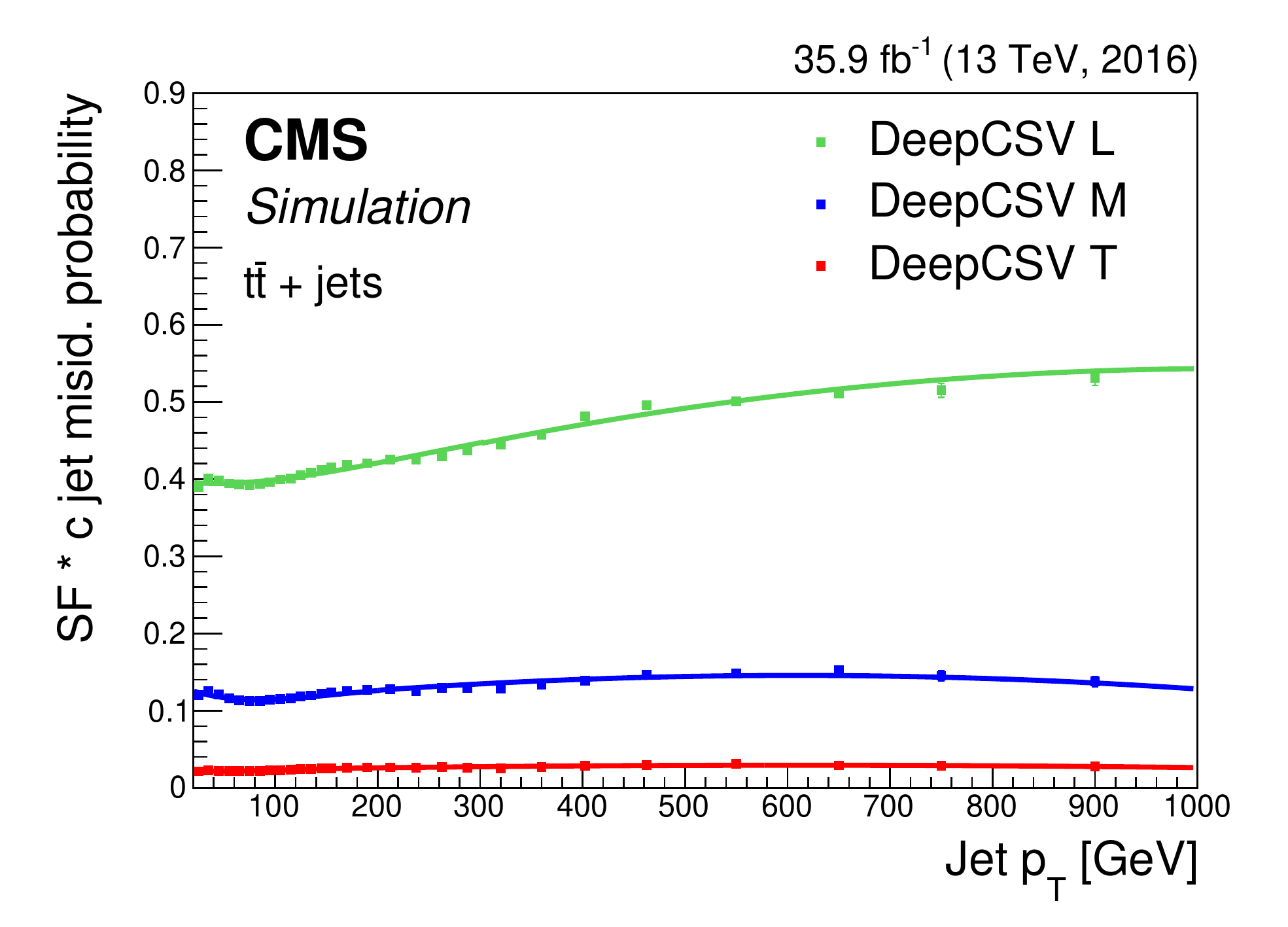}\\
    \includegraphics[width=0.6\textwidth]{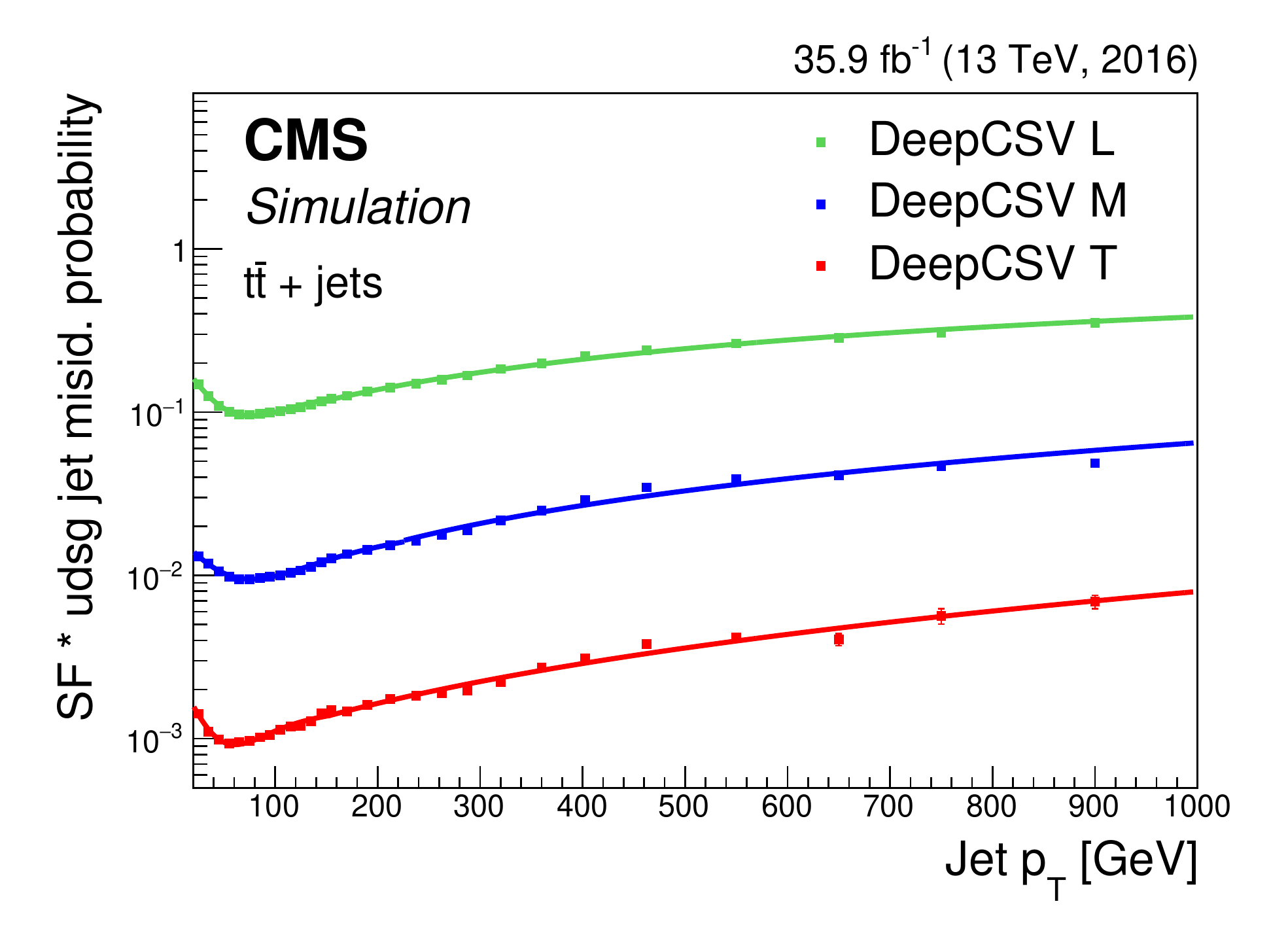}
    \caption{Efficiency for {\cPqb} tagging jets for the three different working points of the DeepCSV algorithm multiplied by the measured data-to-simulation scale factor. The efficiencies are shown as a function of the jet \pt using jets with $\pt > 20\GeV$ in \ttbar events for {\cPqb} jets (upper), {\cPqc} jets (middle), and light-flavour jets (lower). The solid lines represents the functions used to fit the dependence on the jet \pt. The last bin includes the overflow. }
    \label{fig:phenofiteff}
\end{figure}

To facilitate phenomenological studies relying on {\cPqb} jet identification, we provide the {\cPqb} jet identification efficiency as a function of the jet \pt for the three operating points of the DeepCSV algorithm. The efficiency is obtained using jets with $\pt > 20$\GeV in a simulated \ttbar sample and is multiplied by the data-to-simulation scale factor to obtain the tagging efficiency expected in data. This efficiency is shown in Fig.~\ref{fig:phenofiteff} for the three jet flavours. Polynomial functions are used to fit the dependence of the efficiency on the jet \pt for jets with $20<\pt<1000\GeV$. It is worth noting that the parameterization of the fitted functions is not reliable outside this jet \pt range. The parameterizations are summarized in Table~\ref{tab:phenofitfunctions}.

\begin{landscape}
\begin{table*}[htb]
\centering
\topcaption{Polynomial functions used to fit the efficiency of the three working points of the DeepCSV algorithm for the three jet flavours as a function of the jet \pt for jets with $20<\pt<1000\GeV$.}
\label{tab:phenofitfunctions}
\begin{tabular}{cccl}
Flavour & Working point & \pt (\GeVns{}) & Function \\
\hline
{\cPqb} & DeepCSV L & 20--160 & $0.4344  +0.02069 \pt - 0.0004429 {\pt}^{2} +5.137 \times 10^{-6}  {\pt}^{3} -3.406 \times 10^{-8} {\pt}^{4} +1.285 \times 10^{-10} {\pt}^{5}$  \\
        &       &            &  $-2.559 \times 10^{-13} {\pt}^{6} + 2.084 \times 10^{-16} {\pt}^{7} $ \\
        &       &   160--300  &  $0.714 +0.002617 \pt -1.656 \times 10^{-5} {\pt}^{2} +4.767 \times 10^{-8} {\pt}^{3} -6.431 \times 10^{-11}  {\pt}^{4} +3.287 \times 10^{-14} {\pt}^{5}$ \\
        &       &   300--1000 &  $0.872  -6.885 \times 10^{-5} \pt + 4.34 \times 10^{-8} {\pt}^{2} $ \\
        & DeepCSV M &  20--50 & $0.194  +0.0211 \pt -0.000348 {\pt}^{2} +2.761 \times 10^{-6} {\pt}^{3} -1.044 \times 10^{-8} {\pt}^{4} +1.499 \times 10^{-11} {\pt}^{5} $ \\
        &       &  50--250   &  $ 0.557 +0.003417 \pt -3.26 \times 10^{-5} {\pt}^{2} +1.506 \times 10^{-7} {\pt}^{3} -3.63 \times 10^{-10}  {\pt}^{4} + 3.522 \times 10^{-13} {\pt}^{5}$\\
        &       &  250--1000 &  $ 0.768 -0.00055 \pt + 2.876 \times 10^{-7} {\pt}^{2}$ \\
        & DeepCSV T &  20--50 &  $-0.033 +0.0225 \pt -0.00035 {\pt}^{2} +2.586 \times 10^{-6} {\pt}^{3} -9.096 \times 10^{-9} {\pt}^{4} + 1.212 \times 10^{-11} {\pt}^{5}  $ \\
        &       &  50--160 &  $0.169 +0.013 \pt -0.00019 {\pt}^{2} +1.373 \times 10^{-6} {\pt}^{3} -4.923 \times 10^{-9} {\pt}^{4} + 6.87 \times 10^{-12} {\pt}^{5} $ \\
        &       &  160--1000 &  $ 0.62 -0.00083 \pt +4.3078 \times 10^{-7} {\pt}^{2}  $ \\
\hline
{\cPqc} & DeepCSV L & 20--300  & $0.398 -0.000182 \pt +2.53 \times 10^{-6}  {\pt}^{2} -6.796 \times 10^{-9} {\pt}^{3} +8.66 \times 10^{-12} {\pt}^{4} -4.42 \times 10^{-15} {\pt}^{5}  $ \\
        &       &  300--1000 &  $ 0.35 +0.000374 \pt -1.81 \times 10^{-7} {\pt}^{2} $ \\
        & DeepCSV M  & 20--200  & $0.136 -0.000639 \pt +6.188 \times 10^{-6}  {\pt}^{2} -2.26 \times 10^{-8} {\pt}^{3} +3.61 \times 10^{-11} {\pt}^{4} +2.09 \times 10^{-14} {\pt}^{5}  $ \\
        &       &  200--1000 &  $ 0.103 +0.00014 \pt -1.15 \times 10^{-7} {\pt}^{2} $ \\
        & DeepCSV T   & 20--65  & $ 0.0234 -8.417 \times 10^{-5} \pt +1.24 \times 10^{-6}  {\pt}^{2} -5.5 \times 10^{-9} {\pt}^{3} +9.96 \times 10^{-12} {\pt}^{4} -6.32 \times 10^{-15} {\pt}^{5}  $ \\
        &       &  165--1000 &  $ 0.0218  + 2.46 \times 10^{-5}  \pt -2.021 \times 10^{-8}  {\pt}^{2} $ \\
\hline
udsg & DeepCSV L & 20--150 & $ 0.245 -0.0054 \pt +6.92 \times 10^{-5} {\pt}^{2} -3.89 \times 10^{-7} {\pt}^{3} +1.021 \times 10^{-9} {\pt}^{4} -1.007 \times 10^{-12} {\pt}^{5} $ \\
&       & 150--1000 &  $0.0558 +0.000428 \pt -1.0 \times 10^{-7} {\pt}^{2}  $ \\
        & DeepCSV M & 20--225 & $ 0.019 -0.00031 \pt +3.39 \times 10^{-6} {\pt}^{2} -1.47 \times 10^{-8} {\pt}^{3} +2.92 \times 10^{-11} {\pt}^{4} -2.12 \times 10^{-14} {\pt}^{5}  $\\
        &        & 225--1000 & $0.00328 +5.7 \times 10^{-5} \pt +4.7 \times 10^{-9} {\pt}^{2} $ \\
        & DeepCSV T  & 20--150 & $ 0.00284 -8.63 \times 10^{-5} \pt +1.38 \times 10^{-6} {\pt}^{2} -9.69 \times 10^{-9} {\pt}^{3} +3.19 \times 10^{-11} {\pt}^{4} -3.97 \times 10^{-14} {\pt}^{5} $ \\
        &        & 150--1000 & $0.00063 +4.51 \times 10^{-6} \pt +2.83 \times 10^{-9} {\pt}^{2} $ \\
\end{tabular}
\end{table*}
\end{landscape}

\cleardoublepage \section{The CMS Collaboration \label{app:collab}}\begin{sloppypar}\hyphenpenalty=5000\widowpenalty=500\clubpenalty=5000\textbf{Yerevan Physics Institute,  Yerevan,  Armenia}\\*[0pt]
A.M.~Sirunyan, A.~Tumasyan
\vskip\cmsinstskip
\textbf{Institut f\"{u}r Hochenergiephysik,  Wien,  Austria}\\*[0pt]
W.~Adam, F.~Ambrogi, E.~Asilar, T.~Bergauer, J.~Brandstetter, E.~Brondolin, M.~Dragicevic, J.~Er\"{o}, A.~Escalante Del Valle, M.~Flechl, M.~Friedl, R.~Fr\"{u}hwirth\cmsAuthorMark{1}, V.M.~Ghete, J.~Grossmann, J.~Hrubec, M.~Jeitler\cmsAuthorMark{1}, A.~K\"{o}nig, N.~Krammer, I.~Kr\"{a}tschmer, D.~Liko, T.~Madlener, I.~Mikulec, E.~Pree, N.~Rad, H.~Rohringer, J.~Schieck\cmsAuthorMark{1}, R.~Sch\"{o}fbeck, M.~Spanring, D.~Spitzbart, W.~Waltenberger, J.~Wittmann, C.-E.~Wulz\cmsAuthorMark{1}, M.~Zarucki
% ...
\vskip\cmsinstskip
% ...
1:~~Also at Vienna University of Technology, Vienna, Austria\\
% ...

\end{sloppypar}
\end{document}